\documentclass[a4paper,12pt]{book}

\usepackage[twoside,inner=2.8cm,outer=2.8cm,top=3cm,bottom=3cm]{geometry}
\usepackage[utf8x]{inputenc}
\usepackage[english]{babel}
\usepackage{graphicx}
\usepackage{setspace}
\usepackage{braket}
\usepackage{dsfont}
\usepackage{bbm}
\usepackage{amsmath,amssymb}
\usepackage{kantlipsum}
\usepackage{cite}
\allowdisplaybreaks

\input{tex/thesis.sty}

\usepackage[font=small,labelfont=bf]{caption}
\usepackage{fancyhdr}
\newcommand{\normallinespacing}{\renewcommand{\baselinestretch}{1.0} \normalsize}

\DeclareMathOperator{\Tr}{Tr}
\DeclareMathOperator{\B}{B}
\DeclareMathOperator{\R}{Re}
\DeclareMathOperator{\sign}{sign}
\DeclareMathOperator{\perm}{perm}

\begin{document}

\title{Abelian color cycle and abelian color flux dualization methods for non-abelian lattice field theories}
\author{Carlotta Marchis}
\submitdate{\today}

\maketitle
\begin{abstract}
	In this thesis we present the application of the \textit{abelian color cycle} (ACC) and the \textit{abelian color flux} (ACF) methods to several models: the SU(2) principal chiral model, the SU(2) gauge theory with staggered fermions and QCD with staggered fermions. 
	The key step of our approaches consists in decomposing the action of the model one is considering into its minimal units. For gauge theories those minimal terms are complex numbers, which we refer to as abelian color cycles, while for fermions the action is decomposed into Grassmann bilinears, which we called abelian color fluxes. As a result of these decompositions the actions are sums of commuting terms, and thus one can proceed with the dualization of the theory as in the abelian case, by factorizing and expanding the Boltzmann weight. The expansion indices, so-called dual variables, become the new degrees of freedom for the description of the system once the conventional fields are integrated out. The integration over the conventional fields results into weight factors and constraints. The constraints implement the symmetries of the theory in the dual form and imply that the dual configurations which contribute to the long range physics are worldsheets for the gauge degrees of freedom and worldlines for matter fields. On the other hand, the weight factors allow one to organize the dual partition function into a strong coupling series of which all terms are known in closed form. Moreover, the form of the dependence on the chemical potential allows one to identify the net-particle number as the total net temporal winding number of the worldlines in the dual representation. 
\end{abstract}

\cleardoublepage

\body
\chapter{Introduction \label{cha:introduction}}

Quantum Chromodynamics (QCD) is the theory that describes the strong interaction between quarks and gluons, the constituents of hadrons. It is a non-abelian gauge theory with symmetry group SU(3) and six flavors of quarks, two of which are almost massless. The resulting theory is extremely rich, showing properties such as asymptotic freedom, confinement of quarks, spontaneous chiral symmetry breaking and phase transitions. While asymptotic freedom allows the use of perturbation theory at high energies, the strong coupling at low energies requires the use of non-perturbative methods, such as lattice QCD which allows the theoretical study of QCD from first principle calculations. 

The phase diagram of QCD in the $T$ -- $\mu$ plane is thought to have a very rich structure, with a crossover at high temperature and zero chemical potential, which could eventually become a first order phase transition with increasing values of $\mu$, thus signaling the presence of a so-called critical point. More exotic phases, such as color superfluid/superconducting phases, are conjectured at higher values of chemical potential and small temperatures.
As heavy ion collision experiments at LHC and RHIC probe the structure of the phase diagram of QCD at finite baryon chemical potential $\mu$, from the theory side the only region for which we have reliable results is the temperature axes, where numerical simulations predict a rapid crossover transition with the restoration of chiral symmetry at temperatures around 150 MeV. However, non-vanishing values of chemical potential cause the so-called sign problem which prevents the use of standard Monte-Carlo techniques on the lattice. Thus, the need of finding alternative approaches led to the development of a variety of methods over the years. 
Among those, the dual approach has shown to be a powerful tool in solving the sign problem of models with abelian gauge groups. Following that success, the aim of my PhD project was to extend the applicability of the dual approach to non-abelian lattice field theories. The effort we put in that direction resulted in the development of the \textit{abelian color cycle} (ACC) and \textit{abelian color flux} (ACF) dualization methods, which are at the center of the discussion of this thesis. While the ACC method is suitable for the dualization of gauge theories with non-abelian group, the ACF approach can be used for systems with matter. 

The key step of both these methods is the decomposition of the action into its minimal units. For gauge theories those minimal terms are complex numbers, which we refer to as abelian color cycles, while for fermions, the action is decomposed into Grassmann bilinears, which we called abelian color fluxes. As a result of these decompositions the actions are sums of commuting terms, and thus one can proceed with the dualization of the theory as in the abelian case, by factorizing and expanding the Boltzmann weight. The expansion indices, so-called dual variables, become the new degrees of freedom for the description of the system once the conventional fields are integrated out. In general, the integration over the gauge fields results into two contributions: a gauge weight factor and a gauge constraint. The weight factor organizes the partition function into a strong coupling series, where all terms are known in closed form. Moreover, for the specific cases of SU(2) and SU(3) pure gauge theory the weight factor also contain signs, which originate from the explicit signs in the parametrization of the group elements. The constraints, on the other hand, bring information about the original symmetry into the dual formulation, and in particular they allow to determine the configurations that contribute to the long range physics, which are worldsheets of the dual variables. 

Also the Grassmann integration results in a weight factor and a constraint. The fermion constraint implements Pauli's exclusion principle for the fermion dual variables, while the fermion weight factor collects the contributions of the admissible fermion configurations. We find that the long range physics for the fermion degrees of freedom in the dual representation is described by worldlines on the lattice which, however, come with signs. Obviously the appearance of gauge and fermion sign factors in our dual formulations implies that for a Monte Carlo simulation of the ACC and ACF dual representations a strategy for a partial resummation needs to be found. Nevertheless, our dual representations have some features which make them interesting per se, such as the possibility of identifying conserved charges with topological invariants, which thus become much easier to determine in the worldline formulation. In another application the ACF approach solves the complex action problem of the SU(2) principal chiral model in the conventional representation when chemical potentials are coupled to some of the conserved charges. For that system we were actually able to bring the dualization a step further by reformulating it in terms of variables which automatically solve the constraints. Both types of dual representations were successfully implemented in a Monte Carlo simulation.

\newpage
This thesis is organized as follows:
\begin{description}
	\item[Chapter \ref{cha:theoretical}] collects some of the basic concepts of lattice field theories and explains the origin of the sign problem in lattice QCD. After reviewing some of the methods that have been developed in order to overcome the sign problem, we then focus on the dual approach and discuss the dualization of the U(1) Gauge -- Higgs model. 
	
	\item[Chapter \ref{cha:pcm}] presents the SU(2) principal chiral model as first example for the application of the ACF approach. As was mentioned in the introduction, for this model the dual representation solves the complex action problem. At the end of the chapter we also discuss the possibility of completely resolving the constraints by bringing the ACF dualization a step further.
	
	\item[Chapter \ref{cha:su2}] gives the dual representation for SU(2) lattice gauge theory with staggered fermions. We start applying the ACC method to the pure gauge theory. We then show how the generalization of this approach to the theory with fermions leads us to the development of the ACF method.
	
	\item[Chapter \ref{cha:qcd}] shows how to obtain the worldline and worldsheet representation of lattice QCD using the ACC and ACF dualization methods. Also in this case we start with discussing the pure gauge theory, and then present the strong coupling limit before giving the result for full QCD. 
	
	\item[Chapter \ref{cha:conclusions}] ends the thesis by summarizing the findings discussed in the previous chapters and outlining the main features of the dual approaches we developed. 
\end{description}

\newpage
From the projects described in this thesis so far the following papers and conference proceedings were published:

\begin{itemize}
	\item C. Marchis and C. Gattringer,
	\textit{Dual representation of lattice QCD with worldlines and
		worldsheets of abelian color fluxes}, Phys. Rev. D \textbf{97} (2018) 034508, [arXiv:1712.07546 [hep-lat]]
	
	\item C. Gattringer, D. Göschl and C. Marchis,
	\textit{Worldlines and worldsheets for non-abelian lattice field theories: Abelian color fluxes and Abelian color cycles}, EPJ Web Conf. \textbf{175} (2018) 11007, [arXiv:1710.08745 [hep-lat]]
	
	\item C. Gattringer, D. Göschl and C. Marchis,
	\textit{Kramers-Wannier duality and worldline representation for
		the SU(2) principal chiral model}, Phys. Lett. B \textbf{778} (2018) 435, [arXiv:1709.04691 [hep-lat]]
	
	\item C. Gattringer and C. Marchis,
	\textit{Abelian color cycles: a new approach to strong coupling	expansion and dual representations for non-abelian lattice	gauge theory}, Nucl. Phys. B \textbf{916} (2017) 627, [arXiv:1609.00124 [hep-lat]]
	
	\item C. Gattringer and C. Marchis,
	\textit{Dualization of non-abelian lattice gauge theory with Abelian Color Cycles (ACC)}, PoS \textbf{LATTICE2016} (2016) 034,  
	[arXiv:1611.01022 [hep-lat]]
\end{itemize}
All publications can be acquired freely as on-line e-prints at arXiv.org.

\chapter{Theoretical background \label{cha:theoretical}}
The concepts presented in this chapter are at the base of lattice field theory and are explained in much greater detail and clarity in many textbooks, such as \cite{Gattringer:2010zz,Rothe:1992nt,Montvay:1994cy,Peskin:1995ev}. Nonetheless, in Secs.~\ref{sec:teo_path} -- \ref{sec:teo_numerical} we summarize the main concepts of the lattice regularization of quantum field theories with the aim of making this thesis as self-contained as possible. Then, in Sec.~\ref{sec:teo_sign} we explain the technical problem which arises when introducing the quark chemical potential in the discretized QCD action. This is the so-called sign problem, which prevents the study of the QCD phase diagram with first principle lattice calculations and that since the earliest days of lattice field theory have been stimulating the development of a multitude of methods to overcome it. As we outlined in the introduction, this also stands at the very core of the motivation driving our research, and led us to the development of the abelian color cycle and abelian color flux methods as an extension of the applicability of the dual approach to non-abelian lattice field theories. Another important motivation for us is the understanding of general properties of the dual representations, such as the manifestation of the original symmetries in the dual reformulation, or the possibility of identifying conserved quantities as topological invariants.

\section{Path integral quantization of field theories \label{sec:teo_path}}
The quantization of fields on the lattice is done using the path integral approach. Since its introduction by Feynman \cite{Feynman}, the path integral method has become a very important tool for elementary particle physics. In this approach the action, rather than the Hamiltonian, is used as the fundamental quantity. The path integral approach reveals the close analogy between quantum field theories and statistical mechanics. The exploitation of this analogy is at the base of the Monte Carlo simulations performed on the lattice.

In the path integral approach the quantum mechanical amplitude for a particle to travel from a point $x$ to a point $y$ within the time interval $t$ is expressed as an integral over all classical paths weighted by the exponential of $i$ times the classical action $S$ associated with that trajectory:
\begin{equation}
\label{eq:teo_path}
	\langle y | e^{i \hat{H} t}| x \rangle = \int D[x] \, e^{iS} \, ,
\end{equation}
where $\hat{H}$ is the Hamiltonian operator of the quantum mechanical system and $\int D[x]$ denotes the functional integral over all paths. 
In (\ref{eq:teo_path}) quantum mechanical operators have been eliminated in favor of an infinite-dimensional integral. However, a rigorous definition of the path integral $\int D[x]$ is not possible unless a discretization is performed. Moreover, in the form (\ref{eq:teo_path}) the weight $e^{iS}$ is complex and strongly oscillating, which prevents the use of numerical methods to compute the propagation amplitude (\ref{eq:teo_path}). This problem can be bypassed by Wick rotating the time coordinate $t \rightarrow - i x_{4}$, where $x_{4}$ is the Euclidean time. Then, in Euclidean space-time (\ref{eq:teo_path}) reads
\begin{equation}
\label{eq:teo_pathe}
\langle y | e^{- \hat{H} x_{4}}| x \rangle = \int D[x] \, e^{-S_{E}} \, ,
\end{equation}
where $S_{E} = - i S$ is the Euclidean action. In the next section we will see how the Euclidean formalism for quantum field theories makes the analogy with statistical mechanics more transparent, and we will outline the main advantages of the lattice regularization.

\section{Space-time discretization: lattice regularization of quantum field theories \label{sec:teo_lattice}}
Consider a quantum field theory governed by the Euclidean action 
\begin{equation}
\label{eq:teo_action}
	S_{E} = \int d^{4} x \, \mathcal{L} \big(\phi(x), \partial_{\mu} \phi(x) \big) \, ,
\end{equation}
where $\mathcal{L}$ is a general expression of the Euclidean Lagrangian density as a function of the fields $\phi$ and their derivatives $\partial_{\mu} \phi$. In the continuum correlation functions can be computed as functional derivatives of the generating functional $Z[J]$, which is defined as
\begin{equation}
	\label{eq:teo_generating}
	Z [J] = \int D[\phi]\,  e^{- \int d^{4} x (\mathcal{L} - J \phi)} \, ,
\end{equation}
where $\int D[\phi]$ is the infinite-dimensional functional integral over field configurations, and $J \phi$ is a source term.

The lattice counterpart of the partition function $Z = Z[0]$ is
\begin{equation}
	\label{eq:teo_partition}
	Z_{lat} = \int D[\phi]\,  e^{- S_{lat} [\phi]} \, .
\end{equation} 
Eq.~(\ref{eq:teo_partition}) is obtained by discretizing the space-time 
\begin{equation}
\label{eq:teo_discretization}
	x  \rightarrow a n \, , \quad n = (n_1,n_2,n_3,n_4) \quad \text{with } n_{i} = 0,1,\dots, N_i-1 \quad \text{for } i= 1,2,3,4 \, .
\end{equation}
In other words we introduce a four-dimensional lattice $\Lambda$, where the sites $n$ are separated by the lattice spacing $a$, which has the dimension of length.
Then, the space-time integral in Eq.~(\ref{eq:teo_action}) becomes a sum over all lattice sites
\begin{equation}
	\label{eq:teo_latticeintegral}
	\int d^{4} x \, \rightarrow \, a^{4} \sum_{n\in \Lambda} \, ,
\end{equation} 
and the value of the field at a space-time point $x$ is replaced by the value of the field at site $n$, i.e.,
\begin{equation}
\label{eq:teo_latticefield}
\phi (x) \, \rightarrow \, \phi (n) \, .
\end{equation} 
Finally, the derivatives of the fields may be approximated for small lattice spacing $a$ using Taylor expansions as
\begin{equation}
\label{eq:teo_latticederivative}
	\partial_{\mu} \phi (x) \, \rightarrow \, \dfrac{1}{2a} \big(\phi (n + \hat{\mu}) - \phi (n - \hat{\mu})\big)\, .
\end{equation}
Performing the substitutions (\ref{eq:teo_latticeintegral}) -- (\ref{eq:teo_latticederivative}) in (\ref{eq:teo_action}) one obtains the discretized version $S_{lat}[\phi]$ of the Euclidean continuum action $S_E$. As a first advantage, the introduction of the lattice allows for a rigorous definition of the functional integral $\int D[\phi]$ in (\ref{eq:teo_partition}):
\begin{equation}
	\label{eq:teo_measure}
	D[\phi] \equiv \prod_{n \in \Lambda} d \phi(n) \, ,
\end{equation}	
where $d \phi(n)$ is the integration measure for the fields at site $n$ which depends on the nature of the fields $\phi(n)$ one is considering.

Another important advantage of the lattice discretization (\ref{eq:teo_discretization}) is the appearance of an UV cut-off for the discretized field theory. To show this consider, e.g., a function $f(x)$ in one dimension. The Fourier transform $\tilde{f}(k)$ in momentum space of the function $f(x)$ is given by 
\begin{equation}
\label{eq:teo_fourier}
	\tilde{f}(k) \, = \, \int_{-\infty}^{\infty} dx f(x) \, e^{-i k x} \, .
\end{equation} 
If one then restricts $x$ to be a multiple of lattice spacing $a$, i.e., $x = an$, (\ref{eq:teo_fourier}) becomes
\begin{equation}
\label{eq:teo_fourierdiscretized}
\tilde{f}_{a} (k) \, = \, a \sum_{n \in \mathbb{Z}} f(an) \, e^{-i a k n} \, .
\end{equation} 
It is easy to see that the discretized version (\ref{eq:teo_fourierdiscretized}) of the Fourier transform $\tilde{f}(k)$ is invariant under the substitution $k \to k + 2\pi/a$, i.e., it is a periodic function. Hence, in the Fourier representation of the function $f(an)$,
\begin{equation}
\label{eq:teo_fourierdecomposition}
	f(an) \, = \, \int_{-\pi/a}^{+\pi/a} \dfrac{dk}{2\pi} \, \tilde{f}_{a}(k) \, e^{ia k n} \, ,
\end{equation}
the momentum integration is restricted to the so-called Brillouin zone $[-\pi/a, \pi/a]$. Thus we find that the introduction of a lattice $\Lambda$ automatically provides an UV regularization of the discretized field theory thanks to the arising of a momentum cut-off proportional to the inverse lattice spacing $a$. 

We would now like to address the long awaited discussion about the analogy between the path integral formulation of quantum field theories and statistical mechanics. The expression (\ref{eq:teo_partition}) of the discretized partition sum is reminiscent of the partition function of statistical mechanics. It has in fact the same structure of an integral over all possible configurations of an exponential statistical weight. This analogy plays such an important role in lattice field theories that the nomenclatures used in quantum field theory and statistical mechanics are often interchanged or used synonymously. So, for example, throughout this thesis we will call the Feynman path weight $e^{-S_{latt}}$ the Boltzmann weight for a given field configuration on the lattice. A powerful tool that comes with the identification of the lattice regularized quantum field theory with a statistical mechanical system is the Monte Carlo technique for simulation, which we will briefly discuss in Sec.~\ref{sec:teo_numerical}. Before coming to that, in the next two sections we will illustrate in more detail the lattice discretization of fermions and gauge fields for the specific case of lattice QCD.

\section{Lattice discretization of fermions \label{sec:teo_fermions}}
On the lattice the fermionic degrees of freedom are described by Grassmann numbers, which implement the correct Fermi statistic. We therefore start this section by discussing the main properties of Grassmann numbers and giving the rules for integration which we will need in the following chapters of this thesis.

\subsection{Grassmann variables\label{sec:teo_grassmann}}
The Grassmann numbers are numbers that anticommute with each other. So, if we consider the set of Grassmann numbers $\eta_{1}, \dots, \eta_{N}$, the following anticommutation relations hold
\begin{equation}
\label{eq:teo_anticommutation}
	\{\eta_{i}, \eta_{j}\} \, = \, \eta_{i} \eta_{j} + \eta_{j} \eta_{i} \, = \, 0 \, , \qquad i,j = 0,\dots , N \, .
\end{equation}
From Eq.~(\ref{eq:teo_anticommutation}) also follows that Grassmann variables are nilpotent, i.e.,
\begin{equation}
\label{eq:teo_nilpotent}
	\eta_{i}^{2} \, = \, 0 \, .
\end{equation}
General functions of Grassmann numbers are therefore polynomials of the form
\begin{equation}
\label{eq:teo_function}
	f(\eta) = f_{0} + \sum_{i} f_{i} \eta_{i} + \sum_{i \neq j} f_{ij} \eta_{i} \eta_{j} + \dots 
	+  f_{12 \dots N} \eta_{1} \eta_{2} \dots \eta_{N} \, .
\end{equation}
Notice that Eq.~(\ref{eq:teo_nilpotent}) automatically implements Pauli's exclusion principle for fermions.

To compute Grassmann integrals of the form 
\begin{equation}
\label{eq:teo_integral}
	\int \prod_{i=1}^{N} d\eta_{i} \, f(\eta) \, ,
\end{equation}
the two following rules are sufficient
\begin{equation}
\label{eq:teo_integrationrules}
	\int d\eta_{i} = 0 \, , \qquad \int d\eta_{i} \,\eta_{i} = 1 \, ,
\end{equation}
where the integration measures $\{d\eta_{i}\}$ obey the same anticommutation relations as the Grassmann variables $\eta_{i}$:
\begin{equation}
\label{eq:teo_anticommutation2}
	 \{d\eta_{i}, d\eta_{j}\} \, = \, \{d\eta_{i}, \eta_{j}\} \, = \, 0 \, , \qquad \forall i,j \, .
\end{equation}

To illustrate the application of the rules (\ref{eq:teo_integrationrules}), we compute the integral
\begin{equation}
	I[D] \, = \, \int \! \prod_{l = 1}^{N} d\eta_{l} d\overline{\eta}_{l} \ e^{ \, \sum_{i,j = 1}^{N} \overline{\eta}_{i}  D_{i,j} \eta_{j} } \, , 
\label{eq:teo_gaussian}
\end{equation}
where we denoted the $2N$ Grassmann variables as $\eta_{1}, \dots \eta_{N}, \overline{\eta}_{1}, \dots, \overline{\eta}_{N}$. All the terms in the exponent of the integrand in Eq.~(\ref{eq:teo_gaussian}) are quadratic in the Grassmann variables. Hence, they commute among each other, and we can rewrite the sum over the index $i$ in the exponent as a product of exponentials:
\begin{equation}
	I[D] \, 
	= \int \! \prod_{l = 1}^{N} d\eta_{l} d\overline{\eta}_{l} \, \prod_{i=1}^{N} e^{  \overline{\eta}_{i} \sum_{j} D_{i,j} \eta_{j} } \, 
	= \, \int \! \prod_{l = 1}^{N} d\eta_{l} d\overline{\eta}_{l} \, \prod_{i=1}^{N} \Big(1 + \overline{\eta}_{i} \sum_{j} D_{i,j} \eta_{j} \Big) \,.
\label{eq:teo_gaussian2}
\end{equation}
In the second step we exploited the nilpotency of the Grassmann variables $\overline{\eta}_{i}$ and wrote the product of exponentials as a product of binomials. From the integration rules (\ref{eq:teo_integrationrules}) follows that the only terms of the product of binomials that survive after the integration are the ones in which all the Grassmann variables appear exactly once:
\begin{align}
I[D] \, &
= \, \int \! \prod_{l = 1}^{N} d\eta_{l} d\overline{\eta}_{l} \, \prod_{i=1}^{N} \overline{\eta}_{i} \sum_{j} D_{i,j} \eta_{j} \nonumber\\
& = \, \int \! \prod_{l = 1}^{N} d\eta_{l} d\overline{\eta}_{l} \ \overline{\eta}_{1} D_{1,j_{1}} \eta_{j_1} \,\overline{\eta}_{2} D_{2,j_{2}} \eta_{j_2} \dots \overline{\eta}_{N} D_{N,j_{N}} \eta_{j_N}\, ,
\label{eq:teo_gaussian3}
\end{align}
where a sum over repeated indices $j_{k}$, $k=1,\dots,N$ is understood. The product of Grassmann variables in (\ref{eq:teo_gaussian3}) is antisymmetric under the exchange of any pair of indices $j_{l}$ and $j_{l'}$. Moreover, only the terms where all the indices $j_{1},\dots,j_{N}$ are different survive. Hence, we can rewrite (\ref{eq:teo_gaussian3}) using the completely antisymmetric epsilon tensor $\epsilon_{j_{1}j_{2}\dots j_{N}}$ in $N$ dimensions:
\begin{align}
I[D] \, = \, \int \! \prod_{l = 1}^{N} d\eta_{l} d\overline{\eta}_{l} \ \overline{\eta}_{l} \eta_{l}  \sum_{j_{1},\dots,j_{N}} \epsilon_{j_{1}j_{2}\dots j_{N}} D_{1,j_{1}}  D_{2,j_{2}}  \dots D_{N,j_{N}} \, = \, \det D \, ,
\label{eq:teo_gaussian4}
\end{align}
where we used the standard formula for the determinant of a matrix $D$. 

Eq.~(\ref{eq:teo_gaussian4}) and analogous formulas are important when dealing with fermionic systems, as we will see in the next sections.

\subsection{Naive discretization \label{sec:teo_naive}}

We now come to the discussion of the discretization of the fermion action of QCD. In Euclidean space-time the fermionic part of the QCD action is given by:
\begin{equation}
\label{eq:teo_sfcontinuum}
S_{F}[\psi,\overline{\psi},A] \ = \ \int \!\! d^{4}x \ \overline{\psi}(x) \, [\gamma_{\mu} \,(\partial_{\mu} + i A_{\mu}(x)) + m] \, \psi(x) \, , 
\end{equation}
where we used the Einstein summation convention, and matrix-vector notation for the color and Dirac indices. $\overline{\psi}(x)$ and $\psi(x)$ are Dirac 4-spinors, whose entries are Grassmann variables, so that Fermi statistics is enforced for the quarks. The gluons are described by the gauge fields $A_{\mu} (x) \in$ su(3). $\gamma_{\mu}$, $\mu = 1,2,3,4$ are the Dirac $\gamma$-matrices, which in Euclidean space satisfy the Euclidean anti-commutation relations $\{\gamma_{\mu}, \gamma_{\nu}\} = 2 \delta_{\mu \nu} \mathbf{1}$. 
The action (\ref{eq:teo_sfcontinuum}) is invariant under the gauge transformations:
\begin{align}
\label{eq:teo_gaugetransformation1}
	&\overline{\psi}(x) \rightarrow \overline{\psi}'(x) = \overline{\psi} (x) \Omega(x)^{\dagger} \, ,&&\\
\label{eq:teo_gaugetransformation2}
	&\psi(x) \rightarrow \psi'(x) = \Omega(x) \psi (x) \, ,&& \\
\label{eq:teo_gaugetransformation3}
	&A_{\mu}(x) \rightarrow A'_{\mu} (x) = \Omega (x) A_{\mu}(x) \Omega (x)^{\dagger} + i (\partial_{\mu}\Omega (x)) \Omega (x)^{\dagger} \, ,&&
\end{align}
where $\Omega(x) \in $ SU(3).

To carry out the lattice discretization of the fermion action (\ref{eq:teo_sfcontinuum}) we first start with considering the free case, i.e., we set $A_{\mu}(x) = 0$:
\begin{equation}
S_{F}^{free}[\psi,\overline{\psi}] \ = \ \int \! d^{4}x \ \overline{\psi}(x) \, [\gamma_{\mu} \,\partial_{\mu} + m] \, \psi(x) \, . 
\end{equation}
As we outlined in Sec.~\ref{sec:teo_lattice}, the main idea behind the lattice discretization of a field theory is the introduction of a four-dimensional space-time lattice $\Lambda$:
\begin{equation}
	\label{eq:teo_lattice}
	\Lambda \, = \, \{n = (n_1, n_2, n_3, n_4) \, |\,  n_1,n_2,n_3 = 1,\dots, N-1 \, ; \ n_4 = 1,\dots, N_T-1 \} \, .
\end{equation}
The four-vector $n\in \Lambda$ labels the sites of the lattice, which are separated by the lattice spacing $a$. Then, the transition from the continuum to the lattice is effected by making the following substitutions
\begin{flalign}
	&\psi(x) \rightarrow \psi (n) \, , \nonumber \\[0.5em]
	&\int \! d^{4}x \rightarrow a^{4} \sum_{n \in \Lambda} \, , \\
	&\partial_{\mu} \psi (x) \rightarrow \dfrac{1}{2 a} (\psi(n + \hat{\mu}) - \psi(n - \hat{\mu})) \, .\nonumber
\end{flalign}
For the free fermion action on the lattice $\Lambda$ we obtain:
\begin{equation}
\label{eq:teo_sffree}
S_{F}^{free}[\psi,\overline{\psi}] \ = \ a^4 \sum_{n} \overline{\psi}(n)  \left( \sum_{\mu = 1}^{4} \gamma_{\mu} \dfrac{\psi(n + \hat{\mu}) - \psi(n - \hat{\mu}) }{2 a} + m  \psi(x) \right) \, .
\end{equation}
Obviously we want to preserve the gauge invariance under the local transformations (\ref{eq:teo_gaugetransformation1}) and (\ref{eq:teo_gaugetransformation2}), which on the lattice are implemented by choosing $\Omega(n) \in$ SU(3) for each site $n$ and changing the fermion fields according to:
\begin{equation}
\label{eq:teo_gaugelattice}
	\overline{\psi}(n) \rightarrow \overline{\psi}'(n) = \overline{\psi} (n) \Omega(n)^{\dagger} \, , \qquad
	\psi(n) \rightarrow \psi'(n) = \Omega(n) \psi (n) \, .
\end{equation}
While the mass term in (\ref{eq:teo_sffree}) is invariant under (\ref{eq:teo_gaugelattice}), the discretized derivative is not:
\begin{equation*}
	\overline{\psi}'(n) \psi' (n + \hat{\mu}) = \overline{\psi} (n) \Omega(n)^{\dagger} \Omega (n + \hat{\mu}) \psi (n + \hat{\mu}) \, .
\end{equation*}
This problem is overcome by introducing in the expression of the discretized derivative in (\ref{eq:teo_sffree}) the link variables $U_{\mu} (n) \in$ SU(3), which transform as
\begin{equation}
\label{eq:teo_gaugelattice2}
	U_{\mu}(n) \rightarrow 	U'_{\mu}(n) = \Omega(n) U_{\mu}(n) \Omega(n + \hat{\mu})^{\dagger} \, .
\end{equation}
Then, terms of the form $\overline{\psi}(n) U_{\mu}(n) \psi (n + \hat{\mu})$ are invariant under the rotation $\Omega (n)$ of the color indices:
\begin{align*}
	\overline{\psi}'(n) U'_{\mu}(n) \psi' (n + \hat{\mu}) &= \overline{\psi} (n) \Omega(n)^{\dagger} \Omega(n) U_{\mu}(n) \Omega(n + \hat{\mu})^{\dagger} \Omega (n + \hat{\mu}) \psi (n + \hat{\mu}) \\
	&= \overline{\psi}(n) U_{\mu}(n) \psi (n + \hat{\mu}) \, .
\end{align*}

\begin{figure}
	\vspace{2.5mm}
	\centering
	\includegraphics[scale=1.7]{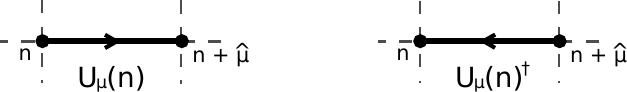}
	\vspace{2.5mm}
	\caption{Graphical representation of the link variable $U_{\mu}(n)$ and its hermitian conjugate $U_{\mu}(n)^{\dagger}$. \label{fig:teo_link}}	
\end{figure}

As we mentioned, the variables $U_{\mu}(n)$ are attached to the links $(n,\mu)$ of the lattice, i.e., they connect the neighboring sites $n$ and $n + \hat{\mu}$. We can then define the hermitian conjugate $U_{\mu} (n)^{\dagger}$ to be the corresponding link variable with negative orientation, namely
\begin{equation*}
	U_{\mu} (n)^{\dagger} \, \equiv \,	U_{-\mu} (n + \hat{\mu}) \, .
\end{equation*} 
For a graphical representation of $U_{\mu}(n)$ and its hermitian conjugate refer to Fig.~\ref{fig:teo_link}.

Putting things together we obtain the following discretized version of the fermion action:
\begin{equation}
S_{F}[\psi,\overline{\psi},U] \ = \ a^4 \sum_{n} \overline{\psi}(n)  \left( \sum_{\mu = 1}^{4} \gamma_{\mu} \dfrac{U_{\mu}(n) \psi(n + \hat{\mu}) - U^{\dagger}_{\mu}(n - \hat{\mu}) \psi(n - \hat{\mu}) }{2 a} + m  \psi(n) \right) \, .
\label{eq:teo_naivesf}
\end{equation}
Eq.~(\ref{eq:teo_naivesf}) is often referred to as the naive discretization of fermions because it actually gives rise to the so-called \textit{doubling problem}, which we will sketch shortly. 
The discretized fermion action (\ref{eq:teo_naivesf}) may be rewritten as
\begin{equation}
	S_{F}[\psi,\overline{\psi},U] \ = \ a^4 \sum_{n,m} \overline{\psi}(n)  D(n|m) \psi(m) \, .
	\label{eq:teo_naivesf2}
\end{equation}
where 
\begin{equation}
\label{eq:teo_naive}
D(n|m)_{_{\alpha\beta}^{a\,b}} \, = \, \sum_{\mu = 1}^{4} (\gamma_{\mu})_{\alpha \beta}  \dfrac{U_{\mu}(n)_{ab} \delta_{n + \hat{\mu},m} - U_{-\mu}(n)_{ab} \delta_{n -\hat{\mu}, m}}{2 a} + m \delta_{ab} \delta_{\alpha \beta} \delta_{n,m} \, ,
\end{equation}
is the naive Dirac operator, with $a,b=1,2,3$ color labels and $\alpha, \beta = 1,2,3,4$ Dirac indices. The inverse $D(n|m)^{-1}$ of the naive Dirac operator (\ref{eq:teo_naive}) is the quark propagator. The quark propagator in momentum space $\tilde{D}(p)^{-1}$ can be computed by inverting the Fourier transform of the discretized Dirac operator (\ref{eq:teo_naive}). In doing so one finds that the momentum space propagator $\tilde{D}(p)^{-1}$ for massless ($m=0$) free fermions has the correct continuum limit when sending $a \to 0$: the propagator has a single pole which corresponds to the massless single fermion described by the continuum Dirac operator. On the lattice, however, the propagator for free fermions has 15 additional unphysical poles, the so-called doublers. 

\subsection{Wilson fermions}
The first solution to the doubling problem was given by Wilson: an extra term, the so-called \textit{Wilson term}, is added to the naive Dirac operator (\ref{eq:teo_naive}). The Wilson term gives a mass to the doublers which is proportional to the inverse lattice spacing $a$. Hence, when $a \to 0$ the doublers become heavy and decouple from the theory. Moreover, in the limit $a \to 0$ the Wilson term vanishes, thus giving the correct continuum limit.

In compact notation the Wilson Dirac operator is given by
\begin{equation}
\label{eq:teo_wilsonfermions}
D(n|m)_{_{\alpha\beta}^{a\,b}} \, = \, \left( m + \dfrac{4}{a}\right) \delta_{ab} \, \delta_{\alpha \beta} \, \delta_{n,m} - \dfrac{1}{2a}\sum_{\mu = \pm 1}^{\pm 4} (\mathbf{1} - \gamma_{\mu})_{\alpha \beta} \, U_{\mu}(n)_{ab} \, \delta_{n + \hat{\mu},m} \, ,
\end{equation}
with
\begin{equation*}
	\gamma_{-\mu} = - \gamma_{\mu} \, , \quad \mu = 1, 2, 3, 4 \, .
\end{equation*}
It is easy to show that the Dirac operator (\ref{eq:teo_wilsonfermions}) is $\gamma_{5}$-hermitian, i.e., it satisfies
\begin{equation}
\label{eq:teo_gamma5}
\gamma_{5} D \gamma_{5} = D^{\dagger} \, .
\end{equation}
Eq.~(\ref{eq:teo_gamma5}) implies that the eigenvalues of the Dirac operator (\ref{eq:teo_wilsonfermions}) are either real or come in complex conjugate pairs.
We will use this property in Sec.~\ref{sec:teo_numerical} to show the feasibility of Monte Carlo simulations of lattice QCD with fermions.

As a remark we stress that the additional term $4/a$ in the expression (\ref{eq:teo_wilsonfermions}) explicitly breaks chiral symmetry. The doubling problem of the naive Dirac operator (\ref{eq:teo_naive}) was therefore traded with the impossibility of studying the effects of the spontaneous chiral symmetry breaking with the Dirac operator (\ref{eq:teo_wilsonfermions}).

\subsection{Staggered fermions}
In this section we introduce another discretization of the fermion action (\ref{eq:teo_sfcontinuum}), so-called staggered fermions first proposed by Kogut and Susskind in \cite{Kogut1975}, which we will use in Chapters \ref{cha:su2} and \ref{cha:qcd}.

The staggered fermion action reads
\begin{equation}
S_{F}[\psi,\overline{\psi},U] \ = \ a^4 \sum_{n} \overline{\psi}(n)  \left( \sum_{\mu = 1}^{4} \eta_{\mu}(n) \dfrac{U_{\mu}(n) \psi(n + \hat{\mu}) - U^{\dagger}_{\mu}(n - \hat{\mu}) \psi(n - \hat{\mu}) }{2 a} + m  \psi(n) \right) ,
\label{eq:teo_staggered}
\end{equation}
where we have introduced the staggered sign functions $\eta_{\mu}(n)$, which are defined as
\begin{equation*}
\eta_{1}(n) = 1 \, , \quad \eta_{2}(n) = (-1)^{n_{1}} \, , \quad \eta_{3}(n) = (-1)^{n_{1} + n_{2}} \, , \quad \eta_{4}(n) = (-1)^{n_{1} + n_{2} + n_{3}} \, .
\end{equation*}
The staggered sign functions $\eta_{\mu}(n)$ play the role of the $\gamma$-matrices: the Dirac indices $\alpha, \beta = 1,2,3,4$ are absent in the staggered formulation, and instead the spinor degrees of freedom are distributed on several sites of the lattice. 

When $m = 0$, the action (\ref{eq:teo_staggered}) is invariant under the global transformation
\begin{equation}
	\label{eq:teo_chiral}
		\psi(n) \rightarrow e^{i \alpha \eta_{5}(n)} \psi(n) \, , \quad
		\overline{\psi}(n) \rightarrow \overline{\psi}(n) \, e^{i \alpha \eta_{5}(n)}\, , \quad
		\text{with }  \eta_{5}(n) = (-1)^{n_{1} + n_{2} + n_{3} + n_{4}} \, ,
\end{equation}
which implements chiral symmetry for staggered fermions ($\eta_{5}(n)$ plays the role of $\gamma_{5}$).

We remark that the form (\ref{eq:teo_staggered}) of the fermion action has a residual doubling problem, since it describes not only one flavor of fermions but four.

\section{Gauge theories on the lattice \label{sec:teo_gauge}}
In Euclidean space-time the continuum gluon action is given by
\begin{equation}
\label{eq:teo_sgcontinuum}
S_{G}[A] \, = \, \dfrac{1}{2g^{2}} \int \! d^{4}x \Tr [F_{\mu\nu}(x) F_{\mu\nu}(x)] \, ,
\end{equation}  
where $g$ is the gauge coupling, $F_{\mu\nu} (x)$ is the field strength tensor defined as
\begin{equation}
\label{eq:teo_fmunu}
F_{\mu\nu} (x) = \partial_{\mu} A_{\nu}(x) - \partial_{\nu} A_{\mu}(x) + i [A_{\mu}(x), A_{\nu}(x)] \, ,
\end{equation} 
and $A_{\mu}(x) \in$ su(3) are the gauge fields.

In Sec.~\ref{sec:teo_naive} we have introduced the link variables $U_{\mu}(n)$ to make the fermion action (\ref{eq:teo_sffree}) invariant under the gauge transformations (\ref{eq:teo_gaugelattice}).  
The variables $U_{\mu}(n)$ are elements of the group SU(3), and absorb the rotations in color space of the fermion variables $\psi (n)$ and $\overline{\psi} (n)$. They thus have the role of the gauge fields on the lattice, and their connection to the continuum gauge fields is given by
\begin{equation}
\label{eq:teo_gaugetransporter}
U_{\mu}(n) = \mathcal{P} \exp \Big( i \int_{n}^{n + \hat{\mu}}dx \, A_{\mu} (x) \Big) \simeq \exp \big( i a \, A_{\mu} (n) \big) \, ,
\end{equation}
i.e., they are the path ordered exponential integrals of the gauge fields $A_{\mu}$ along the link $(n,\mu)$ connecting the neighboring sites $n$ and $n + \hat{\mu}$. In (\ref{eq:teo_gaugetransporter}) we approximated the integral along the link $(n,\mu)$ by the length $a$ of the path and the value $A_{\mu}(n)$ of the field at the starting point.

The gauge transformations of the variables $U_{\mu}(n)$ are given in (\ref{eq:teo_gaugelattice2}). We can then obtain the transformation rules for the ordered product of link variables along a path $\mathcal{L}$ as:
\begin{align}
\label{eq:teo_closedpath}
	\prod_{(n,\mu) \in \mathcal{L}} U'_{\mu}(n) &= U'_{\mu_0}(n_0) U'_{\mu_1}(n_1) \dots U'_{\mu_k}(n_k) \\ \nonumber
	&= \Omega(n_0) U_{\mu_0}(n_0) \Omega(n_1)^{\dagger} \Omega(n_1) U_{\mu_1}(n_1) \Omega(n_2)^{\dagger} \dots \Omega(n_{k}) U_{\mu_k}(n_k) \Omega(n_{k+1})^{\dagger} \, ,
\end{align}
where $(n_{i}, \mu_i)\in \mathcal{L}$, $i = 0,1,\dots, k$ are the links connecting site $n_{i}$ to site $n_{i+1}$ on the path $\mathcal{L}$. Eq.~(\ref{eq:teo_closedpath}) shows that the transformation matrices $\Omega(n_{i})$ and $\Omega(n_{i})^{\dagger}$ cancel for $i = 1, \dots, k$, and the whole product transforms like a gauge transporter:
\begin{equation}
	\prod_{(n,\mu) \in \mathcal{L}} U'_{\mu}(n) = \Omega(n_{0}) \Bigg[ \prod_{(n,\mu) \in \mathcal{L}} U_{\mu}(n) \Bigg] \Omega(n_{k+1})^{\dagger} \, .
\end{equation} 
Therefore we find that the trace of the ordered product of link variables along a closed path on the lattice is gauge invariant.

\begin{figure}
	\vspace{2.5mm}
	\centering
	\includegraphics[scale=1.7]{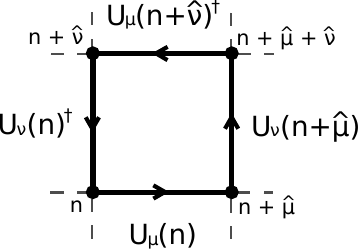}
	\vspace{2.5mm}
	\caption{Graphical illustration of the plaquette variable $U_{\mu\nu}(n)$. \label{fig:teo_plaquette}}	
\end{figure}

Wilson, when giving the first formulation of lattice gauge theories in \cite{Wilson74}, used the smallest gauge-invariant objects one can build on the lattice, i.e., the traces of the plaquette variable, which is the ordered product of link variables around a plaquette (see Fig.~\ref{fig:teo_plaquette} for illustration):
\begin{equation}
\label{eq:teo_plaquette}
	U_{\mu\nu}(n) = U_{\mu}(n) U_{\nu}(n + \hat{\mu}) U_{\mu} (n + \hat{\nu})^{\dagger} U_{\nu} (n)^{\dagger} \, .
\end{equation}
The Wilson gauge action is a sum over all plaquettes, counted with only one orientation:
\begin{equation}
	\label{eq:teo_wilson}
	S_{G}[U] = \dfrac{\beta}{3} \sum_{n \in \Lambda}\sum_{\mu<\nu} \R \Tr \big[\mathbf{1} - U_{\mu\nu}(n) \big] \, .
\end{equation}

We now show that (\ref{eq:teo_wilson}) reproduces the continuum form (\ref{eq:teo_sgcontinuum}) of the gauge action when the continuum limit $a \to 0$ is taken. To do so, we expand the link variables in the form (\ref{eq:teo_gaugetransporter}) for small $a$. Then we use the Baker-Campbell-Hausdorff formula 
\begin{equation}
	\label{eq:teo_bchformula}
	\exp (A) \exp (B) \, = \, \exp \big( A + B + \dfrac{1}{2} [A,B] + \dots \big)
\end{equation}
iteratively to rewrite the plaquette variables in the action (\ref{eq:teo_wilson}) as follows
\begin{align}
\label{eq:teo_plaquettecontinuum}
	U_{\mu\nu}(n) \, = \, \exp \Big(&ia A_{\mu}(n)  + i a A_{\nu}(n + \hat{\mu}) - \frac{a^{2}}{2}[A_{\mu}(n), A_{\nu}(n + \hat{\mu})] \nonumber \\
	&- ia A_{\mu} (n + \hat{\nu}) -i a A_{\nu} (n) - \frac{a^{2}}{2} [A_{\mu}(n + \hat{\nu}), A_{\nu}(n)] \nonumber \\
	&+ \frac{a^{2}}{2} [A_{\mu}(n),A_{\mu}(n + \hat{\nu})] + \frac{a^{2}}{2} [A_{\nu}(n + \hat{\mu}), A_{\nu}(n)] \nonumber \\
	&+ \frac{a^{2}}{2} [A_{\mu}(n),A_{\nu}(n)] + \frac{a^{2}}{2} [A_{\nu}(n + \hat{\mu}), A_{\mu}(n + \hat{\nu})]  + \mathcal{O}(a^{3})\Big) \, .
\end{align}
If one then substitutes in (\ref{eq:teo_plaquettecontinuum}) the Taylor expansion for the shifted fields
\begin{equation*}
	A_{\mu} (n + \hat{\nu}) = A_{\mu}(n) + a \partial_{\nu} A_{\mu}(n) + \mathcal{O}(a^{2}) \, ,
\end{equation*}
many of the terms in Eq.~(\ref{eq:teo_plaquettecontinuum}) cancel and one obtains
\begin{equation}
	\label{eq:teo_plaquettecontinuum2}
	U_{\mu\nu}(n) \, = \, \exp \Big(ia^{2} \big(\partial_{\mu} A_{\nu}(n) - \partial_{\nu} A_{\mu}(n) + i [A_{\mu}(n),A_{\nu}(n)] \big) + \mathcal{O}(a^{3})\Big) \, .
\end{equation}
Now we can insert the form (\ref{eq:teo_plaquettecontinuum2}) in the Wilson action, and expand the exponential to obtain
\begin{equation}
	\label{eq:teo_continuumlimit}
	S_{G}[U] = \dfrac{\beta}{3} \sum_{n \in \Lambda} \sum_{\mu<\nu} \R \Tr [\mathbf{1} - U_{\mu\nu}(n)] = \dfrac{\beta}{6} \sum_{n \in \Lambda} \sum_{\mu,\nu} \Tr [F_{\mu\nu}(n)^{2}] + \mathcal{O}(a^{2}) \, .
\end{equation}
Eq.~(\ref{eq:teo_continuumlimit}) states that the Wilson action (\ref{eq:teo_wilson}) approximates the continuum form (\ref{eq:teo_sgcontinuum}) in the limit $a \to 0$ up to $\mathcal{O}(a^{2})$. Comparing (\ref{eq:teo_continuumlimit}) with (\ref{eq:teo_sgcontinuum}) we also find that the parameter $\beta$, the so-called \textit{inverse coupling}, is related to the gauge coupling $g$ by
\begin{equation}
	\label{eq:teo_beta}
	\beta = \dfrac{6}{g^{2}} \, .
\end{equation}
We remark that this definition of $\beta$ is specific for the SU(3) case, while the general expression for SU(N) groups is $\beta = 2N/g^{2}$.

Another important remark regards the continuum limit. The one we discussed in this chapter is usually referred to as "naive continuum limit" and is used as a guiding principle in the construction of lattice theories. For example in Chapter \ref{cha:pcm} we will use the same steps as described in Sec.~\ref{sec:teo_naive} to discretize the SU(2) principal chiral model with coupled chemical potentials. However, when taking the limit $a \to 0$ the bare parameters of the action, like the gauge coupling $g$ and the quark mass $m$, show a dependence on the lattice spacing $a$. Obviously, when sending $a \to 0$ the so-called \textit{running} of the bare parameters has to reproduce the correct physics. So, a physical observable $P(g(a), m(a), \dots, a)$ has to assume its physical value $P_0$ when taking the continuum limit:
\begin{equation}
	\label{eq:teo_physical}
	\lim_{a \to 0} P(g(a), m(a), \dots, a) = P_0 \, .
\end{equation}
This requirement can be formulated by means of a differential equation \cite{Callan70,Symanzik1970}
\begin{equation}
	\label{eq:teo_callan}
	\dfrac{dP(g,m,\dots,a)}{d\ln a} = 0 \, .
\end{equation}
When using Eq.~(\ref{eq:teo_callan}) for pure gauge theories one finds that the running coupling $g(a)$ decreases when the lattice spacing $a$ decreases, which reproduces the behavior of asymptotic freedom. Then, for pure gauge theories the true continuum limit is performed by sending $\beta \to \infty$, and at the same time changing the number of lattice points $N$ and $N_{T}$ such that the physical extents
$
	L = a N $ and $T = a N_T \,
$ 
of the lattice remain the same.

\section{Numerical simulations \label{sec:teo_numerical}}
The partition function (\ref{eq:teo_partition}) of QCD on the lattice is given by
\begin{equation}
	\label{eq:teo_partitionqcd}
	Z \, = \, \int D[\psi,\overline{\psi}] D[U] \, e^{-S_{F}[\psi,\overline{\psi}, U ] - S_{G}[U]} \, ,
\end{equation}
where $S_{F}[\psi,\overline{\psi}, U]$ and $S_{G}[U]$ are some lattice discretization of the fermion action (\ref{eq:teo_sfcontinuum}) and the gauge action (\ref{eq:teo_sgcontinuum}), e.g., (\ref{eq:teo_naivesf2}) and (\ref{eq:teo_wilson}) respectively. The integrals over the fermion field configurations are the product of Grassmann measures on the sites $n$ of the lattice 
\begin{equation}
	D[\psi,\overline{\psi}] \equiv \prod_{n \in \Lambda} \prod_{\alpha, a} d \psi(n)_{_{a}^{\alpha}} \, d \overline{\psi}(n)_{_{a}^{\alpha}} \, , 
\end{equation} 
while the integrals over the gauge field configurations are the product of Haar measures on the links $(n, \mu)$ of the lattice 
\begin{equation}
\label{eq:teo_haarmeasure}
D[U] \equiv \prod_{n \in \Lambda} \prod_{\mu = 1}^{4} d U_{\mu}(n) \, .
\end{equation} 
Eq.~(\ref{eq:teo_partitionqcd}) can be rewritten as
\begin{equation}
		\label{eq:teo_partitionqcd2}
		Z \, = \, \int D[U] \, e^{- S_{G}[U]}\, Z_{F}[U]\, ,
\end{equation}
where
\begin{equation}
\label{eq:teo_partitionf}
Z_{F} [U] \, = \, \int D[\psi,\overline{\psi}] \, e^{-S_{F}[\psi,\overline{\psi}, U ]} \, .
\end{equation}
When $S_{F}[\psi,\overline{\psi}, U]$ is written as in Eq.~(\ref{eq:teo_naivesf2}), (\ref{eq:teo_partitionf}) has the form of the Grassmann integral (\ref{eq:teo_gaussian}) we computed in Sec.~\ref{sec:teo_grassmann}, and thus
\begin{equation}
\label{eq:teo_fermiondeterminant}
Z_{F} [U] \, = \, \det D(U) \, ,
\end{equation}
which is referred to as \textit{fermion determinant}. The $\gamma_{5}$-hermicity (\ref{eq:teo_gamma5}) of the Wilson Dirac operator (\ref{eq:teo_wilsonfermions}) implies that the fermion determinant (\ref{eq:teo_fermiondeterminant}) is real, a property that is crucial for the Monte Carlo simulations we will discuss shortly. 

Given the form (\ref{eq:teo_partitionqcd2}) of the partition function, expectation values of operators that only depend on the gauge fields may be computed as:
\begin{equation}
\label{eq:teo_expectationU}
\langle O \rangle \, = \, \dfrac{1}{Z} \int D[U] O(U) \det D(U) e^{-S_{G}[U]} \, .
\end{equation}
In general, in the path integral formulation of quantum field theories expectation values are given by:
\begin{equation}
	\label{eq:teo_expectation}
	\langle O \rangle \, = \, \dfrac{1}{Z} \int D[C] O(C) e^{-S[C]} \, ,
\end{equation}
where with the compact notation $C$ we denote the fermion and gauge field configurations. 
The expression (\ref{eq:teo_expectation}) usually cannot be computed analytically. However, Monte Carlo (MC) simulations are very powerful tools to approximate the integral over all the field configurations in (\ref{eq:teo_expectation}).

The key idea behind MC techniques is the interpretation of the Boltzmann weight $e^{-S[C]}/Z$ as a probabilistic weight $P(C)$. It is now clear why the reality of the fermion determinant (\ref{eq:teo_fermiondeterminant}) is so important: only if $P(C)$ is real and positive it can be interpreted as a probability. Then, the MC simulation approximates the expectation value (\ref{eq:teo_expectation}) by the average of the observable evaluated on $M$ sample field configurations $C_{i}$ distributed with probability $P(C_i) = e^{-S[C_{i}]}/Z$:
\begin{equation}
\label{eq:teo_average}
\langle O\rangle_{M} \approx \dfrac{1}{M} \sum_{i=1}^{M} O(C_i) \, .
\end{equation}
From probability theory we know that $\langle O\rangle_{M}$ of Eq.~(\ref{eq:teo_average}) reproduces the correct value $\langle O\rangle$ (\ref{eq:teo_expectation}) when $M \to \infty$, with a statistical error that scales as $1/\sqrt{M}$.

The field configurations $C_i$ distributed with probability $P(C_{i})$ are obtained as a Markov chain: a memoryless stochastic process, i.e., a process for which the probability to get the next configuration $C'$ depends only on the present state $C$ and not on the other configurations $C_{i}$ generated in the Markov chain. This memoryless property goes under the name of Markov property and it requires the transition probability $T(C'|C)$ to depend only on the states $C'$ and $C$ and not on the label $i$. To obtain correct results the Markov chain has to satisfy two important properties. It has to be ergodic, i.e., it must be possible for the process to access all configurations, and the transition probability $T(C'|C)$ has to satisfy the balance equation: 
\begin{equation}
\label{eq:teo_balance}
	\sum_{C} T(C'|C) P(C) = \sum_{C} T(C|C') P(C') \, .
\end{equation}

One way of implementing the Markov Chain in a simulation is the Metropolis algorithm \cite{Metropolis:1953am}, which uses a sufficient condition as a solution of the balance equation (\ref{eq:teo_balance}), called \textit{detailed balance}:
\begin{equation}
\label{eq:teo_detailedbalance}
T(C'|C) P(C) = T(C|C') P(C') \, .
\end{equation}

The Metropolis algorithm consists of the following steps: 
\begin{itemize}
	\item[1:] A candidate configuration $C'$ is chosen according to an a priori selection probability $T_{0}(C'|C)$, with $C = C_{n-1}$.
	\item[2:] The configuration $C'$ is accepted as the new configuration $C_{n}$ of the Markov chain with the acceptance probability
	\begin{equation*}
		T_{A}(C'|C) = \min \Big(1, \dfrac{T_{0}(C'|C) e^{-S[C']}}{T_{0}(C'|C) e^{-S[C]}}\Big) \, .
	\end{equation*}  
	\item[3:] Steps 1 and 2 are repeated.
\end{itemize}
The change from the field configuration $C_n$ to the field configuration $C_{n+1}$ is an \textit{update} of the MC simulation. In general one initializes the simulation with an arbitrary field configuration, and then the observables are calculated as in (\ref{eq:teo_average}) only after a sufficient number of equilibration and decorrelation updates.

In the next section we will see that the introduction of a chemical potential $\mu$ coupled to the net-quark number $\mathcal{N}$ causes the fermion determinant $Z_{F}[U]$ to be complex, thus preventing the use of MC techniques. 

\section{Finite chemical potential and the sign problem \label{sec:teo_sign}}
\subsection{Introducing the chemical potential on the lattice \label{sec:teo_chem}}
In the continuum, the net-quark number $\mathcal{N}$ is given by the spatial volume integral of the temporal component $\overline{\psi} \gamma_{4} \psi$ of the Noether current $\overline{\psi} \gamma_{\mu} \psi$, i.e., it is a conserved charge. As such, it can be coupled to a \textit{quark chemical potential} $\mu$. 
The introduction of the quark chemical potential $\mu$ on the lattice is implemented by replacing the temporal hopping terms in (\ref{eq:teo_wilsonfermions}) with
\begin{equation}
\label{eq:teo_chemical}
 -\dfrac{1}{2a} \left( e^{a \mu} (\mathbf{1} - \gamma_{4})_{\alpha \beta} \, U_{4}(n)_{ab} \, \delta_{n + \hat{4},m} + e^{-a \mu} (\mathbf{1} + \gamma_{4})_{\alpha \beta} \, U_{4}(n - \hat{4})_{ab}^{\dagger} \, \delta_{n - \hat{4},m} \right) \, .
\end{equation}
In (\ref{eq:teo_chemical}) the forward propagation in the time direction is favored by a factor $e^{a \mu}$ while the backward propagation in time is suppressed by a factor $e^{-a \mu}$, thus giving the desired asymmetry between particles and anti-particles.

However, the introduction of the chemical potential on the lattice causes a serious technical problem: for $\mu \neq 0$ the Dirac operator (\ref{eq:teo_wilsonfermions}) is no longer $\gamma_5$-hermitian. It instead satisfies the property
\begin{equation}
	\label{eq:teo_nogamma5}
	D(-\mu^{\star})^{\dagger} \, = \, \gamma_{5} D(\mu) \gamma_{5} \, ,
\end{equation}  
which implies that the eigenvalues of the Dirac operator $D$ no longer come in complex conjugate pairs.
This means that the fermion determinant $\det D$ is now complex, and thus it cannot be used as a probabilistic weight in a Monte Carlo simulation. This is known as the \textit{complex action problem} or \textit{sign problem}, and it has been preventing the study of the QCD phase diagram from first principle lattice calculations in the region of non-vanishing chemical potential.

\subsection{Approaches to the sign problem \label{sec:teo_approaches}}
Over the years a wide variety of methods have been developed in order to overcome the sign problem on the lattice. Here we discuss some of them briefly, and we refer the reader to the reviews at the annual lattice conferences, e.g., \cite{Chandrasekharan:2008gp, deForcrand:2010ys, Wolff:2010zu, Aarts:2013lcm, Gattringer:2014nxa, Sexty:2014dxa, Borsanyi:2015axp} for a more complete overview of the topic.
 
\subsubsection{Reweighting \label{sec:teo_reweighting}}
As we outlined in the previous section, the introduction of the chemical potential causes the fermion determinant $\det D$ to be complex. Using the notation of Sec.~\ref{sec:teo_numerical}, this means that $P(C)$ is now complex, i.e., $P(C) = |P(C)| e^{i \varphi (C)}$ and thus it cannot be interpreted directly as a probability. In reweighting, subsequent configurations $C'$ of the Markov chain are generated using the modulus $|P(C)|$. Then the expectation values (\ref{eq:teo_expectation}) are estimated as
\begin{equation}
	\label{eq:teo_reweighting}
	\langle O \rangle \, = \, \dfrac{\int D[C] |P(C)| e^{i \varphi(C)} O(C)}{\int D[C] |P(C)| e^{i \varphi(C)} } \, = \, \dfrac{\langle e^{i \varphi(C)} O \rangle_{R} }{\langle e^{i \varphi(C)} \rangle_{R}} \, , 
\end{equation}
where with the angular brackets with the subscript $R$ we denote the average (\ref{eq:teo_average}) computed over a set of configurations generated with the real probability distribution $|P(C)|$. When the average $\langle e^{i \varphi(C)} \rangle_{R}$ is vanishing, it indicates strong oscillations of the phases which cause huge cancellations in MC simulations and the sign problem is said to be severe. Moreover, $\langle e^{i \varphi(C)} \rangle_{R}$ can be computed as
\begin{equation}
\label{eq:teo_phase}
	\langle e^{i \varphi(C)} \rangle_{R} = \dfrac{Z}{Z_{R}} = e^{\ln Z - \ln Z_{R}} = e^{- \beta V \Delta f}\, ,
\end{equation}  
where $Z = \int D[C] P(C)$, $Z_{R} = \int D[C] |P(C)|$ and $\Delta f$ is the difference in free energy density of the two partition sums. Eq.~(\ref{eq:teo_phase}) means that, even when $\langle e^{i \varphi(C)} \rangle_{R}$ is finite, it decays exponentially with the volume $V$. Therefore, reweighting can only be applied to systems at small volumes and whose sign problem is mild.

\subsubsection{Density of states \label{sec:teo_density}}
The density of states is formally defined as 
\begin{equation}
	\label{eq:teo_density}
	\rho(E) \, = \, \int D[\phi] \delta \big( S[\phi] - E\big) \, .
\end{equation}
Using (\ref{eq:teo_density}) one can rewrite the partition sum as:
\begin{equation*}
	Z \, = \, \int D[\phi] e^{-S[\phi]} = \int dE \rho (E) e ^{-E} \, ,
\end{equation*}
and expectation values as
\begin{equation*}
	\langle O \rangle \, = \, \dfrac{1}{Z}  \int dE \, O(E) \, \rho (E) \, e ^{-E} \, .
\end{equation*}
Therefore, if $\rho(E)$ is determined numerically, $Z$ and $\langle O \rangle$ can be computed as numerical integrals of a single variable. There exist different methods for the numerical determination of the density, e.g., the LLR method \cite{Langfeld:2012ah}, and the FFA approach \cite{Gattringer:2015lra}. For a more detailed review over the density of states method in the context of the complex action problem see, e.g., \cite{Langfeld:2016kty,Gattringer:2016kco}.

\subsubsection{Complex Langevin \label{sec:teo_langevin}}
The Langevin method, as firstly formulated by Parisi and Wu in \cite{Parisi:1980ys}, treats Euclidean quantum field theory as the equilibrium limit of a statistical system coupled to a thermal reservoir. The system evolves in a fictitious time direction $t$ until it reaches equilibrium as $t\to \infty$. The evolution of the field is described by a stochastic differential equation, the Langevin equation, where the coupling with the heat reservoir is simulated by means of a stochastic noise field $\eta(t)$. In the equilibrium limit stochastic averages become identical to Euclidean expectation values \cite{DAMGAARD1987227}.

The idea of applying this method to system for which MC methods are inapplicable came shortly after, \cite{PARISI1983393,Klauder:1983nn}, and complex Langevin was born. However, also this method faces technical problems, which mainly pertain convergence: sometimes it fails to reach convergence and sometimes it converges to a wrong limit. For a review of the status of complex Langevin see, e.g., \cite{Seiler:2017wvd} and references therein.

\section{The dual approach \label{sec:teo_duality}}
The dual approach consists in exactly rewriting the partition sum of a system in terms of new variables, so-called \textit{dual variables}. The advantages of reformulating field theories in terms of new degrees of freedom are known since long. For example, already almost 80 years ago Kramers and Wannier were able to determine the critical point of the
two-dimensional Ising model using a duality transformation \cite{Kramers:1941zz}. What has been emerging since then is that different representations highlight different aspects of a system, and thus maybe better suited for different scopes (see, e.g., the review on duality \cite{Savit:1979ny}). 

In more recent years, in the framework of lattice field theory, the dual approach has been successful in solving the sign problem of a variety of abelian field theories (see, e.g., \cite{Gattringer:2014nxa}). As a general strategy, the Boltzmann weight is factorized into local exponentials which are then expanded into Taylor series. The expansion coefficients substitute the conventional degrees of freedom in the description of the system once the integration over the conventional fields is performed. The resulting partition function is a sum over the contributions of the admissible configurations of the expansion indices, i.e., the dual variables. Sometimes all those contributions are real and positive and, for those cases, the dual formulation of the partition sum is suitable for Monte-Carlo simulations and the sign problem is solved. 

As an explanatory example for the application of the dual approach to the sign problem of abelian theories we present the U(1) Gauge -- Higgs model \cite{Mercado:2013ola}. We also use this example to outline some important features of the dual approach, such as the emergence of constraints in the dual representation and their connection with the original symmetry of the system.  

\subsection{The U(1) Gauge -- Higgs model \label{sec:teo_u1}}

On the lattice the action of the U(1) Gauge -- Higgs model can be written as the sum of the U(1) Wilson gauge action $S_{G}[U]$ and of the discretized action $S_{M}[U, \phi]$ for the Higgs field:
\begin{equation}
	S[U, \phi] \, = \, S_{G}[U] \, + \, S_{M}[U, \phi] \, .
\end{equation}
The gauge action reads
\begin{equation}
	\label{eq:teo_sg}
	S_{G}[U] \, = \, - \dfrac{\beta}{2} \sum_{x} \sum_{\nu < \rho} \big[U_{x,\nu\rho} + U_{x,\nu\rho}^{\star} \big] \, ,
\end{equation}
where the plaquette variable
\begin{equation}
\label{eq:teo_plaquetteu1}
	U_{x,\nu\rho} \equiv U_{x,\nu} U_{x + \hat{\nu}, \rho} U_{x+ \hat{\rho}, \nu}^{\star} U_{x,\rho}^{\star}\, , \quad
\end{equation}
is the oriented product of the link variables $U_{x,\nu} \in U(1)$ around the plaquette $(x,\nu\rho)$.
The action for the matter fields has the form
\begin{equation}
	\label{eq:teo_sm}
	S_{M}[U,\phi] \, = \, \sum_{x} \Big[ \eta |\phi_{x}|^{2} + \lambda |\phi_{x}|^{4} - \sum_{\nu} \big( e^{\mu \delta_{\nu,4}} \phi_{x}^{\star} U_{x,\nu} \phi_{x+\hat{\nu}} + e^{-\mu \delta_{\nu,4}} \phi_{x+\hat{\nu}}^{\star} U_{x,\nu}^{\star} \phi_{x} \big)\Big] \, ,
\end{equation}
where $\phi_{x} \in \mathbb{C}$ are the charged scalar Higgs fields attached to the sites $x$ of the lattice, the parameter $\eta$ denotes $8 + m^{2}$, where $m$ is the bare mass and $\lambda$ is the coupling of the quartic interaction. In (\ref{eq:teo_sm}) we also coupled a chemical potential $\mu$, which gives different weight to the forward and backward propagation of the $\phi_{x}$ fields, thus making the action $S_{M}[U,\phi]$ complex. Hence, the U(1) Gauge -- Higgs model suffers from the complex action problem in the conventional representation.

The partition function of the model is given by
\begin{equation}
	\label{eq:teo_partitionu1}
	Z \, = \, \int D[U] D[\phi]  e^{-S_{G}[U] {-S_{M}[U,\phi]}} \, ,
\end{equation}
where
\begin{equation}
	\label{eq:teo_haar}
	\int D[U] \equiv \prod_{x,\nu} \int_{U(1)} dU_{x,\nu} 
\end{equation}
is the integration over the U(1) Haar measure, while 
\begin{equation}
	\int D[\phi] \equiv \prod_{x} \int_{\mathbb{C}} \dfrac{d\phi_{x} }{2\pi} = \prod_{x} \int_{0}^{\infty} d r_{x} r_{x} \int_{0}^{2\pi} \dfrac{d\theta_{x}}{2\pi} \, , \qquad \text{with } \ \phi_{x} = r_{x} e^{i \theta_{x}} \, ,
\end{equation}
is the integral over the complex plane for the matter fields. 

We start the dualization of the partition sum (\ref{eq:teo_partitionu1}) considering just the matter fields. The Boltzmann weight for the matter fields can be factorized as follows:
\begin{equation}
\label{eq:teo_esm}
	e^{-S_{M}[U,\phi]} =  \Bigg[\prod_{x} e^{- \eta |\phi_{x}|^{2} - \lambda |\phi_{x}|^{4}} \Bigg] \Bigg[ \prod_{x,\nu} \exp \big( e^{\mu \delta_{\nu,4}} \phi_{x}^{\star} U_{x,\nu} \phi_{x+\hat{\nu}} \big) \exp \big( e^{-\mu \delta_{\nu,4}}  \phi_{x+\hat{\nu}}^{\star} U_{x,\nu}^{\star} \phi_{x} \big) \Bigg] \, ,
\end{equation}
i.e., we rewrite the exponential of sums as a product of exponentials. Then we expand the factors for the nearest neighbor terms:
\begin{align}
&\prod_{x,\nu} \exp \big( e^{\mu \delta_{\nu,4}} \phi_{x}^{\star} U_{x,\nu} \phi_{x+\hat{\nu}} \big) \exp \big( e^{-\mu \delta_{\nu,4}}  \phi_{x+\hat{\nu}}^{\star} U_{x,\nu}^{\star} \phi_{x} \big)\nonumber \\
&\; = \prod_{x,\nu} \sum_{n_{x,\nu} = 0}^{\infty} \dfrac{\big( e^{\mu \delta_{\nu,4}} \phi_{x}^{\star} U_{x,\nu} \phi_{x+\hat{\nu}} \big)^{n_{x,\nu}} }{n_{x,\nu}! }
\sum_{\overline{n}_{x,\nu} = 0}^{\infty} \dfrac{\big( e^{-\mu \delta_{\nu,4}}  \phi_{x+\hat{\nu}}^{\star} U_{x,\nu}^{\star} \phi_{x} \big)^{\overline{n}_{x,\nu}}}{\overline{n}_{x,\nu}!} \nonumber \\
&\; = \sum_{\{n, \overline{n}\}} \Bigg[ \prod_{x, \nu} \dfrac{\big(U_{x,\nu} \big)^{n_{x,\nu}} \big(U_{x,\nu}^{\star} \big)^{\overline{n}_{x,\nu}}}{n_{x,\nu}! \overline{n}_{x,\nu}!} \Bigg]
\Bigg[ \prod_{x} e^{\mu [n_{x,4} - \overline{n}_{x,4}]} \phi_{x}^{\star \sum_{\nu}[n_{x,\nu} + \overline{n}_{x- \hat{\nu},\nu}]}
\phi_{x}^{ \sum_{\nu}[\overline{n}_{x,\nu} + n_{x- \hat{\nu},\nu}]}\Bigg] \, .
\label{eq:teo_linkfactors}
\end{align}
In the first step each local exponential is expanded in a Taylor series. The expansion coefficients are the link variables $n_{x,\nu} \in \mathbb{N}_{0}$ for the forward hops and $\overline{n}_{x,\nu} \in \mathbb{N}_{0}$ for the backward hops. These integer valued variables will be the dynamical degrees of freedom once we will integrate out the conventional fields. In the last line of (\ref{eq:teo_linkfactors}) the terms are simply reorganized, and the notation $\sum_{\{n, \overline{n}\}}$ is introduced to denote the sum over all configurations of the expansion variables, i.e.,
\begin{equation*}
	\sum_{\{n, \overline{n}\}} \equiv \prod_{x,\nu} \sum_{n_{x,\nu} = 0}^{\infty} \sum_{\overline{n}_{x,\nu} = 0}^{\infty} \, .
\end{equation*}
We now insert the expansions (\ref{eq:teo_esm}) and (\ref{eq:teo_linkfactors}) in Eq.~(\ref{eq:teo_partitionu1}) and we write the complex fields as $\phi_{x} = r_{x} e^{i \theta_{x}}$ and the integration measure $\int D[\phi]$ in polar coordinates. For the partition sum we obtain:
\begin{align}
Z &= \sum_{\{n, \overline{n}\}} 
\Bigg[ \prod_{x, \nu} \dfrac{1}{n_{x,\nu}! \overline{n}_{x,\nu}!} \Bigg] 
\int D[U] e^{-S_{G}[U]} \Bigg[ \prod_{x,\nu} \big(U_{x,\nu} \big)^{n_{x,\nu} - \overline{n}_{x,\nu}} \Bigg] \nonumber \\
&\hspace{9mm} \times \Bigg[ \prod_{x} e^{\mu [n_{x,4} - \overline{n}_{x,4}]} \int_{0}^{\infty} d r_{x} r_{x}^{1 + \sum_{\nu}[(n_{x,\nu} + \overline{n}_{x,\nu})+ (n_{x- \hat{\nu},\nu} + \overline{n}_{x- \hat{\nu},\nu})]} e^{- \eta r_{x}^{2} - \lambda r_{x}^{4}}  \Bigg] \nonumber \\
&\hspace{9mm} \times 
\Bigg[ \prod_{x} \int_{0}^{2\pi} \dfrac{d \theta_{x}}{2 \pi} e^{-i \theta_{x} \sum_{\nu}[(n_{x,\nu} - \overline{n}_{x,\nu})- (n_{x- \hat{\nu},\nu} - \overline{n}_{x- \hat{\nu},\nu})]} \Bigg] \, .
\label{eq:teo_partition2}
\end{align}
The integration over the Higgs fields leads to two types of contributions: the integrals over the phases $\theta_{x}$ give rise to Kronecker deltas, which impose constraints on the variables $n_{x,\nu} \in \mathbb{N}_{0}$ and $\overline{n}_{x,\nu} \in \mathbb{N}_{0}$ at all sites $x$. The integrals over the radial coordinates $r_{x}$ together with the factorials in the first line of (\ref{eq:teo_partition2}) give rise to weights for the configurations $\{n,\overline{n}\}$ of the Higgs fields in the dual representation.

To simplify the structure of the constraints it is useful to perform the change of variables:
\begin{equation}
	\label{eq:teo_k}
	\begin{array}{ll}
	n_{x,\nu} - \overline{n}_{x,\nu} = k_{x,\nu} \, ,  \qquad &k_{x,\nu} \in \mathbb{Z} \, , \\
	n_{x,\nu} + \overline{n}_{x,\nu} = |k_{x,\nu}| + 2 l_{x,\nu} \, , \qquad  &l_{x,\nu} \in \mathbb{N}_{0} \, .
	\end{array}
\end{equation}
Then the partition function is given by
\begin{equation}
	Z = \sum_{\{k, l\}} C_{M}[k] \, W_{M}[k,l] \int D[U] e^{-S_{G}[U]} \Bigg[ \prod_{x,\nu} \big(U_{x,\nu} \big)^{k_{x,\nu}} \Bigg]  \, .
\label{eq:teo_partition3}
\end{equation}
The dual variables for the Higgs fields are now the link based variables $k_{x,\nu} \in \mathbb{Z}$ and $l_{x,\nu} \in \mathbb{N}_{0}$ defined in (\ref{eq:teo_k}). The partition function sums over the configurations of those dual variables:
\begin{equation*}
\sum_{\{k, l\}} \equiv \prod_{x,\nu} \sum_{k_{x,\nu} = -\infty}^{+\infty} \sum_{l_{x,\nu} = 0}^{\infty} \, .
\end{equation*}
The configurations of the $k_{x,\nu}$ variables are constrained by the product of Kronecker deltas in the function $C_{M}[k]$:
\begin{equation}
	\label{eq:teo_cm}
	C_{M}[k] = \prod_{x} \delta \Big(  \vec{\nabla} \vec{k}_{x} \Big) \, ,
\end{equation}
where $\vec{\nabla} \vec{k}_{x}$ is the discretized divergence of the link variables $k_{x,\nu}$, which explicitly is given by
\begin{equation}
\label{eq:teo_divergence}
\vec{\nabla} \vec{k}_{x} \equiv \sum_{\nu} [k_{x,\nu} - k_{x- \hat{\nu},\nu}] \, .
\end{equation}
The constraint $C_{M}[k]$ in (\ref{eq:teo_cm}) requires the discretized divergence of the $k_{x,\nu}$ variables to vanish at every site $x$ of the lattice, i.e., it imposes the conservation of the $k$-fluxes. As a result the admissible configurations for the $k_{x,\nu}$ are closed \textit{worldlines}. We will see in Chapter \ref{cha:pcm} that analogous constraints arise from the abelian color flux dualization of the principal chiral model. In general constraints of the type (\ref{eq:teo_cm}) are flux conservation constraints that result from the integration of the phases in the parametrization of the conventional fields.

On the other hand the $l_{x,\nu}$ are unconstrained variables that only contribute to the weight function $W_{M}[k,l]$:
\begin{equation}
	\label{eq:teo_wm}
	W_{M}[k,l] = \prod_{x, \nu} \dfrac{1}{(|k_{x,\nu}|+ l_{x,\nu})! l_{x,\nu}!} \prod_{x} e^{\mu k_{x,4}} P \Big( \sum_{\nu} \big[ |k_{x,\nu}| + |k_{x - \hat{\nu},\nu}| + 2(l_{x,\nu} + l_{x - \hat{\nu},\nu}) \big]\Big) \, ,
\end{equation}
where $P(n)$ are the elementary integrals
\begin{equation}
	P(n) = \int_{0}^{\infty} d x \, x^{1 + n} \, e^{- \eta x^{2} - \lambda x^{4}} \, ,
\end{equation}
which can easily be computed numerically. From Eq.~(\ref{eq:teo_wm}) we can also read off the $\mu$-dependence: the chemical potential $\mu$ multiplies the $k_{x,\nu}$ variables in the temporal direction $\nu = 4$. Since the $k_{x,\nu}$ variables are constrained to form closed loops by $C_{M}[k]$, only temporal winding loops couple to the chemical potential. For those loops $\sum_{x} k_{x,4}$ is equal to the extent of the lattice in the temporal direction $N_{T}$, i.e., the inverse temperature $\beta_{T}$, times the net temporal winding number of the loop $W_{T}$. If one then compares the resulting expression for the $\mu$ dependence $e^{\mu \beta_{T} W_{T}}$ with the usual form $e^{\mu \beta_{T} \mathcal{N}}$ for the coupling with the net-particle number $\mathcal{N}$, we can identify 
\begin{equation}
	W_{T} = \mathcal{N} \, .
\end{equation}
In other words, in the dual representation the net-particle number $\mathcal{N}$ is interpreted as the topological quantity $W_{T}$, which denotes the total net temporal winding of a given configuration. This is one of the most beautiful outcomes of the dualization procedure, as the net temporal winding number is much easier to determine than the net-particle number. We will see that a similar interpretation will arise also in the abelian color flux dualization of the fermion action of QCD. 
  
To complete the reformulation of the partition function $Z$ in (\ref{eq:teo_partition2}) we now focus on the dualization of the gauge fields. The gauge action $S_{G}[U]$ given in (\ref{eq:teo_sg}) is a sum over plaquettes. Hence, analogously to what we have done for the Higgs field, we may rewrite the Boltzmann weight $e^{- S_{G}[U]}$ as a product over plaquettes:
\begin{align}
	e^{-S_{G}[U]} &= \prod_{x, \nu < \rho} e^{\frac{\beta}{2} \big(U_{x,\nu\rho} + U_{x,\nu\rho}^{\star} \big)} = \prod_{x, \nu < \rho} \sum_{p_{x,\nu\rho} = - \infty}^{+\infty} I_{p_{x,\nu\rho}}(\beta) \, U_{x,\nu\rho}^{\ p_{x,\nu\rho}} \nonumber \\
	&= \sum_{\{p\}} \Bigg[\prod_{x, \nu < \rho} I_{p_{x,\nu\rho}}(\beta)\Bigg] 
	\Bigg[\prod_{x, \nu } U_{x,\nu}^{ \ \sum_{\rho > \nu}[p_{x,\nu\rho} - p_{x - \hat{\rho},\nu\rho}] - \sum_{\sigma < \nu}[p_{x,\sigma\nu} - p_{x - \hat{\sigma},\sigma\nu}]} \Bigg] \, .
\label{eq:teo_gauge}
\end{align}
In the second step we used the definition of the generating function of the modified Bessel functions \cite{Olver:2010:NHM:1830479},
\begin{equation*}
	e^{\frac{z}{2} (t + t^{-1})} = \sum_{m = - \infty}^{+\infty} t^{m} I_{m}(z) \, ,
\end{equation*}
thus introducing the expansion variables $p_{x,\nu\rho} \in \mathbb{Z}$ attached to the plaquettes $(x,\nu\rho)$.  We will refer to those variables as \textit{plaquette occupation numbers}. Then we inserted the explicit expression for the plaquettes (\ref{eq:teo_plaquetteu1}) in terms of the link variables and reorganized the product to find the powers of the link variables. We now insert the expression (\ref{eq:teo_gauge}) of the gauge Boltzmann weight inside Eq.~(\ref{eq:teo_partition3}). To obtain the final form of the dualized partition function (\ref{eq:teo_partition3}) we just need to solve the gauge integrals:
\begin{align}
C_{G}[p,k] &= \prod_{x, \nu } \int dU_{x,\nu} \, U_{x,\nu}^{ \ \sum_{\rho > \nu}[p_{x,\nu\rho} - p_{x - \hat{\rho},\nu\rho}] - \sum_{\sigma < \nu}[p_{x,\sigma\nu} - p_{x - \hat{\sigma},\sigma\nu}] + k_{x,\nu}} \nonumber \\
&= \prod_{x, \nu }  \delta \Big( \sum_{\rho > \nu}[p_{x,\nu\rho} - p_{x - \hat{\rho},\nu\rho}] - \sum_{\sigma < \nu}[p_{x,\sigma\nu} - p_{x - \hat{\sigma},\sigma\nu}] + k_{x,\nu}\Big) \, .
\label{eq:teo_cg}
\end{align}
Again the integration over the conventional degrees of freedom leads to constraints for the dual variables in the form of Kronecker deltas. In particular (\ref{eq:teo_cg}) requires the total net link flux to be vanishing at every link $(x,\nu)$. The flux on each link of the lattice receives contributions both from the $k_{x,\nu} \in \mathbb{Z}$ variables, which describe the Higgs fields in the dual representation, and the $p_{x,\nu\rho} \in \mathbb{Z}$ dual variables for the $U(1)$ gauge degrees of freedom. When the Higgs degrees of freedom are absent, i.e., in the pure gauge case, (\ref{eq:teo_cg}) implies that the only admissible gauge configurations are closed surfaces of cycle occupation numbers, also called \textit{worldsheets}. 

Putting things together we get the dual form of the partition function
\begin{equation}
	\label{eq:teo_dualpartitionu1}
	Z = \sum_{\{p,k,l\}} C_{M}[k] \, W_{M}[k,l] \, C_{G} [p,k] \, W_{G}[p] \, .
\end{equation}
The sum $\sum_{\{p,k,l\}}$ runs over the configurations of the plaquette occupation numbers $p_{x,\nu\rho} \in \mathbb{Z}$, the dual variables for the Higgs $k_{x,\nu} \in \mathbb{Z}$ and the auxiliary variables $l_{x,\nu} \in \mathbb{N}_{0}$. The $k_{x,\nu}$ variables must satisfy the flux conservation constraint $C_{M}[k]$ given in (\ref{eq:teo_cm}), as well as the gauge constraint $C_{G}[p,k]$ in (\ref{eq:teo_cm}) together with the $p_{x,\nu\rho}$ variables. These two constraints allow us to interpret the dual variables for matter as worldlines and the dual variables for the gauge degrees of freedom as worldsheets. This kind of interpretation will recur also in the dualized theories we will present in this thesis. The weight factor $W_{M}[k,l]$ for the configurations of the matter flux variables was given in (\ref{eq:teo_wm}) while the weight for the gauge configurations is given by
\begin{equation}
\label{eq:teo_wg}
	W_{G}[p] = \prod_{x, \nu < \rho} I_{p_{x,\nu\rho}}(\beta) \, .
\end{equation}   
Both $W_{M}[k,l]$ and $W_{G}[p]$ are real and positive factors. This means that the dual formulation (\ref{eq:teo_dualpartitionu1}) of the partition function is suitable for Monte Carlo simulations, and thus the sign problem is solved. Obviously, in order for the simulation to be efficient one must use simulation strategies that take into account the constraints (\ref{eq:teo_cm}) and (\ref{eq:teo_cg}). We refer the interested reader to the original paper \cite{Mercado:2013yta}, where the numerics are described in detail. 
\chapter{The SU(2) principal chiral model \label{cha:pcm}}

In this chapter we present the application of the abelian color flux (ACF) approach to the SU(2) principal chiral model. The study of this model is motivated by its analogies with non-abelian field theories, like asymptotic freedom and dynamic mass generation. Also, this model can be used as a chiral effective model for low energy QCD with isospin chemical potential \cite{Rindlisbacher:2015xku}. 

However, here our interest lies on more theoretical aspects. We want to demonstrate the applicability of the abelian color flux (ACF) approach to general non-abelian theories, other than fermion theories with a gauge background. Furthermore, we want to address the question of how the symmetry of the conventional representation manifests itself in the dual worldline representation. In doing so we figured out that the ACF approach easily solves the complex action problem that this model has in the conventional representation when chemical potentials are coupled to some of the conserved currents. Furthermore, we were able to find yet another dual representation, as we will discuss in the last section of this chapter. 

\section{The continuum theory \label{sec:pcm_continuum}}

The Euclidean continuum action of the SU(2) principal chiral model in $d$ dimensions is given by
\begin{equation}
\label{eq:pcm_continuumaction}
S = \dfrac{J}{2} \int \! d^{d}x \, \Tr \left[ ( \partial_{\nu} U(x) )^{\dagger} (\partial_{\nu} U(x))\right] \, ,
\end{equation}
where $J$ is the coupling and $U(x)\in$ SU(2) are the dynamical degrees of freedom. The labels $\nu = 1,2,\dots,d$ are space-time indices which are summed over in the expression for the action (\ref{eq:pcm_continuumaction}). The model enjoys the global symmetry $\text{SU(2)}_{L} \times \text{SU(2)}_{R}$ which corresponds to the left and right multiplications of $U(x)$ by arbitrary constant SU(2) matrices
\begin{equation}
\label{eq:pcm_symmetries}
	U(x) \, \rightarrow \, V_{L} \, U(x) \, V^{\dagger}_{R} \, , \qquad \qquad V_{L}, V_{R} \in \text{SU(2)} \, .
\end{equation}
Let us now consider one of the symmetries (\ref{eq:pcm_symmetries}) written in the infinitesimal form
\begin{equation}
\label{eq:pcm_infinitesimaltransformation}
	U(x) \, \rightarrow \, U^{\prime}(x) \, = \, U(x) + \alpha \, \Delta U(x) \, ,
\end{equation}
where $\alpha$ is an infinitesimal parameter and $\Delta U(x)$ is the linearized deformation of the field $U(x)$ at order $\alpha$. Noether's theorem implies the existence of the conserved currents
\begin{equation}
\label{eq:pcm_conservedcurrents}
	j_{\nu} (x) = J \, \Tr \left[(\partial_{\nu} U^{\dagger} (x)) \Delta U(x)\right] \, , \qquad \text{with } \,
	\partial_{\nu} j_{\nu} (x) = 0 \, .
\end{equation}
The corresponding conserved charge then is
\begin{equation}
\label{eq:pcm_conservedcharge}
		Q = \int \!d^{d-1}x \, j_{d}(x) = J \int \!d^{d-1}x \Tr \left[(\partial_{d} U^{\dagger} (x)) \Delta U(x)\right] \, ,
\end{equation}
where the integration runs over the $d-1$ dimensional space and $d$ denotes the temporal direction. 
In the Euclidean formalism this charge can be coupled to a chemical potential $\mu$ as follows
\begin{equation}
\label{eq:pcm_coupling}
	S_{\mu} = S + i \mu \beta Q = J \int d^{d} x \left\{ \frac{1}{2} \Tr \left[ ( \partial_{\nu} U(x) )^{\dagger} (\partial_{\nu} U(x))\right] + i \mu \Tr \left[\partial_{d} U^{\dagger} (x) \Delta U(x)\right] \right\} \, ,
\end{equation}
where in the second step we used the fact that the size of the compactified time direction is equivalent to the inverse temperature: $\int d x_{4} = \beta$.
 
For our purposes we consider two of the symmetries (\ref{eq:pcm_symmetries}),
\begin{equation}
\label{eq:pcm_symmetries12}
		U(x) \, \rightarrow \, e^{i \frac{\alpha_{1}}{2} \sigma_{3}} \, U(x) \, e^{i \frac{\alpha_{1}}{2} \sigma_{3}} \, , \qquad
		U(x) \, \rightarrow \, e^{i \frac{\alpha_{2}}{2} \sigma_{3}} \, U(x) \, e^{-i \frac{\alpha_{2}}{2} \sigma_{3}} \, ,
\end{equation} 
where $\alpha_{1} \in \mathbb{R}$ and $\alpha_{2} \in \mathbb{R}$ are independent parameters and $\sigma_{3}$ denotes the third Pauli matrix. Following the discussion above we can compute the Noether charges corresponding to these two symmetries and obtain
\begin{align}
	\nonumber
	&Q_{1} = \dfrac{i J}{4}  \int \!d^{d-1}x \Tr \left[(\partial_{d} U^{\dagger} (x)) [ \sigma_{3} U(x) + U(x) \sigma_{3} ] - U^{\dagger} (x) [ \sigma_{3} (\partial_{d} U(x)) + (\partial_{d} U(x)) \sigma_{3} ]\right] \, ,\\
	&Q_{2} = \dfrac{i J}{4}  \int \!d^{d-1}x \Tr \left[(\partial_{d} U^{\dagger} (x)) [ \sigma_{3} U(x) - U(x) \sigma_{3} ] - U^{\dagger} (x) [ \sigma_{3} (\partial_{d} U(x)) - (\partial_{d} U(x)) \sigma_{3} ]\right] \, .
\label{eq:pcm_charges12}
\end{align}
In the next section we will obtain the lattice version of the principal chiral model with chemical potentials $\mu_{1}$ and $\mu_{2}$ coupled to the two charges.

\section{Lattice discretization \label{sec:pcm_lattice}}

The lattice discretization is carried out following the standard procedure, as described in Section \ref{sec:teo_lattice}: the continuous Euclidean space-time is replaced by a finite lattice, derivatives are discretized using Taylor expansion and the space-time integral is replaced by a sum over all the links of the lattice. The lattice version of the model is then defined by the action
\begin{align}
\nonumber
	S = - \dfrac{J}{2} \sum_{x,\nu} &\left(\Tr \left[e^{\,  \delta_{\nu,d} \sigma_{3} \frac{\mu_{1} + \mu_{2}}{2}} \, U_{x} \, e^{\,\delta_{\nu,d}\sigma_{3}\frac{\mu_{1} - \mu_{2}}{2}} \, U_{x + \hat{\nu}}^{\dagger}\right] \right.\\
	& + \left. \Tr \left[e^{-\delta_{\nu,d}\sigma_{3}\frac{\mu_{1} - \mu_{2}}{2}} \, U_{x}^{\dagger} \, e^{-\delta_{\nu,d}\sigma_{3}\frac{\mu_{1} + \mu_{2}}{2}} \, U_{x + \hat{\nu}} \right]\right) \, .
\label{eq:pcm_action}
\end{align}
The matrices $U_{x} \in$ SU(2) live on the sites $x$ of a $d$-dimensional $N^{d-1} \times N_{t}$ lattice with periodic boundary conditions and lattice constant $a$ set to $1$. From the explicit expression (\ref{eq:pcm_action}) of the lattice action it is clear that the chemical potentials $\mu_{1}$ and $\mu_{2}$ give different weights to temporal ($\nu = d$) forward and backward nearest neighbor terms. In other words, they favor matter over anti-matter of the charge they couple to. Furthermore, it is important to notice that the introduction of the chemical potentials $\mu_{\lambda}$, $\lambda = 1,2$ also causes the action $S$ to become complex. Therefore in the conventional representation the model suffers from the complex action problem.

As usual, the partition function is obtained integrating the Boltzmann weight $e^{-S}$ over the product of SU(2) Haar measures $\int \! D[U] = \prod_{x} \int_{SU(2)} dU_{x}$, giving rise to
\begin{equation}
\label{eq:pcm_partition}
	Z = \int \! D[U] \, e^{-S} \, .
\end{equation}
In what follows we will use the explicit parametrization of the SU(2) matrices
\begin{equation}
\label{eq:pcm_su2parametrization}
U_{x} = \left[
\begin{array}{cc}
\cos \theta_{x} \, e^{\,i \alpha_{x}} & \sin \theta_{x} \, e^{\, i \beta_{x}} \\
-\sin \theta_{x} \, e^{-i \beta_{x}} & \cos \theta_{x} \, e^{-i \alpha_{x}}
\end{array}\right]	\quad , \quad dU_{x} = 2 \sin \theta_{x} \cos \theta_{x} \, d\theta_{x} \dfrac{d\alpha_{x}}{2 \pi} \, \dfrac{d\beta_{x}}{2 \pi} \, ,
\end{equation} 
with $\theta_{x} \in [0, \pi/2]$, $\alpha_{x} \in [-\pi, \pi]$ and $\beta_{x} \in [-\pi, \pi]$.

In the next section we will show how to obtain a worldline representation of the model described by the action (\ref{eq:pcm_action}) using the abelian color flux (ACF) method. Our dualization will result in an expression of the partition sum (\ref{eq:pcm_partition}) that only has real and positive terms. It will thus solve the complex action problem. 

\section{Worldline formulation \label{sec:pcm_worldline}}

For the ACF dualization of the SU(2) principal chiral model with chemical potentials $\mu_{\lambda}$, $\lambda = 1,2$, we start by making all the traces and matrix products in the lattice action (\ref{eq:pcm_action}) explicit, i.e., we rewrite them as sums over the indices $a$, $b$ of the matrices $U_{x}^{ab}$:
\begin{equation}
\label{eq:pcm_action2}
	S = - J \sum_{x,\nu} \Big[ e^{\,\mu_{1} \delta_{\nu,d}} \, U_{x}^{11} U_{x + \hat{\nu}}^{11 \, \star} + e^{\,\mu_{2} \delta_{\nu,d}} \, U_{x}^{12} U_{x + \hat{\nu}}^{12 \, \star} + e^{-\mu_{2} \delta_{\nu,d}} \, U_{x}^{21} U_{x + \hat{\nu}}^{21 \, \star} + e^{-\mu_{1} \delta_{\nu,d}} \, U_{x}^{22} U_{x + \hat{\nu}}^{22 \, \star} \Big] \, .
\end{equation}
This expression of the action makes evident that non-zero values of the chemical potentials give rise to a complex action. In fact, from the parametrization (\ref{eq:pcm_su2parametrization}), the following identities between products of SU(2) matrix elements hold
\begin{equation*}
	U_{x}^{22} U_{x + \hat{\nu}}^{22 \, \star} = (U_{x}^{11} U_{x + \hat{\nu}}^{11 \, \star})^{\star}\, , \qquad \qquad
	U_{x}^{21} U_{x + \hat{\nu}}^{21 \, \star} = (U_{x}^{12} U_{x + \hat{\nu}}^{12 \, \star})^{\star}\, .
\end{equation*}
Hence, when either of the chemical potentials $\mu_{1}$ or $\mu_{2}$ are finite, the imaginary parts of the terms $U_{x}^{11} U_{x + \hat{\nu}}^{11 \, \star}$ and $U_{x}^{12} U_{x + \hat{\nu}}^{12 \, \star}$ respectively do not cancel, consequently the action $S$ becomes complex. We will see at the end of our dualization program that the worldline formulation we will obtain completely solves the complex action problem. 

We continue the ACF dualization with introducing the matrices $M_{\nu}$ with entries
\begin{equation}
\label{eq:Mentries}
M_{\nu}^{11} = e^{\mu_{1} \delta_{\nu,d}} \, , \qquad 
M_{\nu}^{22} = e^{-\mu_{1} \delta_{\nu,d}} \, , \qquad 
M_{\nu}^{12} = e^{\mu_{2} \delta_{\nu,d}} \, , \qquad 
M_{\nu}^{21} = e^{-\mu_{2} \delta_{\nu,d}} \, ,
\end{equation}
and rewriting the partition sum as
\begin{equation}
	Z = \int \! D[U] \exp \left( J \sum_{x,\nu} \sum_{a,b = 1}^{2} M_{\nu}^{ab} U_{x}^{ab} U_{x + \hat{\nu}}^{ab \, \star} \right) = \int \! D[U] \prod_{x,\nu} \prod_{a,b = 1}^{2} e^{\,J  M_{\nu}^{ab} U_{x}^{ab} U_{x + \hat{\nu}}^{ab \, \star}} \, .
\label{eq:pcm_partition2}
\end{equation}
In the second step of (\ref{eq:pcm_partition2}) the exponential of sums was factorized into a product of exponentials. Then, we expand each of the exponentials $e^{J  M_{\nu}^{ab} U_{x}^{ab} U_{x + \hat{\nu}}^{ab \, \star}}$ in a power series, thus introducing summation variables $j_{x,\nu}^{ab}$,
\begin{align}
\nonumber	
	Z& = \int \!\! D[U] \prod_{x,\nu} \prod_{a,b = 1}^{2} \sum_{j_{x,\nu}^{ab} = 0}^{\infty} \dfrac{(J M_{\nu}^{ab})^{j_{x,\nu}^{ab}}}{j_{x,\nu}^{ab}!} (U_{x}^{ab} U_{x + \hat{\nu}}^{ab \, \star})^{j_{x,\nu}^{ab}} \\
	&= \sum_{\{ j \}} W_{J,\mu}[j] 
	\int \!\! D[U] \prod_{x,\nu} \prod_{a,b}  (U_{x}^{ab})^{j_{x,\nu}^{ab}} (U_{x}^{ab \, \star})^{j_{x - \hat{\nu},\nu}^{ab}}  \, .
	\label{eq:pcm_partition3}
\end{align}
From the first line of this equation we can already observe that, since every matrix element $U_{x}^{ab}$ is multiplied by the same matrix element on the neighboring site $U_{x + \hat{\nu}}^{ab}$, there is no complex action problem left after the expansion. 
In the second step of (\ref{eq:pcm_partition3}) we have reordered the factors $U_{x}^{ab}$ and introduced the sum $\sum_{\{j\}}$ over all configurations of the summation variables $j_{x,\nu}^{ab}$, as well as the weight factors $W_{J,\mu}[j]$:
\begin{align}
\nonumber
	&\sum_{\{j\}} \equiv \prod_{x,\nu} \prod_{ab} \sum_{j_{x,\nu}^{ab} = 0}^{\infty} \, , \\
	&W_{J,\mu}[j] \equiv \prod_{x,\nu} \prod_{ab} \dfrac{(J M_{\nu}^{ab})^{j_{x,\nu}^{ab}}}{j_{x,\nu}^{ab}!} = e^{\, \mu_{1} \sum_{x}[j_{x,d}^{11} - j_{x,d}^{22}]} \, e^{\, \mu_{2} \sum_{x}[j_{x,d}^{12} - j_{x,d}^{21}]} \prod_{x,\nu} \prod_{ab} \dfrac{J^{j_{x,\nu}^{ab}}}{j_{x,\nu}^{ab}!} \, .
\label{eq:pcm_wjmu}
\end{align}
The integration over the product Haar measure $\int \! D[U] = \prod_{x} \int_{SU(2)} dU_{x}$ in (\ref{eq:pcm_partition3}) can be carried out analytically by inserting the explicit parametrization (\ref{eq:pcm_su2parametrization}) for the $U_{x}$ matrices and the Haar integration measure in Eq.~(\ref{eq:pcm_partition3}):
\begin{align}
\nonumber	
	Z= &\sum_{\{ j \}} W_{J,\mu}[j] \prod_{x} 2 \int_{0}^{\frac{\pi}{2}} \! d\theta_{x} 
	(\cos \theta_{x})^{1 + \sum_{\nu}[j_{x,\nu}^{11} + j_{x,\nu}^{22} + j_{x-\hat{\nu},\nu}^{11} + j_{x-\hat{\nu},\nu}^{22}]} \\ \nonumber
	&\times 
	(\sin \theta_{x})^{1 + \sum_{\nu}[j_{x,\nu}^{12} + j_{x,\nu}^{21} + j_{x-\hat{\nu},\nu}^{12} + j_{x-\hat{\nu},\nu}^{21}]} \int_{-\pi}^{\pi} \! \dfrac{d\alpha_{x}}{2\pi} e^{i\alpha_{x} \sum_{\nu}[(j_{x,\nu}^{11} - j_{x,\nu}^{22}) - (j_{x-\hat{\nu},\nu}^{11} - j_{x-\hat{\nu},\nu}^{22})]}\\ 
	& \times 
	\int_{-\pi}^{\pi} \! \dfrac{d\beta_{x}}{2\pi} e^{i\beta_{x} \sum_{\nu}[(j_{x,\nu}^{12} - j_{x,\nu}^{21}) - (j_{x-\hat{\nu},\nu}^{12} - j_{x-\hat{\nu},\nu}^{21})]}
  \, .
\label{eq:pcm_partition4}
\end{align}
The integrals over the angles $\alpha$ and $\beta$ give rise to Kronecker deltas for the integer valued combinations of $j_{x,\nu}^{ab}$ in the respective exponents. They enforce constraints for the $j$-fluxes at every site $x$. The integrals over the $\theta$ angles can be solved using the following representation of the beta function \cite{Olver:2010:NHM:1830479}
\begin{equation}
	2 \int_{0}^{\frac{\pi}{2}} \! d\theta \, (\cos \theta)^{1 + n}	\,(\sin \theta)^{1 + m} \, = \, \B	\bigg(\frac{n}{2} + 1 \bigg| \frac{m}{2} + 1 \bigg) = 
	\dfrac{\frac{n}{2}! \frac{m}{2}!}{\big(\frac{n + m}{2} + 1 \big)!} \, ,
	\label{eq:pcm_betafunction}
\end{equation}
where the last equality holds only if both $n$ and $m$ are even. It can be easily shown that this is the case in Eq.~(\ref{eq:pcm_partition4}). In fact, as we have already said, the $\alpha$ and $\beta$ integrals result in two constraints at every site $x$:
\begin{align*}
\sum_{\nu}[(j_{x,\nu}^{11} - j_{x,\nu}^{22}) - (j_{x-\hat{\nu},\nu}^{11} - j_{x-\hat{\nu},\nu}^{22})] = 0 \, , \quad
\sum_{\nu}[(j_{x,\nu}^{12} - j_{x,\nu}^{21}) - (j_{x-\hat{\nu},\nu}^{12} - j_{x-\hat{\nu},\nu}^{21})] = 0 \, . 
\end{align*}
The expressions on the left hand side of these two equations can be turned into the combinations of $j$-fluxes at the exponent of the cosine and the sine in Eq. (\ref{eq:pcm_partition4}) by adding on both sides of the equations even quantities. Hence, these exponents are even and the $\theta$ integral results in simple combinatorial factors, that are fractions of factorials, as given in (\ref{eq:pcm_betafunction}).

In order to make the constraints more transparent it is favorable to introduce new \textit{flux variables} $k_{x,\nu}^{\lambda} \in \mathbb{Z}$, $\lambda = 1, 2$ and \textit{auxiliary variables} $m_{x,\nu}^{\lambda} \in \mathbb{N}_{0}$, $\lambda = 1, 2$, defined as linear combinations of the $j_{x,\nu}^{ab}$:
\begin{align}
\nonumber
&k_{x,\nu}^{1} = j_{x,\nu}^{11} + j_{x,\nu}^{22} \, , \qquad
k_{x,\nu}^{2} = j_{x,\nu}^{12} + j_{x,\nu}^{21} \, , \\
&m_{x,\nu}^{1} = \dfrac{j_{x,\nu}^{11} + j_{x,\nu}^{22} - |k_{x,\nu}^{1}|}{2} \, , \qquad
m_{x,\nu}^{2} = \dfrac{j_{x,\nu}^{12} + j_{x,\nu}^{21} - |k_{x,\nu}^{2}|}{2} \, .
\label{eq:pcm_dualvariables}
\end{align}
Using these the constraints turn into 
\begin{equation}
\label{eq:pcm_constraints}
	\vec{\nabla} \vec{k}_{x}^{1} = 0 \quad \forall x \, , \qquad \vec{\nabla} \vec{k}_{x}^{2} = 0 \quad \forall x \, ,
\end{equation}
where we introduced the discretized divergence
\begin{equation}
\label{eq:pcm_divergence}
\vec{\nabla} \vec{k}_{x}^{\lambda} \equiv \sum_{\nu} [k_{x,\nu}^{\lambda} - k_{x- \hat{\nu},\nu}^{\lambda}] \, .
\end{equation}
We can now obtain the final form of the worldline representation of the partition sum just by inserting (\ref{eq:pcm_dualvariables}) into (\ref{eq:pcm_partition4}) after all integrals were solved
\begin{equation}
\label{eq:pcm_partitionworldline}
Z = \sum_{\{k,m\}} \, W_{J} [k,m] \, W_{H} [k,m] \, W_{\mu}[k] \prod_{x} \, \prod_{\lambda=1}^{2} \, \delta \left(\vec{\nabla} \vec{k}_{x}^{\lambda}\right) \, .
\end{equation}
The partition function is a sum $\sum_{\{k,m\}}$ over all possible configurations of the dual variables $k_{x,\nu}^{\lambda} \in \mathbb{Z}$, $\lambda = 1, 2$ and $m_{x,\nu}^{\lambda} \in \mathbb{N}_{0}$, $\lambda = 1, 2$. The admissible configurations of the flux variables $k_{x,\nu}^{\lambda}$ are specified by the product of Kronecker deltas (we use the notation $\delta(n) \equiv \delta_{n,0}$) in (\ref{eq:pcm_partitionworldline}). These Kronecker deltas require the discretized divergence of both $k_{x,\nu}^{1}$ and $k_{x,\nu}^{2}$ to vanish at every site $x$. Therefore, the condition $\vec{\nabla} \vec{k}_{x}^{\lambda} = 0$ $\forall x$ implemented by the product of the Kronecker deltas in (\ref{eq:pcm_partitionworldline}) implies that at each site $x$ the total flux of $k_{x,\nu}^{\lambda}$ has to vanish. In other words, the fluxes of $k_{x,\nu}^{1}$ and of $k_{x,\nu}^{2}$ must form closed worldlines.
The auxiliary variables $m_{x,\nu}^{\lambda}$ are unconstrained.
Hence, in the worldline representation, the principal chiral model 
consists of two species of worldlines ($\lambda = 1$ and $\lambda = 2$) that are constrained to form closed loops, and two additional species of auxiliary variables which are unconstrained. 

\begin{figure}
	\centering
	\includegraphics[scale=1.7]{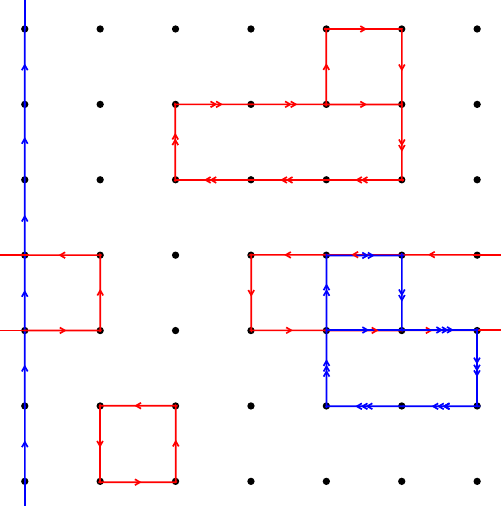}
	\caption{Example of an admissible configuration of the flux variables. $k_{x,\nu}^{1}$ is represented in red and $k_{x,\nu}^{2}$ in blue. The number of arrows denotes $|k_{x,\nu}^{\lambda}|$, i.e., the units of flux on the link $(x,\nu)$. The direction of the arrow reflects the sign of $k_{x,\nu}^{\lambda}$: it has positive (negative) orientation when $k_{x,\nu}^{\lambda}$ is positive (negative). \label{fig:pcm_config}}
\end{figure}

An example of an admissible configuration of the $k$-flux variables is represented in Fig.~\ref{fig:pcm_config}.
There we depicted $k_{x,\nu}^{1}$ in red and $k_{x,\nu}^{2}$ in blue. The number of arrows denotes $|k_{x,\nu}^{\lambda}|$, which has the interpretation of the units of flux on the link $(x,\nu)$, while the direction of the arrow depends on the sign of $k_{x,\nu}^{\lambda}$: it has positive (negative) orientation when $k_{x,\nu}^{\lambda}$ is positive (negative). Since the constraints are independent for the two species, the two kinds of worldlines must form closed loops or winding loops independently from each other. Nevertheless, as we will see later, they interact with each other, as well as with the auxiliary variables, through the weight factors $W$. In fact, the configurations of the $k_{x,\nu}^{\lambda}$ and $m_{x,\nu}^{\lambda}$ come with three different weights: $W_{J} [k, m]$, $W_{H} [k, m]$ and $W_{\mu} [k]$. $W_{J} [k, m]$ is the $J$-dependent combinatorial weight factor that arises from the Taylor expansion of the exponential factors in (\ref{eq:pcm_partition2}). It is given by
\begin{equation}
	W_{J} [k, m] = \prod_{x,\nu} \prod_{\lambda = 1}^{2}\dfrac{J^{D_{x,\nu}^{\lambda}}}{(D_{x,\nu}^{\lambda} - m_{x,\nu}^{\lambda})!m_{x,\nu}^{\lambda}!} \, ,
\label{eq:pcm_wj}
\end{equation} 
where we have introduced the abbreviation
\begin{equation}
D_{x,\nu}^{\lambda} \equiv |k_{x,\nu}^{\lambda}| + 2 m_{x,\nu}^{\lambda} \, .
\label{eq:pcm_d}
\end{equation}
$W_{H} [k, m]$ collects the fractions of factorials (\ref{eq:pcm_betafunction}) resulting from the integration of the $\theta$ angle in the Haar measure integrals. Also in this case we use the abbreviation (\ref{eq:pcm_d}) and obtain 
\begin{equation}
	W_{H} [k, m] = \prod_{x} \dfrac{\prod_{\lambda = 1}^{2} \left(\frac{1}{2} \sum_{\nu} [D_{x,\nu}^{\lambda} + D_{x-\hat{\nu},\nu}^{\lambda}]\right)!}{\left(1 + \frac{1}{2} \sum_{\nu} \sum_{\lambda} [D_{x,\nu}^{\lambda} + D_{x-\hat{\nu},\nu}^{\lambda}]\right)!} \, .
\label{eq:pcm_wh}
\end{equation}
Finally, $W_{\mu} [k]$ gives the $\mu$-dependence for the coupling with the chemical potentials
\begin{equation}
\label{eq:pcm_wmu}
W_{\mu}[k] = \prod_{\lambda = 1}^{2} \prod_{x} e^{\ \mu_{\lambda} k_{x,d}^{\lambda}} = \prod_{\lambda = 1}^{2} e^{\ \mu_{\lambda} \sum_{x} k_{x,d}^{\lambda}} = e^{\ \mu_{1} \, \beta \, \omega_{1}[k]}\ e^{\ \mu_{2} \, \beta \, \omega_{2}[k]} \, .
\end{equation}
In the last step of (\ref{eq:pcm_wmu}) we used the identity $\sum_{x} k_{x,d}^{\lambda} = N_{t} \omega_{\lambda}[k]$ where $\omega_{\lambda}[k]$ is the temporal net winding number of the $k^{\lambda}$-flux and $N_{t}$ the extent of the lattice in time ($\nu = d$) direction. The identity holds because the admissible configurations of the $k_{x,\nu}^{\lambda}$ are closed worldlines. In (\ref{eq:pcm_wmu}) we also used the fact that $N_{t}$ is the inverse temperature in lattice units and replaced $N_{t}$ by the more conventional symbol $\beta$. From Eq.~(\ref{eq:pcm_wmu}) we can also deduce that the conserved charges we computed in Section \ref{sec:pcm_kw} in the worldline formulation have the topological interpretation of the integer valued net winding numbers of the respective worldlines. This feature is of great advantage, because it allows one to identify the conserved charges easily, and thus it also opens the possibility to perform canonical simulations (compare \cite{Orasch:2017niz}).

Since all the weight factors (\ref{eq:pcm_wj}) -- (\ref{eq:pcm_wmu}) are real and positive, numerical simulations of this model in the worldline representation (\ref{eq:pcm_partitionworldline}) are indeed possible. In \cite{Gattringer:2017hhn} an exploratory numerical test was done with this representation. Here we just remark that, due to the fact that the $k$ variables are subject to constraints,  simulation strategies have to involve techniques that implement updates that do not violate the constraints. Furthermore these techniques have to be set up such that the updates are still efficient. In \cite{Gattringer:2017hhn} we performed two types of simulations. One in which the $k$ variables are updated via a generalization of the worm algorithm \cite{PhysRevLett.87.160601} (taking into account the site terms using the variant described in \cite{Giuliani:2017mxu}), the other using local updates. The auxiliary $m$ variables were always updated with local Metropolis sweeps.
We remark that our worldline representation is not unique and other alternative dual representations can be found \cite{Rindlisbacher:2015xku,Kovacs:1993ta,PhysRevD.49.6072}. Furthermore, the SU(2) principal chiral model is equivalent to the O(4) nonlinear sigma model. In this form dual formulations were presented in \cite{Bruckmann:2015sua,Wolff:2009kp}.

As we outlined at the beginning of this chapter, one of our main goals was to identify how the original symmetry of the model translates in the dual representation. It is clear since quite some time now that the constraints arising from the integration of the conventional degrees of freedom carry information about the original symmetry of the system. Often for matter fields the constraints enforce site based flux conservation laws.
In the specific case of the principal chiral model, constraints over the $k$-fluxes arise from the integration over the U(1) phases. The whole SU(2) description is recovered with the weight factors, that tie together the two worldlines $k^{1}$ and $k^{2}$, and the auxiliary variables $m^{\lambda}_{x,\nu}$, $\lambda=1,2$.

\section{Kramers-Wannier dual representation \label{sec:pcm_kw}}

Having discussed the ACF dualization of the principal chiral model with two chemical potentials, here we present a second reformulation of the same model. Starting from the worldline representation of the partition function given by Eqs.~(\ref{eq:pcm_partitionworldline})--(\ref{eq:pcm_wmu}), we show how it is possible to take the dualization a step further by rewriting the partition sum in terms of yet another set of dual variables. These variables live on what is called \textit{dual lattice}, and they automatically solve the constraints in Eq.~(\ref{eq:pcm_partitionworldline}). In this sense we can say that the system is fully dualized, \textit{a la} Kramers and Wannier \cite{Kramers:1941zz}. 

\begin{figure}
	\centering
	\includegraphics[scale=1.5]{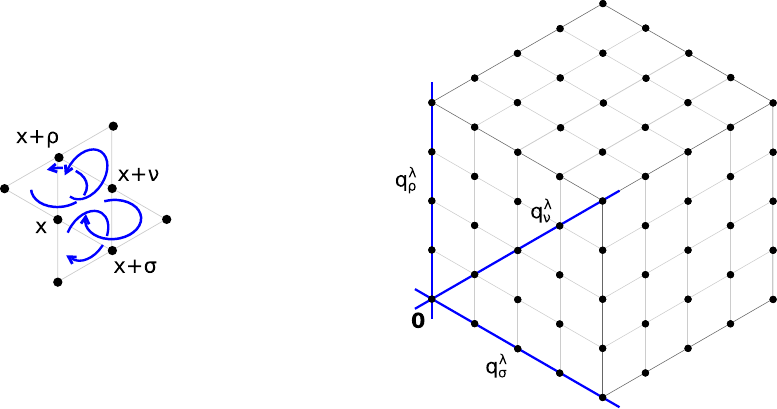}
	\caption{Graphical representation of the plaquette fluxes $n_{x,\nu\rho}^{\lambda}$ (left), and the disorder loops $q_{\nu}^{\lambda}$ (right). In both plots $\sigma < \nu < \rho$. On the left we show the graphical illustration of the two sums in Eq.~(\ref{eq:pcm_hodge}). We depict the four plaquette fluxes that contribute positively to the $k_{x,\nu}^{\lambda}$ in 3 dimensions (there would be six contributions in 4 dimensions). On the right we show the disorder loops, which are forced to be on the coordinate axes by the support functions $\Theta_{x,\nu}^{(\nu)}$, and that implement closed winding loops due to the periodic boundary conditions. \label{fig:pcm_kw}}
\end{figure}

The first step for the Kramers-Wannier (KW) dualization program consists in finding new variables that can generate all the admissible configurations of the $k^{\lambda}$-fluxes and that, at the same time, resolve the flux conservation constraints enforced by the Kronecker deltas in Eq.~(\ref{eq:pcm_partitionworldline}). In our case these conditions can be fulfilled by introducing two kinds of new dual variables: plaquette fluxes and disorder loops. For a graphical illustration we refer to Fig.~\ref{fig:pcm_kw}.
The plaquette variables are denoted by $n_{x,\nu\rho}^{\lambda} \in \mathbb{Z}$, $\lambda = 1,2$. They generate $|n_{x,\nu\rho}^{\lambda}|$ units of $k^{\lambda}$-flux around the plaquette $(x,\nu\rho)$, $\nu<\rho$, with mathematically positive orientation if $n_{x,\nu\rho}^{\lambda} > 0$ and negative orientation if $n_{x,\nu\rho}^{\lambda} < 0$. The disorder flux is written as $\Theta_{x,\rho}^{(\nu)} \, q_{\nu}^{\lambda}$, $q_{\nu}^{\lambda}\in\mathbb{Z}$, $\lambda = 1,2$ where
\begin{equation}
	\Theta_{x,\rho}^{(\nu)} = \left\{
	\begin{array}{l}
	1 \quad \text{if } (x,\rho)\in \{ (\mathbf{0} + n \, \hat{\nu},\nu), n = 0, 1, \dots, N_{\nu}-1\} \\
	0 \quad \text{if } (x,\rho)\notin \{ (\mathbf{0} + n \, \hat{\nu},\nu), n = 0, 1, \dots, N_{\nu}-1\} 
	\end{array}
	\right.
\label{eq:pcm_supportfunction}
\end{equation}
is the support function for the coordinate axis in the $\nu$ direction. In the definition (\ref{eq:pcm_supportfunction}), $\mathbf{0}$ denotes the origin of lattice, defined as the corner with site $(0,0,\dots,0)$. $N_{\nu}$, $\nu = 1, \dots, d-1$ are the spatial extents of the lattice, and $N_{d}$ the temporal extent $N_{d} = N_{t}$. $q_{\nu}^{\lambda} \in \mathbb{Z}$ therefore introduces $|q_{\nu}^{\lambda}|$ units of $k^{\lambda}$-flux on the $\nu$ coordinate axis which is oriented in positive $\nu$-direction for $q_{\nu}^{\lambda} > 0$ and has negative orientation for $q_{\nu}^{\lambda} < 0$. Due to the periodic boundary conditions, the term $\Theta_{x,\nu}^{(\nu)} \, q_{\nu}^{\lambda}$ actually introduces a closed winding loop of $|q_{\nu}^{\lambda}|$ units of flux placed on the coordinate axes through the origin of our $d$-dimensional lattice. Since both the plaquette fluxes and the disorder variables implement only closed worldlines of $k^{\lambda}$-fluxes, they are unconstrained.

The flux $k_{x,\nu}^{\lambda}$ receives contributions from all the plaquettes that contain the link $(x,\nu)$, as well as the disorder lines if $(x,\nu)$ sits on the $\nu$-coordinate axis: 
\begin{equation}
\label{eq:pcm_hodge}
k_{x,\nu}^{\lambda} = \sum_{\rho: \nu < \rho} \left[ n_{x,\nu\rho}^{\lambda} - n_{x - \hat{\rho},\nu\rho}^{\lambda}\right] - \sum_{\sigma: \nu > \sigma} \left[ n_{x,\sigma\nu}^{\lambda} - n_{x - \hat{\sigma},\sigma\nu}^{\lambda}\right] + \Theta_{x,\nu}^{(\nu)} q_{\nu}^{\lambda} \ .
\end{equation}
This transformation is a so-called Hodge decomposition and generates all possible configurations of $k_{x,\nu}^{\lambda}$, $\lambda = 1,2$ that obey the zero-divergence constraints. 

Substituting (\ref{eq:pcm_hodge}) in (\ref{eq:pcm_partitionworldline}) one obtains the following form of the partition function,
\begin{equation}
\label{eq:pcm_partition5}
Z = \sum_{\{q,n,m\}} \, W_{J} [q,n,m] \, W_{H} [q,n,m] \, W_{\mu}[q]  \, .
\end{equation}
The sum is now over all configurations of the variables $q_{\nu}^{\lambda}\in\mathbb{Z}$, $n_{x,\nu\rho}^{\lambda}\in\mathbb{Z}$ and the auxiliary variables $m_{x,\nu}^{\lambda}\in\mathbb{N}_{0}$, $\lambda=1,2$. The product of Kronecker deltas has disappeared, since the constraints are automatically satisfied by the configurations of the new variables $\{q,n\}$. The weight factors $W_{J} [q,n,m]$ and $W_{H} [q,n,m]$ are still given by (\ref{eq:pcm_wj}) and (\ref{eq:pcm_wh}) respectively, but now the combinations $D_{x,\nu}^{\lambda} \in \mathbb{N}_{0}$ read
\begin{equation}
D_{x,\nu}^{\lambda} \equiv \bigg|\sum_{\rho: \nu < \rho} \left[ n_{x,\nu\rho}^{\lambda} - n_{x - \hat{\rho},\nu\rho}^{\lambda}\right] - \sum_{\sigma: \nu > \sigma} \left[ n_{x,\sigma\nu}^{\lambda} - n_{x - \hat{\sigma},\sigma\nu}^{\lambda}\right] + \Theta_{x,\nu}^{(\nu)} q_{\nu}^{\lambda}\bigg| + 2 m_{x,\nu}^{\lambda} \, .
\label{eq:pcm_d2}
\end{equation}
The $\mu$-dependent weight factor is 
\begin{equation}
\label{eq:pcm_wmu2}
W_{\mu}[q] =  e^{\ \beta (\mu_{1} \, q_{d}^{1}\ + \mu_{2} \, q_{d}^{2} ) } \, .
\end{equation}
We obtained this expression by observing that, in terms of the new dual variables introduced in Eq.~(\ref{eq:pcm_hodge}), the net temporal winding number $\omega_{\lambda}[k]$ is indeed represented by the disorder variables $q_{d}^{\lambda}$ (compare with the definition of $q_{\nu}^{\lambda}$ we gave earlier).

The last step to obtain the KW-dual of the principal chiral model consists in switching to the dual lattice. In general, the construction of the lattice dual to a $d$-dimensional hypercubical lattice can be done by shifting the lattice by half a lattice spacing in every direction. More mathematically we may introduce the notion of a \textit{simplex} of dimension $s$ as an $s$-dimensional element of the hypercubical lattice: a simplex of dimension 0 is a site of the lattice, a simplex of dimension 1 is a link on the lattice, a simplex of dimension 2 is a plaquette, a simplex of dimension 3 an elementary cube, and so on. Obviously, in a $d$-dimensional lattice there are simplices of dimension $0 \leq s \leq d$. From the way the dual lattice is constructed follows that each simplex of dimension $s$ of the original lattice is in one-to-one correspondence with one simplex of dimension $\tilde{s} = d - s$ of the dual lattice. This correspondence will become clearer in the next subsections, where we will discuss in detail the mapping of the worldline representation of the principal chiral model to the dual lattice for the special cases of two and four dimensions. The generalization to any other dimensions is straightforward. 

\subsection{Fully dualized form in two dimensions}
\begin{figure}
	\centering
	\includegraphics[scale=1]{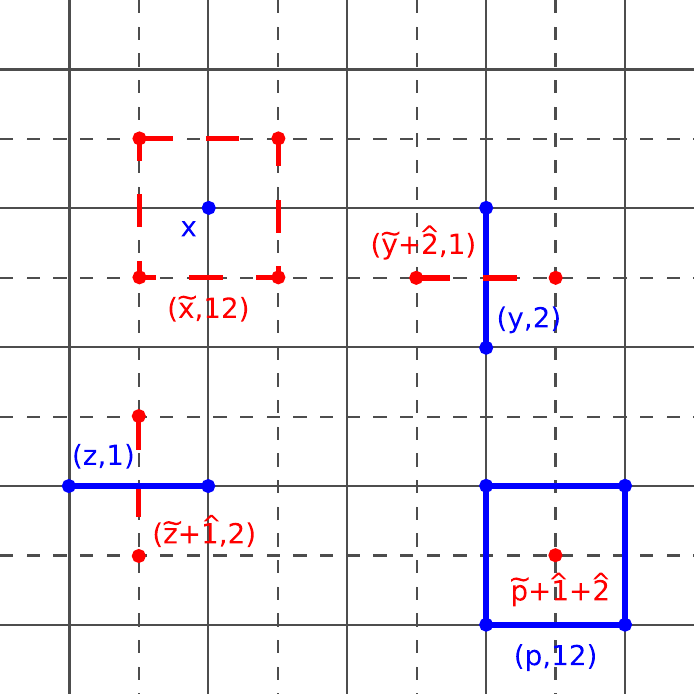}
	\caption{In 2 dimensions the dual lattice is built by placing a new site in the center of each plaquette of the original lattice. Here we represent the dual lattice with dashed lines and the original lattice with solid lines. We graphically show the one-to-one correspondence discussed in the text: simplices of dimension $s$, $0 \leq s \leq 2$, of the original lattice (illustrated in blue) are mapped to simplices of dimension $\tilde{s} = 2 - s$ of the dual lattice (illustrated in red). \label{fig:pcm_dual}}
\end{figure}

Following the above discussion, in two dimensions the dual lattice is built by placing a new site in the center of each plaquette of the original lattice. We will denote the sites of the dual lattice, as well as the variables living on it, with a tilde, e.g., $\tilde{x}$, $\tilde{n}_{\tilde{x}}^{\lambda}$. For the two-dimensional case it is easy to identify the one-to-one correspondence we mentioned earlier: each site $x$ of the original lattice lies in the center of one plaquette $(\tilde{x},12)$ of the dual lattice; likewise, each site $(\tilde{x}+\hat{1}+\hat{2})$ of the dual lattice is placed at the center of one plaquette $(x,12)$ of the original lattice; finally, each link $(x,1)$ in the spatial direction of the original lattice crosses one link in the temporal direction $(\tilde{x}+\hat{1},2)$ of the dual lattice and vice versa (refer to Fig.~\ref{fig:pcm_dual} for a graphical illustration of this mapping).
The dual variables $m_{x,\nu}^{\lambda}$, $n_{x,\nu\rho}^{\lambda}$ and $q_{\nu}^{\lambda}$, $\lambda = 1,2$ are mapped to the dual lattice accordingly:
\begin{equation}
\label{eq:pcm_2dmaptodual}
n_{x,12}^{\lambda} \rightarrow \tilde{n}_{\tilde{x} + \hat{1} + \hat{2}}^{\lambda} \, , \qquad
m_{x,\nu}^{\lambda} \rightarrow \tilde{m}_{\tilde{x} + \hat{\nu}, \rho}^{\lambda} \, , \qquad 
\Theta_{x,\nu}^{(\nu)}\, q_{\nu}^{\lambda} \rightarrow \tilde{\Theta}_{\tilde{x} + \hat{\nu}, \rho }^{(\tilde{\nu})}\, \tilde{q}_{\tilde{\nu}}^{\lambda} \, , \quad \text{with } \rho\neq \nu \, .        
\end{equation}
In (\ref{eq:pcm_2dmaptodual}) the support function $ \tilde{\Theta}_{\tilde{x}, \rho }^{(\tilde{\nu})}$ is nonzero only on the links $ (\tilde{x}, \rho )$ of the dual lattice that are dual to the $\nu$ coordinate axis.
In the KW-dual formulation of the principal chiral model in 2 dimensions the dynamical degrees of freedom are the dual variables $\tilde{n}_{\tilde{x}}^{\lambda} \in \mathbb{Z}$ assigned to the sites of the dual lattice, the dual auxiliary variables $\tilde{m}_{\tilde{x}, \rho}^{\lambda} \in \mathbb{N}_{0}$ on the links of the dual lattice, and the dual disorder variables $\tilde{q}_{\tilde{\nu}}^{\lambda}\in\mathbb{Z}$ for the flux through the links $(\tilde{x},\rho)$ dual to the $\nu$ coordinate axes ($\rho \neq \nu$).

At this point, to obtain the fully dualized form of the partition sum it is sufficient to apply the mapping (\ref{eq:pcm_2dmaptodual}) to the expression (\ref{eq:pcm_partition5}) of the partition function,
\begin{equation}
\label{eq:pcm_2dpartitionkw}
Z = \sum_{\tilde{q}_{\tilde{2}}^{1}, \tilde{q}_{\tilde{2}}^{2} \in \mathbb{Z}} e^{\, \beta \, \left( \mu_{1} \tilde{q}_{\tilde{2}}^{1} + \mu_{2} \tilde{q}_{\tilde{2}}^{2}\right)} \sum_{\tilde{q}_{\tilde{1}}^{1}, \tilde{q}_{\tilde{1}}^{2} \in \mathbb{Z}} \, \sum_{\{\tilde{n}, \tilde{m}\}} W_{J} [\tilde{q}, \tilde{n}, \tilde{m}]  W_{H} [\tilde{q}, \tilde{n}, \tilde{m}] \, .
\end{equation}
In (\ref{eq:pcm_2dpartitionkw}) we wrote explicitly the sum $\sum_{\{\tilde{q}\}}$ over the configurations of the disorder variables. In this way, we are able to factorize the $\mu$-dependence, and the partition sum has the form of a double fugacity expansion. Thus, the form (\ref{eq:pcm_2dpartitionkw}) of the partition function is suitable for canonical simulations of the system. The sum over the configurations of the dual site variables  $\tilde{n}_{\tilde{x}}^{\lambda} \in \mathbb{Z}$ and the dual auxiliary variables $\tilde{m}_{\tilde{x}, \rho}^{\lambda} \in \mathbb{N}_{0}$ is expressed by  $\sum_{\{\tilde{n}, \tilde{m}\}}$, and the weights of the configurations are
\begin{equation}
\label{eq:pcm_2dwj}
W_{J} [\tilde{q}, \tilde{n}, \tilde{m}] = \prod_{\tilde{x}, \nu} \prod_{\lambda = 1}^{2}\dfrac{J^{\tilde{D}_{\tilde{x}, \nu}^{\lambda}}}{(\tilde{D}_{\tilde{x}, \nu}^{\lambda} - \tilde{m}_{\tilde{x}, \nu}^{\lambda})!\, \tilde{m}_{\tilde{x}, \nu}^{\lambda}!},
\end{equation}
\begin{equation}
\label{eq:pcm_2dwh}
W_{H} [\tilde{n}, \tilde{q}, \tilde{m}] = \prod_{\tilde{p}}  \dfrac{\prod_{\lambda} \left( \frac{1}{2} \sum_{(\tilde{x}, \nu) \in \partial \tilde{p}} \tilde{D}_{\tilde{x}, \nu}^{\lambda} \right) !}{\left(1 + \frac{1}{2} \sum_{\lambda} \sum_{(\tilde{x}, \nu) \in \partial \tilde{p}} \tilde{D}_{\tilde{x}, \nu}^{\lambda} \right) !} \, ,
\end{equation}
where the combinations $\tilde{D}_{\tilde{x}, \nu}^{\lambda} \in \mathbb{N}_{0}$ assigned to the links of the dual lattice are
\begin{equation}
\tilde{D}_{\tilde{x}, 1}^{\lambda} = \left| \tilde{n}_{\tilde{x}}^{\lambda} - \tilde{n}_{\tilde{x} + \hat{1}}^{\lambda} + \tilde{\Theta}_{\tilde{x}, 1}^{(\tilde{2})} \, q_{\tilde{2}}^{\lambda}\, \right| + 2 m_{\tilde{x},1}^{\lambda} \, ,\quad 
\tilde{D}_{\tilde{x}, 2}^{\lambda} = \left| \tilde{n}_{\tilde{x} + \hat{2}}^{\lambda} - \tilde{n}_{\tilde{x}}^{\lambda} + \tilde{\Theta}_{\tilde{x}, 2}^{(\tilde{1})} \, q_{\tilde{1}}^{\lambda}\, \right| + 2 m_{\tilde{x},2}^{\lambda} \, .
\end{equation}
In Eq.~(\ref{eq:pcm_2dwh}) the product $\prod_{\tilde{p}}$ runs over all plaquettes $\tilde{p}$ of the dual lattice, and the sum $\sum_{(\tilde{x}, \nu) \in \partial \tilde{p}}$ is over all links $(\tilde{x},\nu)$ on the boundary $\partial \tilde{p}$ of $\tilde{p}$.

Concluding, we found that in two dimensions the KW-dual of the principal chiral model with two chemical potentials is given by a sum over the configurations of the dual disorder variables $\tilde{q}_{\tilde{\nu}}^{\lambda} \in \mathbb{Z}$, the dual site variables $\tilde{n}_{\tilde{x}}^{\lambda} \in \mathbb{Z}$ and the dual auxiliary variables $\tilde{m}_{\tilde{x}, \nu}^{\lambda} \in \mathbb{N}_{0}$. All the degrees of freedom are unconstrained in the KW-formulation of the model. The only variables that couple to the chemical potentials are the disorder variables on the links dual to the temporal coordinate axis of the original lattice, i.e., $\tilde{q}_{\tilde{2}}^{\lambda} \in \mathbb{Z}$. Therefore, in this representation it is possible to completely factorize the $\mu$-dependence and write the partition sum as a double fugacity expansion. The canonical partition function then sums over the configurations of the disorder variables $\tilde{q}_{\tilde{1}}^{\lambda} \in \mathbb{Z}$, the site variables and the auxiliary variables, and it collects the weights $W_{J} [\tilde{q}, \tilde{n}, \tilde{m}]$ and $W_{H} [\tilde{q}, \tilde{n}, \tilde{m}]$ for those configurations. Notice that both $W_{J} [\tilde{q}, \tilde{n}, \tilde{m}]$ and $W_{H} [\tilde{q}, \tilde{n}, \tilde{m}]$ are real and positive, therefore also the KW-dual formulation (\ref{eq:pcm_2dpartitionkw}) is free of the sign problem, and it can be used for simulations at finite $\mu_{\lambda}$.

\subsection{Fully dualized form in four dimensions}
 
Having discussed the two-dimensional case, it is now straightforward to generalize the results to four dimensions. The dual lattice is constructed by placing the new sites in the center of each hypercube of the original lattice. Explicitly the map to the dual lattice reads
\begin{equation*}
\begin{array}{cccccc}
	\text{site }& x &\leftrightarrow & (\tilde{x}, 1234)& \text{hypercube } \, ,& \\
	\text{link } &(x, \nu) &\leftrightarrow & (\tilde{x} + \hat{\nu}, \rho \sigma \tau) &\text{cube }& \rho < \sigma < \tau \, ,\\
	\text{plaquette } &(x, \nu \rho) &\leftrightarrow & (\tilde{x} + \hat{\nu} + \hat{\rho}, \sigma \tau) &\text{plaquette } & \nu < \rho \text{ and } \sigma < \tau \, ,\\
	\text{cube } &(x, \nu\rho\sigma) &\leftrightarrow & (\tilde{x} + \hat{\nu} + \hat{\rho} + \hat{\sigma}, \tau) &\text{link }& \nu < \rho < \sigma \, ,\\
	\text{hypercube }& (x, 1234) &\leftrightarrow &\tilde{x} + \hat{1} + \hat{2} + \hat{3} + \hat{4}& \text{site } \, .&
\end{array}	
\end{equation*}
The variables are then mapped to the dual lattice as follows
\begin{align}
\nonumber
&n_{x,\nu\rho}^{\lambda} \rightarrow \tilde{n}_{\tilde{x} + \hat{\nu} + \hat{\rho}, \sigma \tau}^{\lambda}\, , \qquad \text{with } \nu < \rho \text{ and } \sigma < \tau \, , \\ \nonumber
&m_{x,\nu}^{\lambda} \rightarrow \tilde{m}_{\tilde{x} + \hat{\nu}, \rho \sigma \tau}^{\lambda}\, , \qquad \ \text{with } \rho < \sigma < \tau \, ,\\
&\Theta_{x,\nu}^{(\nu)}\, q_{\nu}^{\lambda} \rightarrow \tilde{\Theta}_{\tilde{x} +\hat{\nu}, \tilde{\nu} }^{(\tilde{\nu})}\, \tilde{q}_{\tilde{\nu}}^{\lambda} \, .
\label{eq:pcm_4dmaptodual}
\end{align}
In the last line of (\ref{eq:pcm_4dmaptodual}) we use the notation $\tilde{\nu}$ to label the cubes of the dual lattice which are dual to links in direction $\nu$ of the original lattice, i.e., $(\tilde{x},\tilde{\nu}) \equiv (\tilde{x}, \sigma \tau \omega )$, $\sigma < \tau < \omega$. The support function $\tilde{\Theta}_{\tilde{x}, \tilde{\nu} }^{(\tilde{\nu})}$ on the dual lattice then is nonvanishing only on the cubes $\tilde{\nu}$ dual to the $\nu$ coordinate axis.

In the KW-dual formulation we thus use the dual dynamical variables $\tilde{n}_{\tilde{x}, \sigma \tau}^{\lambda} \in \mathbb{Z}$ assigned to the plaquettes of the dual lattice, the dual auxiliary variables $\tilde{m}_{\tilde{x}, \tilde{\nu}}^{\lambda} \in \mathbb{N}_{0}$ on the cubes of the dual lattice, and the dual disorder variables $\tilde{q}_{\tilde{x}, \tilde{\nu}}^{\lambda} \in \mathbb{Z}$ for the flux through the dual cubes dual to the coordinate axes. The fully KW-dual form of the partition sum in four dimensions is
\begin{equation}
\label{eq:pcm_4dpartitionkw}
Z = \sum_{\tilde{q}_{\tilde{4}}^{1}, \tilde{q}_{\tilde{4}}^{2} \in \mathbb{Z}} e^{\, \beta \, \left( \mu_{1} \tilde{q}_{\tilde{4}}^{1} + \mu_{2} \tilde{q}_{\tilde{4}}^{2}\right)} \left( \prod_{\nu = 1}^{3} \sum_{\tilde{q}_{\tilde{\nu}}^{1}, \tilde{q}_{\tilde{\nu}}^{2} \in \mathbb{Z}} \right) \, \sum_{\{\tilde{n}, \tilde{m}\}} W_{J} [\tilde{n}, \tilde{q}, \tilde{m}]  W_{H} [\tilde{n}, \tilde{q}, \tilde{m}] \, ,
\end{equation}
where the sum $\sum_{\{\tilde{n}, \tilde{m}\}}$ now runs over all configurations of the dual $\tilde{n}$- and $\tilde{m}$-variables on the dual lattice. In (\ref{eq:pcm_4dpartitionkw}) the sums over the dual disorder variables $\tilde{q}_{\tilde{\nu}}^{\lambda}$ were written explicitly up front. They are ordered such that the first double sum is over the temporal disorder variables $\tilde{q}_{\tilde{4}}^{\lambda}$ which carry the dependence on the chemical potentials $\mu_{\lambda}$. Thus, also in this case, the KW-dual partition sum (\ref{eq:pcm_4dpartitionkw}) is already organized in the form of a double fugacity expansion. The weight factor $W_{J} [\tilde{n}, \tilde{q}, \tilde{m}]$ from the Taylor expansion of the original Boltzmann factors reads
\begin{equation}
\label{eq:weightJPCMKW}
W_{J} [\tilde{n}, \tilde{q}, \tilde{m}] = \prod_{\tilde{x}, \tilde{\nu}} \prod_{\lambda = 1}^{2}\dfrac{J^{\tilde{D}_{\tilde{x}, \tilde{\nu}}^{\lambda}}}{(\tilde{D}_{\tilde{x}, \tilde{\nu}}^{\lambda} - \tilde{m}_{\tilde{x}, \tilde{\nu}}^{\lambda})!\, \tilde{m}_{\tilde{x}, \tilde{\nu}}^{\lambda}!},
\end{equation}
where the combinations $\tilde{D}_{\tilde{x}, \tilde{\nu}}^{\lambda} \in \mathbb{N}_{0}$ assigned to the dual cubes $(\tilde{x}, \tilde{\nu})$ are given by
\begin{equation}
\tilde{D}_{\tilde{x}, \tilde{\nu}}^{\lambda} = \left| \sum_{\rho: \nu < \rho} \left[  \tilde{n}_{\tilde{x} + \hat{\rho}, \sigma \tau}^{\lambda} - \tilde{n}_{\tilde{x}, \sigma \tau}^{\lambda} \right] - \sum_{\omega: \nu > \omega}  \left[  \tilde{n}_{\tilde{x} + \hat{\omega}, \sigma \tau}^{\lambda} - \tilde{n}_{\tilde{x}, \sigma \tau}^{\lambda} \right] + \tilde{\Theta}_{\tilde{x}, \tilde{\nu}}^{(\tilde{\nu})} \, q_{\tilde{\nu}}^{\lambda}\, \right| + 2 m_{\tilde{x},\tilde{\nu}}^{\lambda} \, .
\end{equation}
The weight $W_{H} [\tilde{n}, \tilde{q}, \tilde{m}]$ that originates from the Haar measure integration and implements the SU(2) symmetry of the conventional representation in the KW-dual formulation is given by
\begin{equation}
\label{eq:pcm_4dwh}
W_{H} [\tilde{n}, \tilde{q}, \tilde{m}] = \prod_{\tilde{h}}  \dfrac{\prod_{\lambda} \left( \frac{1}{2} \sum_{(\tilde{x}, \tilde{\nu}) \in \partial \tilde{h}} \tilde{D}_{\tilde{x}, \tilde{\nu}}^{\lambda} \right) !}{\left(1 + \frac{1}{2} \sum_{\lambda} \sum_{(\tilde{x}, \tilde{\nu}) \in \partial \tilde{h}} \tilde{D}_{\tilde{x}, \tilde{\nu}}^{\lambda} \right) !} \, ,
\end{equation}
where the product $\prod_{\tilde{h}}$ runs over all hypercubes $\tilde{h}$ of the dual lattice and the sum $\sum_{(\tilde{x}, \tilde{\nu}) \in \partial \tilde{h}}$ is over all dual cubes $(\tilde{x}, \tilde{\nu})$ in the boundary $\partial \tilde{h}$ of $\tilde{h}$. In the KW-dual form all constraints have disappeared and again all weight factors are real and positive, such that a simulation is possible at finite $\mu_{\lambda}$.

The KW-dual form (\ref{eq:pcm_4dpartitionkw}) -- (\ref{eq:pcm_4dwh}) was tested in an exploratory numerical simulation in \cite{Gattringer:2017hhn}. A very good agreement was found between results for bulk variables from simulations in the conventional
representation (for vanishing chemical potentials), in the worldline representation (both results from local updates and worm simulation results), as well as in the KW-dual form. The preliminary numerical findings presented there indicate that in some coupling regions the worm update is inefficient despite fine tuning the
worm amplitude parameter. In these cases switching to a completely KW-dual form is a good choice for efficient simulations, and the absence of constraints in the KW-dual form might even allow for using strategies such as Swendsen–Wang type algorithms.

\chapter{Dual representation for SU(2) lattice gauge theory with staggered fermions \label{cha:su2}}

As we outlined in Sec.~\ref{sec:teo_sign}, QCD at finite quark density is not amenable to lattice Monte Carlo simulations because of the sign problem. One of the tactics that has been used to circumvent this problem is to consider non-abelian gauge theory with fermions, in which the color group is SU(2) rather than SU(3). The obvious advantage of studying this theory, known as \textit{two color QCD} or \textit{QC$_{2}$D}, is the absence of the complex action problem, due to the pseudo-reality of SU(2). Moreover, QC$_2$D is of particular interest since it shares most of the salient features of QCD, such as confinement and dynamical chiral symmetry breaking.
Consequently, the phase structure of QC$_2$D has been extensively studied using lattice methods \cite{NAKAMURA1984391, DAGOTTO1986421, Kogut:2001na, Hands:2006ve, Braguta:2016cpw}. 

Our motivation for studying the SU(2) lattice gauge theory lies on a slightly different ground with respect to the studies we just mentioned, but it shares the same final goal of studying the phase diagram of lattice QCD from first principle calculations. 
Our aim was to develop a dual approach that would enable us to reformulate non-abelian gauge theories in terms of new dual variables, and SU(2) looked like the perfect playground to create and refine our techniques. 

Duality transformations are important tools for understanding quantum field theories, since often physical phenomena are better described by changing the representation. In particular, when using numerical simulations in the framework of lattice field theories, novel representations may give rise to new simulation strategies which allow one to explore parameter regions that were not accessible before (see, e.g., the reviews on worldline and dual representations in lattice field theories \cite{Chandrasekharan:2008gp,deForcrand:2010ys,Gattringer:2014nxa}).
In the past the dualization of non-abelian lattice field theories has often involved the use of strong coupling expansion techniques based on the character expansion \cite{DROUFFE1978133,Savit:1979ny}. In the specific case of SU(2), exact dualizations were found, e.g., in 3 dimensions in terms of $6j$ symbols \cite{PhysRevLett.65.813,Anishetty:1992xa} or in the framework of spin foams \cite{Oeckl:2000hs,Cherrington:2007ax,Cherrington:2009ak,Cherrington:2009am}.

In this chapter we present the \textit{abelian color cycle} dualization method, using four-dimensional SU(2) lattice gauge theory as a first example of application. We then show how the generalization of this approach to the theory with fermions leads us to the development of the \textit{abelian color flux} (ACF) method. Both these methods are versatile, as shown by the fact that we already applied the ACF approach to the SU(2) principal chiral model in the previous chapter. Nonetheless, the technicalities and the mathematical tools needed vary, depending on the theory one focuses on.

\section{Abelian color cycle dualization of the pure SU(2) gauge theory \label{sec:su2_acc}}

The Wilson action for SU(2) lattice gauge theory reads:
\begin{equation}
S_G[U] \; = \; -\dfrac{\beta}{2} \sum_{x,\mu < \nu} 
\Tr U_{x,\mu} \; U_{x+\hat{\mu},\nu} \, U_{x+\hat{\nu},\mu}^{\dagger} \, U_{x,\nu}^\dagger \, ,
\label{eq:su2_gaugeaction}
\end{equation}
where $U_{x,\mu} \in$ SU(2) are the dynamical degrees of freedom, which live on the links $(x,\mu)$ of a four-dimensional lattice, where we impose periodic boundary conditions. The partition sum $Z$ is then given by
\begin{equation}
\label{eq:su2_partition}
Z \, = \, \int \! D[U] \, e^{-S_{G}[U]} \, = \, \prod_{x,\mu} \int_{SU(2)} \! dU_{x,\mu} \, e^{-S_{G}[U]}\, ,
\end{equation}
where in the second step we have explicitly written the product over the links $(x,\mu)$ of the SU(2) Haar measures $\int \! D[U] = \prod_{x,\mu} \int_{SU(2)} dU_{x,\mu}$.
As in the previous chapter, the explicit parametrization of the SU(2) matrices we adopt is
\begin{equation}
\label{eq:su2_parametrization}
U_{x,\mu} = \left[
\begin{array}{cc}
\cos \theta_{x,\mu} \, e^{\,i \alpha_{x,\mu}} & \sin \theta_{x,\mu} \, e^{\, i \beta_{x,\mu}} \\
-\sin \theta_{x,\mu} \, e^{-i \beta_{x,\mu}} & \cos \theta_{x,\mu} \, e^{-i \alpha_{x,\mu}}
\end{array}\right]	\, , 
\end{equation} 
with $\theta_{x,\mu} \in [0, \pi/2]$, $\alpha_{x,\mu}, \beta_{x,\mu} \in [-\pi, \pi]$ and Haar measure given by
\begin{equation}
\label{eq:su2_haar}
dU_{x,\mu} = 2 \sin \theta_{x,\mu} \cos \theta_{x,\mu} \, d\theta_{x,\mu} \dfrac{d\alpha_{x,\mu}}{2 \pi} \, \dfrac{d\beta_{x,\mu}}{2 \pi} \, .
\end{equation} 

The crucial step to dualize this theory consists in decomposing the action into its minimal terms, which we refer to as \textit{"Abelian Color Cycles"} (ACCs). This decomposition is carried out by writing explicitly both the trace and the matrix multiplications in the expression (\ref{eq:su2_gaugeaction}) of the gauge action:
\begin{equation}
\label{eq:su2_accaction}
S_G[U] \; = \; -\dfrac{\beta}{2} \sum_{x,\mu < \nu} \sum_{a,b,c,d=1}^{2} 
U_{x,\mu}^{ab} U_{x+\hat{\mu},\nu}^{bc} U_{x+\hat{\nu},\mu}^{dc \ \star} U_{x,\nu}^{ad \ \star}.
\end{equation} 
The products of the four link elements $U_{x,\mu}^{ab} U_{x+\hat{\mu},\nu}^{bc} U_{x+\hat{\nu},\mu}^{dc \, \star} U_{x,\nu}^{ad \, \star} \in \mathbb{C}$ are the ACCs. They are complex numbers and, as such, abelian. We will see later that this is a key property, because it enables the factorization and the expansion of the Boltzmann weight $e^{-S_{G}[U]}$, and thus it allows us to proceed with the dualization of the partition sum $Z$ as in the well-known case of abelian theories (in Sec.~\ref{sec:teo_u1} we discussed the example of dualizing U(1) pure gauge theory). 

Before doing so, let us spend a few words describing the geometrical interpretation we give to the ACCs on the lattice. First of all, notice that each abelian color cycle $U_{x,\mu}^{ab} U_{x+\hat{\mu},\nu}^{bc} U_{x+\hat{\nu},\mu}^{dc \, \star} U_{x,\nu}^{ad \, \star}$ carries two sets of indicies: the usual space-time position indices $(x,\mu\nu)$, and the color indices $a,b,c,d=1,2$. Clearly, $(x,\mu\nu)$ $\mu<\nu$ denotes the space-time position of the plaquette on which the ACC sits. The color indicies $a$, $b$, $c$ and $d$ come from the labeling of the four matrix entries that constitute the ACC (compare Eq.~(\ref{eq:su2_accaction})). Each of these four indices labels a corner of the plaquette $(x,\mu\nu)$, starting with $a$ at site $x$ and continuing in the mathematically positive direction for the other three color labels. If we now imagine the lattice to have two layers, each layer corresponding to a different color, then the four-tuple $(a,b,c,d)$ specifies the color path described by the ACC around the plaquette $(x,\mu\nu)$. 

\begin{figure}
	\begin{center}
		\includegraphics[scale=0.6,clip]{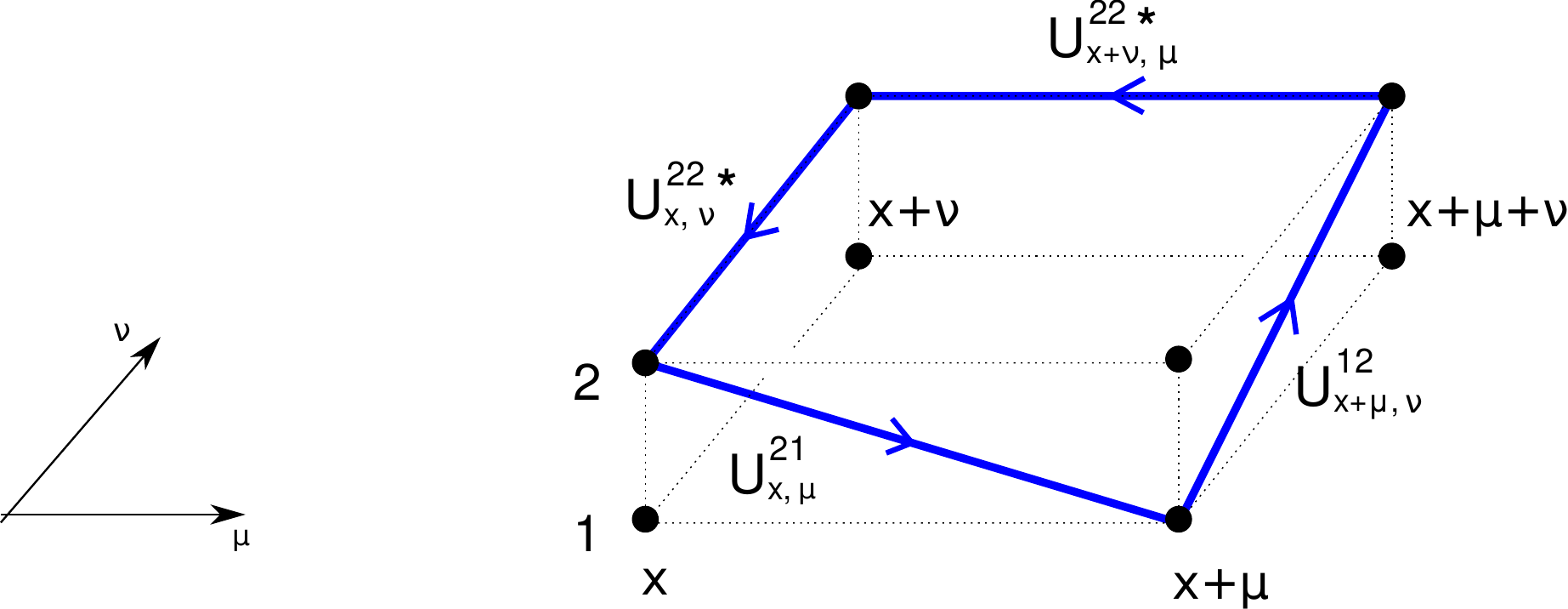}
	\end{center}
	\caption{Geometrical representation of an abelian color cycle (ACC) in the $\mu$--$\nu$ plane for the example $U_{x,\mu}^{21} U_{x+\hat{\mu},\nu}^{12} U_{x+\hat{\nu},\mu}^{22 \, \star} U_{x,\nu}^{22 \, \star}$. Each link element $U_{x,\mu}^{ab}$ is represented as an arrow connecting color $a$ at site $x$ to color $b$ at site $x+\hat{\mu}$ and we use two layers of the lattice to represent the two colors. The ACC shown here corresponds to the cycle occupation number $p_{x,\mu\nu}^{2122}\in\mathbb{C}$. \label{fig:su2_acc}}		
\end{figure}

For a better understanding of this interpretation, in Fig.~\ref{fig:su2_acc} we show the abelian color cycle $U_{x,\mu}^{21} U_{x+\hat{\mu},\nu}^{12} U_{x+\hat{\nu},\mu}^{22 \, \star} U_{x,\nu}^{22 \, \star}$ as an example.
The two colors degrees of freedom correspond to the two layers of the lattice which, in the figure, we sketch in light grey as two copies of the plaquette we consider. 
The first link element of the $2122$ ACC is $U_{x,\mu}^{21}$. It is represented with a positively oriented arrow and it connects color $2$ on site $x$ to color $1$ on site $x+\hat{\mu}$. The next factor $U_{x+\hat{\mu},\nu}^{12}$ subsequently connects color $1$ on site $x+\hat{\mu}$ to color $2$ on site $x+\hat{\mu} + \hat{\nu}$, also with a  positively oriented arrow. Then, since in our representation complex conjugation corresponds to a negative orientation of the arrow, the last two link elements $U_{x+\hat{\nu},\mu}^{22 \, \star} U_{x,\nu}^{22 \, \star}$ close the ACC around the plaquette. 

\begin{figure}
	\begin{center}
		\includegraphics[scale=1.10,clip]{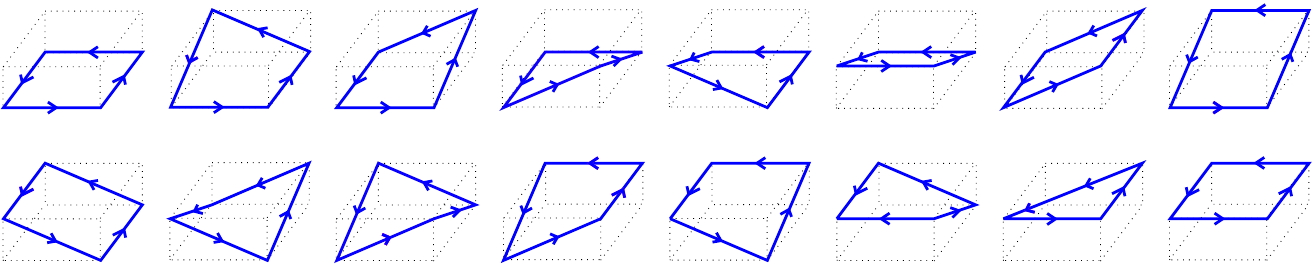}
	\end{center}
	\caption{The 16 possible abelian color cycles which are attached to a given plaquette. In the dual representation their occupation is given by the corresponding cycle occupation number $p_{x,\mu\nu}^{abcd} \in \mathds{N}_0$.\label{fig:su2_allcycles}}		
\end{figure}

Summarizing, we interpret the ACCs $U_{x,\mu}^{ab} U_{x+\hat{\mu},\nu}^{bc} U_{x+\hat{\nu},\mu}^{dc \ \star} U_{x,\nu}^{ad \ \star}$ as paths in color space closing around plaquettes. The path in color is described by the labels $(a,b,c,d)$, one for every corner of the plaquette $(x,\mu\nu)$. This results in a total of $2^4 = 16$ possible ACCs for every plaquette of the lattice. We show all of these possibilities in Fig.~\ref{fig:su2_allcycles}. Notice that, in the case of SU(2), only ACCs with mathematically positive orientations are needed. This fact can be traced back to the pseudo-reality of SU(2), that guaranties the reality of the traces of SU(2) matrices.  

We may now continue with the dualization of the partition sum (\ref{eq:su2_partition}):
\begin{align}
	Z  & =  \int \! D[U] \, e^{-S_G[U]} \; = \; \int \! D[U] \prod_{x,\mu<\nu} \prod_{a,b,c,d = 1}^{2} 
	e^{\frac{\beta}{2} U_{x,\mu}^{ab} U_{x+\hat{\mu},\nu}^{bc} U_{x+\hat{\nu},\mu}^{dc \ \star} U_{x,\nu}^{ad \ \star}} 
	\nonumber \\
	& = \int \! D[U] \prod_{x,\mu<\nu} \prod_{a,b,c,d = 1}^{2} \sum_{p_{x,\mu\nu}^{abcd} = 0}^{\infty} 
	\dfrac{ \left( \beta/2 \right)^{p_{x,\mu\nu}^{abcd}}}{p_{x,\mu\nu}^{abcd}\, !} 
	\left( U_{x,\mu}^{ab} U_{x+\hat{\mu},\nu}^{bc} U_{x+\hat{\nu},\mu}^{dc \ \star} 
	U_{x,\nu}^{ad \ \star} \right)^{p_{x,\mu\nu}^{abcd}} \, .
	\label{eq:su2_partition2}
\end{align}
In the first line we rewrote the sums in the exponent as a product over all ACCs and all plaquettes of the lattice with the local exponentials $e^{\frac{\beta}{2} U_{x,\mu}^{ab} U_{x+\hat{\mu},\nu}^{bc} U_{x+\hat{\nu},\mu}^{dc \ \star} U_{x,\nu}^{ad \ \star}} $ as factors. We stress once more that this factorization is possible thanks to the ACC decomposition (\ref{eq:su2_accaction}), in which every term of the gauge action is a complex number and therefore commutes. In the second step we then expanded each local factor in a Taylor series, thus introducing the expansion coefficients $p_{x,\mu\nu}^{abcd} \in \mathbb{N}_{0}$, which will turn out to be our dual variables. Since each expansion coefficient $p_{x,\mu\nu}^{abcd} \in \mathbb{N}_{0}$ is related to one ACC $U_{x,\mu}^{ab} U_{x+\hat{\mu},\nu}^{bc} U_{x+\hat{\nu},\mu}^{dc \ \star} 
U_{x,\nu}^{ad \ \star}$, we refer to them as \textit{cycle occupation numbers}. 

We now rearrange the product in the last line of Eq.~(\ref{eq:su2_partition2}) so that all factors of $U_{x,\mu}^{ab}$ and $U_{x,\mu}^{ab \, \star}$ associated with a given link element are grouped together:
\begin{equation}
	Z  = \sum_{\{p\}} \left[ \prod_{x,\mu<\nu} \prod_{a,b,c,d}  
	\dfrac{ \left(\beta/2 \right)^{p_{x,\mu\nu}^{abcd}}}{p_{x,\mu\nu}^{abcd}\, !} \right] 
	\prod_{x,\mu} \int \! \! dU_{x,\mu} \; \prod_{a,b} \left( U_{x,\mu}^{ab} \right) ^{N_{x,\mu}^{ab}} 
	\left( U_{x,\mu}^{ab \ \star} \right) ^{\overline{N}_{x,\mu}^{ab}} \, .
\label{eq:su2_partition3}
\end{equation} 
Written in this form the partition function is suitable for the Haar integration, as we shall discuss later. 
In (\ref{eq:su2_partition3}) we use the short-hand notation 
\begin{equation}
	\sum_{\{p\}} \; = \; \prod_{x,\mu<\nu} \prod_{a,b,c,d = 1}^{2} \sum_{p_{x,\mu\nu}^{abcd} = 0}^{\infty}
\label{eq:su2_cyclesum}
\end{equation}
to denote the sum over all the configurations of the cycle occupation numbers $p_{x,\mu\nu}^{abcd} \in \mathbb{N}_{0}$. Moreover, we introduced $N_{x,\mu}^{ab}\in \mathbb{N}_{0}$ and $\overline{N}_{x,\mu}^{ab}\in \mathbb{N}_{0}$ for the exponents of the link variable elements $U_{x,\mu}^{ab}$ and $U_{x,\mu}^{ab \ \star}$ respectively. Explicitly they are given by the following linear combinations of cycle occupation numbers:
\begin{equation}
	N_{x,\mu}^{ab} \; = \; 
	\sum_{\nu:\mu<\nu}p_{x,\mu\nu}^{abss} + \sum_{\rho:\mu>\rho} p_{x-\hat{\rho},\rho\mu}^{sabs}\, , \qquad \overline{N}_{x,\mu}^{ab} \; = \; 
	\sum_{\nu:\mu<\nu}p_{x-\hat{\nu},\mu\nu}^{ssba} + \sum_{\rho:\mu>\rho}p_{x,\rho\mu}^{assb} \, ,
\label{eq:su2_nnbar}
\end{equation}
where the label $s$ stands for the independent summation of the color indices replaced by it, e.g., $p_{x,\mu\nu}^{abss} = \sum_{c,d =1}^{2} p_{x,\mu\nu}^{abcd}$ or $p_{x,\mu\nu}^{sabs} = \sum_{c,d =1}^{2} p_{x,\mu\nu}^{cabd}$.

\begin{figure}
	\begin{center}
		\includegraphics[scale=0.5,clip]{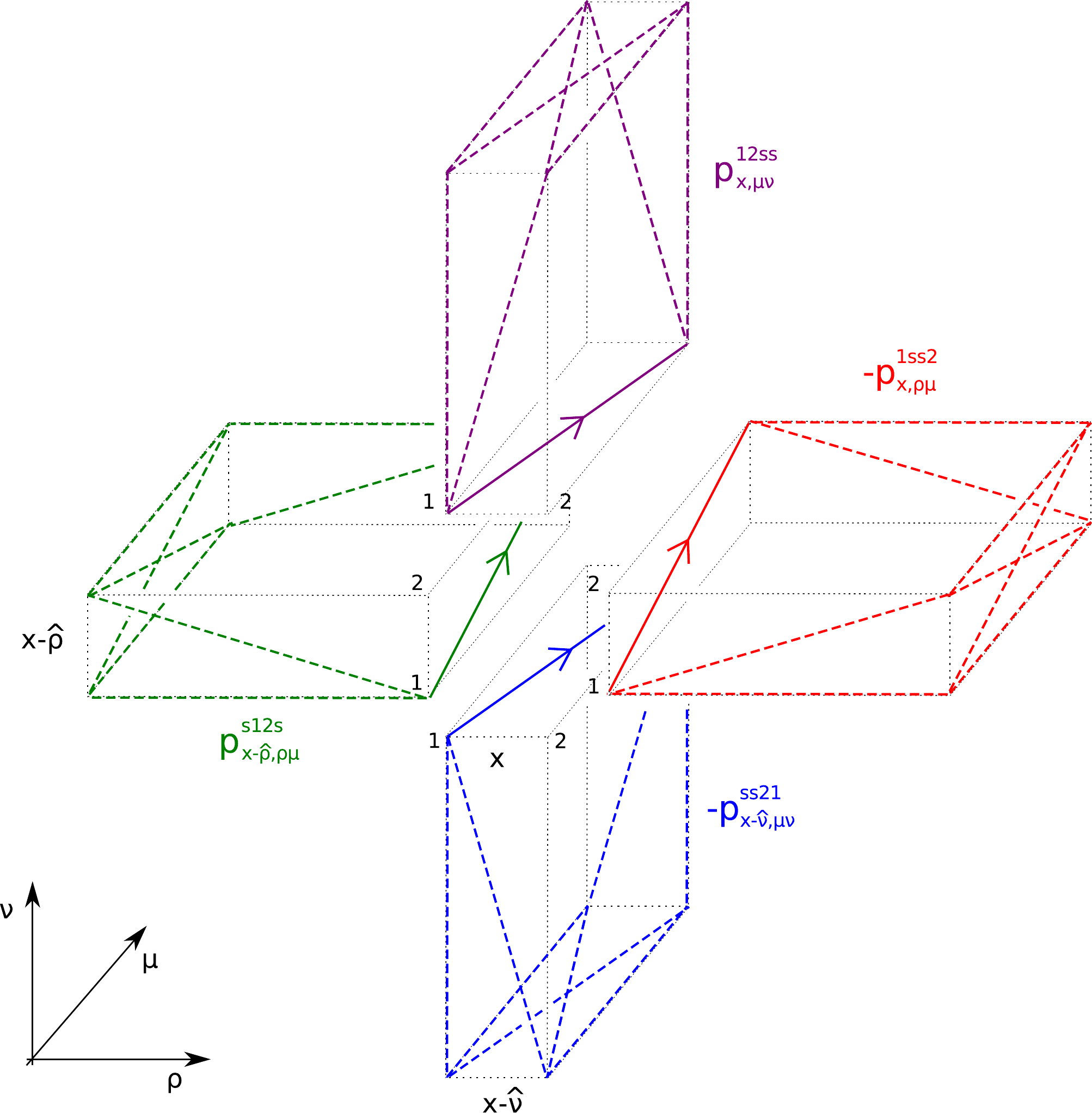}
	\end{center}
	\caption{Graphical illustration of the contributions from the cycle occupation numbers to the $J$-flux using the example of the $J_{x,\mu}^{12}$ element. For a description of the figure see the text.\label{fig:su2_j12}}		
\end{figure}

At this point we substitute the parametrization (\ref{eq:su2_parametrization}) of the link elements and the expression (\ref{eq:su2_haar}) of the Haar measure in Eq.~(\ref{eq:su2_partition3}) and obtain:
\begin{align}
Z \; = &\; \sum_{\{p\}} \! \left[ \prod_{x,\mu<\nu} \prod_{a,b,c,d} 
\dfrac{ \left( \beta/2 \right)^{p_{x,\mu\nu}^{abcd}}}{p_{x,\mu\nu}^{abcd}!} \right] \nonumber
\\
& 
\times \prod_{x,\mu} (-1)^{J_{x,\mu}^{21}} \; \,
2 \! \int_{0}^{\pi/2} \!\!\!\!  d\theta_{x,\mu} \, (\cos\theta_{x,\mu})^{1 + S_{x,\mu}^{11} + S_{x,\mu}^{22}} 
\; (\sin\theta_{x,\mu})^{1 + S_{x,\mu}^{12} + S_{x,\mu}^{21}} 
\nonumber
\\
& \hspace{20mm} \times \; 
\int_{0}^{2\pi} \! \dfrac{d\alpha_{x,\mu}}{2\pi} \; e^{i\alpha_{x,\mu}[J_{x,\mu}^{11}-J_{x,\mu}^{22}]} \; 
\int_{0}^{2\pi} \! \dfrac{d\beta_{x,\mu}}{2\pi} \; e^{i\beta_{x,\mu}[J_{x,\mu}^{12}-J_{x,\mu}^{21}]} 	\; .
\label{eq:su2_partition4}
\end{align} 
For a convenient notation we introduced the integer valued fluxes $J_{x,\mu}^{ab}\in \mathbb{Z}$ and $S_{x,\mu}^{ab}\in \mathbb{N}_{0}$,
\begin{equation}
\label{eq:su2_jfluxes}
J_{x,\mu}^{ab} \; = \; N_{x,\mu}^{ab} - \overline{N}_{x,\mu}^{ab} \; = \; 
\sum_{\nu:\mu<\nu}[\, p_{x,\mu\nu}^{abss} - p_{x-\hat{\nu},\mu\nu}^{ssba} \, ] - 
\sum_{\rho:\mu>\rho}[\, p_{x,\rho\mu}^{assb} - p_{x-\hat{\rho},\rho\mu}^{sabs} \, ] \; ,	
\end{equation}
\begin{equation}
\label{eq:su2_sfluxes}
S_{x,\mu}^{ab} \; = \; N_{x,\mu}^{ab} + \overline{N}_{x,\mu}^{ab} \; = \; 
\sum_{\nu:\mu<\nu}[\, p_{x,\mu\nu}^{abss} + p_{x-\hat{\nu},\mu\nu}^{ssba}\, ] + 
\sum_{\rho:\mu>\rho}[\, p_{x,\rho\mu}^{assb} + p_{x-\hat{\rho},\rho\mu}^{sabs}\, ] \; .
\end{equation}	
It is now easy to see that the link integrals in Eq.~(\ref{eq:su2_partition4}) can be solved in closed form. The integrals over the phases $\alpha_{x,\mu}$ and $\beta_{x,\mu}$ give rise to Kronecker deltas that impose constraints over the color link fluxes $J_{x,\mu}^{ab}$ we have just introduced. In particular, the constraints enforce the following equalities between components of the $J$-fluxes at all links $(x,\mu)$:
\begin{equation}
	J_{x,\mu}^{11} \, - \, J_{x,\mu}^{22} \; = \; 0 \quad \forall x,\mu \qquad \text{and} \qquad J_{x,\mu}^{12} \, - \, J_{x,\mu}^{21} \; = \; 0  \quad \forall x,\mu \, .
\label{eq:su2_constraints}
\end{equation}
These constraints can be easily understood after we discuss the geometrical interpretation of the currents $J_{x,\mu}^{ab}$. From their definition in Eq.~(\ref{eq:su2_jfluxes}), we know that the $J_{x,\mu}^{ab}$ sum over every cycle occupation number that contributes to the flux from color $a$ on site $x$ to color $b$ on site $x+\hat{\mu}$. In Fig.~\ref{fig:su2_j12} we show four of the plaquettes attached to the link $(x,\mu)$ and illustrate how they contribute to $J_{x,\mu}^{12}$ as an example. On the link $(x,\mu)$ the flux from color 1 to 2 is kept fixed and represented with solid arrows. For every plaquette attached to the link this flux gets contributions from four different cycle occupation numbers, which are summed over in the definition (\ref{eq:su2_jfluxes}), and illustrated with dotted lines in the figure.
Hence, $J_{x,\mu}^{ab}$ is the total flux from color $a$ on site $x$ to color $b$ on site $x+\hat{\mu}$.
\begin{figure}
	\begin{center}
		\includegraphics[scale=0.8,clip]{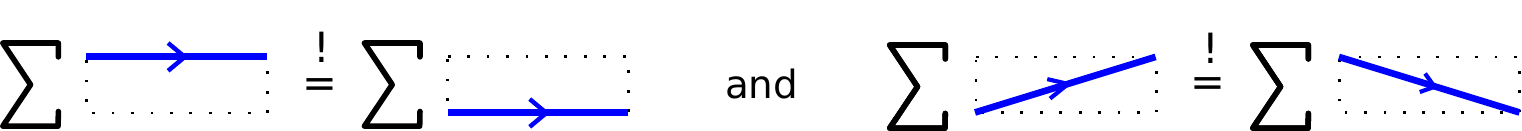}
	\end{center}
	\caption{Geometrical illustration of the two constraints in Eq.~\protect(\ref{eq:su2_constraints}) for the fluxes $J_{x,\mu}^{ab}$ on all links $(x,\mu)$. The first constraint (lhs.~plot) requires the sum over all 1-1 fluxes to equal the sum over all 2-2	fluxes. The second constraint requires the sum over 1-2 fluxes to equal the sum over 2-1 fluxes.\label{fig:su2_constraints}}		
\end{figure}

Now that we have given the interpretation of the $J_{x,\mu}^{ab}$ fluxes, it is straightforward to understand the meaning of the constraints in Eq.~(\ref{eq:su2_constraints}): for every link of the lattice the total fluxes on the two color layers have to be equal, and the total fluxes between the two layers have to match. We represent these conditions schematically in Fig.~\ref{fig:su2_constraints}. 

The constraints (\ref{eq:su2_constraints}) can also be used to simplify the result of the last link integral in Eq.~(\ref{eq:su2_partition4}). The integrals over the $\theta_{x,\mu}$ angles have the form of the well-known integral representation of the beta function $\B$
\begin{align}
2 \int_{0}^{\frac{\pi}{2}} \! d\theta \, (\cos \theta)^{1 + n}	\,(\sin \theta)^{1 + m} \, &= \, \B	\bigg(\frac{n}{2} + 1 \bigg| \frac{m}{2} + 1 \bigg) = \dfrac{\Gamma\big(\frac{n}{2} + 1 \big) \Gamma \big( \frac{m}{2} + 1 \big) }{\Gamma\big(\frac{n + m}{2} + 2 \big)} \nonumber \\
&= \, \dfrac{\frac{n}{2}! \frac{m}{2}!}{\big(\frac{n + m}{2} + 1 \big)!} \, .
\label{eq:su2_betafunction}
\end{align}
In the second step we represent the beta function $\B$ in terms of gamma functions $\Gamma$, while the last simplified expression in terms of factorials holds only if $n$ and $m$ are even. In our case $n$ and $m$ are given by the sums 
\begin{equation}
S_{x,\mu}^{11} \, + \, S_{x,\mu}^{22} \; = \; \text{even} \quad \forall x,\mu \qquad \text{and} \qquad S_{x,\mu}^{12} \, + \, S_{x,\mu}^{21} \; = \; \text{even}  \quad \forall x,\mu \, 
\label{eq:su2_even}
\end{equation}
respectively. It is easy to prove the evenness of these combinations using the constraints (\ref{eq:su2_constraints}) and recalling the definitions (\ref{eq:su2_jfluxes}) and (\ref{eq:su2_sfluxes}) for the $J$- and $S$-fluxes: the expressions on the left hand sides of the equations in (\ref{eq:su2_even}) can be obtained from the expressions on the left hand sides of (\ref{eq:su2_constraints}) by adding even quantities on both sides of the equations. Therefore, the integrals over the $\theta_{x,\mu}$ angles in Eq.~(\ref{eq:su2_partition4}) result in the simple combinatorial factors
\begin{equation}
W_{H}[p] \; = \; 
\prod_{x,\mu} \dfrac{\left(\frac{S_{x,\mu}^{11} + S_{x,\mu}^{22}}{2}\right)! 
	\left(\frac{S_{x,\mu}^{12} + S_{x,\mu}^{21}}{2}\right)!}{\left(\frac{S_{x,\mu}^{11} + S_{x,\mu}^{22} + S_{x,\mu}^{12} + S_{x,\mu}^{21}}{2} + 1\right)!} \; .
\label{eq:su2_wh}	
\end{equation}
We then introduce the $\beta$-dependent weight factor $W_{\beta}[p]$, which collects the powers and factorials from the expansion of the exponentials
\begin{equation}
\label{eq:su2_wbeta}	
W_{\beta}[p] \; = \; 
\prod_{x,\mu<\nu} \prod_{a,b,c,d}  \dfrac{ \left( \frac{\beta}{2} \right)^{p_{x,\mu\nu}^{abcd}}}{p_{x,\mu\nu}^{abcd}!}  \; .
\end{equation}
This weight organizes the partition function in the form of a power series of the inverse lattice coupling $\beta$, i.e., in a strong coupling expansion.

\begin{figure}
	\begin{center}
		\includegraphics[scale=0.5,clip]{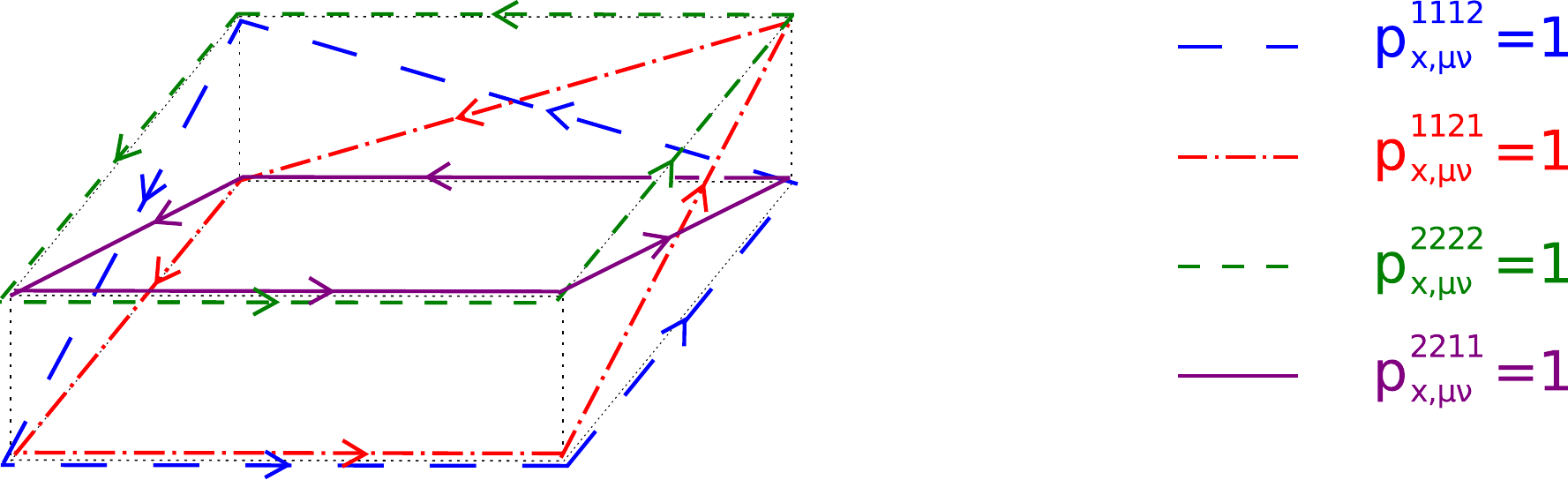}
	\end{center}
	\caption{Example of a configuration that contributes to the partition function with a negative sign. We show the four cycle occupation numbers that are set to 1. Obviously all constraints are obeyed. The configuration has three flux crossings and thus a negative sign. \label{fig:su2_unhappy}}		
\end{figure}

Putting things together we obtain for the partition sum
\begin{equation}
Z \; = \; \sum_{\{p\}} W_{\beta}[p] \; W_H[p] \; (-1)^{\sum_{x,\mu}J_{x,\mu}^{21}} \; 
\prod_{x,\mu}  \delta(J_{x,\mu}^{11}-J_{x,\mu}^{22}) \; \delta(J_{x,\mu}^{12}-J_{x,\mu}^{21}) \; .
\label{eq:su2_dualgaugepartition}
\end{equation}
In its dual form (\ref{eq:su2_dualgaugepartition}) the partition function is a sum over configurations of cycle occupation numbers $p_{x,\mu\nu}^{abcd} \in \mathbb{N}_{0}$ attached to the plaquettes $(x, \mu < \nu)$ of the lattice. At each link $(x,\mu)$ the $p_{x,\mu\nu}^{abcd}$ have to  obey constraints which are expressed in terms of the two Kronecker deltas (we use the notation $\delta(n) \equiv \delta_{n,0}$) in (\ref{eq:su2_dualgaugepartition}) which relate different components of the link fluxes $J_{x,\mu}^{ab}$, given in Eq.~(\ref{eq:su2_jfluxes}). Each configuration comes with two weight factors $W_{\beta}[p]$ and $W_{H}[p]$ that collect the coefficients of the Taylor expansion and the combinatorial factors from the Haar measure integrals respectively. Both these weight factors are real and positive (compare the explicit expressions (\ref{eq:su2_wh}) and (\ref{eq:su2_wbeta})). Nonetheless, the partition sum (\ref{eq:su2_dualgaugepartition}) also contains the explicit sign factor $(-1)^{\sum_{x,\mu}J_{x,\mu}^{21}}$. This minus sign origins from the minus sign in 
the (2,1) matrix element in the parametrization (\ref{eq:su2_parametrization}) of our SU(2) link variables. Using the constraint in Eq.~(\ref{eq:su2_dualgaugepartition}) we can give a simple interpretation of the sign factor: since by the constraints (\ref{eq:su2_constraints}) the $J_{x,\mu}^{21}$ flux equals the $J_{x,\mu}^{12}$ flux, the configurations that contribute to the partition function with a negative sign are the ones that have an odd number of flux crossings:
\begin{equation*}
	(-1)^{\sum_{x,\mu} J_{x,\mu}^{12}} = (-1)^{\text{\# of flux crossings}} \, .
\end{equation*}
An example of a configuration with an odd number of crossing is given by
\begin{equation*}
	p_{x,\mu\nu}^{1112} = 1 \, , \qquad p_{x,\mu\nu}^{1121} = 1 \, , \qquad 
	p_{x,\mu\nu}^{2222} = 1 \, , \qquad p_{x,\mu\nu}^{2211} = 1 \, .
\end{equation*}
We illustrate this contribution in Fig.~\ref{fig:su2_unhappy}. 
Obviously this configuration is admissible since it fulfils both the constraints in (\ref{eq:su2_constraints}) on each link of the plaquette but, since it has an odd number of crossings, it contributes to the partition function with a negative sign. This configuration is one of the lowest negative sign configurations which one can construct. In fact, the first negative configurations appear at order $\beta^{4}$. At this order negative configurations are local, i.e., they are built from four non-zero cycle occupation numbers on the same plaquette.

A large class of admissible dual pure gauge configurations contributing to the partition sum with a positive weight are closed orientable and non-orientable surfaces of plaquettes where certain cycle occupation numbers are occupied. While the examples of negative configurations we found are local, the surfaces are non-local and contribute to the long distance properties of the theory. It is an interesting open question if all the configurations that contribute at long distance are positive.

\section{Abelian color flux dualization of fermions \label{sec:su2_acf}}
Having discussed the pure gauge part of the theory, we now focus on fermions. We consider one flavor of staggered fermions described by the action
\begin{equation}
	S_F[\,\overline{\psi}, \psi,U]   =   \sum_x \Big[ m \, \overline{\psi}_x \psi_x + \sum_{\mu} \frac{\gamma_{x,\mu}}{2}
	\Big( \, \overline{\psi}_x U_{x,\mu} \psi_{x + \hat\mu} - \overline{\psi}_{x+\hat\mu} U_{x,\mu}^\dagger \psi_{x} \Big) \Big]  \, ,
\label{eq:su2_actionf}
\end{equation}
where $m$ is the mass and $\gamma_{x,\mu}$ is the usual staggered sign factor defined as
\begin{equation*}
	\gamma_{x,1} = 1 \, , \ \quad \gamma_{x,2} = (-1)^{x_{1}}  \, , \
	\quad \gamma_{x,3} = (-1)^{x_{1} + x_{2}}  \, , \ \quad \gamma_{x,4} = (-1)^{x_{1} + x_{2} + x_{3}}\, .
\end{equation*}
$\psi_{x}$ and $\overline{\psi}_{x}$ are two-component Grassmann vectors in color space
\begin{equation}
\psi_{x} = \left(
\begin{array}{c}
\psi_{x}^{1} \\
\psi_{x}^{2} \\
\end{array} \right) \ ,\qquad
\overline{\psi}_{x} = \left( \overline{\psi}_{x}^{1} \ , \ \overline{\psi}_{x}^{2} \right) \, ,
\label{eq:su2_grassmann}
\end{equation}
that live on the sites $x$ of our four-dimensional $N^{3} \times N_{t}$ lattice, with anti-periodic boundary conditions in Euclidean time ($\nu = 4$) for the fermions and periodic in the spatial directions ($\nu=1,2,3$).
The fermionic partition function in a gauge background is given by
\begin{equation}
Z_F[U] \; = \; \int \!\! D[\, \overline{\psi}, \psi] \; e^{-S_F[\,\overline{\psi}, \psi,U]} \; ,
\label{eq:su2_partitionf}
\end{equation}
where the measure is a product over Grassmann measures, 
$D[\, \overline{\psi}, \psi] = \prod_{x,a} d \psi^a_x d \overline{\psi}^a_x$. The full partition sum $Z$ is then obtained as 
$Z = \int D[U]  e^{-S_G[U]} Z_F[U]$.

The key step in the \textit{abelian color flux} (ACF) dualization of the fermionic partition sum (\ref{eq:su2_partitionf}) is the decomposition of the staggered action (\ref{eq:su2_actionf}) into Grassmann bilinears. This is done by making the sums over the color indices explicit:  
\begin{equation}
S_F[\,\overline{\psi}, \psi,U]  = \sum_x \Big[ \, m \!\sum_{a} 
\overline{\psi}_x^{\,a} \psi_x^{a} + \sum_{\mu} \frac{\gamma_{x,\mu}}{2} \sum_{a,b}
\Big( \, \overline{\psi}^{\,a}_x U_{x,\mu}^{\,ab} 
\psi_{x + \hat\mu}^{b} - \overline{\psi}_{x+\hat\mu}^{\,b} U_{x,\mu}^{\, ab \, \star} \psi_{x}^{a} 
\Big) \Big] \; .
\label{eq:su2_acfactionf}
\end{equation}
As a result, every term in the decomposition (\ref{eq:su2_acfactionf}) of the staggered action commutes and, again, this is what enables us to proceed with the dualization. We find for the fermionic partition function
\begin{align}
Z_F[U] \, &= \int \!\!\! D[\, \overline{\psi}, \psi]  \prod_x \prod_a e^{-m \overline{\psi}_x^{\,a} \psi_x^a} \; 
\prod_{x,\mu} \prod_{a,b} e^{- \,\frac{\gamma_{x,\mu}}{2} \overline{\psi}^{\,a}_x U_{x,\mu}^{\,ab} \psi_{x + \hat\mu}^b} \;
e^{ \, \frac{\gamma_{x,\mu}}{2} \overline{\psi}_{x+\hat\mu}^{\,b} U_{x,\mu}^{\, ab\, \star} \psi_{x}^a} 
\nonumber \\ 
&= \int \!\!\!D[\, \overline{\psi}, \psi]  \prod_x \prod_a \sum_{s_x^a = \, 0}^{1} \! 
( -m \overline{\psi}_x^{\,a} \psi_x^a )^{s_x^a} 
 \nonumber \\
& \qquad \qquad \times
\prod_{x,\mu} \prod_{a,b} 
\sum_{k_{x,\mu}^{\;ab} = \, 0 }^{1} \!\!\! 
\Big( \!- \frac{\gamma_{x,\mu}}{2} \, \overline{\psi}^{\,a}_x U_{x,\mu}^{\,ab} \psi_{x + \hat\mu}^b \Big)^{k_{x,\mu}^{\; ab}} \!
\sum_{ \overline{k}_{x,\mu}^{\;ab} = \, 0}^{1}\!\!\! 
\Big( \frac{\gamma_{x,\mu}}{2} \, \overline{\psi}_{x+\hat\mu}^{\,b} U_{x,\mu}^{\, ab\, \star} 
\psi_{x}^a \Big)^{\overline{k}_{x,\mu}^{\; ab}}
\nonumber \\
&= \, \frac{1}{2^{2V}} \! \sum_{\{s,k,\overline{k}\}} \!\! (2m)^{\, \sum_{x,a} s_x^a} \; \prod_{x,\mu} \prod_{a,b} 
( U_{x,\mu}^{\,ab} )^{k_{x,\mu}^{\; ab}} \, 
( U_{x,\mu}^{\, ab\, \star} )^{\overline{k}_{x,\mu}^{\; ab}}  \; (-1)^{k_{x,\mu}^{\; ab}} \; 
( \, \gamma_{x,\mu}\,) ^{k_{x,\mu}^{\; ab} + \overline{k}_{x,\mu}^{\; ab} } 
\nonumber \\
& \qquad \qquad \times \!\!
\int \!\! D[\, \overline{\psi}, \psi] \! \prod_x \prod_a  (\overline{\psi}_x^{\,a} \psi_x^a )^{s_x^a} \prod_{x,\mu} \prod_{a,b}
(\overline{\psi}^{\,a}_x \psi_{x + \hat\mu}^b)^{k_{x,\mu}^{\; ab}} 
(\overline{\psi}_{x+\hat\mu}^{\,b} \psi_{x}^a)^{\overline{k}_{x,\mu}^{\; ab}} \; .
\label{eq:su2_partitionf2}
\end{align}
In the first step we have factorized the Boltzmann weight into the local color site factors $e^{-m \overline{\psi}_x^{\,a} \psi_x^a}$, the forward hops $e^{-\frac{\gamma_{x,\mu}}{2} \overline{\psi}_x^{\,a} U_{x,\mu}^{ab} \psi_{x+\hat{\mu}}^b}$ and the backward hops $e^{\frac{\gamma_{x,\mu}}{2} \overline{\psi}_{x+\hat{\mu}}^{\,b} U_{x,\mu}^{ab\star} \psi_{x}^a}$. Afterwards, we expanded each single exponential in a power series which, in the case of fermions, terminates after the linear term due to the nilpotency of the Grassmann variables. The expansion indices, $s_{x}^{a} = 0,1$, $k_{x,\mu}^{ab} = 0,1$ and $\overline{k}_{x,\mu}^{ab} = 0,1$ are the new dual variables for fermions. 
\begin{figure}
	\begin{center}
		\vspace*{0.2cm}
		\includegraphics[scale=0.45,clip]{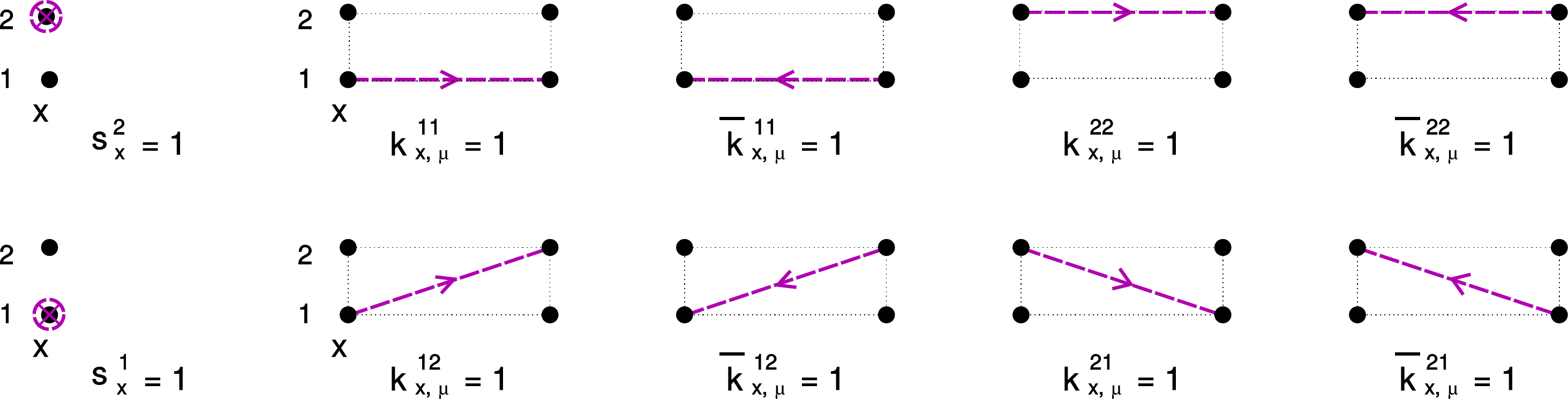}
	\end{center}
	\caption{Graphical representation of the dual variables for the fermions. 
		The first two diagrams on the very left represent the monomers $s_x^a$, 
		the arrows are for the dual variables $k_{x,\mu}^{\,ab}, \overline{k}_{x,\mu}^{\,ab}$. \label{fig:su2_grassmann}}		
\end{figure}
Their graphical representation, shown in Fig.~\ref{fig:su2_grassmann}, is obtained as an extension of the interpretation we gave to the abelian color cycle in the previous section. Whenever one of the dual variables for the fermions is equal to 1, we say that it \textit{activates} the Grassmann bilinear it is associated to. So, for example, $s_{x}^{a}$ activates the color $a$ component of the mass term on site $x$. We refer to these objects as \textit{monomers}, and we represent them with a circle around the corresponding site and color, as done in the two diagrams on the very left of Fig.~\ref{fig:su2_grassmann}. $k_{x,\mu}^{ab}$ is used for the forward hop that connects color $a$ to color $b$ on the link $(x,\mu)$ and, analogously, $\overline{k}_{x,\mu}^{ab}$ represents the backward hop that connects color $b$ to color $a$ on the link $(x,\mu)$. They are illustrated as oriented arrows connecting color layers on neighboring sites. As we will see shortly, these flux variables can be used to build up either \textit{dimers}, by setting $k_{x,\mu}^{ab} = \overline{k}_{x,\mu}^{ab} = 1$, or \textit{loops}, by setting closed chains of $k$ and $\overline{k}$ to 1.

In the last step of Eq.~(\ref{eq:su2_partitionf2}) we reorganized the terms by collecting all the factors that do not depend on the Grassmann variables in front of the Grassmann integral. The Grassmann integral will give as a result either $0$ or $\pm 1$, depending on the values of the fermion dual variables $s_{x}^{a},k_{x,\mu}^{ab},\overline{k}_{x,\mu}^{ab} = 0,1$. Obviously, for the Grassmann integral to be non-vanishing each Grassmann variable $\overline{\psi}_{x}^{a} \psi_{x}^{a}$ has to appear exactly once. This requirement can be enforced by means of a Kronecker delta at every site and color layer:
\begin{equation}
\label{eq:su2_cf}
C_{F}[s,k,\overline{k}] = \prod_{x,a} \delta \Big( 1 - s_{x}^{a} - \dfrac{1}{2} \sum_{\mu,b} \big[k_{x,\mu}^{ab} + \overline{k}_{x,\mu}^{ab} + k_{x - \hat{\mu},\mu}^{ba} + \overline{k}_{x - \hat{\mu},\mu}^{ba} \big]\Big) \, .
\end{equation} 
The way of satisfying this fermion constraint is to completely fill the two layers of our four-dimensional lattice with monomers, dimers and fermionic loops. For this reason, when the conditions imposed by the product of Kronecker deltas (\ref{eq:su2_cf}) are satisfied, we say that the Grassmann integrals are \textit{saturated}. 

We now continue the discussion of (\ref{eq:su2_partitionf2}) focusing on the sign factors. The sources of signs in the case of fermions are various, and therefore a detailed discussion about them is mandatory. Distinctly visible from Eq.~(\ref{eq:su2_partitionf2}) are the negative sign factors associated to the forward hops $(-1)^{k_{x,\mu}^{ab}}$, as well as the staggered sign factors $(\gamma_{x,\mu})^{k_{x,\mu}^{ab} + \overline{k}_{x,\mu}^{ab}}$. Other sources of signs are the anti-periodic boundary conditions along the time direction ($\mu=4$), and the Grassmann integral itself. As we mentioned earlier, the Grassmann integral is non-vanishing only if it is saturated, namely if the whole lattice is filled up with monomers, dimers and loops. It is clear from Eq.~(\ref{eq:su2_partitionf2}) that monomers, which are activated by setting $s_{x}^{a} = 1$, contribute just with a factor of $2m$. Moreover, the corresponding Grassmann variables $\overline{\psi}_{x}^{a} \psi_{x}^{a}$ are already in the canonical order for the Grassmann integration, which means that the corresponding Grassmann integral gives a factor of $+1$. We conclude that monomers are not a source of signs. 
Let us now consider a dimer, activated by setting $k_{x,\mu}^{ab} = \overline{k}_{x,\mu}^{ab} = 1$:
\begin{equation}
\overline{\psi}^{\,a}_x \, \psi_{x + \hat\mu}^b \, \overline{\psi}_{x+\hat\mu}^{\,b}  \, \psi_{x}^a \; = \; 
- \; \overline{\psi}^{\,a}_x  \, \psi_{x}^a \, \overline{\psi}_{x+\hat\mu}^{\,b} \, \psi_{x + \hat\mu}^b \; .
\label{dimer}
\end{equation}
On the right hand side of the equality we reordered the Grassmann variables in the canonical way for the integration. By doing so we pick up a minus sign, which anyway is compensated by the minus sign coming from the forward hop ($(-1)^{k_{x,\mu}^{ab}} = (-1)^{1} = -1$). Moreover, the staggered sign factor is always positive for dimers ($(\gamma_{x,\mu})^{k_{x,\mu}^{ab} + \overline{k}_{x,\mu}^{ab}} = (-1)^{2} = +1$) and, ultimately, since dimers are backtracking loops of length 2, they always have vanishing net winding number. We conclude that dimers are not a source of signs either. 
The discussion about loops is a bit more complicated. Let us start with considering the product of the staggered sign factors along a plaquette: 
\begin{equation}
\sigma_{x,\mu\nu} \; = \; 
\gamma_{x,\mu} \,  \gamma_{x+\hat{\mu},\nu} \, \gamma_{x+\hat{\nu},\mu} \, \gamma_{x,\nu} \; = \; -\, 1 \,.
\end{equation}
This result always holds, independently of the plane $\mu,\nu$ on which the plaquette sits, the position $x$, or the orientation of the loop. If we then consider a loop closing around two adjacent plaquettes, the product of the staggered sing factors along that loop can be equivalently computed by multiplying the plaquette factors $\sigma_{x,\mu\nu}$ corresponding to the two plaquettes bordered by the loop, e.g.:
\begin{align*}
	\sigma_{x,\mu\nu} \, \sigma_{x + \hat{\mu},\mu\nu}\; &\equiv \; 
	\gamma_{x,\mu} \,  \gamma_{x+\hat{\mu},\nu} \, \gamma_{x+\hat{\nu},\mu} \, \gamma_{x,\nu} \, \gamma_{x + \hat{\mu},\mu} \,  \gamma_{x+2\hat{\mu},\nu} \, \gamma_{x+ \hat{\mu}+\hat{\nu},\mu} \, \gamma_{x + \hat{\mu},\nu} \; \\
	&= \; \gamma_{x,\mu} \, \gamma_{x + \hat{\mu},\mu} \,  \gamma_{x+2\hat{\mu},\nu} \, \gamma_{x+ \hat{\mu}+\hat{\nu},\mu} \, \gamma_{x+\hat{\nu},\mu} \, \gamma_{x,\nu} \,.
\end{align*}
The last equivalence holds because the staggered sign factor on the link common to the two adjacent plaquettes is squared, i.e., $(\gamma_{x + \hat{\mu},\nu})^{2} = 1$. This procedure can be iterated to form any kind of loops, and the resulting staggered sign factor can therefore be computed simply as:
\begin{equation}
	\prod_{(x,\mu)\in \mathcal{L}} \gamma_{x,\mu} \; = \; (\, -1 \,)^{P_{\mathcal{L}}} \, ,	
\label{eq:su2_signpl}
\end{equation}
where $P_{\mathcal{L}}$ stands for the number of plaquettes necessary to cover the surface bounded by the loop $\mathcal{L}$. Note that there is an ambiguity in this definition caused by the fact that, in more than two dimensions, the surface bounded by a loop is not unique. Nevertheless, different surfaces always differ by an even number of plaquettes, and the result in (\ref{eq:su2_signpl}) is the same also if different surfaces $P_{\mathcal{L}}$ are used. 

The loops pick up further signs from the sign factors $\prod_{x,\mu,a,b} (-1)^{k_{x,\mu}^{ab}}$ introduced by the forward hops of the loop. For trivially closing loops, where the number of forward hops is half the length of the loop, this sign factor can be expressed as $(-1)^{|\mathcal{L}|/2}$, where $|\mathcal{L}|$ is the length of the loop, i.e., the number of links that constitute the loop
$\mathcal{L}$. Loops that wind around the compact time or space direction do not share the property of equal numbers of forward and backward hops and one would have to distinguish different cases. For simplicity we here assume that the temporal and spatial extents of the lattice are all multiples of 4 and it is easy to see that $(-1)^{|\mathcal{L}|/2}$ then always correctly takes into account the signs from forward hopping. 

Loops that wind in the temporal direction also pick up a minus sign due to the anti-periodic boundary conditions, and we express this sign as $(-1)^{\mathcal{W}_\mathcal{L}}$, where $\mathcal{W}_\mathcal{L}$ denotes the net temporal winding number of the loop $\mathcal{L}$. Finally, each loop also picks up a minus sign from the reordering of the Grassmann variables. Summarizing, we obtained that the sign of a loop $\mathcal{L}$ only depends on its geometry, and is given by
\begin{equation}
\mbox{sign}\,(\mathcal{L}) \; = \; (-1)^{ \, |\mathcal{L}|/2 \, + \, \mathcal{W}_\mathcal{L} \, + \,  P_\mathcal{L} \, + \, 1 } \; ,
\label{eq:su2_loopsign}
\end{equation}
where $|\mathcal{L}|$ is the length of the loop, $\mathcal{W}_\mathcal{L}$ is the number of windings around the compact time direction, and $P_{\mathcal{L}}$ the number of plaquettes of a surface with $\mathcal{L}$ as its boundary.

Putting things together we find the following expression for the fermionic partition sum,
\begin{equation}
Z_F[U] \, =   \, 
\frac{1}{2^{2V}} \! \sum_{\{s,k,\overline{k}\}}  C_{F}[s,k,\overline{k}\,] \; W_M[s] \; \prod_{\mathcal{L}} \, \mbox{sign} \,(\mathcal{L}) 
\prod_{x,\mu} \prod_{a,b} 
( U_{x,\mu}^{\,ab} )^{k_{x,\mu}^{\; ab}} \, 
( U_{x,\mu}^{\, ab\, \star} )^{\overline{k}_{x,\mu}^{\; ab}}  \; ,
\label{eq:su2_partitionf3}
\end{equation}
where we collected the contributions of the monomers into the weight factor $W_M[s]$:
\begin{equation}
\label{eq:su2_wm}
	W_M[s] = \prod_{x,a}(2m)^{s_{x}^{a}}  \, .
\end{equation}
The fermionic partition function (\ref{eq:su2_partitionf3}) is a sum over all the configurations of the fermion variables $s_{x}^{a}, k_{x,\mu}^{ab}, \overline{k}_{x,\mu}^{ab} \in \{0,1\}$. The admissible configurations must satisfy the fermion constraint $C_{F}[s,k,\overline{k}]$, which implement Pauli's exclusion principle for our dual representation. The only sources of sign for fermions are loops, which come with the sign function $\sign(\mathcal{L})$ given in (\ref{eq:su2_loopsign}). Monomers come with the weight (\ref{eq:su2_wm}), while dimers and loops activate factors $U_{x,\mu}^{ab}$ and $U_{x,\mu}^{ab \,\star}$. In the next section we discuss the integration of the gauge fields and obtain the final result for the partition function of the SU(2) lattice gauge theory with one flavor of staggered fermions.

\section{The full theory \label{sec:su2_full}}

\begin{figure}[t]
	\begin{center}
		\vskip5mm
		\includegraphics[scale=0.8,clip]{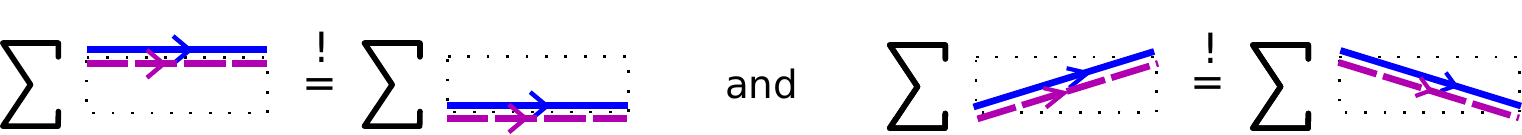}
	\end{center}
	\caption{Geometrical illustration of the generalized constraints that combine
		the gauge fluxes $J_{x,\mu}^{ab}$ (full lines) and the matter fluxes $k_{x,\mu}^{\,ab} - \overline{k}_{x,\mu}^{\,ab}$
		(dashed lines). Similar to the pure gauge case shown in Fig.~\ref{fig:su2_constraints} the sum over all 1-1 
		fluxes must equal the sum over all 2-2 fluxes and the sum over 1-2 fluxes must equal the sum over 2-1 fluxes. \label{fig:su2_fullconstraints}}	
\end{figure}

The last step necessary in order to obtain the final dual form of the partition function $Z = \int \! D[U] \, e^{-S_{G}[U]} Z_{F}[U]$ consists in performing the link integrals 
\begin{align}
&\int\!  D[U] \, e^{-S_{G}[U]} \, \prod_{x,\mu} \prod_{a,b} 
( U_{x,\mu}^{\,ab} )^{k_{x,\mu}^{\; ab}} \, 
( U_{x,\mu}^{\, ab\, \star} )^{\overline{k}_{x,\mu}^{\; ab}} \nonumber \\
&= \, \prod_{x,\mu} \int \!  dU_{x,\mu} \prod_{a,b} 
( U_{x,\mu}^{\,ab} )^{N_{x,\mu}^{\; ab} + k_{x,\mu}^{\; ab}} \, 
( U_{x,\mu}^{\, ab\, \star} )^{\overline{N}_{x,\mu}^{\; ab} + \overline{k}_{x,\mu}^{\; ab} }  \, .
\end{align}
The solution to these integrals can be derived from the one obtained in Section \ref{sec:su2_acc}, substituting $N_{x,\mu}^{\; ab}$ with $N_{x,\mu}^{\; ab} + k_{x,\mu}^{\; ab}$ and $\overline{N}_{x,\mu}^{\; ab}$ with $\overline{N}_{x,\mu}^{\; ab} + \overline{k}_{x,\mu}^{\; ab}$. Thus we find that the integral over the $\theta_{x,\mu}$ angles results in the combinatorial factor
\begin{equation}
\label{su2_whfull}
W_H [p,k,\overline{k}]  \; = \; 
\prod_{x,\mu} \dfrac{\!\left(\!\frac{S_{x,\mu}^{11} +k_{x,\mu}^{\; 11}+\overline{k}_{x,\mu}^{\; 11}+ 
		S_{x,\mu}^{22} +k_{x,\mu}^{\; 22}+\overline{k}_{x,\mu}^{\; 22}}{2}\!\right) ! \; 
	\left(\frac{S_{x,\mu}^{12} +k_{x,\mu}^{\; 12}+\overline{k}_{x,\mu}^{\; 12} + 
		S_{x,\mu}^{21}+k_{x,\mu}^{\; 21}+\overline{k}_{x,\mu}^{\; 21}}{2}\right) !}{\!\left(\frac{
		S_{x,\mu}^{11} + k_{x,\mu}^{\; 11}+\overline{k}_{x,\mu}^{\; 11} + 
		S_{x,\mu}^{22} + k_{x,\mu}^{\; 22}+\overline{k}_{x,\mu}^{\; 22} + 
		S_{x,\mu}^{12} + k_{x,\mu}^{\; 12}+\overline{k}_{x,\mu}^{\; 12} + 
		S_{x,\mu}^{21} + k_{x,\mu}^{\; 21}+\overline{k}_{x,\mu}^{\; 21}  }{2} + 1\!\right) !} \; ,
\end{equation}
while the integrals over the phases $\alpha_{x,\mu}$ and $\beta_{x,\mu}$ give rise to Kronecker deltas that enforce constraints over the color fluxes on each link $(x,\mu)$ of the lattice:
\begin{gather}
J_{x,\mu}^{11} \, + k_{x,\mu}^{11} \, - \overline{k}_{x,\mu}^{11} \; = \; J_{x,\mu}^{22} \, + k_{x,\mu}^{22} \, - \overline{k}_{x,\mu}^{22} \qquad \forall x,\mu \, , \label{eq:su2_fullconstraints1} \\
J_{x,\mu}^{12} \, + k_{x,\mu}^{12} \, - \overline{k}_{x,\mu}^{12} \; = \; J_{x,\mu}^{21} \, + k_{x,\mu}^{21} \, - \overline{k}_{x,\mu}^{21} \qquad \forall x,\mu  \, .
\label{eq:su2_fullconstraints2}
\end{gather}
In these constraints, which are illustrated schematically in Fig.~\ref{fig:su2_fullconstraints}, the color link fluxes receive contributions both from the gauge and the fermion fields. In our reformulation the gauge fields are represented by the cycle occupation numbers $p_{x,\mu\nu}^{abcd} \in \mathbb{N}_{0}$, which are variables assigned to the plaquettes. We write the contribution they give to the color link fluxes by means of the $J_{x,\mu}^{ab}$ fluxes (\ref{eq:su2_jfluxes}), which sum up all the cycle occupation numbers that are attached to the link $(x,\mu)$ and that have color path $(a,b)$ on that link. They are depicted in blue full lines in Fig.~\ref{fig:su2_fullconstraints}.
The contribution of the fermions comes from the link variables $k_{x,\mu}^{ab} \in \{0,1\}$ and $\overline{k}_{x,\mu}^{ab} \in \{0,1\}$, and is shown with dashed purple lines in the figure.

The final expression for the partition sum in our dual representation then is: 
\begin{align}
Z = &\frac{1}{2^{2V}} \!\!\sum_{\{p,k,\overline{k},s \}} \!\! C_{F}[s,k,\overline{k}\,] \;  W_M[s] \, W_\beta [p] \, 
W_H [p,k,\overline{k}]\; 
\prod_{x,\mu} (-1)^{\, J_{x,\mu}^{\,21} + k_{x,\mu}^{\,21} + \overline{k}_{x,\mu}^{\,21}} \; 
\prod_{\mathcal{L}} \, \sign \,(\mathcal{L}) 
\nonumber \\
 \times & 
\prod_{x, \mu} 
\delta \!\left(\! J_{x,\mu}^{\,11} \!+\! k_{x,\mu}^{\,11} \!-\! \overline{k}_{x,\mu}^{\,11} - 
[J_{x,\mu}^{\,22} \!+\! k_{x,\mu}^{\,22} \!-\! \overline{k}_{x,\mu}^{\,22} ]\! \right)
\delta \!\left(\! J_{x,\mu}^{\,12} \!+\! k_{x,\mu}^{\,12} \!-\! \overline{k}_{x,\mu}^{\,12} - 
[J_{x,\mu}^{\,21} \!+\! k_{x,\mu}^{\,21} \!-\! \overline{k}_{x,\mu}^{\,21} ]\! \right).
\label{eq:su2_partitiondualfull}
\end{align}
It sums over all the possible configurations of the cycle occupation numbers $p_{x,\mu\nu}^{abcd} \in \mathbb{N}_{0}$, and the fermion dual variables $s_{x}^{a}, k_{x,\mu}^{ab}, \overline{k}_{x,\mu}^{ab}\in \{0,1\}$. The fermionic variables have to satisfy the constraint $C_{F}[s,k,\overline{k}]$ given in Eq.~(\ref{eq:su2_cf}), which ensures that the Grassmann integrals are saturated by requiring the two layers of our four-dimensional lattice to be filled with monomers, dimers and loops. Monomers come with the weight factor $W_{M}[s]$ (\ref{eq:su2_wm}), while dimers and loops, along with the cycle occupation numbers, give contributions to the combinatorial factor $W_{H}[p,k,\overline{k}]$. Loops, together with the cycle occupation numbers, are further constrained by the product of Kronecker deltas in Eq.~(\ref{eq:su2_partitiondualfull}) which impose equalities between different components of the total color flux at each link $(x,\mu)$. Notice that the dual representation we obtain suffers from the sign problem, which has two different origins: $(-1)^{\, J_{x,\mu}^{\,21} + k_{x,\mu}^{\,21} + \overline{k}_{x,\mu}^{\,21}}$ is the gauge sign, that can be traced back to the negative sign of the (2,1) entry in the parametrization (\ref{eq:su2_parametrization}) of the SU(2) matrices, and $\sign(\mathcal{L})$ is the fermion sign, which depends only on the geometry of the fermion loop $\mathcal{L}$. 

In the next and final section of this chapter we will show how the interplay of these two sources of sign makes the first terms of the joint hopping and strong coupling expansion positive.

\section{The strong coupling limit \label{sec:su2_sc}}
The aim of this section is to analyse the contributions to the partition function in its dual form (\ref{eq:su2_partitiondualfull}). We initially focus on the strong coupling limit \cite{PhysRevD.10.2445}, i.e., $\beta = 0$, and then add finite $\beta$ corrections. 

\begin{figure}[t]
	\begin{center}
		\vskip5mm
		\includegraphics[scale=0.53,clip]{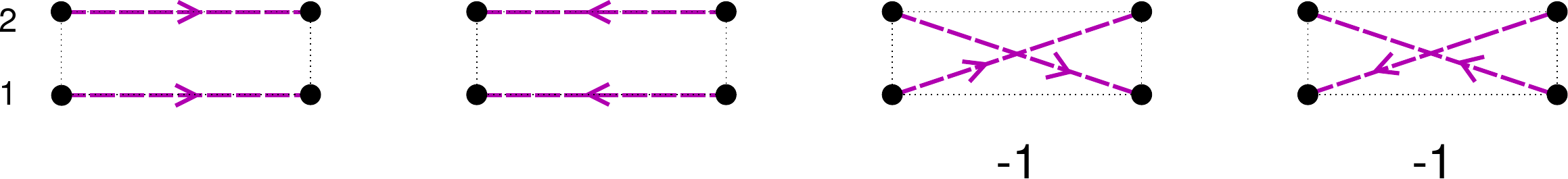}
	\end{center}
	\caption{Loop elements in the strong coupling limit. Only the four double fluxes shown here are admissible
		for loops in the strong coupling limit. These strong coupling loops, together with monomers and the 
		dimers shown in Fig.~\ref{fig:su2_dimers} have to completely 
		fill the lattice. The elements with crossing color flux come with an explicit minus sign.	\label{fig:su2_strongloops}}
\end{figure}

In the strong coupling limit all the cycle occupation numbers must be 0, as shown by the fact that $\lim_{\beta \to 0} W_{\beta}[p] = 0$ unless $p_{x,\mu\nu}^{abcd} = 0$ $\forall x,\mu,\nu,a,b,c,d$. Consequently, also the $J$- and $S$-fluxes vanish and, in particular, the constraints (\ref{eq:su2_fullconstraints1}) and (\ref{eq:su2_fullconstraints2}) reduce to:
\begin{equation}
k_{x,\mu}^{\,11} \!-\! \overline{k}_{x,\mu}^{\,11} \; = \; k_{x,\mu}^{\,22} \!-\! \overline{k}_{x,\mu}^{\,22}
\quad \forall x, \mu \qquad \mbox{and} \qquad 
k_{x,\mu}^{\,12} \!-\! \overline{k}_{x,\mu}^{\,12} \; = \; k_{x,\mu}^{\,21} \!-\! \overline{k}_{x,\mu}^{\,21}
\quad \forall x, \mu \; .
\label{strongconstraints}
\end{equation}
From these constraints one infers that only special types of loops can be used to fill the lattice together with monomers and dimes. The four combinations of $k$-fluxes that give rise to admissible loop segments are (refer to Fig.\ref{fig:su2_strongloops} for a graphical illustration):
\begin{equation*}
	k_{x,\mu}^{\,11} \, = \, k_{x,\mu}^{\,22} \, ,\quad 
	\overline{k}_{x,\mu}^{\,11} \, = \, \overline{k}_{x,\mu}^{\,22} \, , \quad 
	k_{x,\mu}^{\,12} \, = \, k_{x,\mu}^{\,21} \, ,\quad 
	\overline{k}_{x,\mu}^{\,12} \, = \, \overline{k}_{x,\mu}^{\,21} \, .
\end{equation*}
This means that admissible strong coupling loops are built from segments of two units of flux that run in the same direction. The fluxes can either run parallel in color space, or cross. The flux crossings come with negative sign, as we discussed in Sec.~\ref{sec:su2_acc}.

\begin{figure}[t]
	\begin{center}
		\vskip5mm
		\includegraphics[scale=0.53,clip]{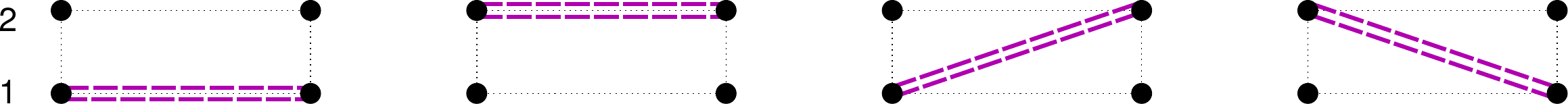}
	\end{center}
	\caption{Graphical representation of the possible dimers. Dimers saturate the Grassmann integrals for one of the two colors of two neighboring sites. Four different color combinations are possible. \label{fig:su2_dimers}}
\end{figure}

\begin{figure}[t]
	\begin{center}
		\vskip5mm
		\includegraphics[scale=0.6,clip]{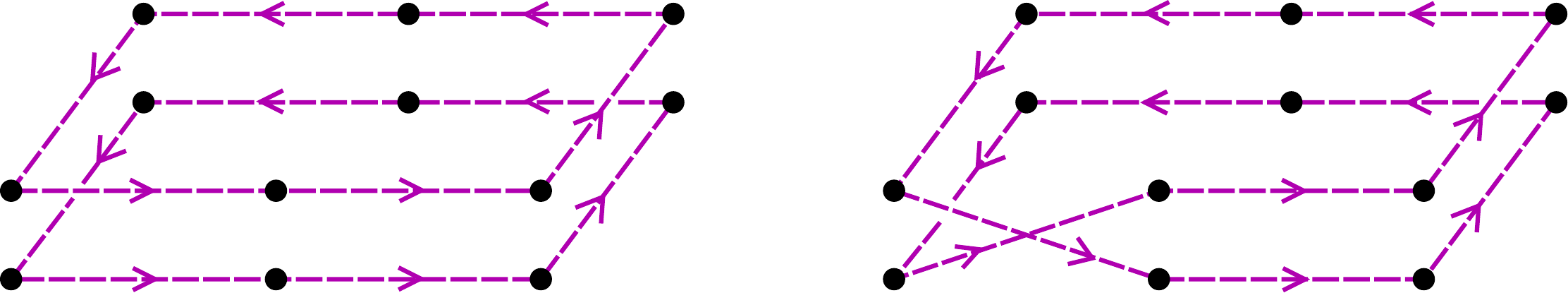}
	\end{center}
	\caption{Examples of fermion loops in the strong coupling limit. The double loop on the lhs.\ has a positive sign
		since here the signs of the individual loops in the color 1 and the color 2 layers are the same and thus this sign 
		squares to $+1$. The loop on the rhs.\ differs from the lhs.\ by one link element with a color crossing which comes with 
		an explicit minus sign. On the other hand the color crossing also connects the two loops from the lhs.\ into a single 
		loop such that there is one less minus sign from the number of loops. 
		In total the loop on the rhs.\ thus also has a positive sign. \label{fig:su2_loops}}
\end{figure}

Summarizing, at strong coupling the admissible configurations are the ones that completely fill the lattice with monomers, dimers (represented in Fig.~\ref{fig:su2_dimers}), and loops made of the strong coupling loop elements depicted in Fig.~\ref{fig:su2_strongloops}. 
Examples of such loops are shown in Fig.~\ref{fig:su2_loops}. 
The configuration on the lhs.\ of the plot has evidently a positive sign, since it is made of two identical loops $\mathcal{L}$ running parallel to each other on the two different color layers ($(\sign \mathcal{L})^2 = 1$). The configuration on the rhs.\ has a flux crossing, which gives rise to a factor of $-1$. On the other hand, the crossing connects the previously disconnected loops into a single loop, such that we have one minus sign less, while all other signs from the number of forward hops and from the staggered sign factors remain the same (compare with the definition of the loop sign factor (\ref{eq:su2_loopsign})). Thus the overall sign of the new loop with a single flux crossing is again $+1$. Any loop configuration at strong coupling can be built following this procedure. One starts with two units of flux running parallel along the path chosen for the loops (like in the plot on the lhs.\ of Fig.~\ref{fig:su2_loops}). Then, more general loop configurations can be obtained by interchanging loop elements of parallel flux (the two elements shown on the lhs.\ of Fig.~\ref{fig:su2_strongloops}) with flux crossings (the two elements on the rhs.\ of Fig.~\ref{fig:su2_strongloops}). Each interchange has the effect we just described: it adds a minus sign factor for the crossing and, at the same time, changes the total number of loops by $\pm 1$, thus leaving the overall sign unchanged. We conclude that loops in the strong coupling limit have positive weights. This result is remarkable because it implies that, at strong coupling, all the terms in the partition function are real and positive (recall that, for fermions, only loops come with signs).  

The first $\beta$ corrections can be derived easily and here, in order to illustrate such a calculation, we display the leading terms of a coupled expansion of the partition sum in $1/m$ and in $\beta$:
\begin{align}
Z  \; = \; m^{2V} \bigg[& 1 \; + \; \left( \!\frac{1}{2m}\! \right)^{\!2} \!\! \times \frac{1}{2!} \times 16 V \nonumber \\
&+ \;   
\left( \! \frac{1}{2m}\! \right)^{\!4} \! \!\times \left( \!  \frac{1}{2!} \! \right)^{\! 2} \!\! \times [ 128 V^2 - 264 V] 
\; + \; \left(\! \frac{1}{2m}\! \right)^4 \!\!\times \frac{2! + 1}{3!} \times 8 V 
\nonumber
\\
& + 
\left(\! \frac{1}{2m}\! \right)^4 \!\!\times \frac{\beta}{2} \times  
\left( \!  \frac{1}{2!} \! \right)^{\! 4} \!\! \times 192 V \; + \; 
\left(\! \frac{\beta}{2}\! \right)^2 \!\! \times  
\left( \!  \frac{1}{2!} \! \right)^{\! 4} \!\! \times 48 V \nonumber \\ 
&+ \; 
{\cal O} \left(\!\left(\! \frac{1}{2m}\! \right)^{\!6\,}\!\right) +
{\cal O} \left(\beta^4\right) \bigg] .
\label{eq:su2_expansion}
\end{align}
The leading term of this expansion is the contribution of the configuration in which the lattice is saturated placing a monomer on all the $2V$ sites. The corresponding weight factor is $(2m)^{2V}$, which together with the overall factor $(1/2)^{2V}$ gives rise to the contribution $m^{2V}$. Notice that on the rhs.\ of Eq.~(\ref{eq:su2_expansion}) we wrote the factor $m^{2V}$ up front. Therefore, all further terms in the expansion are relative to the lattice completely filled with monomers.

The next term corresponds to the configuration where we substitute two adjacent monomers with a dimer. We thus account for a suppressing factor of $(1/2m)^2$. The combinatorial factor $1/2!$ comes from the weight $W_{H}[p,k,\overline{k}]$, where we set $k_{x,\mu}^{ab} = \overline{k}_{x,\mu}^{ab} = 1$ on a single link $(x,\mu)$ for some color combination $a,b$. Lastly, the factor $16V$ is the degeneracy of the configuration we are considering, i.e., the number of ways to place a single dimer on our double layer lattice: on every link we have 4 ways to place a dimer (compare Fig.~\ref{fig:su2_dimers}) and, on a four-dimensional lattice, there are a total of $4V$ links. This results in $16V$ ways to place a single dimer.

The following contribution, in the second line of Eq.~(\ref{eq:su2_expansion}), is the one where we place two dimers on different links. The suppressing factor this time is $(1/2m)^4$, since we interchanged four monomers with two dimers. The weight factor $W_{H}[p,k,\overline{k}] = (1/2!)^2$ is the square of the one for a single dimer. The number of ways to place two dimers on the lattice is the square of the degeneracy for a single dimer divided by two: $(16V)^2/2 = 128V^2$. To this number we have to subtract the $264V$ configurations in which the two dimers are either on the same link, or touch (not admissible).

The second term in the second line of Eq.~(\ref{eq:su2_expansion}) accounts for the configurations where two dimers sit on the same link. There are only two ways of placing two dimers on one link: they must be either parallel ($k_{x,\mu}^{11} = \overline{k}_{x,\mu}^{11} = k_{x,\mu}^{22} = \overline{k}_{x,\mu}^{22} = 1$) or crossing ($k_{x,\mu}^{12} = \overline{k}_{x,\mu}^{12} = k_{x,\mu}^{21} = \overline{k}_{x,\mu}^{21} = 1$). Together with the number of links this gives $2\times 4V = 8V$ possibilities. $W_{H}[p,k,\overline{k}] = 2!/3!$ is the combinatorial weight for each of those configurations. The weight $1/3!$ accounts for self-crossing loops sitting on a single link. 

\begin{figure}[t]
	\begin{center}
		\vskip5mm
		\includegraphics[scale=0.5,clip]{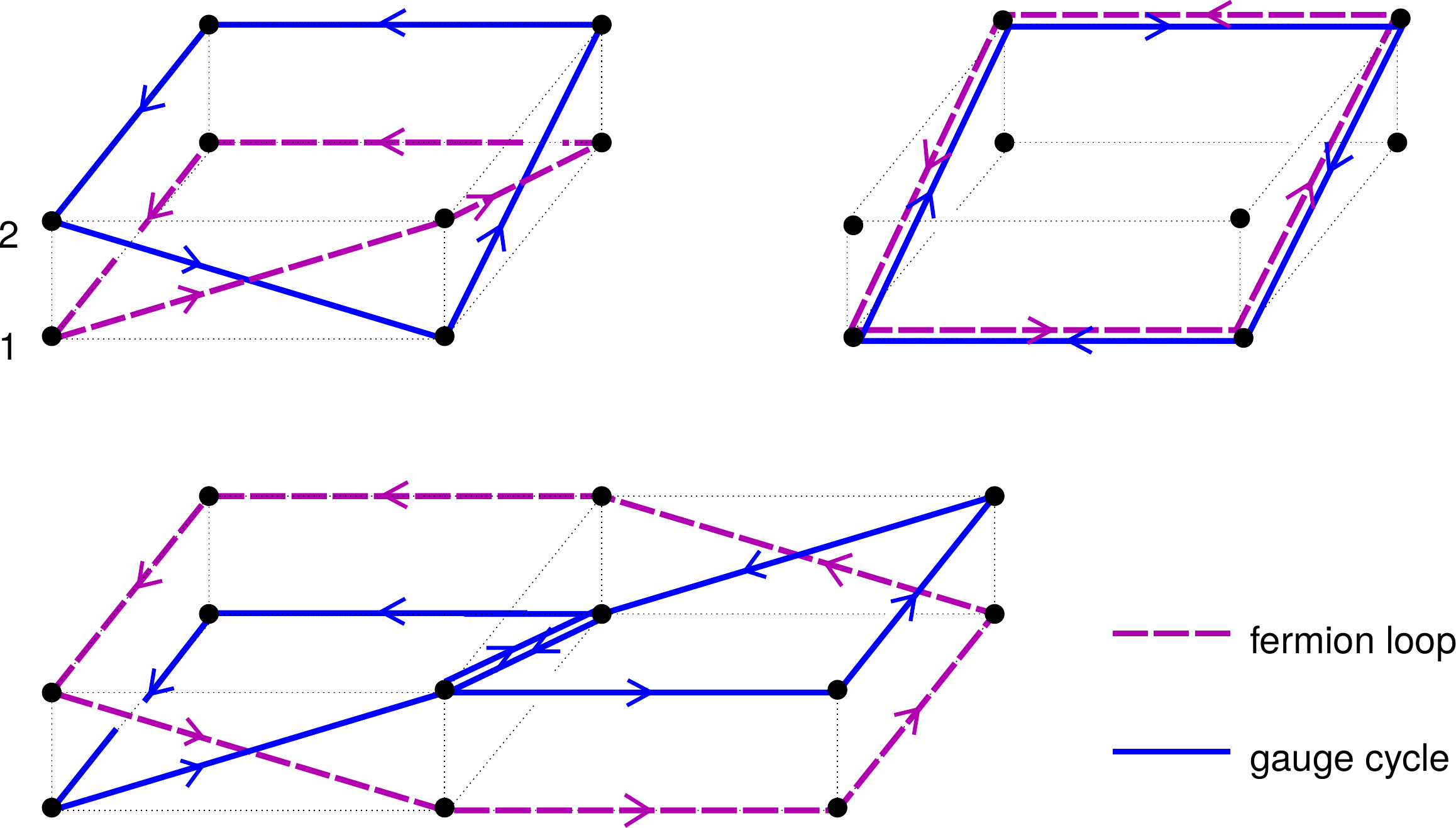}
	\end{center}
	\caption{Example for the saturation of a fermion loop with ACCs. \label{fig:su2_saturateloops}}
\end{figure}

The subsequent term, i.e., the first term in the third line in Eq.~(\ref{eq:su2_expansion}), is the contribution of the configurations with a single fermion loop around a plaquette for which the gauge constraints (\ref{eq:su2_fullconstraints1}) and (\ref{eq:su2_fullconstraints2}) are satisfied with an ACC, thus coming with a factor $\beta/2$ for the non-trivial value of one cycle occupation number. The factor $(1/2m)^4$ is there because we replaced the monomers on the four corners of a plaquette with a fermion loop. For every plaquette there are $16$ different fermion loops we can place (compare with Fig.~\ref{fig:su2_allcycles}), each with two orientations. Together with the number of plaquettes, this gives $2 \times 16 \times 6V = 192V$. Then we must adequately place an ACC such that the gauge constraints are satisfied at every link of the plaquette. Since the ACCs only have positive orientation, there is just one way of doing so for every fermion loop: for positively oriented fermion loops (see the example on the lhs.\ in the top row of Fig.~\ref{fig:su2_saturateloops}) the ACC has the opposite color at each corner, while for negatively oriented fermion loops the ACC runs alongside the fermion loop but with opposite orientation (rhs.\ top of Fig.~\ref{fig:su2_saturateloops}). In both cases we have two units of flux on each link of the plaquette, such that we have the additional factor $W_H [p, k, \overline{k}] = (1/2!)^4$. 

Finally, the last term is for the configurations where the fermion constraint is fulfilled just with monomers, but where we place two ACCs on top of each other, such that they satisfy the gauge constraints. Since the ACCs only have positive orientation, the admissible configurations are the ones for which the matching ACCs have opposite color at each corner, thus we can form 8 matching pairs from the 16 ACCs. Counting the $6V$ plaquettes of the lattice, this results in $48V$ possibilities. The two non-trivial cycle occupation numbers come with a weight of $(\beta/2)^2$ and, since they generate two units of flux on the four links of the plaquette, $W_H [p, k, \overline{k}] = (1/2!)^4$ also for these configurations. 

Subsequent terms of the coupled strong coupling and hopping expansion can be computed following the same steps we discussed for the leading terms. The plot at the bottom of Fig.~\ref{fig:su2_saturateloops} shows an example of a configuration contributing at $\mathcal{O}\Big(\big(\frac{1}{2m}\big)^6 \beta^2\Big)$.

We remark that we cross-checked some of the terms of the series (\ref{eq:su2_expansion}) with previous results: the term depending only on $\beta$ matches the corresponding term of conventional strong coupling expansion \cite{PhysRevD.10.2445}, while the leading $(2m)^{-2}$ and $(2m)^{-4}$ contributions of the $\beta$-independent terms were verified by comparison to the free case. We also stress that, in our dual formulation, the leading contributions of the expansion (\ref{eq:su2_expansion}) for the partition sum $Z$ are real and positive. Negative configurations only appear at $\mathcal{O}(\beta^4)$ (see Fig.~\ref{fig:su2_unhappy}) or at $\mathcal{O}\Big(\big(\frac{1}{2m}\big)^4 \beta^3\Big)$ when a fermion loop is included.

\chapter{Dual representation for lattice QCD with one flavor of staggered fermions \label{cha:qcd}}
As a natural extension of the methods we developed in Chapter \ref{cha:su2} for QC$_2$D, we here discuss the generalization of the abelian color cycle (ACC) and the abelian color flux (ACF) dual approaches to lattice QCD with one flavor of staggered fermions. 

The study of the phase diagram of QCD on the lattice has been tremendously slowed down by the so-called sign problem. This problem arises when a non-vanishing chemical potential is introduced in the action. As a result, the fermion determinant becomes complex and thus the applicability of standard Monte Carlo methods, which requires the interpretation of the Boltzmann weight as a probability, is prevented. 

As we learned in Sec.~\ref{sec:teo_u1} as well as in Chapter \ref{cha:pcm}, the sign problem is, in general, representation dependent. Unfortunately, up to this date no representation of lattice QCD has been found that solves the sign problem (and the one we present in this chapter makes no exception). Nevertheless, already more than 30 years ago, a monomer-dimer-loop representation of QCD at strong coupling was found that presented a sign problem much milder than the one of QCD itself \cite{Rossi:1984cv,Karsch:1988zx}, so that the standard technique of reweighting can be used for simulations \cite{deForcrand:2009dh}. In the strong coupling limit the inverse lattice coupling $\beta$ is set to 0. Thus, in this regime it is impossible to take the continuum limit and one expects large lattice artifacts. Nonetheless, strong coupling QCD shares some key features of QCD as, for examples, it confines colored objects and, for vanishing quark masses, it presents a chiral spontaneously broken phase in the vacuum and a restored chiral symmetric phase at high temperature or high chemical potential. These properties explain the abundance of literature about this subject. Older publications \cite{Adams:2003cca, Chandrasekharan:2003im, Chandrasekharan:2004kd, Chandrasekharan:2006tz, deForcrand:2009dh, Unger:2011in, deForcrand:2012vh, Kim:2016izx, deForcrand:2017fky} use the monomer-dimer-loop representation derived in \cite{Rossi:1984cv,Karsch:1988zx} to study the phase diagram of strong coupling QCD, using adaptation of the worm algorithm \cite{PhysRevLett.87.160601} or continuous time Monte Carlo as simulation techniques. In \cite{deForcrand:2014tha,deForcrand:2015daa} the first order correction in $\beta$ is included as a first step towards the continuum limit, while in \cite{Gagliardi:2017uag} a method to include corrections at higher orders of $\beta$ is proposed. Other diagrammatic representations for QCD and QCD-like lattice field theories can be found in \cite{Bruckmann:2017vri,Borisenko:2017gql,Vairinhos:2014uxa,Gattringer:2018mrg}.

Differently from what was done in the past, the reformulation of lattice QCD we present in this chapter has the form of a strong coupling expansion in which all the terms are known in closed form. The long range physics, relevant for the continuum limit, has the structure of worldsheets for the gauge degrees of freedom, and worldlines for the matter fields. We also study in details the strong coupling limit. We find that in this regime baryons propagate as free fermion loops in a background of monomers and dimers. As a future challenge, we propose the use of fermion bags \cite{PhysRevD.82.025007,PhysRevD.85.091502,Chandrasekharan2013,Gattringer:2018mrg} as an updating strategy for our form of strong coupling QCD.

\section{Abelian color cycle dualization of the SU(3) pure gauge theory \label{sec:qcd_acc}}

We start the dualization of lattice QCD focusing on the pure gauge theory. The Wilson action for SU(3) reads
\begin{equation}
	S_G[U] \; = \; -\dfrac{\beta}{3} \sum_{x,\nu < \rho} \R \Tr U_{x,\nu} \; U_{x+\hat{\nu},\rho} \, U_{x+\hat{\rho},\nu}^{\dagger} \, U_{x,\rho}^\dagger \; ,
\label{eq:qcd_sg}
\end{equation}
where $U_{x,\nu} \in$ SU(3) are the gauge degrees of freedom. They sit on the links $(x,\nu)$ of a four-dimensional lattice with periodic boundary conditions. The partition function $Z$ is obtained by integrating the Boltzmann weight $e^{-S_G[U]}$ over the product of Haar measures $\int \! D[U] = \prod_{x,\nu} \int_{\text{SU(3)}} dU_{x,\nu}$:
\begin{equation}
	Z \, = \, \int \! D[U] \; e^{-S_G[U]} \, .
\label{eq:qcd_partition}
\end{equation}
As we have seen in the previous chapter, the key step of the abelian color cycle (ACC) method consists in decomposing the action in its minimal terms. This is achieved by making explicit the real part, the trace and the matrix multiplications in Eq.~(\ref{eq:qcd_sg}). As a result, the action is rewritten as a sum of products of four link matrix elements $U_{x,\nu}^{ab}$
\begin{equation}
	S_G[U] \; = \; -\dfrac{\beta}{6} \sum_{x,\nu < \rho}  \sum_{a,b,c,d=1}^{3} \Big[
	U_{x,\nu}^{ab} U_{x+\hat{\nu},\rho}^{bc} U_{x+\hat{\rho},\nu}^{dc \, \star} U_{x,\rho}^{ad \, \star} + U_{x,\nu}^{ab \, \star} U_{x+\hat{\nu},\rho}^{bc \, \star} U_{x+\hat{\rho},\nu}^{dc} U_{x,\rho}^{ad} \Big]\, .
	\label{eq:qcd_accaction}
\end{equation} 
The products $U_{x,\nu}^{ab} U_{x+\hat{\nu},\rho}^{bc} U_{x+\hat{\rho},\nu}^{dc \ \star} U_{x,\rho}^{ad \ \star}$ are what we call \textit{Abelian Color Cycles} (ACCs) \cite{Gattringer:2016lml}. These objects are complex numbers, and we interpret them as paths in color space closing around plaquettes, analogously to what we have done in Section \ref{sec:su2_acc} for the case of SU(2). The four color labels $a,b,c,d = 1,2,3$ denote the colors on each of the four corners of the plaquette $(x,\nu\rho)$. On the first site $x$, which is at the bottom left of the plaquette, the ACC runs trough color $a$, on site $x + \hat{\nu}$ trough color $b$, and so on, closing around the plaquette with mathematically positive orientation. Notice that, differently from what we found for the SU(2) gauge theory, also the complex conjugate ACCs $U_{x,\nu}^{ab \ \star} U_{x+\hat{\nu},\rho}^{bc \ \star} U_{x+\hat{\rho},\nu}^{dc} U_{x,\rho}^{ad}$ appear in the decomposition (\ref{eq:qcd_accaction}). This is a result of having to take the real part in the Wilson action (\ref{eq:qcd_sg}), an operation that was unnecessary in Section \ref{sec:su2_acc} because of the pseudo-reality of SU(2). Geometrically the complex conjugate ACCs will be interpreted as running through the plaquette with mathematically negative orientation.

\begin{figure}[t]
	\begin{center}
		\vskip5mm
		\includegraphics[scale=0.6,clip]{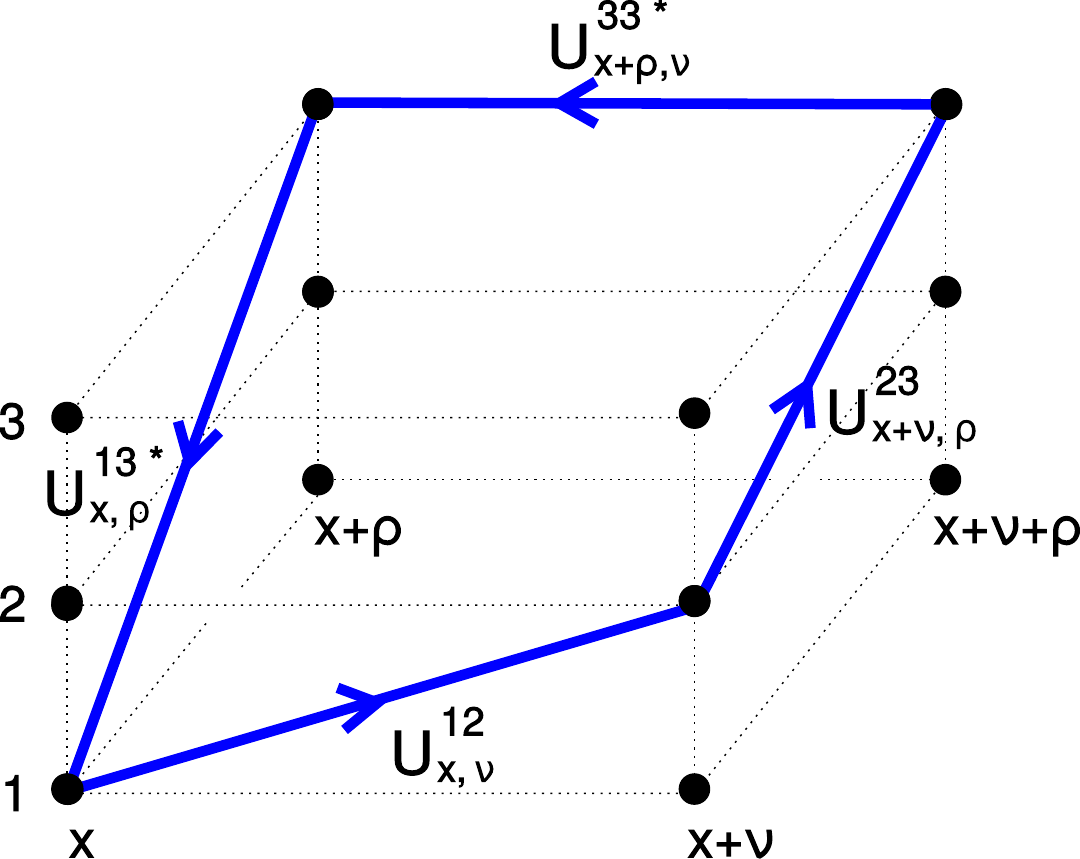}
	\end{center}
	\caption{Graphical illustration of the (1,2,3,3)-ACC, which explicitly is given by $U_{x,\nu}^{12} U_{x+\hat{\nu},\rho}^{23} U_{x+\hat{\rho},\nu}^{33 \ \star} U_{x,\rho}^{13 \ \star}$. This ACC closes around the plaquette $(x,\nu\rho)$ running through the sequence $(1,2,3,3)$ of color indices at the corners of the plaquette. In the	graphical representation the color degrees of freedom are shown	as three distinct layers of the space-time lattice, labeled with 1,2 and 3 on the lhs.\ of the plot. Each of the matrix elements $U_{x,\nu}^{ab}$ constituting the ACC is represented by an arrow along the corresponding link $(x,\nu)$ connecting color $a$ with color $b$. For	complex conjugate matrix elements the link is run through with negative orientation. \label{fig:qcd_acc}}
\end{figure}

To give an example of the geometrical interpretation, in Fig.~\ref{fig:qcd_acc} we illustrate the $(1,2,3,3)$-ACC. It closes around the plaquette $(x,\nu\rho)$, which has three color layers represented in the figure with grey dashed lines. The four link matrix elements $U_{x,\nu}^{ab}$ constituting the ACC are represented with oriented arrows which connect color layers on neighboring sites, e.g., the matrix element $U_{x,\nu}^{12}$ links color $1$ on site $x$ to color $2$ on site $x + \hat{\nu}$. As we already mentioned, complex conjugation reverts the orientation of the arrows, so that, e.g., the link element $U_{x,\nu}^{13 \ \star}$ connects color $3$ on site $x + \hat{\rho}$ to color $1$ on site $x$, thus closing the cycle around the plaquette. Obviously, since for each of the four corners of a plaquette we have three different possible choices of color, then for SU(3) there are a total of $3^4=81$ different ACCs on every plaquette, each of which can have either positive or negative mathematical orientation. 

The ACC decomposition (\ref{eq:qcd_accaction}) of the action (\ref{eq:qcd_sg}) allows us to proceed with the dualization of the theory as follows: 
\begin{align}
	Z \, &= \, \int \! D[U] \prod_{x,\nu<\rho} \prod_{a,b,c,d = 1}^{3} 
	e^{\frac{\beta}{6} U_{x,\nu}^{ab} U_{x+\hat{\nu},\rho}^{bc} U_{x+\hat{\rho},\nu}^{dc \, \star} U_{x,\rho}^{ad \, \star}} \,e^{\frac{\beta}{6} U_{x,\nu}^{ab\, \star} U_{x+\hat{\nu},\rho}^{bc \, \star} U_{x+\hat{\rho},\nu}^{dc} U_{x,\rho}^{ad}}  \nonumber
	\\
	& = \, \int \! D[U] \prod_{x,\nu<\rho} \prod_{a,b,c,d = 1}^{3} \sum_{n_{x,\nu\rho}^{abcd} = 0}^{\infty} \sum_{\overline{n}_{x,\nu\rho}^{abcd} = 0}^{\infty} 
	\dfrac{ \left( \beta/6 \right)^{n_{x,\nu\rho}^{abcd} + \overline{n}_{x,\nu\rho}^{abcd}}}{n_{x,\nu\rho}^{abcd}\, ! \overline{n}_{x,\nu\rho}^{abcd}\, !} \nonumber \\
	& \hspace{2.5cm} \times 
	\left( U_{x,\nu}^{ab} U_{x+\hat{\nu},\rho}^{bc} U_{x+\hat{\rho},\nu}^{dc \, \star} 
	U_{x,\rho}^{ad \, \star} \right)^{n_{x,\nu\rho}^{abcd}}
	\left( U_{x,\nu}^{ab\, \star} U_{x+\hat{\nu},\rho}^{bc \, \star} U_{x+\hat{\rho},\nu}^{dc} 
	U_{x,\rho}^{ad} \right)^{\overline{n}_{x,\nu\rho}^{abcd}} \, .
\label{eq:qcd_partition2}
\end{align}
In the first step we wrote the Boltzmann weight as a product over all the plaquettes, and all the ACCs on that plaquette of the two local factors $e^{\frac{\beta}{6} U_{x,\nu}^{ab} U_{x+\hat{\nu},\rho}^{bc} U_{x+\hat{\rho},\nu}^{dc \ \star} U_{x,\rho}^{ad \ \star}}$ and $e^{\frac{\beta}{6} U_{x,\nu}^{ab \ \star} U_{x+\hat{\nu},\rho}^{bc \ \star} U_{x+\hat{\rho},\nu}^{dc} U_{x,\rho}^{ad}}$. These exponentials are the Boltzmann weights for the ACCs with positive and negative orientation respectively. In the second step of (\ref{eq:qcd_partition2}) we expanded each factor in a Taylor series, thus introducing two sets of expansion indices assigned to the plaquettes: $n_{x,\nu\rho}^{abcd} \in \mathbb{N}_{0}$ and $\overline{n}_{x,\nu\rho}^{abcd} \in \mathbb{N}_{0}$, where $a,b,c,d=1,2,3$ are the color indices. The variables $n_{x,\nu\rho}^{abcd}$ correspond to the units of flux around the plaquette $(x,\nu\rho)$ in positive orientation, with color indices $(a,b,c,d)$, while $\overline{n}_{x,\nu\rho}^{abcd}$ is used for flux with negative orientation.

All the factors in the last line of (\ref{eq:qcd_partition2}) are complex numbers, such that we can freely commute them and reorganize the products to determine the integer valued powers for the link elements $U_{x,\nu}^{ab}$ and theirs complex conjugate $U_{x,\nu}^{ab\, \star}$. The partition sum becomes
\begin{equation}
\label{eq:qcd_partition3}
	Z \; = \; \sum_{\{n, \overline{n}\}} \left[ \prod_{x,\nu<\rho} \prod_{a,b,c,d}  
	\dfrac{ \left( \beta/6 \right)^{n_{x,\nu\rho}^{abcd} + \overline{n}_{x,\nu\rho}^{abcd}}}{n_{x,\nu\rho}^{abcd}\, !\,  \overline{n}_{x,\nu\rho}^{abcd}\, !} \right] \prod_{x,\nu} \int \! \! dU_{x,\nu} \; \prod_{a,b} \left( U_{x,\nu}^{ab} \right) ^{N_{x,\nu}^{ab}} 
	\left( U_{x,\nu}^{ab \ \star} \right) ^{\overline{N}_{x,\nu}^{ab}} \; ,
\end{equation}
where we introduced the short-hand notation 
\begin{equation*}
	\sum_{\{n, \overline{n}\}} = \prod_{x,\nu<\rho} \prod_{a,b,c,d = 1}^{3} \sum_{n_{x,\nu\rho}^{abcd} = 0}^{\infty} \sum_{\overline{n}_{x,\nu\rho}^{abcd} = 0}^{\infty} \, ,
\end{equation*}
to express the sum over the configurations of the variables $n_{x,\nu\rho}^{abcd} \in \mathbb{N}_{0}$ and $\overline{n}_{x,\nu\rho}^{abcd} \in \mathbb{N}_{0}$. The powers $N_{x,\nu}^{ab} \in \mathbb{N}_{0}$ and $\overline{N}_{x,\nu}^{ab} \in \mathbb{N}_{0}$ for the matrix elements $U_{x,\nu}^{ab}$ and $U_{x,\nu}^{ab \, \star}$ are given by
\begin{gather} 
\label{eq:qcd_N}
N_{x,\nu}^{ab} \; = \; 
\sum_{\rho:\nu<\rho}n_{x,\nu\rho}^{abss} + \overline{n}_{x-\hat{\rho},\nu\rho}^{ssba} + \sum_{\sigma:\nu>\sigma}\overline{n}_{x,\sigma\nu}^{assb} + n_{x-\hat{\sigma},\sigma\nu}^{sabs},
\\
\label{eq:qcd_Nbar}
\overline{N}_{x,\nu}^{ab} \; = \; 
\sum_{\rho:\nu<\rho}\overline{n}_{x,\nu\rho}^{abss} + n_{x-\hat{\rho},\nu\rho}^{ssba} + \sum_{\sigma:\nu>\sigma}n_{x,\sigma\nu}^{assb} + \overline{n}_{x-\hat{\sigma},\sigma\nu}^{sabs} \; .
\end{gather} 
The label $s$ introduced in the last two equations stands for the independent sum over the color indices that are replaced by $s$, e.g., $n_{x,\nu\rho}^{assb} \equiv \sum_{c,d} n_{x,\nu\rho}^{acdb}$.

To compute the Haar measure integrals in Eq.~(\ref{eq:qcd_partition3}) we use the following explicit parametrization for the SU(3) matrices \cite{Bronzan:1988wa}
\begin{align}
	\label{eq:qcd_parametrization}
	&U_{x,\nu} =\\[1em] \nonumber 
	&\left( \! \!
	\begin{array}{ccc}
	c_1 c_2 \, e^{i\phi_1}  & s_1 \, e^{i\phi_3} & c_1 s_2 \, e^{i\phi_4}\\
	s_2 s_3 \, e^{-i\phi_4 -i\phi_5} - s_1 c_2 c_3 \, e^{i\phi_1 +i\phi_2 -i\phi_3} & c_1 c_3 \, e^{i\phi_2} & - c_2 s_3 \, e^{-i\phi_1 -i\phi_5} - s_1 s_2 c_3 \, e^{i\phi_2 -i\phi_3 +i\phi_4}\\
	- s_2 c_3 \, e^{-i\phi_2 -i\phi_4} - s_1 c_2 s_3 \, e^{i\phi_1 -i\phi_3 +i\phi_5} & c_1 s_3 \, e^{i\phi_5} & c_2 c_3 \, e^{-i\phi_1 -i\phi_2} - s_1 s_2 s_3 \, e^{-i\phi_3 +i\phi_4 +i\phi_5}
	\end{array} \!\!\right) ,
\end{align}
where $c_{i} = \cos \theta_{x,\nu}^{(i)}$, $s_{i} = \sin \theta_{x,\nu}^{(i)}$, with $\theta_{x,\nu}^{(i)} \in [0, \pi/2]$, $i = 1,2,3$ and $\phi_{j} = \phi_{x,\nu}^{(j)}$, with $\phi_{x,\nu}^{(j)} \in [-\pi,\pi]$, $j=1,2,\dots,5$. The normalized Haar measure is
\begin{equation}
	dU_{x,\nu}  =  \frac{1}{2\pi^5} \, d\theta_1 c_1^3 s_1 \, d\theta_2 c_2 s_2 \, d\theta_3 c_3 s_3 \, d\phi_1 \, d\phi_2 \, d\phi_3 \, d\phi_4 \, d\phi_5 \, .
	\label{eq:qcd_haarmeasure}
\end{equation}
In what follows it will prove convenient to perform the changes of variables:
\begin{gather}
\label{eq:qcd_cycleoccupationnumbers}
	n_{x,\nu\rho}^{abcd} - \overline{n}_{x,\nu\rho}^{abcd} = p_{x,\nu\rho}^{abcd} \, , \qquad p_{x,\nu\rho}^{abcd} \in \mathbb{Z}\, ; \\
	\label{eq:qcd_auxiliary}
	n_{x,\nu\rho}^{abcd} + \overline{n}_{x,\nu\rho}^{abcd} = |p_{x,\nu\rho}^{abcd}| + 2 l_{x,\nu\rho}^{abcd} \, , \qquad l_{x,\nu\rho}^{abcd} \in \mathbb{N}_0 \, .
\end{gather}
Eqs.~(\ref{eq:qcd_cycleoccupationnumbers}) and (\ref{eq:qcd_auxiliary}) define the dual variables $p_{x,\nu\rho}^{abcd} \in \mathbb{Z}$ and $l_{x,\nu\rho}^{abcd} \in \mathbb{N}_0$, which will be the new dynamical degrees of freedom replacing the gauge fields once the conventional fields $U_{x,\nu}$ are integrated out. 
From the definition (\ref{eq:qcd_cycleoccupationnumbers}) and the interpretation of the $n_{x,\nu\rho}^{abcd}$ ($\overline{n}_{x,\nu\rho}^{abcd}$) as the activation numbers for the $(a,b,c,d)$-ACCs with positive (negative) orientation, it is clear that the new variables $p_{x,\nu\rho}^{abcd}$ activate $|p_{x,\nu\rho}^{abcd}|$ units of flux for the $(a,b,c,d)$-ACC on the plaquette $(x,\nu\rho)$, with the orientation of the flux given by the sign of the $p_{x,\nu\rho}^{abcd}$. We refer to the $p_{x,\nu\rho}^{abcd}$ as \textit{cycle occupation numbers}. As we will see later, the $p_{x,\nu\rho}^{abcd}$ will be subject to constraints, while the variables $l_{x,\nu\rho}^{abcd} \in \mathbb{N}_{0}$ will be unconstrained, thus we simply refer to them as \textit{auxiliary plaquette variables}.
  
We also introduce the fluxes
\begin{equation}
\label{eq:qcd_Jfluxes}
	J_{x,\nu}^{ab} \; =  
	\sum_{\rho:\nu<\rho}  [\, p_{x,\nu\rho}^{abss} - p_{x-\hat{\rho},\nu\rho}^{ssba} \, ] - \!\!
	\sum_{\sigma:\nu>\sigma}[\, p_{x,\sigma\nu}^{assb} - p_{x-\hat{\sigma},\sigma\nu}^{sabs} \, ] \; ,
\end{equation}
\begin{align}
	\nonumber
	S_{x,\nu}^{ab} &=  
	\sum_{\rho:\nu<\rho}  [ |p_{x,\nu\rho}^{abss}| + |p_{x-\hat{\rho},\nu\rho}^{ssba}| + 
	2(l_{x,\nu\rho}^{abss} + l_{x-\hat{\rho},\nu\rho}^{ssba}) ] 
	\\
	&\, + \sum_{\sigma:\nu>\sigma} [ |p_{x,\sigma\nu}^{assb}| + |p_{x-\hat{\sigma},\sigma\nu}^{sabs}| + 
	2(l_{x,\sigma\nu}^{assb} + l_{x-\hat{\sigma},\sigma\nu}^{sabs}) ] \; .
\label{eq:qcd_Sfluxes}
\end{align}
Since the $J$-fluxes will enter the constraints, it is important to discuss their geometrical interpretation on the lattice. The $J_{x,\nu}^{ab}$ represents the total net amount of color flux on the link $(x,\nu)$ connecting color $a$ and $b$ of the neighboring sites $x$ and $x + \hat{\nu}$. This color flux is the sum of the contributions of all the ACCs attached to the link $(x,\nu)$ that have a path from color $a$ to color $b$ on that link. So, if we consider the plaquette $(x,\nu\rho)$, with $\nu < \rho$, we have 9 different ACCs that contribute to that flux, namely the ones corresponding to the cycle occupation numbers $p_{x,\nu\rho}^{abef}$, where $a$ and $b$ are the color indices which we fix at $x$ and $x + \hat{\nu}$. The colors $e$ and $f$, chosen independently from the set $\{1,2,3\}$, determine the ACC at the remaining two corners of the plaquette $(x,\nu\rho)$. We thus have $3^2 = 9$ possibilities. Since the flux of these ACCs has a positive orientation along the link $(x,\nu)$, the 9 ACCs contribute with a positive sign in the definition (\ref{eq:qcd_Jfluxes}) of the $J_{x,\nu}^{ab}$ fluxes. 
If we then consider another plaquette attached to the link $(x,\nu)$, e.g., the plaquette $(x,\sigma\nu)$ with $\sigma < \nu$, we find that also in this case there are 9 ACCs which contribute to the flux $J_{x,\nu}^{ab}$, corresponding to the 9 cycle occupation numbers $p_{x,\sigma\nu}^{aefb}$. Since these ACCs have negative flux along the link $(x,\nu)$, the $p_{x,\sigma\nu}^{aefb}$ contribute with a negative sign in the definition (\ref{eq:qcd_Jfluxes}) of the $J_{x,\nu}^{ab}$. For the remaining four plaquettes attached to the link $(x,\nu)$ an analogous discussion holds.

\begin{figure}[t]
	\begin{center}
 		\vskip5mm
 		\includegraphics[width=6.5cm]{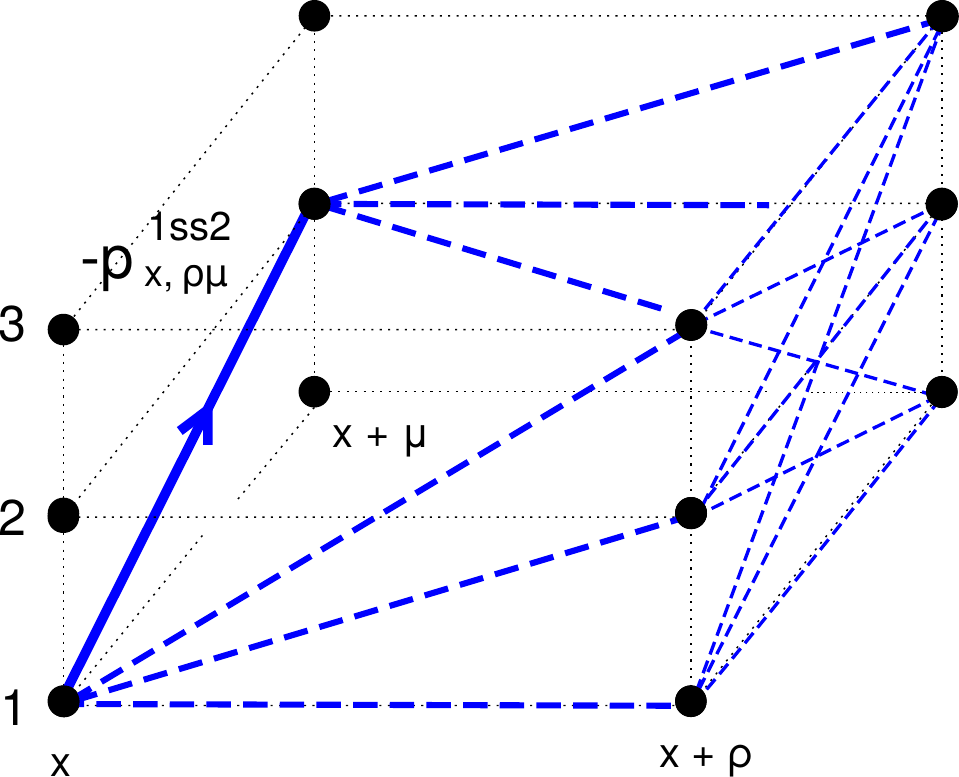}%
 	\end{center}
 	\caption{Graphical illustration of the sum $- p_{x,\sigma\nu}^{1ss2}$ contributing to the $J_{x,\nu}^{12}$ flux. \label{fig:qcd_colorflux}}
\end{figure}

As an example, in Fig.~\ref{fig:qcd_colorflux} we illustrate the sum of the contributions to the flux $J_{x,\nu}^{12}$ from the plaquette $(x,\sigma\nu)$, with $\sigma < \nu$. The flux from color 1 to 2 on the link $(x,\nu)$ is fixed and represented with an arrow oriented towards the positive $\nu$ direction. The 9 ACCs on the plaquette $(x,\sigma\nu)$ which contribute to this flux are summed over in the definition (\ref{eq:qcd_Jfluxes}) and we represent them with dashed lines in the figure. Since the ACCs on the $(x,\sigma\nu)$ plaquette have a negative orientation of the flux on the link we are considering, they contribute with a negative sign to the flux $J_{x,\nu}^{12}$. 

Having discussed the interpretation we give to the $J$-fluxes, we can now continue with the derivation of the dual representation for the partition function $Z$. Using the change of variables (\ref{eq:qcd_cycleoccupationnumbers}) and (\ref{eq:qcd_auxiliary}) we obtain the following expression:
\begin{equation}
\label{eq:qcd_partition4}
Z \; = \; \sum_{\{p, l\}} \left[ \prod_{x,\nu<\rho} \prod_{a,b,c,d}  
\dfrac{ \left( \beta/6 \right)^{|p_{x,\nu\rho}^{abcd}| + 2l_{x,\nu\rho}^{abcd}}}{(|p_{x,\nu\rho}^{abcd}| + l_{x,\nu\rho}^{abcd})\, !\,  l_{x,\nu\rho}^{abcd}\, !} \right] \prod_{x,\nu} \int \! \! dU_{x,\nu} \; \prod_{a,b} \left( U_{x,\nu}^{ab} \right) ^{N_{x,\nu}^{ab}} 
\left( U_{x,\nu}^{ab \ \star} \right) ^{\overline{N}_{x,\nu}^{ab}} \; .
\end{equation}
Hence, the partition function is now a sum over the configurations of the cycle occupation numbers $p_{x,\nu\rho}^{abcd} \in \mathbb{Z}$ and the auxiliary plaquette variables $l_{x,\nu\rho}^{abcd} \in \mathbb{N}_{0}$. In terms of these dual variables, $N_{x,\nu}^{ab}$ and $\overline{N}_{x,\nu}^{ab}$ can be written as
\begin{equation}
\label{eq:qcd_N2}
N_{x,\nu}^{ab} = \frac{S_{x,\nu}^{ab} + J_{x,\nu}^{ab}}{2} \, , \qquad \qquad
\overline{N}_{x,\nu}^{ab} = \frac{S_{x,\nu}^{ab} - J_{x,\nu}^{ab}}{2} \, .
\end{equation}

To compute the Haar measure integrals in Eq.~(\ref{eq:qcd_partition4}) we have to substitute the parametrization (\ref{eq:qcd_parametrization}) for the matrix elements $U_{x,\nu}^{ab}$ in (\ref{eq:qcd_partition3}). 
Notice, however, that some of the elements $U_{x,\nu}^{ab}$ of the matrix parametrization (\ref{eq:qcd_parametrization}) are not in the simple form $U_{x,\nu}^{ab} = r_{x,\nu}^{ab} \, e^{i \varphi_{x,\nu}^{ab}}$ (like in the case of SU(2)), but are sums $U_{x,\nu}^{ab} = \rho_{x,\nu}^{ab} \, e^{i \alpha_{x,\nu}^{ab}} + \omega_{x,\nu}^{ab} \, e^{i \beta_{x,\nu}^{ab}}$. Therefore an additional step is still required in order to perform  analytically the Haar integration in (\ref{eq:qcd_partition4}). For those elements that are sums we make use of the binomial theorem $(x + y)^{N} = \sum_{m = 0}^{N} \binom{N}{m} x^{N-m} y^{m}$ and rewrite the integrand in (\ref{eq:qcd_partition4}) as
\begin{align}
	\nonumber
	&\left( U_{x,\nu}^{ab} \right) ^{N_{x,\nu}^{ab}} 
	\left( U_{x,\nu}^{ab \ \star} \right) ^{\overline{N}_{x,\nu}^{ab}} \, = \,
	\left( \rho_{x,\nu}^{ab}\,  e^{i \alpha_{x,\nu}^{ab}} 
	+ \omega_{x,\nu}^{ab} \, e^{i \beta_{x,\nu}^{ab}} \right)^{N_{x,\nu}^{ab}} 
	\left( \rho_{x,\nu}^{ab}\,  e^{-i \alpha_{x,\nu}^{ab}} 
	+ \omega_{x,\nu}^{ab} \, e^{-i \beta_{x,\nu}^{ab}} \right)^{\overline{N}_{x,\nu}^{ab}} \\ 
	& \hspace{1.5cm}= \, \sum_{m_{x,\nu}^{ab} = 0}^{N_{x,\nu}^{ab}} \sum_{\overline{m}_{x,\nu}^{ab} = 0}^{\overline{N}_{x,\nu}^{ab}} \binom{N^{ab}_{x,\nu}}{m^{ab}_{x,\nu}}
	\binom{\overline{N}^{ab}_{x,\nu}}{\overline{m}^{ab}_{x,\nu}} \left( \rho_{x,\nu}^{ab} \right)^{s_{x,\nu}^{ab}} \left( \omega_{x,\nu}^{ab} \right)^{S_{x,\nu}^{ab} - s_{x,\nu}^{ab}} e^{i \alpha_{x,\nu}^{ab} j_{x,\nu}^{ab}} \, e^{i \beta_{x,\nu}^{ab} \left(J_{x,\nu}^{ab} - j_{x,\nu}^{ab}\right)}
	\nonumber \\
	&\text{with} \quad m_{x,\nu}^{ab} = 0,1\dots N_{x,\nu}^{ab} \, , \qquad \overline{m}_{x,\nu}^{ab} = 0,1 \dots \overline{N}_{x,\nu}^{ab} \, , \nonumber \\
	&\text{and } \quad j_{x,\nu}^{ab} \equiv m_{x,\nu}^{ab} - \overline{m}_{x,\nu}^{ab} \, , \qquad \, \overline{s}_{x,\nu}^{ab} \equiv m_{x,\nu}^{ab} + \overline{m}_{x,\nu}^{ab} \, . 
\label{eq:qcd_binomial}
\end{align}
The new auxiliary variables $m_{x,\nu}^{ab}$ and $\overline{m}_{x,\nu}^{ab}$ that we use for the binomial decomposition (\ref{eq:qcd_binomial}) of the matrix elements with $(a,b) = (2,1),(2,3),(3,1)$ and $(3,3)$ live on the links of the lattice. 
	
To obtain the final result for the partition function we substitute the parametrization (\ref{eq:qcd_parametrization}) and the Haar measure (\ref{eq:qcd_haarmeasure}) in (\ref{eq:qcd_partition4}), and we use the binomial decomposition (\ref{eq:qcd_binomial}). For the partition function we obtain
\begin{align}
\nonumber
	&Z \; = \; 2^{4V} \sum_{\{p,l\}} \sum_{\{m,\overline{m}\}} \! 
	\Bigg[ \prod_{x,\nu<\rho} \prod_{a,b,c,d} 
	\dfrac{ \left( \beta/2 \right)^{|p_{x,\nu\rho}^{abcd}| + 2l_{x,\nu\rho}^{abcd}}}{\left(|p_{x,\nu\rho}^{abcd}| + l_{x,\nu\rho}^{abcd}\right)!l_{x,\nu\rho}^{abcd}!} \Bigg] 
	\Bigg[ \prod_{x,\nu} (-1)^{S_{x,\nu}^{23} + S_{x,\nu}^{31} + s_{x,\nu}^{21} + s_{x,\nu}^{33}} \Bigg]
	\\
	& \hspace{28mm} \times \; \Bigg[
	\prod_{x,\nu} \prod_{a = 2,3} \prod_{b=1,3}
	\binom{N_{x,\nu}^{ab}}{m_{x,\nu}^{ab}} \binom{\overline{N}_{x,\nu}^{ab}}{\overline{m}_{x,\nu}^{ab}} \Bigg]
	\nonumber
	\\
	&  \hspace{23mm} \times  \prod_{x,\nu}  
	2 \! \int_{0}^{\pi/2} \!\!\!\!  d\theta_{x,\nu}^{(1)} \, (\cos\theta_{x,\nu}^{(1)})^{3 + S_{x,\nu}^{11} + S_{x,\nu}^{13} + S_{x,\nu}^{22} + S_{x,\nu}^{32}}
	\nonumber
	\\
	&  \hspace{53mm} 
	(\sin\theta_{x,\nu}^{(1)})^{1 + S_{x,\nu}^{12} + s_{x,\nu}^{21} +s_{x,\nu}^{23} + s_{x,\nu}^{31} + s_{x,\nu}^{33}} 
	\nonumber
	\\
	& \hspace{30mm} \times \!  
	2 \! \int_{0}^{\pi/2} \!\!\!\!  d\theta_{x,\nu}^{(2)} \, (\cos\theta_{x,\nu}^{(2)})^{1 + S_{x,\nu}^{11} + s_{x,\nu}^{21} + S_{x,\nu}^{23} - s_{x,\nu}^{23} + s_{x,\nu}^{31} + S_{x,\nu}^{33} - s_{x,\nu}^{33}} 
	\nonumber
	\\
	& \hspace{53mm} (\sin\theta_{x,\nu}^{(2)})^{1 + S_{x,\nu}^{13} + S_{x,\nu}^{21} - s_{x,\nu}^{21} + s_{x,\nu}^{23} + S_{x,\nu}^{31} - s_{x,\nu}^{31} + s_{x,\nu}^{33}} 
	\nonumber
	\\
	& \hspace{30mm} \times \! 
	2 \! \int_{0}^{\pi/2} \!\!\!\!  d\theta_{x,\nu}^{(3)} \, (\cos\theta_{x,\nu}^{(3)})^{1 + s_{x,\nu}^{21} + S_{x,\nu}^{22} + s_{x,\nu}^{23} + S_{x,\nu}^{31} - s_{x,\nu}^{31} + S_{x,\nu}^{33} - s_{x,\nu}^{33}} 
	\nonumber
	\\
	& \hspace{53mm} (\sin\theta_{x,\nu}^{(3)})^{1 + S_{x,\nu}^{21} - s_{x,\nu}^{21} + S_{x,\nu}^{23} - s_{x,\nu}^{23} + s_{x,\nu}^{31} + S_{x,\nu}^{32} + s_{x,\nu}^{33}} 
	\nonumber
	\\
	&  \hspace{31mm} \times \! 
	\int_{0}^{2\pi} \! \dfrac{d\phi_{x,\nu}^{(1)}}{2\pi} \; e^{i\phi_{x,\nu}^{(1)}[J_{x,\nu}^{11}-J_{x,\nu}^{23}-J_{x,\nu}^{33}+j_{x,\nu}^{21}+j_{x,\nu}^{23}+j_{x,\nu}^{31}+j_{x,\nu}^{33}]}
	\nonumber
	\\
	&  \hspace{31mm} \times \! 
	\int_{0}^{2\pi} \! \dfrac{d\phi_{x,\nu}^{(2)}}{2\pi} \; e^{i\phi_{x,\nu}^{(2)}[J_{x,\nu}^{22}-J_{x,\nu}^{31}-J_{x,\nu}^{33}+j_{x,\nu}^{21}+j_{x,\nu}^{23}+j_{x,\nu}^{31}+j_{x,\nu}^{33}]} \; 
	\nonumber
	\\
	&  \hspace{31mm} \times \! 
	\int_{0}^{2\pi} \! \dfrac{d\phi_{x,\nu}^{(3)}}{2\pi} \; e^{i\phi_{x,\nu}^{(3)}[J_{x,\nu}^{12}-j_{x,\nu}^{21}-j_{x,\nu}^{23}-j_{x,\nu}^{31}-j_{x,\nu}^{33}]}
	\nonumber
	\\
	&  \hspace{31mm} \times \! 
	\int_{0}^{2\pi} \! \dfrac{d\phi_{x,\nu}^{(4)}}{2\pi} \; e^{i\phi_{x,\nu}^{(4)}[J_{x,\nu}^{13}-J_{x,\nu}^{21}-J_{x,\nu}^{31}+j_{x,\nu}^{21}+j_{x,\nu}^{23}+j_{x,\nu}^{31}+j_{x,\nu}^{33}]} \; 
	\nonumber
	\\
	&  \hspace{31mm} \times \! 
	\int_{0}^{2\pi} \! \dfrac{d\phi_{x,\nu}^{(5)}}{2\pi} \; e^{i\phi_{x,\nu}^{(5)}[J_{x,\nu}^{32}-J_{x,\nu}^{21}-J_{x,\nu}^{23}+j_{x,\nu}^{21}+j_{x,\nu}^{23}+j_{x,\nu}^{31}+j_{x,\nu}^{33}]} \; ,
\label{eq:qcd_partition5}
\end{align} 
where we introduced the short hand notations 
\begin{equation*}
	\sum_{\{p\}} \, = \, \prod_{x,\nu<\rho} \, \prod_{a,b,c,d = 1}^{3} \, \sum_{p_{x,\nu\rho}^{abcd} = -\infty}^{\infty} \, , \qquad  \sum_{\{l\}} \, = \, \prod_{x,\nu<\rho} \, \prod_{a,b,c,d = 1}^{3} \, \sum_{l_{x,\nu\rho}^{abcd} = 0}^{\infty} \, ,
\end{equation*} 
to express the sums over configurations of the cycle occupation numbers $p_{x,\nu\rho}^{abcd} \in \mathbb{Z}$ and the auxiliary plaquette variables $l_{x,\nu\rho}^{abcd} \in \mathbb{N}_{0}$, and
\begin{equation}
	\sum_{\{m, \overline{m}\}} \, = \, \prod_{x,\nu} \, \prod_{a=2,3} \, \prod_{b=1,3} \, 
	\sum_{m_{x,\nu}^{ab} = 0}^{N_{x,\nu}^{ab}} \,
	\sum_{\overline{m}_{x,\nu}^{ab} = 0}^{\overline{N}_{x,\nu}^{ab}} \, ,
\end{equation}
for the sums over the configurations of the link-based auxiliary variables $m_{x,\nu}^{ab}$ and $\overline{m}_{x,\nu}^{ab}$ used in the binomial decomposition (\ref{eq:qcd_binomial}).
		
The remarkable outcome of our approach is that, written as in (\ref{eq:qcd_partition5}), the Haar measure integrals can be solved in closed form. More precisely, the integrals over the $\theta_{x,\nu}^{(i)}$ angles $i=1,2,3$ give rise to beta functions \cite{Olver:2010:NHM:1830479}
\begin{equation}
\label{eq:qcd_betafunction}
	2 \int_{0}^{\pi/2} \! d\theta (\cos \theta)^{n + 1} (\sin \theta)^{m + 1} = \B\left(\dfrac{n}{2} + 1\right| \left. \dfrac{m}{2} + 1\right) \, ,
\end{equation}
whereas the integrals over the phases $\phi_{x,\nu}^{(j)}$ $j=1,2,\dots,5$ give rise to Kronecker deltas (we use the notation $\delta(n) \equiv \delta_{n,0}$), which impose constraints on the dual variables.

Performing the gauge field integration we find
\begin{equation}
\label{eq:qcd_dualpartitiong}
	Z \, = \, \sum_{\{p\}} \, W_{G}[p] \, C_{G}[p] \, .
\end{equation}
The partition function is a sum over the configurations of the cycle occupation numbers $p_{x,\nu\rho}^{abcd}\in \mathbb{Z}$. The factor $C_{G}[p]$ collects the four constraints which arise from the integration of the four phases $\phi_{x,\nu}^{(j)}$, $j=1,2,4,5$: 
\begin{align}			
	C_G[p] \; =& \; \prod_{x,\nu} 
	\delta(J_{x,\nu}^{11} + J_{x,\nu}^{12} - J_{x,\nu}^{33} - J_{x,\nu}^{23}) \;
	\delta(J_{x,\nu}^{22} + J_{x,\nu}^{12} - J_{x,\nu}^{33} - J_{x,\nu}^{31}) 
	\nonumber \\ 
	& \hspace{3mm}\times  	
	\delta(J_{x,\nu}^{13} + J_{x,\nu}^{12} - J_{x,\nu}^{31} - J_{x,\nu}^{21}) \; 
	\delta(J_{x,\nu}^{32} + J_{x,\nu}^{12} - J_{x,\nu}^{23} - J_{x,\nu}^{21}) \; .
	\label{eq:qcd_cg}
\end{align}
These Kronecker deltas enforce relations between different color components of the fluxes $J_{x,\nu}^{ab}$ (\ref{eq:qcd_Jfluxes}) at every link $(x,\nu)$, thus limiting the number of admissible configurations $\{p\}$ of the cycle occupation numbers $p_{x,\mu\rho}^{abcd}$. In Eq.~(\ref{eq:qcd_cg}) we have already taken into account another constraint, which origins from the $\phi_{x,\nu}^{(3)}$ integral in (\ref{eq:qcd_partition5}):
\begin{equation}
\label{eq:qcd_constrain}
	j_{x,\nu}^{21} + j_{x,\nu}^{23} + j_{x,\nu}^{31} + j_{x,\nu}^{33} \, = \, J_{x,\nu}^{12} \, .
\end{equation}
This constraint relates $J_{x,\nu}^{12}$ to the auxiliary fluxes $j_{x,\nu}^{ab} = m_{x,\nu}^{ab} - \overline{m}_{x,\nu}^{ab}$ for the variables $m_{x,\nu}^{ab}$ and $\overline{m}_{x,\nu}^{ab}$, introduced in (\ref{eq:qcd_binomial}) for the binomial decomposition of the $(2,1)$, $(2,3)$, $(3,1)$ and $(3,3)$ matrix elements of the parametrization (\ref{eq:qcd_parametrization}). To obtain (\ref{eq:qcd_cg}) we used (\ref{eq:qcd_constrain}) to substitute the sum $j_{x,\nu}^{21} + j_{x,\nu}^{23} + j_{x,\nu}^{31} + j_{x,\nu}^{33}$ with $J_{x,\nu}^{21}$ in the integrals over $\phi_{x,\nu}^{(j)}$, $j=1,2,4,5$ in (\ref{eq:qcd_partition5}). We included the resulting four Kronecker deltas in the expression (\ref{eq:qcd_cg}) for the gauge constraint $C_{G}[p]$, while (\ref{eq:qcd_constrain}) is incorporated in the weight $W_{G}[p]$.

The weight factor $W_{G}[p]$ is itself a sum $\sum_{\{l,m,\overline{m}\}}$ over the auxiliary plaquette variables $l_{x,\nu\rho}^{abcd}\in \mathbb{N}_{0}$ and the auxiliary link variables $m_{x,\nu}^{ab} \in \{0,1 \dots N_{x,\nu}^{ab} \}$ and $\overline{m}_{x,\nu}^{ab} \in \{0,1 \dots \overline{N}_{x,\nu}^{ab} \}$:
\small
\begin{align}
	&W_{G} [p] \, = \, 2^{4V} \!\! \sum_{\{l,m,\overline{m}\}} 
	\Bigg[\prod_{x,\nu} \delta(J_{x,\nu}^{12} - j_{x,\nu}^{21} - j_{x,\nu}^{23} - j_{x,\nu}^{31} - j_{x,\nu}^{33})\Bigg]
	\Bigg[\prod_{x,\nu} (-1)^{J_{x,\nu}^{12} + J_{x,\nu}^{23} + J_{x,\nu}^{31} - j_{x,\nu}^{23} - j_{x,\nu}^{31}}  \Bigg] \nonumber \\
	&\hspace{29mm} \times
	\Bigg[ \prod_{x,\nu} \prod_{a=2,3} \prod_{b=1,3}
	\binom{N_{x,\nu}^{ab}}{m_{x,\nu}^{ab}} \binom{\overline{N}_{x,\nu}^{ab}}{\overline{m}_{x,\nu}^{ab}}\Bigg] 
	\Bigg[\prod_{x,\nu<\rho} \prod_{a,b,c,d} 
	\dfrac{ \left( \beta/2 \right)^{|p_{x,\nu\rho}^{abcd}| + 2l_{x,\nu\rho}^{abcd}}}{\left(|p_{x,\nu\rho}^{abcd}| + l_{x,\nu\rho}^{abcd}\right)!l_{x,\nu\rho}^{abcd}!}\Bigg] \nonumber \\
	&\!\times \!\Bigg[ \prod_{x,\nu} \B\left(\dfrac{S_{x,\nu}^{11} + S_{x,\nu}^{13} + S_{x,\nu}^{22} + S_{x,\nu}^{32}}{2} + 2\right.\left|\dfrac{S_{x,\nu}^{12} + s_{x,\nu}^{21} +s_{x,\nu}^{23} + s_{x,\nu}^{31} + s_{x,\nu}^{33}}{2} + 1\right)
	\nonumber \\
	&\!\times\!\! \B\!\left(\!\!\dfrac{S_{x,\nu}^{11} \! + \! s^{21}_{x,\rho} \! + \! S_{x,\nu}^{23}\! -\! s_{x,\nu}^{23}\! + s_{x,\nu}^{31}\! + \!S_{x,\nu}^{33}\! - \!s_{x,\nu}^{33}}{2} \!+\! 1\right.\!\!\left|\dfrac{\!S_{x,\nu}^{13}\! +\! S_{x,\nu}^{21}\! - \!s_{x,\nu}^{21}\! +\! s_{x,\nu}^{23} \!+\! S_{x,\nu}^{31}\! -\! s_{x,\nu}^{31}\!  +\! s_{x,\nu}^{33}}{2}\! + \!\!1\!\right) \nonumber
	\\
	&\times\!\! \B\!\left(\!\!\dfrac{s_{x,\nu}^{21} \!+ \!S^{22}_{x,\rho}\! + \!s_{x,\nu}^{23} \!+\! S_{x,\nu}^{31}\! -\! s_{x,\nu}^{31}\! +\! S_{x,\nu}^{33}\! -\! s_{x,\nu}^{33}}{2}\! +\! 1\right.\!\!\left|\dfrac{\!S_{x,\nu}^{21} \!- \!s_{x,\nu}^{21} \!+ \!S_{x,\nu}^{23} \!- \!s_{x,\nu}^{23} \!+ \!s_{x,\nu}^{31} \!+ \!S_{x,\nu}^{32} \!+ \!s_{x,\nu}^{33}}{2}\! \!+\! 1\! \!\right) \!\!\!\Bigg].
\label{eq:qcd_wg}
\end{align}
\normalsize
The configurations of the auxiliary link variables $m_{x,\nu}^{ab}$ and $\overline{m}_{x,\nu}^{ab}$ are constrained by the value of $J_{x,\nu}^{12}$ at every link $(x,\nu)$, as expressed by the Kronecker delta in the first line in (\ref{eq:qcd_wg}). The two factors in the second line in (\ref{eq:qcd_wg}) are the weights arising from the binomial decomposition and the Taylor expansion respectively. Finally, the beta functions arise from the $\theta_{x,\nu}^{(j)}$, $j=1,2,3$, integrals in (\ref{eq:qcd_partition5}). Notice that in (\ref{eq:qcd_wg}) there is an explicit sign factor. It origins from the minus signs in the parametrization (\ref{eq:qcd_parametrization}) of the SU(3) group elements.

\begin{figure}
	\begin{center}
		\vskip5mm
		\includegraphics[width=15cm]{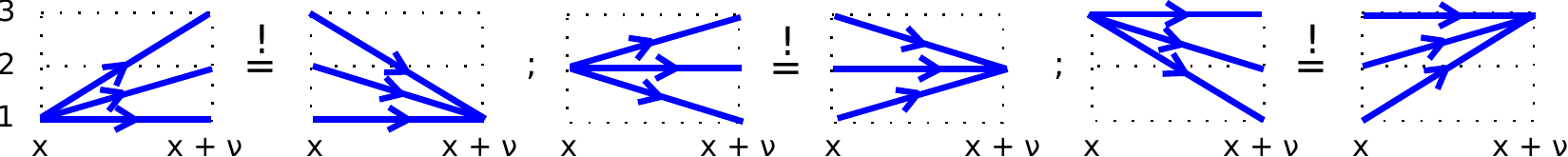}\\ \vskip10mm
		\includegraphics[width=8cm]{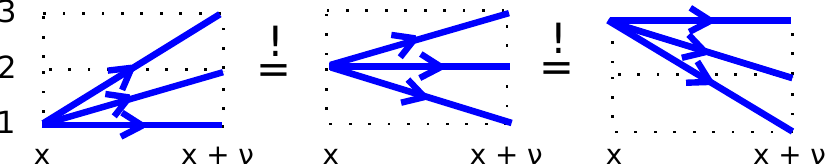}%
	\end{center}
	\caption{Schematic representation of the three \textit{color conservation constraints} (\ref{eq:qcd_fluxconservation1}) -- (\ref{eq:qcd_fluxconservation3}) (the three plots at the top) and the \textit{color exchange constraint} (\ref{eq:qcd_fluxexchange}) (bottom plot). They impose relations between the fluxes $J_{x,\nu}^{ab}$ and admissible configurations of the cycle	occupation numbers $p_{x,\nu\rho}^{abcd}$ have to respect these constraints. \label{fig:qcd_constraints}}
\end{figure}

Let us now spend a few words discussing the gauge constraints in Eq.~(\ref{eq:qcd_cg}). Understanding the constraints is of crucial importance because, as we mentioned already several times in this thesis, they are the primary manifestation of the symmetry of the original theory in the dual representation. Explicitly the Kronecker deltas in (\ref{eq:qcd_cg}) require the following equalities to hold at every link $(x,\nu)$ of the lattice:
\begin{gather}
\label{eq:qcd_constraint1}
	J_{x,\nu}^{11} + J_{x,\nu}^{12} = J_{x,\nu}^{23} + J_{x,\nu}^{33} \; ,\\
\label{eq:qcd_constraint2}
	J_{x,\nu}^{22} + J_{x,\nu}^{12} = J_{x,\nu}^{33} + J_{x,\nu}^{31} \; ,\\
\label{eq:qcd_constraint4}
	J_{x,\nu}^{13} + J_{x,\nu}^{12} = J_{x,\nu}^{31} + J_{x,\nu}^{21} \; ,\\
\label{eq:qcd_constraint5}
	J_{x,\nu}^{23} + J_{x,\nu}^{21} =  J_{x,\nu}^{32} + J_{x,\nu}^{12}\; .
\end{gather} 
The requirements imposed by these constraints can be better understood by using linear combinations of (\ref{eq:qcd_constraint1})--(\ref{eq:qcd_constraint5}). So, for example, if we add $J_{x,\nu}^{11}$ on both sides of Eq.~(\ref{eq:qcd_constraint4}) we obtain
\begin{equation*}
	J_{x,\nu}^{11} + J_{x,\nu}^{12} + J_{x,\nu}^{13} \, = \, J_{x,\nu}^{11} + J_{x,\nu}^{21} + J_{x,\nu}^{31} \, .
\end{equation*}
This relation is represented schematically in the top left plot of Fig~\ref{fig:qcd_constraints}. From the figure it is easy to see that the constraint requires the sum of all the fluxes coming out of color 1 at site $x$ on the link $(x,\nu)$ to be equal to the sum of all the color fluxes going into color 1 at site $x + \hat{\nu}$. Therefore, we can interpret this relation as a flux conservation constraint for the color layer 1. The analogous relation for color 2 can be obtained by adding $J_{x,\nu}^{22}$ on both sides of Eq.~(\ref{eq:qcd_constraint5}), while adding Eq.~(\ref{eq:qcd_constraint4}) to Eq.~(\ref{eq:qcd_constraint5}) and then $J_{x,\nu}^{33}$ on both sides of the resulting equation gives the relation for color 3. Thus we obtain the following three constraints
\begin{gather}
\label{eq:qcd_fluxconservation1}
	J_{x,\nu}^{11} + J_{x,\nu}^{12} + J_{x,\nu}^{13} \, = \, J_{x,\nu}^{11} + J_{x,\nu}^{21} + J_{x,\nu}^{31} \, , \\
\label{eq:qcd_fluxconservation2}
	J_{x,\nu}^{22} + J_{x,\nu}^{21} + J_{x,\nu}^{23} \, = \, J_{x,\nu}^{22} + J_{x,\nu}^{12} + J_{x,\nu}^{32} \, , \\
\label{eq:qcd_fluxconservation3}
	J_{x,\nu}^{33} + J_{x,\nu}^{32} + J_{x,\nu}^{31} \, = \, J_{x,\nu}^{33} + J_{x,\nu}^{23} + J_{x,\nu}^{13} \, ,
\end{gather}
which we graphically illustrate in the plots in the first line of Fig.~\ref{fig:qcd_constraints}. Since (\ref{eq:qcd_fluxconservation1}) -- (\ref{eq:qcd_fluxconservation3}) impose the conservation of fluxes going through each of the 3 color layers at every link, we refer to them as \textit{color flux conservation} constraints. 

Another set of constraints that results from linearly combining Eqs.~(\ref{eq:qcd_constraint1})--(\ref{eq:qcd_constraint5}) is:
\begin{equation}
	\label{eq:qcd_fluxexchange}
	J_{x,\nu}^{11} + J_{x,\nu}^{12} + J_{x,\nu}^{13} \, = \, J_{x,\nu}^{22} + J_{x,\nu}^{21} + J_{x,\nu}^{23} \,  = \,
	J_{x,\nu}^{33} + J_{x,\nu}^{32} + J_{x,\nu}^{31} \, . 
\end{equation}
The first equality in (\ref{eq:qcd_fluxexchange}) is obtained by adding Eq.~(\ref{eq:qcd_constraint1}) to Eq.~(\ref{eq:qcd_constraint4}) and using (\ref{eq:qcd_constraint2}) to substitute $J_{x,\nu}^{33} + J_{x,\nu}^{31}$ with $J_{x,\nu}^{22} + J_{x,\nu}^{21}$. The last is obtained by adding Eq.~(\ref{eq:qcd_constraint2}) to Eq.~(\ref{eq:qcd_constraint5}). The chain of equalities in (\ref{eq:qcd_fluxexchange}) requires the magnitude of the flux to be the same on each color layer at all links $(x,\nu)$. We refer to the set of constraints (\ref{eq:qcd_fluxexchange}) as \textit{color exchange} constraints, and we illustrate them in the plot at the bottom of Fig.~(\ref{fig:qcd_constraints}). We stress at this point that Eqs.~(\ref{eq:qcd_fluxconservation1}) -- (\ref{eq:qcd_fluxexchange}) are over-complete. Nonetheless, the gauge constraints $C_{G}[p]$ are easier to understand in the form (\ref{eq:qcd_fluxconservation1}) -- (\ref{eq:qcd_fluxexchange}).  

As a result of applying the ACC method to pure SU(3) gauge theory, we obtained a representation of the partition function where the dynamical degrees of freedom are integer valued. The cycle occupation numbers $p_{x,\mu\rho}^{abcd} \in \mathbb{Z}$, which are attached to plaquettes, have to satisfy the constraints encoded in the function $C_{G}[p]$. These constraints impose relations between the link color fluxes $J_{x,\nu}^{ab}$ generated by the cycle occupation numbers attached to the link $(x,\nu)$. We categorized the constraints into two sets: the color flux conservation constraints (\ref{eq:qcd_fluxconservation1}) -- (\ref{eq:qcd_fluxconservation3}), which force the flux going through a color to be the same at all sites, and the color flux exchange constraint (\ref{eq:qcd_fluxexchange}), which requires the fluxes going through the three colors to match. 

The admissible configurations, i.e., the configurations that obey the gauge constraint $C_{G}[p]$, come with the weight factor $W_{G}[p]$, whose explicit expression is given in (\ref{eq:qcd_wg}). $W_{G}[p]$ is itself a sum over the configurations of the auxiliary plaquette variables $l_{x,\nu\rho}^{abcd} \in \mathbb{N}_{0}$, and the auxiliary link variables $m_{x,\nu}^{ab} \in \{0,1\dots N_{x,\nu}^{ab}\}$ and $\overline{m}_{x,\nu}^{ab} \in \{0,1\dots \overline{N}_{x,\nu}^{ab}\}$, which come from the binomial decomposition of the matrix elements $(2,1)$, $(2,3)$, $(3,1)$ and $(3,3)$ of the parametrization (\ref{eq:qcd_parametrization}) of the SU(3) elements. The auxiliary plaquette variables $l_{x,\nu\rho}^{abcd} \in \mathbb{N}_{0}$ are unconstrained and act as a background field, while the link variables $m_{x,\nu}^{ab} \in \{0,1\dots N_{x,\nu}^{ab}\}$ and $\overline{m}_{x,\nu}^{ab} \in \{0,1\dots \overline{N}_{x,\nu}^{ab}\}$ are constrained by the value of the $J_{x,\nu}^{12}$ flux at every link. 

Admissible configurations of the cycle occupation numbers can be built in a simple way: one starts setting one cycle occupation number to one. This causes the constraints (\ref{eq:qcd_cg}) to be violated on all the four links that contour the plaquette. To obviate this problem one has two possibilities. The first one consists in activating at least two other cycle occupation numbers on the same plaquette, so that the constraints are fulfilled. However, the configurations obtained in this way are not relevant for the long range physics. Alternatively, one can activate a cycle occupation number on an adjacent plaquette, so that the constraints are satisfied on the common link. One must then proceed similarly for the other six links that contour the two plaquettes. For the resulting 2D surface the constraints will still be violated along the boundary, unless the surface is closed. Therefore, in our dual representation, the long range physics, which is relevant for the continuum limit, is described by closed \textit{worldsheets}. 

Before continuing with the discussion about fermions, we stress that in our dual representation (\ref{eq:qcd_dualpartitiong}) there is an explicit sign factor $(-1)^{\sum_{x,\mu}J_{x,\mu}^{12} + J_{x,\mu}^{23} + J_{x,\mu}^{31} - j_{x,\mu}^{23} - j_{x,\mu}^{31}}$. This sign factor origins from the explicit minus signs in the parametrization of the SU(3) matrices (\ref{eq:qcd_parametrization}). This implies that for a Monte Carlo simulation of the ACC dual form of the partition sum (\ref{eq:qcd_dualpartitiong}) a strategy for a partial resummation needs to be found.

\section{Abelian color flux dualization of fermions: the strong coupling limit \label{sec:qcd_sc}}

In this section we discuss the abelian color flux (ACF) dualization of fermions, focusing first on the strong coupling limit. In this limit $\beta = 0$, therefore the gauge action is absent and the continuum limit cannot be performed. Nevertheless, the strong coupling limit of lattice QCD shares some non-perturbative properties with QCD in the continuum, like confinement and the spontaneous breaking of chiral symmetry.

For our purposes the worldline representation of strong coupling QCD is interesting because, even if the constraints have the same structure of those for the gauge degrees of freedom, they are simpler to interpret thanks to an additional constraint coming from the Pauli principle. 

The fermionic partition function is given by
\begin{equation}
Z_{F}[U] = \int \! D[\overline{\psi},\psi] \, e^{-S_{F}[U,\psi,\overline{\psi}]} \, ,
\label{eq:qcd_partitionf}
\end{equation}
where $\psi_{x}$ and $\overline{\psi}_{x}$ are 3-component Grassmann vectors
\begin{equation}
\psi_{x} = \left(
\begin{array}{c}
\psi_{x}^{1} \\
\psi_{x}^{2} \\
\psi_{x}^{3} \\
\end{array} \right) \ ,\qquad
\overline{\psi}_{x} = \left( \overline{\psi}_{x}^{1} \ \ \overline{\psi}_{x}^{2} \ \ \overline{\psi}_{x}^{3} \right) \, ,
\label{eq:qcd_grassmann}
\end{equation}
that live on the sites of the four-dimensional lattice, with anti-periodic boundary conditions in Euclidean time, i.e., $\nu = 4$ and periodic boundary conditions in the spatial directions, i.e., $\nu =1,2,3$. The measure is a product over Grassmann measures $\int \! D[\overline{\psi},\psi] = \prod_{x}\prod_{a = 1}^{3} \int d\psi_{x}^{a} d\overline{\psi}^{a}_{x}$. The staggered fermion action is given by
\begin{align}
\nonumber
	&S_{F}[U,\psi,\overline{\psi}] = \sum_{x}\biggl[ m \overline{\psi}_{x} \psi_{x} 
	+ \sum_{\nu} \dfrac{\gamma_{x,\nu}}{2} \left( \overline{\psi}_x U_{x,\nu} \psi_{x + \hat{\nu}}\, e^{\mu \delta_{\nu,4}} -\, \overline{\psi}_{x + \hat{\nu}} U_{x,\nu}^{\dagger} \psi_{x}\, e^{-\mu \delta_{\nu,4}} \right) \biggr]\\
	&= \sum_{x}\biggl[ m \sum_{a = 1}^{3} \overline{\psi}_{x}^{a} \psi_{x}^{a} 
	+ \sum_{\nu} \dfrac{\gamma_{x,\nu}}{2} \sum_{a, b = 1}^{3} \left( \overline{\psi}_{x}^{a} U_{x,\nu}^{ab} \psi_{x + \hat{\nu}}^{b}\, e^{\mu \delta_{\nu,4}} -\, \overline{\psi}_{x + \hat{\nu}}^{b} U_{x,\nu}^{ab \, \star} \psi_{x}^{a}\, e^{-\mu \delta_{\nu,4}} \right) \biggr] \, ,
\label{eq:qcd_actionf}
\end{align}
where $m$ is the fermion's mass, $\gamma_{x,\nu}$ are the staggered sign factors defined as
\begin{equation*}
	\gamma_{x,1} = 1 \, , \quad \gamma_{x,2} = (-1)^{x_{1}} \, , \quad \gamma_{x,3} = (-1)^{x_{1} + x_{2}} \, , \quad \gamma_{x,4} = (-1)^{x_{1} + x_{2} + x_{3}} \, ,
\end{equation*}
and $U_{x,\nu}$ are the gauge degrees of freedom, which live on the links $(x,\nu)$. To obtain the partition function in the strong coupling limit it is sufficient to integrate the fermionic partition function (\ref{eq:qcd_partitionf}):
\begin{equation*}
	Z = \int \! D[U] \, Z_{F}[U] \, .
\end{equation*}
The measure $\int \! D[U]$ is the product of SU(3) Haar measures on the links of the lattice, $\int \! D[U] = \prod_{x,\nu} \int_{SU(3)} \! dU_{x,\nu}$ whose explicit expression is given in (\ref{eq:qcd_haarmeasure}). 
Notice that in Eq.~(\ref{eq:qcd_actionf}) we introduced a chemical potential $\mu$, which couples to the temporal hopping terms in the canonical way. 

\begin{figure}
	\includegraphics[width=15cm]{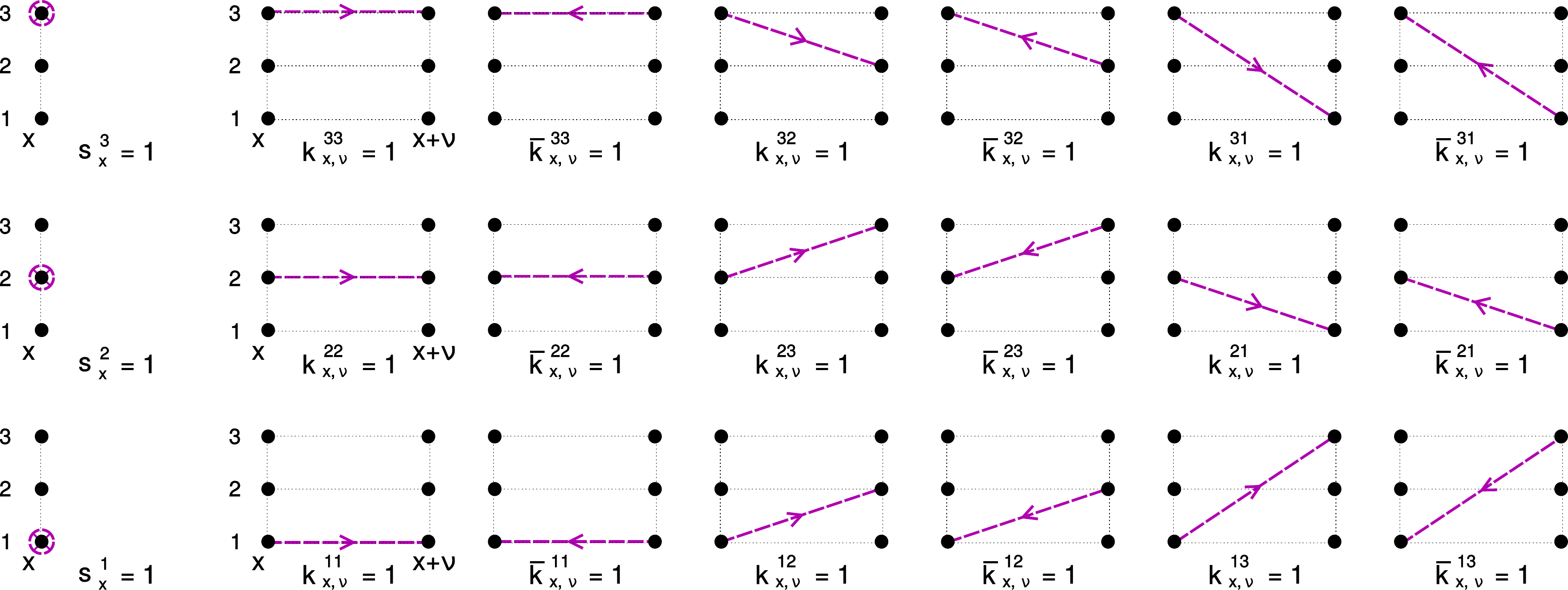}%
	\caption{Graphical representation of the fermion's dual variables. In the first column we show the monomers $s_{x}^{a}$, while the arrows represent the dual variables $k_{x,\nu}^{ab}$ and $\overline{k}_{x,\nu}^{ab}$ for the forward and backward hopping respectively. With these link variables it is possible to build dimers and oriented loops that, together with monomers constitute the admissible configurations for fermions. \label{fig:qcd_grassmann}}
\end{figure}

In the first line of Eq.~(\ref{eq:qcd_actionf}) we used matrix-vector notation for gauge links and fermions, while in the second line the sums over color indices are made explicit, using the color labels $a,b = 1,2,3$. In this decomposition all terms of the action are single Grassmann bilinears, which therefore commute. Once again the rewriting of the action into its minimal units is what allows us to proceed further with the dualization. For the fermionic partition function we obtain
\begin{align}
	&Z_{F}[U] =  \nonumber \\
	&= \int \! D[\overline{\psi},\psi] \prod_{x} \prod_{a = 1}^{3} e^{- m \overline{\psi}_{x}^{a} \psi_{x}^{a}} \prod_{x, \nu} \prod_{a, b = 1}^{3} e^{- \frac{\gamma_{x,\nu}}{2} \overline{\psi}_{x}^{a} U_{x,\nu}^{ab} \psi_{x + \hat{\nu}}^{b} e^{\mu \delta_{\nu,4}}} 
	e^{\frac{\gamma_{x,\nu}}{2} \overline{\psi}_{x + \hat{\nu}}^{b} U_{x,\nu}^{ab \, \star} \psi_{x}^{a} e^{-\mu \delta_{\nu,4}}} \nonumber \\
	&= \int \! D[\overline{\psi},\psi] \prod_{x} \prod_{a = 1}^{3} \sum_{s^a_{x} = 0}^{1} \left( m \overline{\psi}_{x}^{a} \psi_{x}^{a} \right)^{s_{x}^{a}}  \prod_{x, \nu} \prod_{a, b = 1}^{3} \sum_{k_{x,\nu}^{ab} = 0}^{1} \left(- \frac{\gamma_{x,\nu}}{2}\, \overline{\psi}_{x}^{a} U_{x,\nu}^{ab} \psi_{x + \hat{\nu}}^{b}\, e^{\mu \delta_{\nu,4}} \right)^{k_{x,\nu}^{ab}} 
	\nonumber\\
	&\hspace*{6.1cm} \times  \prod_{x, \nu} \prod_{a, b = 1}^{3} \sum_{\overline{k}_{x,\nu}^{ab} = 0}^{1} \left( \frac{\gamma_{x,\nu}}{2}\, \overline{\psi}_{x + \hat{\nu}}^{b} U_{x,\nu}^{ab \, \star} \psi_{x}^{a}\, e^{-\mu \delta_{\nu,4}} \right)^{\overline{k}_{x,\nu}^{ab}} \nonumber \\
	\nonumber
	&= \left(\dfrac{1}{2}\right)^{\!\!3V} \! \! \! \sum_{\{s,k,\overline{k}\}} (2m)^{\sum_{x,a} s_{x}^{a}} \, e^{\mu \sum_{x,ab} [k_{x,\hat{4}}^{ab} - \overline{k}_{x,\hat{4}}^{ab}]} 
	\prod_{x,\nu} \prod_{a, b} \left( U_{x,\nu}^{ab} \right)^{k_{x,\nu}^{ab}}
	\left( U_{x,\nu}^{ab \, \star} \right)^{\overline{k}_{x,\nu}^{ab}} \\
	& \times \prod_{x,\nu} \prod_{a, b} (-1)^{k_{x,\nu}^{ab}} (\gamma_{x,\nu})^{k_{x,\nu}^{ab} + \overline{k}_{x,\nu}^{ab}} \int \! D[\overline{\psi},\psi] \prod_{x,a} (\overline{\psi}_{x}^{a} \psi_{x}^{a})^{s_{x}^{a}}
	\prod_{x, \nu} \prod_{a, b} ( \overline{\psi}_{x}^{a}  \psi_{x + \hat{\nu}}^{b} )^{k_{x,\nu}^{ab}}
	( \overline{\psi}_{x + \hat{\nu}}^{b}  \psi_{x}^{a})^{\overline{k}_{x,\nu}^{ab}} \, .
\label{eq:qcd_partitionf2}
\end{align}
In the first step we rewrote the exponential of sums as products of exponentials. In the second step we then Taylor-expanded each of these exponentials. Notice that the Taylor series terminate after the linear term. This is due to the nilpotency of the Grassmann variables that causes any higher order of the series to vanish. The three expansion indices that we introduced, one for every bilinear of the action, are the new dual variables for fermions: $s_{x}^{a} = 0, 1$ is the dual variable corresponding to the color component $a$ of the mass term on site $x$, $k_{x,\nu}^{ab} = 0, 1$ represents the forward hop from color $a$ to color $b$ on the link $(x,\nu)$, and $\overline{k}_{x,\nu}^{ab} = 0, 1$ is the respective backward hop on the same link. 
The graphical representation of the dual variables for fermions is shown in Fig. \ref{fig:qcd_grassmann}. The monomer variables $s_{x}^{a}$ are illustrated as circles around the color component $a$ of the site $x$, while the forward and backward hops $k_{x,\nu}^{ab}, \overline{k}_{x,\nu}^{ab}$ are shown as oriented arrows connecting color $a$ and $b$ on the link $(x,\nu)$.

In the last step of Eq.~(\ref{eq:qcd_partitionf2}) we reorganized the terms: we collected an overall factor $(1/2)^{3V}$ and brought all factors independent of the Grassmann variables in front of the Grassmann integrals. We also introduced the notation 
\begin{equation}
	\sum_{\{s,k,\overline{k}\}} = \Bigg[\prod_{x,a} \sum_{s_{x}^{a} = 0}^{1} \Bigg] \Bigg[\prod_{x,\nu} \prod_{a,b} \sum_{k_{x,\nu}^{ab} = 0}^{1} \sum_{\overline{k}_{x,\nu}^{ab} = 0}^{1} \Bigg] \, ,
\end{equation}
to denote the sum over all the possible configurations of the fermion dual variables.

The Grassmann integral in the last line of (\ref{eq:qcd_partitionf2}) is non-vanishing only if each Grassmann variable $\overline{\psi}_{x}^{a}\psi_{x}^{a}$ appears exactly once. This condition, which implements Pauli's exclusion principle, can be expressed by means of the following fermion constraint 
\begin{equation}
\label{eq:qcd_cf}
C_{F}[s,k,\overline{k}] \, = \, \prod_{x,a} \delta \bigg( 1 - s_{x}^{a} - \dfrac{1}{2} \sum_{\nu,b} \big[k_{x,\nu}^{ab} + \overline{k}_{x,\nu}^{ab} + k_{x - \hat{\nu},\nu}^{ba} + \overline{k}_{x - \hat{\nu},\nu}^{ba} \big] \bigg) \, .
\end{equation} 
The configurations that satisfy the fermion constraint (\ref{eq:qcd_cf}) are the ones that completely saturate the four-dimensional lattice: the three color layers of every site must be either occupied by a monomer ($s_{x}^{a} = 1$), be the endpoint of a dimer ($k_{x,\nu}^{ab} = \overline{k}_{x,\nu}^{ab} = 1$), or run through by a loop (closed chains of $k_{x,\nu}^{ab} = 1$ and $\overline{k}_{x,\nu}^{ab} = 1$). Notice that Eq.~(\ref{eq:qcd_cf}) is the analogous of the triality constraint in \cite{Karsch:1988zx}.

When $C_{F}[s,k,\overline{k}]$ is fulfilled, the Grassmann integral in the last line of Eq.~(\ref{eq:qcd_partitionf2}) gives either $+1$ or $-1$, depending on the values of the dual variables $s_{x}^{a}$, $k_{x,\nu}^{ab}$ and $\overline{k}_{x,\nu}^{ab}$. Forward and backward hops also contribute with signs, since they activate the staggered sign factor $(\gamma_{x,\nu})^{k_{x,\nu}^{ab} + \overline{k}_{x,\nu}^{ab}}$, as well as the factor $(-1)^{k_{x,\nu}^{ab}}$. Moreover, loops winding in the temporal direction also produce signs, caused by the anti-periodic boundary conditions for fermions. Because of the multitude of sign sources, we now discuss in detail the sign contributions coming from the monomers, the dimers and the loops. 

Monomers $\overline{\psi}_{x}^{a} \psi_{x}^{a}$ are activated by setting $s_{x}^{a} = 1$, which leads to the contribution of a factor $2m$. Furthermore, the Grassmann variables $\overline{\psi}_{x}^{a} \psi_{x}^{a}$ are already in the canonical order for the Grassmann integral and do not introduce any signs. Hence, we conclude that monomers are sign free objects in our representation.

Dimers are built by setting $k_{x,\nu}^{ab} = \overline{k}_{x,\nu}^{ab} = 1$. This corresponds to activating the factor
\begin{equation}
\label{eq:qcd_dimer}
\overline{\psi}_{x}^{a}\psi_{x + \hat{\nu}}^{b} \overline{\psi}_{x + \hat{\nu}}^{b}\psi_{x}^{a} \, =
\, - \overline{\psi}_{x}^{a} \psi_{x}^{a}  \overline{\psi}_{x + \hat{\nu}}^{b} \psi_{x + \hat{\nu}}^{b}
\end{equation}
in the Grassmann integral. The minus sign on the right hand-side of (\ref{eq:qcd_dimer}) results from the reordering of the Grassmann variables. Anyway it is compensated by the explicit minus sign arising from the activation of the forward hop $(-1)^{k_{x,\nu}^{ab}} = (-1)^{1} = -1$. Finally, the staggered sign factor for dimers is always positive, i.e., $(\gamma_{x,\nu})^{k_{x,\nu}^{ab} + \overline{k}_{x,\nu}^{ab}} = (\gamma_{x,\nu})^{2} = 1$, therefore also dimers do not contribute with negative signs.

Loops are built by setting closed chains of $k_{x,\nu}^{ab}$ and $\overline{k}_{x,\nu}^{ab}$ variables to 1. Each loop $\mathcal{L}$ picks up an overall minus sign from the necessary interchanges to bring the Grassmann variables into the canonical order for the integration. Moreover, each forward hop of the loop will contribute with a minus sign. Hence, for trivially closing loops, if $|\mathcal{L}|$ is the length of the loop $\mathcal{L}$, the sign coming from the forward hops is $(-1)^{|\mathcal{L}|/2}$. On the other hand, the forward hops sign for loops that wind in one of the spatial directions $(\nu = 1,2,3)$ or in the temporal direction $(\nu = 4)$ depends on the size $N_{\nu}$ of the lattice in that direction. We choose to restrict our attention to lattices in which $N_{\nu}$ are multiples of 4, so that we do not need to distinguish different cases. Loops winding in the temporal direction pick up an additional minus sign because of the anti-periodic boundary conditions. We denote this sign as $(-1)^{W_{\mathcal{L}}}$, where $W_{\mathcal{L}}$ is the net number of windings of the loop $\mathcal{L}$ around the compactified time. 

Finally we have to work out the contributions from the staggered sign factors. Let us first consider a loop closing around a single plaquette $(x,\nu\rho)$ with $\nu < \rho$. For this simple case the staggered sign is given by
\begin{equation}
\label{eq:qcd_staggered}
	\gamma_{x,\nu} \gamma_{x + \hat{\nu},\rho} \gamma_{x + \hat{\rho},\nu} \gamma_{x,\rho} = (-1)^{\sum_{i=1}^{\nu-1} x_{i}} (-1)^{\sum_{i=1}^{\rho-1} x_{i} + 1} (-1)^{\sum_{i=1}^{\nu-1} x_{i}} (-1)^{\sum_{i=1}^{\rho-1}x_{i}}= -1 \,,
\end{equation}
where we used the compact definition of the staggered sign 
\begin{equation*}
	\gamma_{x,\nu} = (-1)^{\sum_{i=1}^{\nu-1} x_{i}}\, , \qquad x = (x_1, \ x_2, \ x_3, x_4) \, .
\end{equation*}
The result (\ref{eq:qcd_staggered}) holds independently from the position $x$, and from the plane $\nu,\rho$ on which the plaquette lies. 
We can then elongate the loop by one plaquette. The product of the six staggered signs corresponding to the six links of the loop can equivalently be computed by multiplying the staggered signs for the two plaquettes bounded by the loop, i.e., $(-1)(-1) = +1$. The equivalence holds because in the product the staggered sign on the link common to the two plaquettes gets squared. Obviously this procedure can be iterated to build loops of any shape, and the staggered sign factor can be expressed as $(-1)^{P_\mathcal{L}}$, where $P_\mathcal{L}$ is the number of plaquettes necessary to cover the surface bounded by the loop $\mathcal{L}$. Since we have three layers of colors, the admissible configurations may also contain loops that wind around the same contour up to three times (see Fig.~\ref{fig:qcd_loop} for a simple example of such loop). For these cases we need a multiple covered surface (e.g., a surface covered 3 times for the example in the bottom plot of Fig.~\ref{fig:qcd_loop}) and the total number $P_{\mathcal{L}}$ of plaquettes in the surface spanned by the loop is understood in the sense that it also takes into account multiple coverings. We finally remark that, even if the surface that has the loop $\mathcal{L}$ as its boundary is not unique, those surfaces that share the same boundary always differ by an even number of plaquettes. Hence, the result $(-1)^{P_\mathcal{L}}$ is valid independently from the surface chosen to compute $P_\mathcal{L}$.

Summarizing we found that in our representation, while monomers and dimers are sign free configurations, loops come with non-trivial signs which can be expressed as:
\begin{equation}
\label{eq:qcd_signl}
	\sign(\mathcal{L}) = (-1)^{1 + |\mathcal{L}|/2 + P_\mathcal{L} + W_\mathcal{L}} \, ,
\end{equation}
where $|\mathcal{L}|$ is the length of the loop $\mathcal{L}$, $P_\mathcal{L}$ is the number of plaquettes necessary to cover the surface bounded by the loop $\mathcal{L}$ and $W_\mathcal{L}$ is the net winding number in the compactified temporal direction. 

To obtain the full partition sum at strong coupling $Z = \int \! D[U] Z_{F}[U]$ we still have to integrate the fermionic partition function $Z_{F}[U]$ over the product of SU(3) Haar measures. Putting things together the partition function reads
\begin{equation}
	Z = \sum_{\{s,k,\overline{k}\}} C_{F}[s,k,\overline{k}] \, W_{F}[s,k,\overline{k}] \int \! D[U] \prod_{x, \nu} \prod_{a, b} 	\left( U_{x,\nu}^{ab} \right)^{k_{x,\nu}^{ab}}
	\left( U_{x,\nu}^{ab \, \star} \right)^{\overline{k}_{x,\nu}^{ab}} \, .
\label{eq:qcd_partitionsc}
\end{equation}
$C_{F}[s,k,\overline{k}]$ is the fermion constraint given in Eq.~(\ref{eq:qcd_cf}) and $W_{F}[s,k,\overline{k}]$ is the weight for the fermion configurations defined as: 
\begin{align}
	W_{F}[s,k,\overline{k}] &= \dfrac{1}{2^{3V}} \prod_{\mathcal{L}} \sign (\mathcal{L}) \prod_{x} \bigg[\prod_{a} (2m)^{s_{x}^{a}}\bigg] \bigg[\prod_{ab} e^{\mu[k_{x,4}^{ab} - \overline{k}_{x,4}^{ab}]}\bigg] \nonumber \\
	 &= \dfrac{1}{2^{3V}} \prod_{\mathcal{L}} \sign (\mathcal{L}) \, e^{\mu \beta W_{\mathcal{L}}} \bigg[\prod_{x,a} (2m)^{s_{x}^{a}}\bigg] \, .
\label{eq:qcd_wf}
\end{align}
The product $\prod_{\mathcal{L}}$ runs over all loops $\mathcal{L}$ that are formed by the $k_{x,\nu}^{ab}$ and $\overline{k}_{x,\nu}^{ab}$ variables.
In the second step we reorganized the factor that gives the $\mu$-dependence. The chemical potential $\mu$ multiplies the difference $k_{x,4}^{ab} - \overline{k}_{x,4}^{ab}$ between the forward and backward hops on the links in the temporal direction. That quantity is obviously vanishing for dimers, since they are activated by setting $k_{x,\nu}^{ab} = \overline{k}_{x,\nu}^{ab} = 1$. Also trivially closing loop do not contribute to the difference $k_{x,4}^{ab} - \overline{k}_{x,4}^{ab}$, since they have the same amount of forward and backward hops. Therefore, only loops that wind in the temporal direction couple with the chemical potential. For them
\begin{equation*}
	\sum_{x\in \mathcal{L}} \sum_{a,b} [k_{x,4}^{ab} - \overline{k}_{x,4}^{ab}] = N_{t} W_{\mathcal{L}} = \beta W_{\mathcal{L}}\, ,
\end{equation*}
where $N_{t} = \beta$ is the temporal extent of the lattice and $W_{\mathcal{L}}$ is the temporal winding number of the loop $\mathcal{L}$. The chemical potential then enhances loops winding around the temporal direction with positive orientation $(W_{\mathcal{L}} > 0)$, while it suppresses loops that have a negative orientation of the winding $(W_{\mathcal{L}} < 0)$. If we then compare the $\mu$-dependence in the last line of (\ref{eq:qcd_wf}) with the usual form $e^{\mu \beta \mathcal{N}}$ for the coupling of the chemical potential with the net-particle number $\mathcal{N}$, it is straightforward to identify the equality
\begin{equation}
	\sum_{\mathcal{L}} W_{\mathcal{L}} \, = \, \mathcal{N} \, .
\label{eq:qcd_wl}
\end{equation} 
The geometrical interpretation (\ref{eq:qcd_wl}) of the net-particle number $\mathcal{N}$ as the total temporal net-winding number of all fermion loops $\sum_{\mathcal{L}} W_{\mathcal{L}}$ is one of the most beautiful features of our worldline formulation of QCD. A clear advantage coming from (\ref{eq:qcd_wl}) is that in the dual representation the net-particle number is easily determined as a topological quantity, i.e., the total net-winding number for all the loops of a configuration. On the contrary, in the conventional representation the net-particle number can be quite challenging to compute, since it is given by the discretized integral over the zero component of the conserved vector current. Moreover, the interpretation (\ref{eq:qcd_wl}) opens the door to the implementation of simulations of the canonical ensemble \cite{Orasch:2017niz,Giuliani:2017qeo}.

To obtain the final result for the strong coupling partition sum $Z$, we still have to perform the integrals over the SU(3) Haar measures in (\ref{eq:qcd_partitionsc}). Similarly to what we have done in the last section, we insert the explicit expressions (\ref{eq:qcd_parametrization}) and (\ref{eq:qcd_haarmeasure}) for the matrix elements $U_{x,\nu}^{ab}$ and the path integral measure $D[U]$ in the integral in (\ref{eq:qcd_partitionsc}). For the matrix elements $(2,1)$, $(2,3)$, $(3,1)$ and $(3,3)$, which are sums of complex numbers, we again use the binomial decomposition (\ref{eq:qcd_binomial}). In this way we introduce the link variables $m_{x,\nu}^{ab} \in \{0, k_{x,\nu}^{ab}\}$ and $\overline{m}_{x,\nu}^{ab} \in \{0, \overline{k}_{x,\nu}^{ab}\}$. Since $k_{x,\nu}^{ab}, \overline{k}_{x,\nu}^{ab}, m_{x,\nu}^{ab},\overline{m}_{x,\nu}^{ab} \in \{0,1\}$, all binomial factors $\binom{k_{x,\nu}}{m_{x,\nu}}, \binom{\overline{k}_{x,\nu}}{\overline{m}_{x,\nu}}$ are equal to 1 and we can therefore drop them here. We can then solve the gauge integrals in closed form and for the partition function we obtain
\begin{equation}
	Z \, = \, \sum_{\{s,k,\overline{k}\}}  C_{F}[s,k,\overline{k}] \, W_{F}[s,k,\overline{k}] \, C_{G}[k,\overline{k}] \, W_{G}[k,\overline{k}] \, .
\label{eq:qcd_partitionsc2}
\end{equation}
The gauge integration in (\ref{eq:qcd_partitionsc}) has generated a gauge constraint $C_{G}[k,\overline{k}]$ and a weight factor $W_{G}[k,\overline{k}]$. To represent the constraints and the weight factor in a transparent way, we introduce combinations of the dual variables $k_{x,\nu}^{ab}$, $\overline{k}_{x,\nu}^{ab}$ and the auxiliary variables $m_{x,\nu}^{ab}$, $\overline{m}_{x,\nu}^{ab}$ for $(a,b) = (2,1), (2,3), (3,1), (3,3)$ as follows:
\begin{align}
\label{eq:qcd_shortsc}
	&K_{x,\nu}^{ab} = k_{x,\nu}^{ab} - \overline{k}_{x,\nu}^{ab} \, ,  \qquad \ P_{x,\nu}^{ab} = k_{x,\nu}^{ab} + \overline{k}_{x,\nu}^{ab} \, , \\
	&j_{x,\nu}^{ab} = m_{x,\nu}^{ab} - \overline{m}_{x,\nu}^{ab} \, , \qquad s_{x,\nu}^{ab} = m_{x,\nu}^{ab} + \overline{m}_{x,\nu}^{ab} \, .
	\nonumber
\end{align}
As for the pure gauge case the constraints are generated by the integration over the four phases $\phi_{x,\nu}^{(j)}$, $j=1,2,4,5$ of the parametrization (\ref{eq:qcd_parametrization}) of the SU(3) matrices. We collect the corresponding four Kronecker deltas in the gauge constraint 
\begin{align}			
	C_G[k,\overline{k}] \; =& \; \prod_{x,\nu} 
	\delta(K_{x,\nu}^{11} + K_{x,\nu}^{12} - K_{x,\nu}^{33} - K_{x,\nu}^{23}) \;
	\delta(K_{x,\nu}^{22} + K_{x,\nu}^{12} - K_{x,\nu}^{33} - K_{x,\nu}^{31}) 
	\nonumber \\ 
	& \hspace{3mm}\times  	
	\delta(K_{x,\nu}^{13} + K_{x,\nu}^{12} - K_{x,\nu}^{31} - K_{x,\nu}^{21}) \; 
	\delta(K_{x,\nu}^{32} + K_{x,\nu}^{12} - K_{x,\nu}^{23} - K_{x,\nu}^{21}) \; .
\label{eq:qcd_cgsc}
\end{align}
For writing Eq.~(\ref{eq:qcd_cgsc}) we actually made use of the additional constraint which results from the $\phi_{x,\nu}^{(3)}$ integration
\begin{equation}
\label{eq:qcd_scconstraint3}
	J_{x,\nu}^{12} = j_{x,\nu}^{21} + j_{x,\nu}^{23} + j_{x,\nu}^{31} + j_{x,\nu}^{33} \, ,
\end{equation}
to substitute the sum of the auxiliary fluxes $j_{x,\nu}^{ab} = m_{x,\nu}^{ab} - \overline{m}_{x,\nu}^{ab}$ with the fermion color flux $K_{x,\nu}^{12} = k_{x,\nu}^{12} - \overline{k}_{x,\nu}^{12}$. Eq.~(\ref{eq:qcd_scconstraint3}) relates the value of the auxiliary variables $m_{x,\nu}^{ab}$ and $\overline{m}_{x,\nu}^{ab}$ to the $k_{x,\nu}^{12}, \overline{k}_{x,\nu}^{12}$ fermion color fluxes, thus limiting the amount of admissible configurations. Later we will discuss the implications of the constraint (\ref{eq:qcd_scconstraint3}) in more detail.  

The weight factor $W_{G}[k,\overline{k}]$ has the following form
\small
\begin{align}
	&W_{G} [k,\overline{k}]  =  2^{4V} \! \! \sum_{\{m,\overline{m}\}} \! \! 
	\Bigg[\prod_{x,\nu} \delta(K_{x,\nu}^{12} - j_{x,\nu}^{21} - j_{x,\nu}^{23} - j_{x,\nu}^{31} - j_{x,\nu}^{33})\Bigg]
	\Bigg[\prod_{x,\nu} (-1)^{K_{x,\nu}^{12} + K_{x,\nu}^{23} + K_{x,\nu}^{31} - j_{x,\nu}^{23} - j_{x,\nu}^{31}}  \Bigg] \nonumber \\
	&\times \Bigg[ \prod_{x,\nu} \B\left(\dfrac{P_{x,\nu}^{11} + P_{x,\nu}^{13} + P_{x,\nu}^{22} + P_{x,\nu}^{32}}{2} + 2\right.\left|\dfrac{P_{x,\nu}^{12} + s_{x,\nu}^{21} +s_{x,\nu}^{23} + s_{x,\nu}^{31} + s_{x,\nu}^{33}}{2} + 1\right)
	\nonumber \\
	&\! \times\! \B\!\left(\!\!\dfrac{P_{x,\nu}^{11} \! + \! s^{21}_{x,\rho} \! + \! P_{x,\nu}^{23}\! -\! s_{x,\nu}^{23}\! + s_{x,\nu}^{31}\! + \!P_{x,\nu}^{33}\! - \!s_{x,\nu}^{33}}{2} \!+\! 1\right.\!\left|\dfrac{\!P_{x,\nu}^{13}\! +\! P_{x,\nu}^{21}\! - \!s_{x,\nu}^{21}\! +\! s_{x,\nu}^{23} \!+\! P_{x,\nu}^{31}\! -\! s_{x,\nu}^{31}\! +\! s_{x,\nu}^{33}}{2}\! + \!1\!\right) \nonumber
	\\
	&\! \times\! \B\!\left(\!\!\dfrac{s_{x,\nu}^{21}\!  \!+ \!P^{22}_{x,\rho}\! + \!s_{x,\nu}^{23}\!  \!+\! P_{x,\nu}^{31}\! -\! s_{x,\nu}^{31}\! \! +\! P_{x,\nu}^{33}\! -\! s_{x,\nu}^{33}}{2}\! \! +\! 1\right.\!\left|\dfrac{\!P_{x,\nu}^{21} \!- \!s_{x,\nu}^{21}\!  \!+ \!P_{x,\nu}^{23} \!- \!s_{x,\nu}^{23} \! \!+ \!s_{x,\nu}^{31} \! \!+ \!P_{x,\nu}^{32} \!+ \!s_{x,\nu}^{33}}{2}\! \! +\! 1\!\right)\! \! \Bigg].
\label{eq:qcd_wgsc}
\end{align}
\normalsize
It sums over the configurations of the auxiliary variables $m_{x,\nu}^{ab}$, $\overline{m}_{x,\nu}^{ab}$, which we introduced for the binomial decomposition (\ref{eq:qcd_binomial}) of the matrix elements $(2,1)$, $(2,3)$, $(3,1)$ and $(3,3)$:
\begin{equation}
	\sum_{\{m,\overline{m}\}} \, = \, \prod_{x,\nu} \prod_{a =2,3} \prod_{b=1,3}
	\sum_{m_{x,\nu}^{ab} = 0}^{k_{x,\nu}^{ab}} \sum_{\overline{m}_{x,\nu}^{ab} = 0}^{\overline{k}_{x,\nu}^{ab}} \, .
\end{equation}
The admissible configurations of the dual variables $m_{x,\nu}^{ab}$ and $\overline{m}_{x,\nu}^{ab}$ must satisfy the constraint (\ref{eq:qcd_scconstraint3}), which is enforced at every link $(x,\nu)$ of the lattice by the product of Kronecker deltas in (\ref{eq:qcd_wgsc}). The three beta functions $\B$ in Eq.~(\ref{eq:qcd_wgsc}) are the result of the three $\theta_{x,\nu}^{(i)}$ integrals, $i=1,2,3$ in (\ref{eq:qcd_partitionsc}). They have the same structure as the beta functions we found for the pure gauge case (compare with (\ref{eq:qcd_wg})).

Summarizing the discussion about the dual representation of strong coupling QCD, we found that the partition function is a sum over the configurations of the dual variables for fermions: $s_{x}^{a} = 0,1$ for the monomers and $k_{x,\nu}^{ab}, \overline{k}_{x,\nu}^{ab} = 0,1$ for the forward and backward hops respectively. The configurations of the fermion dual variables must completely fill the lattice, as imposed by the fermion constraint $C_{F}[s,k,\overline{k}]$ given in (\ref{eq:qcd_cf}). Hence, the admissible configurations are those for which every color layer of every site of the lattice is either occupied my a monomer, is the endpoint of a dimer or is run through by a loop. The fermion configurations come with the weight factor $W_{F}[s,k,\overline{k}]$ in (\ref{eq:qcd_wf}): monomers contribute with a factor $2m$, while loops come with the sign function $\sign (\mathcal{L})$ given in (\ref{eq:qcd_signl}). Only loops winding in the compactified temporal direction couple with the chemical potential $\mu$, and in Eq.~(\ref{eq:qcd_wf}) we were able to find a nice geometrical interpretation for the net-particle number as the total net-winding number of a loop configuration.
Dimers and loops are further restricted by the gauge constraint $C_{G}[k,\overline{k}]$ in (\ref{eq:qcd_cf}). In the next section we will discuss the structure of the admissible configurations determined by the gauge constraint. Moreover, dimers and loops give contributions to the gauge weight $W_{G}[k,\overline{k}]$. The sign factor $(-1)^{K_{x,\nu}^{12} + K_{x,\nu}^{23} + K_{x,\nu}^{31} - j_{x,\nu}^{23} - j_{x,\nu}^{31}}$ in (\ref{eq:qcd_wgsc}) has a simple interpretation in the strong coupling limit, as we will see later. 

Before coming to the detailed discussion of the structure of the strong coupling configurations, we would like to compare the strong coupling results for the gauge integration with the ones we obtained in the last section for the pure gauge case. 
The gauge constraint (\ref{eq:qcd_cgsc}) in the strong coupling limit imposes the same relations between color link fluxes as the gauge constraint (\ref{eq:qcd_cg}) in the pure gauge theory. This similarity in the structure of the constraints reflects the fact that both systems are SU(3) symmetric. However, the variables that generate the color link fluxes are different in the two cases. In the pure gauge case the $J_{x,\nu}^{ab}$ are generated by the cycle occupation numbers $p_{x,\nu\rho}^{abcd} \in \mathbb{Z}$. Since the $p_{x,\nu\rho}^{abcd}$ live on the plaquettes $(x,\nu\rho)$ of the lattice, each $J_{x,\nu}^{ab}$ receives contributions from the cycle occupation numbers on the 6 plaquettes that are attached to the link $(x,\nu)$ (compare with the definition (\ref{eq:qcd_Jfluxes})). On the other hand, the fermion fluxes $K_{x,\nu}^{ab}$ are generated from the fermion dual variables $k_{x,\nu}^{ab}$ and $\overline{k}_{x,\nu}^{ab}$ for the forward and backward hops on the link $(x,\nu)$. Another main difference is that while the $J$-fluxes take values in $\mathbb{Z}$, the fermion fluxes $K_{x,\nu}^{ab} \in \{-1, 0 , 1\}$. This will make the interpretation of the gauge constraint (\ref{eq:qcd_cgsc}) together with the constraint (\ref{eq:qcd_scconstraint3}) for the auxiliary variables $m_{x,\nu}^{ab}, \overline{m}_{x,\nu}^{ab}$ much easier than in the pure gauge case. 

Comparing $W_{G}[p]$ of the pure gauge theory in Eq.~(\ref{eq:qcd_wg}) with the weight factor $W_{G}[k,\overline{k}]$ for strong coupling QCD in (\ref{eq:qcd_wgsc}) we also notice a structural similarity. Both are sums over configurations of the auxiliary variables $m_{x,\nu}^{ab}$ and $\overline{m}_{x,\nu}^{ab}$ needed for the binomial decomposition. In both cases the sum of the auxiliary fluxes $j_{x,\nu}^{ab} = m_{x,\nu}^{ab} - \overline{m}_{x,\nu}^{ab}$ is constrained by the value of the $K_{x,\nu}^{12}$ ($J_{x,\nu}^{12}$ for the gauge case) color flux at every link. Furthermore, the same sign factors appear in the summands of both weight factors. Because of the Pauli principle, the fluxes in the strong coupling case are restricted to the values 0,1 and -1, such that all binomial coefficients are equal to 1, while in the pure gauge theory weight $W_{G}[p]$ in Eq.~(\ref{eq:qcd_wg}) the binomial coefficients can have non-trivial values. The pure gauge theory weight $W_{G}[p]$ additionally have plaquette based weight factors from the expansion of the gauge action which also depends on the auxiliary plaquette variables $l_{x,\nu\rho}^{abcd}$. Clearly these terms are absent in strong coupling QCD where we have no gauge action. Nonetheless the weight factors that come from the Haar measure integration are the same in both cases, key signal of the fact that not only the constraints carry the information about the SU(3) symmetry in our dual representation.

\subsection{Strong coupling baryon loops}
In this section we focus on the interpretation of the constraints (\ref{eq:qcd_cgsc}) and (\ref{eq:qcd_scconstraint3}), and we discuss the consequences that they have on the admissible configurations of QCD at strong coupling. 

The gauge constraint $C_{G}[k,\overline{k}]$ imposes relations between different components of the color fluxes $K_{x,\nu}^{ab} \equiv k_{x,\nu}^{ab} - \overline{k}_{x,\nu}^{ab}$ we defined in (\ref{eq:qcd_shortsc}). Explicitly those constraints are:
\begin{gather}
\label{eq:qcd_scconstraint1}
K_{x,\nu}^{11} + K_{x,\nu}^{12} = K_{x,\nu}^{23} + K_{x,\nu}^{33} \, ,\\	
\label{eq:qcd_scconstraint2}
K_{x,\nu}^{22} + K_{x,\nu}^{12} = K_{x,\nu}^{33} + K_{x,\nu}^{31} \, ,\\
\label{eq:qcd_scconstraint4}
K_{x,\nu}^{13} + K_{x,\nu}^{12} = K_{x,\nu}^{31} + K_{x,\nu}^{21} \, ,\\
\label{eq:qcd_scconstraint5}
K_{x,\nu}^{32} + K_{x,\nu}^{12} = K_{x,\nu}^{23} + K_{x,\nu}^{21} \, .
\end{gather} 
Comparing Eqs.~(\ref{eq:qcd_scconstraint1}) -- (\ref{eq:qcd_scconstraint5}) with the constraints (\ref{eq:qcd_constraint1}) -- (\ref{eq:qcd_constraint5}) in Sec.~\ref{sec:qcd_acc} for the pure gauge theory we notice the structural similarity between the two sets of equations. As a consequence also (\ref{eq:qcd_scconstraint1}) -- (\ref{eq:qcd_scconstraint5}) can be organized into the three \textit{color flux conservation} constraints (\ref{eq:qcd_fluxconservation1}) -- (\ref{eq:qcd_fluxconservation3}) (plots in the top row of Fig.~\ref{fig:qcd_constraints}) and the three  \textit{color flux exchange} constraints (\ref{eq:qcd_fluxexchange}) (plots at the bottom of Fig.~\ref{fig:qcd_constraints}) as we did in Sec.~\ref{sec:qcd_acc}, where we substitute the color fluxes $J_{x,\nu}^{ab} \in \mathbb{Z}$ of the pure gauge theory with the fermion color fluxes $K_{x,\nu}^{ab} \in \{-1,0,1\}$. The fact that the $K_{x,\nu}^{ab}$ fluxes take values from such a small set, together with the further restrictions imposed by the constraints (\ref{eq:qcd_scconstraint1}) -- (\ref{eq:qcd_scconstraint5}), makes the determination of the admissible configurations at strong coupling much easier than in the pure gauge case.
Moreover, the additional constraint 
\begin{equation}
\label{eq:qcd_scconstraint3b}
K_{x,\nu}^{12} = j_{x,\nu}^{21} + j_{x,\nu}^{23} + j_{x,\nu}^{31} + j_{x,\nu}^{33} \, ,
\end{equation}
which relates the fermion flux $K_{x,\nu}^{12}$ to the auxiliary fluxes $j^{ab}_{x,\nu} \equiv m^{ab}_{x,\nu} - \overline{m}^{ab}_{x,\nu}$ of the variables $m^{ab}_{x,\nu} \in \{0,k^{ab}_{x,\nu}\}$ and $\overline{m}^{ab}_{x,\nu} \in \{0,\overline{k}^{ab}_{x,\nu}\}$ will be crucial for the determination of the gauge sign 
\begin{equation}
	\label{eq:qcd_gaugesign}
	(-1)^{K_{x,\nu}^{12} + K_{x,\nu}^{23} + K_{x,\nu}^{31} - j_{x,\nu}^{23} - j_{x,\nu}^{31} } \, .
\end{equation}

\begin{figure}
	\centering
	\includegraphics[width=7.5cm]{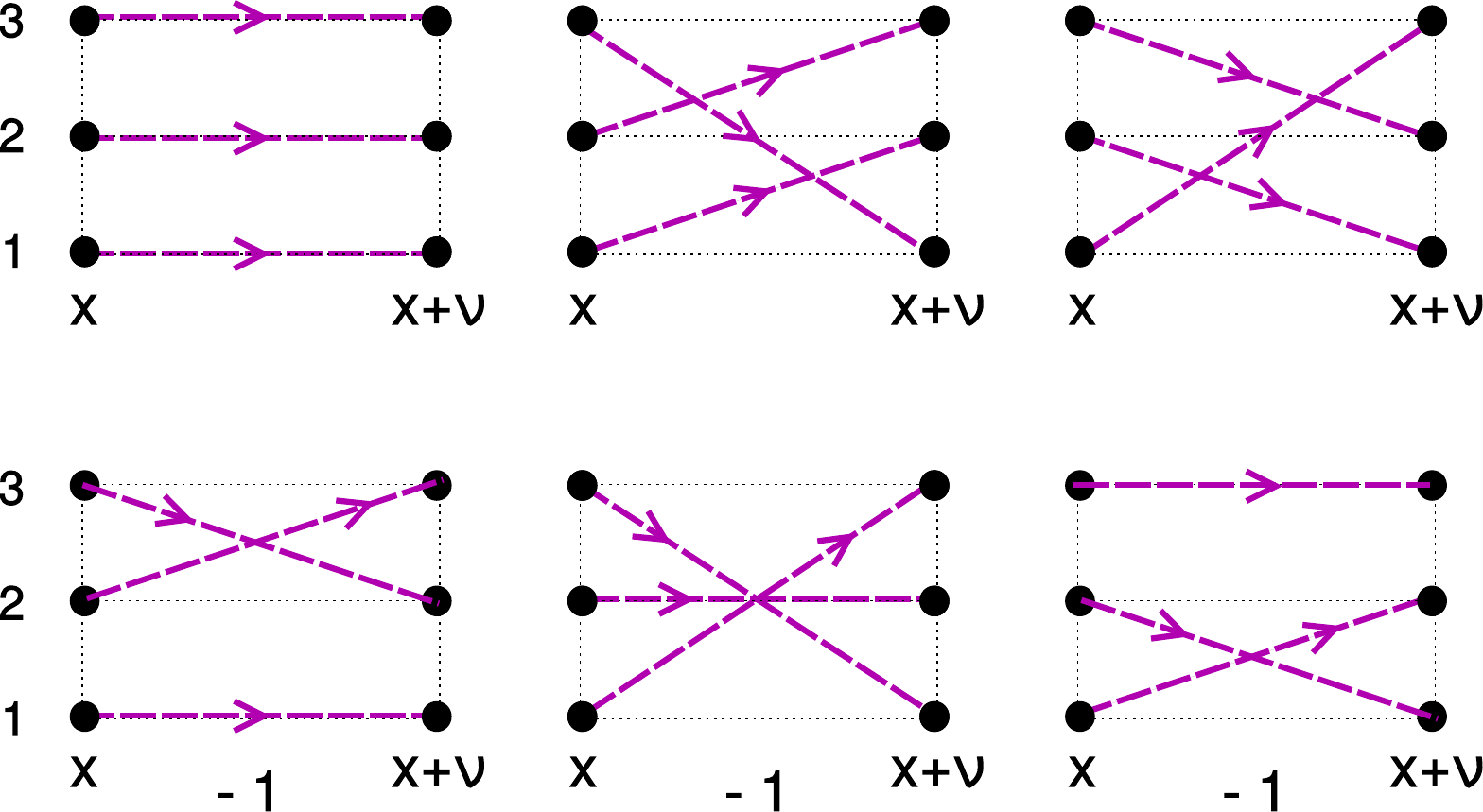}
	\caption{Baryon loop elements in the strong coupling limit. Only the six combinations shown here are admissible for the propagation of fluxes in the strong coupling limit. The elements with an odd number of color flux crossings come with an explicit minus sign. For the negative direction the same fluxes are admissible and	have the same signs. The corresponding diagrams are obtained by reverting the arrows. \label{fig:qcd_strongloops}}
\end{figure}

\begin{figure}
	\centering
	\includegraphics[width=9.5cm]{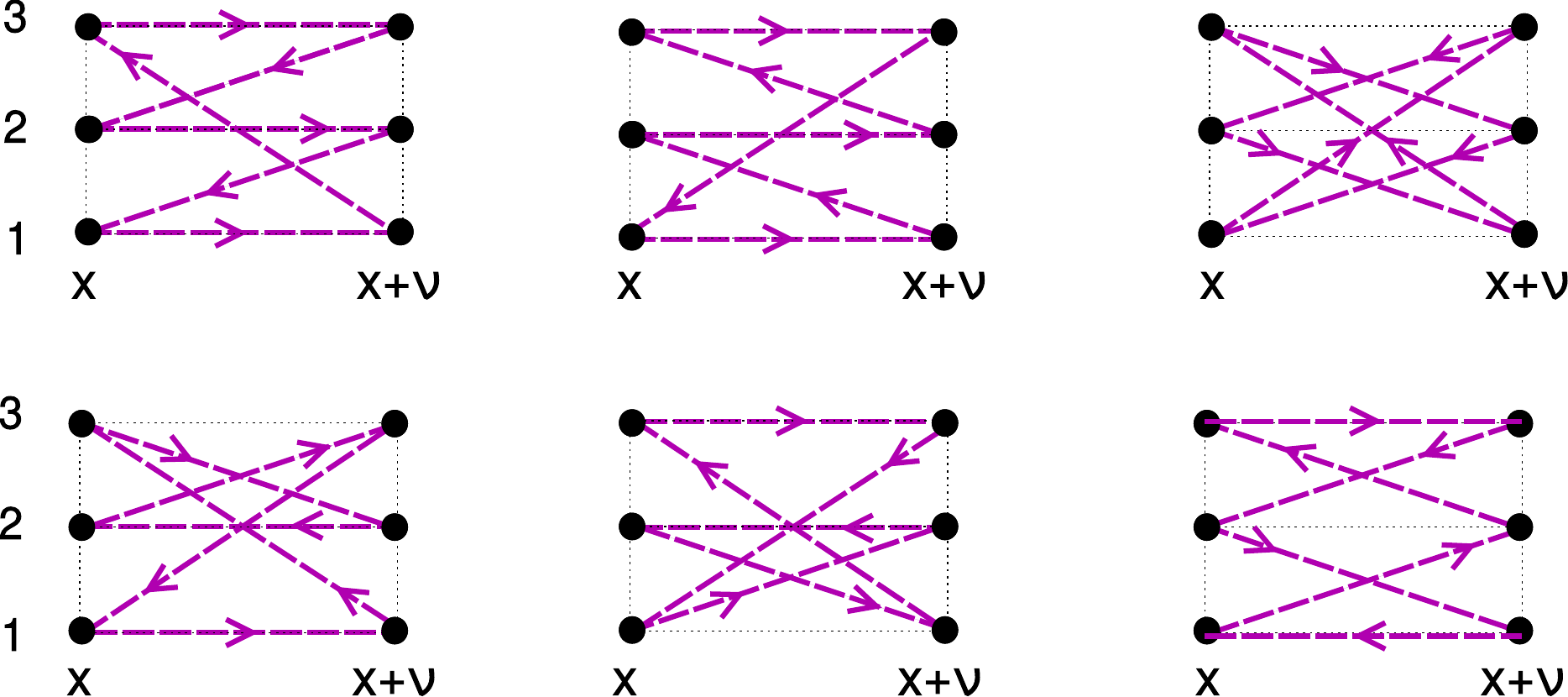}
	\caption{Closed, non-propagating one-link loops at strong coupling. All of these loops come with a positive weight. Also the opposite orientation is possible, which is obtained by reverting all arrows. \label{fig:qcd_linkloops}}
\end{figure}

The admissible combinations of the strong coupling fluxes $K_{x,\nu}^{ab}$ at a single link come in two types: three lines of flux that run in the same direction (see Fig.~\ref{fig:qcd_strongloops}), or six lines of flux that form a closed loop on a single link (Fig.~\ref{fig:qcd_linkloops}).
Obviously only the first type allows for long distance propagation and we refer to these strong coupling elements as \textit{strong coupling baryon fluxes}. The locally closing ones are referred to as \textit{one-link loops}.

For the discussion of the complete list of strong coupling baryon fluxes we start with solutions of the constraint equations (\ref{eq:qcd_scconstraint1}) -- (\ref{eq:qcd_scconstraint5}) where we allow only the values $K_{x,\nu}^{ab}= 0,1$, i.e., we consider forward propagation. In addition to the gauge constraint $C_{G}[k,\overline{k}]$ also the fermion constraint $C_{F}[k,\overline{k}]$ has to be obeyed. This implies that, if we consider the link $(x,\nu)$, at site $x$ only a single line of flux can originate from each color $a$, and the three lines must end on different colors at site $x + \hat{\nu}$. For the forward propagation of the strong coupling baryon fluxes one finds exactly six solutions, which we represent in Fig.~\ref{fig:qcd_strongloops}. The six solutions for the backward propagation of the strong coupling baryon fluxes are obtained reverting the arrows in Fig.~\ref{fig:qcd_strongloops}, which corresponds to $K_{x,\nu}^{ab} \rightarrow - K_{x,\nu}^{ab}$. 

The gauge sign (\ref{eq:qcd_gaugesign}) for the strong coupling baryon fluxes in Fig.~\ref{fig:qcd_strongloops} is easily determined. We here discuss two examples to illustrate the method. The strong coupling baryon flux at the top left of Fig.~\ref{fig:qcd_strongloops} is built by setting $K_{x,\nu}^{11} = K_{x,\nu}^{22} = K_{x,\nu}^{33} = 1$. Therefore, for this example the fermion fluxes $K_{x,\nu}^{12}$, $K_{x,\nu}^{23}$ and $K_{x,\nu}^{31}$ that appear at the exponent of the sign factor (\ref{eq:qcd_gaugesign}) are vanishing. Moreover, since $K_{x,\nu}^{23} = K_{x,\nu}^{31} = 0$, also $j_{x,\nu}^{23} = j_{x,\nu}^{31} = 0$ (recall that $j_{x,\nu}^{23} = m_{x,\nu}^{ab} - \overline{m}_{x,\nu}^{ab}$ and that $m_{x,\nu}^{ab} = 0, k_{x,\nu}^{ab}$ and $\overline{m}_{x,\nu}^{ab} = 0, \overline{k}_{x,\nu}^{ab}$, with $(a,b) = (2,1), (2,3), (3,1), (3,3)$). Hence, the strong coupling baryon flux at the top left of Fig.~\ref{fig:qcd_strongloops} has positive gauge sign. 

Let us now consider the top center example of Fig.~\ref{fig:qcd_strongloops}. This strong coupling baryon flux has non-vanishing fluxes $K_{x,\nu}^{12} = K_{x,\nu}^{23} = K_{x,\nu}^{31} = 1$, which means that the sum $K_{x,\nu}^{12} + K_{x,\nu}^{23} + K_{x,\nu}^{31}$ at the exponent of the sign factor (\ref{eq:qcd_gaugesign}) is odd. However, the constraint (\ref{eq:qcd_scconstraint3b}) forces the sum of the auxiliary fluxes $j_{x,\nu}^{ab}$ to be equal to $K_{x,\nu}^{12} = 1$. Since $K_{x,\nu}^{21}$ and $K_{x,\nu}^{33}$ are vanishing, $j_{x,\nu}^{21} = j_{x,\nu}^{33} = 0$ and the constraint (\ref{eq:qcd_scconstraint3b}) is satisfied only if $j_{x,\nu}^{23} + j_{x,\nu}^{31} = 1$. As a result, the gauge sign for the top center example of Fig.~\ref{fig:qcd_strongloops} is also positive. 

The gauge signs for all the other strong coupling baryon fluxes in Fig.~\ref{fig:qcd_strongloops} can be determined in a similar way. In particular we find:
\begin{equation}
\label{eq:qcd_gaugesignsc}
	(-1)^{K_{x,\nu}^{12} + K_{x,\nu}^{23} + K_{x,\nu}^{31} - j_{x,\nu}^{23} - j_{x,\nu}^{31} } = (-1)^{\text{\# crossings of K-flux}}\, .
\end{equation}
In other words, at strong coupling the exponent of the gauge sign in (\ref{eq:qcd_gaugesign}) has the simple interpretation of the total number of $K$-flux crossings. In Fig.~\ref{fig:qcd_strongloops} we marked the strong coupling baryon fluxes where the sign (\ref{eq:qcd_gaugesignsc}) is negative with $-1$.

The second class of solutions of (\ref{eq:qcd_scconstraint1}) -- (\ref{eq:qcd_scconstraint5}) are the one-link loops depicted in Fig.~\ref{fig:qcd_linkloops}. They are obtained by allowing all values $K_{x,\nu}^{ab} = -1,0,+1$ and enforcing the fermion constraints $C_{F}[s,k,\overline{k}]$, such that each node is run through by a loop. It is easy to see that the six solutions in Fig.~\ref{fig:qcd_linkloops} are obtained by combining one of the forward baryon fluxes from Fig.~\ref{fig:qcd_strongloops} with a matching backward baryon flux such that the fermion constraints are obeyed. One finds that only fluxes with the same sign in Fig.~\ref{fig:qcd_strongloops} can be combined among each other, such that the total sign from (\ref{eq:qcd_gaugesign}) is always +1. Recalling the definition (\ref{eq:qcd_signl}) of $\sign (\mathcal{L})$, we find that also the fermion loop sign is positive for the six one-link loops in Fig.~\ref{fig:qcd_linkloops}. In fact, for each of them there is an overall minus sign and then a factor $(-1)^{3}$ for the three forward hops. Thus one-link loops always come with a positive weight which is given by the products of beta functions in (\ref{eq:qcd_wgsc}). These weights can be summed, and all possible one-link loops may be combined into a dual element that plays a similar role as the monomers and dimers: they are all local fermionic monomials that can be used to saturate the fermion constraints on the sites and links that are not occupied by a strong coupling baryon loop.

We now demonstrate that the interplay between the gauge sign (\ref{eq:qcd_gaugesignsc}) for the strong coupling baryon fluxes in Fig.~\ref{fig:qcd_strongloops} and the fermion loop sign makes the overall sign of baryon loops equivalent to the fermion loop sign (\ref{eq:qcd_signl}) itself. 

\begin{figure}
	\begin{center}
			\includegraphics[width=7cm]{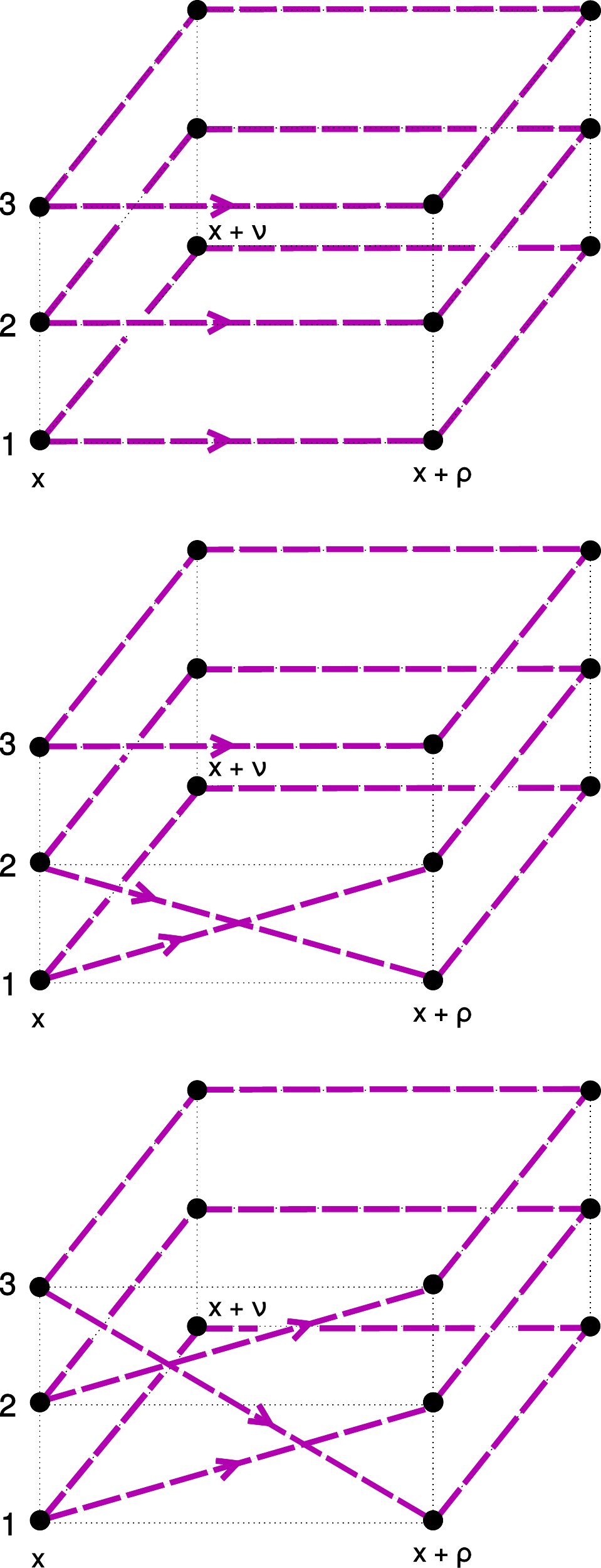}
	\end{center}
	\caption{Examples of simple strong coupling baryon loops with different connectivity properties. \label{fig:qcd_loop}}
\end{figure}

A general baryon loop is a closed non-intersecting path on the lattice made out of the strong coupling baryon fluxes represented in Fig.~\ref{fig:qcd_strongloops}. Let us start considering the simple strong coupling baryon loop shown in the top plot of Fig.~\ref{fig:qcd_loop}: a baryon loop made out of just parallel fluxes closing around a single plaquette. For this loop the gauge sign is evidently positive because there are no crossings of $K$-fluxes. Moreover, since the baryon loop is made out of three identical fermion loops $\mathcal{L}$ running parallel on the three different color layers, the fermion sign for the baryon loop can be determined as the third power of the sign function $\sign (\mathcal{L})$ of a single loop:
\begin{equation}
	\big(\sign (\mathcal{L})\big)^3 = \sign (\mathcal{L}) \, .
	\label{eq:qcd_signb}
\end{equation}  
Hence, we proved that for this simple example the baryon loops again obey the sign formula (\ref{eq:qcd_signl}) for staggered fermions. 

We then replace one of the parallel fluxes of the baryon loop with a strong coupling flux that has just one crossing (see the middle example in Fig.~\ref{fig:qcd_loop}). The resulting baryon loop have an overall negative gauge sign. However, the crossing also changes the connectivity properties of the loop: inserting one crossing either connects two fermion loops into one, as in the example in the middle of Fig.~\ref{fig:qcd_loop}, or splits a loop into two loops. Thus every crossing changes the number of loops by one, and since every loop comes with an overall minus sign, the crossing also changes the fermion sign. However, the overall sign of the baryon loop remains the same, because the change in the fermion sign is compensated by the change of the gauge sign. Consequently the sign of a baryon loop is always given by (\ref{eq:qcd_signb}), i.e., the signs of the strong coupling baryon loops are equivalent to the signs for loops of a single free staggered fermion. 

In our dual representation it is possible to bring the identification of the strong coupling baryon loops with the loops of a free staggered fermion a step further: when computing the link weight factor for the six strong coupling flux elements in Fig.~\ref{fig:qcd_strongloops}, one finds that they all come with the same weight of $1/12$. Thus for every link of the loop we can sum over all six possible fluxes and obtain a total link weight of $1/2$. The resulting weight for a baryon loop $\mathcal{L}_{B}$ is then given by:
\begin{equation}
\label{eq:qcd_wb}
	W(\mathcal{L}_{B}) = \sign (\mathcal{L}_{B}) \left(\dfrac{1}{2}\right)^{|\mathcal{L}_{B}|} \, ,
\end{equation}
where $|\mathcal{L}_{B}|$ denotes the length of the loop $\mathcal{L}_{B}$.

We conclude that the dual form of strong coupling QCD is a gas of free staggered fermion loops that come with the weight $W(\mathcal{L}_{B})$. These loops describe baryons and are embedded in a background of monomers, dimers and local link loops, such that the fermion constraints are obeyed at all sites of the lattice. We think that the interpretation of the strong coupling baryon loops as free staggered fermions could open the possibility of updating our form of strong coupling QCD with fermion bags \cite{PhysRevD.82.025007}. This idea has also led to a new reformulation of strong coupling QCD in terms of baryon bags, presented in \cite{Gattringer:2018mrg}.

\section{Worldlines and worldsheets representation of the full theory \label{sec:qcd_full}}

We complete the presentation of the dual representation of QCD discussing the formulation in terms of worldsheets and worldlines of the full partition function. The partition sum of full QCD can be written as
\begin{equation}
\label{eq:qcd_partitionfull}
Z = \int \! D[U] Z_{F} [U] e^{-S_{G}[U]} \, .
\end{equation}
In other words, we integrate the fermionic partition sum $Z_{F} [U]$ given in (\ref{eq:qcd_partitionf}) over the Haar measure $\int \! D[U]$, weighted with the gauge Boltzmann factor $e^{-S_{G}[U]}$, where $S_{G}[U]$ is the Wilson action (\ref{eq:qcd_sg}). We can therefore use the intermediate result (\ref{eq:qcd_partitionsc}) we obtained in the last section for the partition sum at strong coupling, and simply multiply the integrand by the Boltzmann weight $e^{-S_{G}[U]}$. What we are left with is the following gauge integral
\begin{equation}
\label{eq:qcd_gaugeintegral}
	\Bigg[\prod_{x,\nu} \int \! dU_{x,\nu} \Bigg] e^{-S_{G}[U]} \,
	\prod_{x, \nu} \prod_{a, b} 	\left( U_{x,\nu}^{ab} \right)^{k_{x,\nu}^{ab}}
	\left( U_{x,\nu}^{ab \, \star} \right)^{\overline{k}_{x,\nu}^{ab}} \, .
\end{equation} 
We can now treat the Boltzmann weight $e^{-S_{G}[U]}$ as we did in Sec.~\ref{sec:qcd_acc}: we decompose the action into a sum over abelian color cycles as in Eq.~(\ref{eq:qcd_accaction}), and then we factorize and Taylor-expand the Boltzmann weight. As a result the integrand in (\ref{eq:qcd_gaugeintegral}) takes the following form:
\begin{equation}
\label{eq:qcd_gaugeintegral2}
	\Bigg[\prod_{x,\nu} \int \! dU_{x,\nu} \Bigg]
	\prod_{x, \nu} \prod_{a, b} 	\left( U_{x,\nu}^{ab} \right)^{N_{x,\nu}^{ab} + k_{x,\nu}^{ab}}
	\left( U_{x,\nu}^{ab \, \star} \right)^{\overline{N}_{x,\nu}^{ab} + \overline{k}_{x,\nu}^{ab}} \, .
\end{equation} 
Eq.~(\ref{eq:qcd_gaugeintegral2}) can be solved in closed form following exactly the same steps we used in Sec.~\ref{sec:qcd_acc}, just replacing $N_{x,\nu}^{ab}$ with $N_{x,\nu}^{ab} + k_{x,\nu}^{ab}$ and $\overline{N}_{x,\nu}^{ab}$ with $\overline{N}_{x,\nu}^{ab} + \overline{k}_{x,\nu}^{ab}$. Again we rewrite
\begin{equation}
	N_{x,\nu}^{ab} + k_{x,\nu}^{ab} = \dfrac{Q_{x,\nu}^{ab} + L_{x,\nu}^{ab}}{2} \, , \qquad
	\overline{N}_{x,\nu}^{ab} + \overline{k}_{x,\nu}^{ab} = \dfrac{Q_{x,\nu}^{ab} - L_{x,\nu}^{ab}}{2} \, ,
\end{equation}
where
\begin{equation}
	L_{x,\nu}^{ab} = J_{x,\nu}^{ab} + K_{x,\nu}^{ab} \, , \qquad
	Q_{x,\nu}^{ab} = S_{x,\nu}^{ab} + P_{x,\nu}^{ab} \, .
\end{equation}
$J_{x,\nu}^{ab}$ and $S_{x,\nu}^{ab}$ are the linear combinations of the cycle occupation numbers $p_{x,\nu\rho}^{abcd}$ given in (\ref{eq:qcd_Jfluxes}) and (\ref{eq:qcd_Sfluxes}), while $K_{x,\nu}^{ab}$ and $P_{x,\nu}^{ab}$ are the combinations of the fermion dual variables $k_{x,\nu}^{ab}$ and $\overline{k}_{x,\nu}^{ab}$ given in (\ref{eq:qcd_shortsc}). 

Putting things together we obtain the following dual form for the partition sum of full QCD:
\begin{equation}
\label{eq:qcd_partitionfull2}
	Z \; = \; \sum_{\{p,s,k,\overline{k}\}} C_{F}[s,k,\overline{k}] \; W_{F}[s,k,\overline{k}] \; C_{G}[p,k,\overline{k}] \; W_{G}[p,k,\overline{k}] \, .
\end{equation}
$C_{F}[s,k,\overline{k}]$ and $W_{F}[s,k,\overline{k}]$ are the fermion constraint (\ref{eq:qcd_cf}) and the fermion weight (\ref{eq:qcd_wf}) we discussed in the previous section. The gauge constraint $C_{G}[p,k,\overline{k}]$ and the gauge weight $W_{G}[p,k,\overline{k}]$ originate from the gauge integrals in (\ref{eq:qcd_gaugeintegral2}). They therefore have the same structure as the gauge constraints and weights we obtained in the pure gauge case and the strong coupling limit. In particular,  
\begin{align}			
	C_G[p,k,\overline{k}] \; =& \; \prod_{x,\nu} 
	\delta(L_{x,\nu}^{11} + L_{x,\nu}^{12} - L_{x,\nu}^{33} - L_{x,\nu}^{23}) \;
	\delta(L_{x,\nu}^{22} + L_{x,\nu}^{12} - L_{x,\nu}^{33} - L_{x,\nu}^{31}) 
	\nonumber \\ 
	& \hspace{3mm}\times  	
	\delta(L_{x,\nu}^{13} + L_{x,\nu}^{12} - L_{x,\nu}^{31} - L_{x,\nu}^{21}) \; 
	\delta(L_{x,\nu}^{32} + L_{x,\nu}^{12} - L_{x,\nu}^{23} - L_{x,\nu}^{21}) \; ,
\label{eq:qcd_cgfull}
\end{align}
is again a product of four Kronecker deltas which impose relations between different color components of the link-based fluxes $L_{x,\nu}^{ab}$. In full QCD these fluxes take contributions both from the gauge degrees of freedom, represented by the cycle occupation numbers $p_{x,\nu\rho}^{abcd}$ attached to the link $(x,\nu)$, and the fermionic degrees of freedom, in the form of the hopping variables $k_{x,\nu}^{ab}$ and $\overline{k}_{x,\nu}^{ab}$.

The weight factor 
\small
\begin{align}
	&W_{G} [p,k,\overline{k}]  = 2^{4V}\!\! \sum_{\{l,m,\overline{m}\}} \!
	\Bigg[\prod_{x,\nu} \delta(L_{x,\nu}^{12} - j_{x,\nu}^{21}\! - j_{x,\nu}^{23} \!- j_{x,\nu}^{31}\! - j_{x,\nu}^{33})\Bigg] \!
	\Bigg[\prod_{x,\nu} (-1)^{L_{x,\nu}^{12} + L_{x,\nu}^{23} + L_{x,\nu}^{31} - j_{x,\nu}^{23} - j_{x,\nu}^{31}}  \Bigg] \nonumber \\
	&\hspace{18mm} \times
	\Bigg[ \prod_{x,\nu} \prod_{a=2,3} \prod_{b=1,3}
	\binom{N_{x,\nu}^{ab} + k_{x,\nu}^{ab}}{m_{x,\nu}^{ab}} \binom{\overline{N}_{x,\nu}^{ab} + \overline{k}_{x,\nu}^{ab}}{\overline{m}_{x,\nu}^{ab}}\Bigg] 
	\Bigg[\prod_{x,\nu<\rho} \prod_{a,b,c,d} 
	\dfrac{ \left( \beta/2 \right)^{|p_{x,\nu\rho}^{abcd}| + 2l_{x,\nu\rho}^{abcd}}}{\left(|p_{x,\nu\rho}^{abcd}| + l_{x,\nu\rho}^{abcd}\right)!l_{x,\nu\rho}^{abcd}!}\Bigg] \nonumber \\
	&\times \Bigg[ \prod_{x,\nu} \B\left(\dfrac{Q_{x,\nu}^{11} + Q_{x,\nu}^{13} + Q_{x,\nu}^{22} + Q_{x,\nu}^{32}}{2} + 2\right.\left|\dfrac{Q_{x,\nu}^{12} + s_{x,\nu}^{21} +s_{x,\nu}^{23} + s_{x,\nu}^{31} + s_{x,\nu}^{33}}{2} + 1\right)
	\nonumber \\
	&\!\times\!\! \B\!\left(\!\!\dfrac{Q_{x,\nu}^{11} \! + \! s^{21}_{x,\rho}\! \! + \! Q_{x,\nu}^{23}\! -\! s_{x,\nu}^{23}\!\! + s_{x,\nu}^{31}\!\! + \!Q_{x,\nu}^{33}\! - \!s_{x,\nu}^{33}}{2} \!\!+\! 1\right.\!\left|\dfrac{\!Q_{x,\nu}^{13}\! +\! Q_{x,\nu}^{21}\! - \!s_{x,\nu}^{21}\!\! +\! s_{x,\nu}^{23}\! \!+\! Q_{x,\nu}^{31}\! -\! s_{x,\nu}^{31}\!\! +\! s_{x,\nu}^{33}}{2}\!\! + \!1\!\right) \nonumber
	\\
	&\!\times\! \!\B\!\left(\!\!\dfrac{s_{x,\nu}^{21} \!\!+ \!Q^{22}_{x,\rho}\! + \!s_{x,\nu}^{23} \!\!+\! Q_{x,\nu}^{31}\! -\! s_{x,\nu}^{31}\!\! +\! Q_{x,\nu}^{33}\! -\! s_{x,\nu}^{33}}{2}\!\! +\! 1\right.\!\left|\dfrac{\!Q_{x,\nu}^{21} \!- \!s_{x,\nu}^{21} \!\!+ \!Q_{x,\nu}^{23} \!- \!s_{x,\nu}^{23} \!\!+ \!s_{x,\nu}^{31} \!\!+ \!Q_{x,\nu}^{32} \!+ \!s_{x,\nu}^{33}}{2}\!\! +\! 1\!\right)\!\!\Bigg],
\label{eq:qcd_wgfull}
\end{align}
\normalsize
contains the additional constraint 
\begin{equation}
	L_{x,\nu}^{12}  = j_{x,\nu}^{21} + j_{x,\nu}^{23} +  j_{x,\nu}^{31} + j_{x,\nu}^{33} \, ,
\label{eq:qcd_fullconstraint3}
\end{equation}
which relates the link flux $L_{x,\nu}^{12}$ with the fluxes $j_{x,\nu}^{ab} \equiv m_{x,\nu}^{ab} - \overline{m}_{x,\nu}^{ab}$ of the auxiliary variables $m_{x,\nu}^{ab}$ and $\overline{m}_{x,\nu}^{ab}$ we introduced for the binomial decomposition of the matrix elements $(2,1)$, $(2,3)$, $(3,1)$ and $(3,3)$. In full QCD $W_{G}[p,k,\overline{k}]$ sums over the configurations of the auxiliary plaquettes variables $l_{x,\nu\rho}^{abcd} \in \mathbb{N}_{0}$ as well as the link-based auxiliary variables $m_{x,\nu}^{ab} \in \{0,1\dots N_{x,\nu}^{ab} + k_{x,\nu}^{ab}\}$ and $\overline{m}_{x,\nu}^{ab} \in \{0,1\dots \overline{N}_{x,\nu}^{ab} + \overline{k}_{x,\nu}^{ab}\}$:
\begin{equation}
	\sum_{\{l,m,\overline{m}\}} = \Bigg[\prod_{x,\nu<\rho} \prod_{a,b,c,d} \sum_{l_{x,\nu\rho}^{abcd} = 0}^{\infty}\Bigg] \Bigg[\prod_{x,\nu} \prod_{a=2,3} \prod_{b=1,3} \sum_{m_{x,\nu}^{ab} = 0}^{N_{x,\nu}^{ab} + k_{x,\nu}^{ab}} \sum_{\overline{m}_{x,\nu}^{ab} = 0}^{\overline{N}_{x,\nu}^{ab} + \overline{k}_{x,\nu}^{ab}}\Bigg] \, .
\end{equation} 
The beta functions and the sign factor resulting from the Haar measure integration have the same dependence on the component of the link fluxes which, also for $W_{G}[p,k,\overline{k}]$ depend on both the cycle occupation numbers and the fermion dual variables. Moreover, $W_{G}[p,k,\overline{k}]$ collects the coefficients from the Taylor expansion of the local exponentials (\ref{eq:qcd_partition2}), as well as the binomial factors from the binomial decomposition (\ref{eq:qcd_binomial}).

Summarizing, in the dual formulation the partition function is a sum over the dual variables for fermions $s_{x}^{a},k_{x,\nu}^{ab},\overline{k}_{x,\nu}^{ab} \in \{0,1\}$ and the cycle occupation numbers $p_{x,\nu\rho}^{abcd} \in \mathbb{Z}$ which represent the gauge degrees of freedom. The configurations of the fermion dual variables must satisfy the fermion constraint $C_{F}[s,k,\overline{k}]$ given in (\ref{eq:qcd_cf}), which enforces Pauli's exclusion principle. As a result, the admissible fermion configurations must saturate the lattice with monomers dimers and loops. Each fermion configuration comes with a weight given by $W_{F}[s,k,\overline{k}]$ in (\ref{eq:qcd_wf}): monomers contribute a factor of $2m$, loops come with a sign $\sign (\mathcal{L})$ given in (\ref{eq:qcd_signl}) and the chemical potential couples to the temporal winding number $W_{\mathcal{L}}$ of the loops. Dimers and loops, together with the gauge degrees of freedom, enter in the expressions of the gauge constraint $C_{G}[p,k,\overline{k}]$ and the gauge weight $W_{G}[p,k,\overline{k}]$ that we just discussed. Structurally these are the same constraints and weights as for pure gauge theory and strong coupling QCD, since they are generated by integrating the SU(3) link matrices. However, in full QCD they link the color flux contributions from both the gauge fields, via the cycle occupation numbers $p_{x,\nu\rho}^{abcd}$, and the fermion loops, via $k_{x,\nu}^{ab}$ and $\overline{k}_{x,\nu}^{ab}$.

We conclude this section on full QCD with addressing two important aspects of the new representation: as in the case of pure SU(3) lattice gauge theory, our dual form of the partition sum (\ref{eq:qcd_partitionfull2}) has the structure of a strong coupling expansion, and again, our approach allows one to compute all coefficients of this expansion in closed form. Furthermore, it is obvious how to generalize the construction to several flavors: one simply uses multiple sets of dual fermion variables, which all couple in the same way to the gauge fields. Thus instead of the variables $k_{x,\nu}^{ab}$ and $\overline{k}_{x,\nu}^{ab}$ one has flavor sums over such variables, and the color fluxes at each link have contributions from all flavors. These flavor sums over the dual fermion variables enter the constraints and weights, which otherwise have the same form as presented in this section.

\chapter{Summary \label{cha:conclusions}}

In this thesis we developed two methods, the \textit{abelian color cycle} (ACC) and \textit{abelian color flux} (ACF) methods, with the aim of extending the applicability of the dual approach to non-abelian lattice field theories. We achieved this by making explicit the sums over the color indices in the action of the model we were considering. As a result, the action is decomposed into a sum over commuting terms, and thus the Boltzmann weight may be factorized into local factors. Then the dualization program can be carried out as in the abelian case, by expanding the single exponentials and integrating out the conventional fields. As a result the partition function is exactly rewritten in terms of new variables, the so-called dual variables, that are integer valued variables attached either to the links or the plaquettes of the lattice. The final partition function is a sum over the configurations of those dual variables, which come with weights and must satisfy constraints. The emergence of the constraints comes from the integration over the phases in the parametrization of the conventional fields. In general, those constraints impose flux conservation for the dual variables which imply that the long range physics is described by worldlines for the matter fields and worldsheets for the gauge degrees of freedom. When integrating fermion fields we see the emergence of additional fermionic constraints which implement Pauli's exclusion principle for the fermionic dual variables. 

As a first example for the application of the ACF approach, in Chapter \ref{cha:pcm} we presented the dualization of the SU(2) principal chiral model with coupled chemical potentials. In the dual formulation this model consists of two species of worldlines that are constrained to form closed loops, and two additional species of auxiliary variables which are unconstrained. As we outlined several times, the constraints implement the original symmetry of a theory in the dual representation. However, this is not the only way the symmetry is represented in the dual formulation. Also key are the types of dual variables, and their role in the final form of the partition function, as well as the form of the weight factors. In fact, in the principal chiral model the two constraints for the two species of flux variables are the same as the constraint for the flux variable we found in Sec.~\ref{sec:teo_u1} for the case of the U(1) Gauge -- Higgs model. Hence, the constraints alone would hint at a $U(1) \times U(1)$ symmetry. Then the auxiliary variables as well as the structure of the weight tie together the two species of worldlines and give the full SU(2) symmetry in the dual representation.

Our worldline representation of the dual partition function of the SU(2) principal chiral model sums over the admissible configurations of the flux variables and the auxiliary variables which come with real and positive weights. Hence, in this particular case, the ACF method solves the sign problem that this model has in the conventional representation. 
Moreover, thanks to the simple structure of the constraints, in a second instant we were able to reformulate the model in terms of yet another set of dual variables which automatically fulfill the flux conservation constraints, thus obtaining a complete dualization of the partition function in the sense of Kramers and Wannier. 

In Chapters \ref{cha:su2} and \ref{cha:qcd} we derived our dual formulation for QC$_{2}$D and QCD respectively. In both cases we started with the ACC dualization of the pure gauge theories. The minimal terms the gauge actions are decomposed into are the so-called abelian color cycles, which we interpreted as paths in color space closing around plaquettes. In SU(2) there are a total of 16 different ACCs for each plaquette of the lattice, which can only have positive orientation. For SU(3) the number of the ACC increases to 81 for each plaquette, because we have 3 color layers instead of 2. Furthermore, they can have both positive and negative orientation. In the dualization those abelian color cycles are in one to one correspondence with a set of dual variables called cycle occupation numbers, which are positive integers for SU(2) and integer valued variables for SU(3). For SU(3) we additionally have plaquette auxiliary variables, which result from having to take the real part in the Wilson gauge action.

Our dualization process factorizes the integrals over the link variables, thus allowing us to perform the Haar measure integrals analytically.  
The Haar measure integrations give rise to gauge constraints and gauge weight factors. The gauge constraints arise when integrating over the phases of the parametrization chosen for the conventional gauge fields, hence we find two constraints for the SU(2) cycle occupation numbers, and five constraints for the SU(3) cycle occupation numbers. Those constraints impose relations between different color components of the link fluxes generated by the cycle occupation numbers attached to that link. As a result we find that the admissible configurations which contribute to the long range physics are worldsheets of cycle occupation numbers. The weight factors collect the combinatorial factors arising from the integrals over the angles of the parametrization of the elements of the group, as well as the power series in the inverse gauge coupling which come from the Taylor expansion of the Boltzmann weight. Hence, our worldsheet representations for pure gauge theories correspond to a strong coupling series, where all terms are known in closed form. We stress at this point that the gauge weight factors contain explicit signs which originate from the negative signs in the parametrization of the group elements. 

To add the matter fields we consider staggered fermions. The minimal terms the fermionic actions are decomposed into are the so-called abelian color fluxes. The dual variables for fermions are then monomer variables, and forward and backward hopping variables which we illustrated in Fig.~\ref{fig:su2_grassmann} and Fig.~\ref{fig:qcd_grassmann}. The Grassmann integrals give rise to a fermion constraints which enforce Pauli's exclusion principle for the dual variables, and a fermion weight. The fermion constraints imply that the admissible fermion configurations are the ones that completely fill the lattice with monomers, dimers and loops. Loops come with a sign factor. Moreover, when considering non-vanishing values of the quark chemical potential, in the dual representation the chemical potential only couples with loops winding in the temporal direction. We thus can identify the net-particle number as the total net temporal winding number of the worldline configuration, which from a geometrical point of view is a very elegant representation. In fact, the identification of conserved quantities as topological invariant, not only makes the determination of such quantities easier in the dual representation, but also opens the possibility of performing simulations in the canonical ensemble. 
Finally, the Haar measure integration leads to the same constraints and weights discussed for the pure gauge case, with the difference that now the link fluxes receive contributions not only from the cycle occupation numbers, but also from the forward and backward hopping terms for the fermions.  

Obviously the appearance of gauge and fermion sign factors in our dual formulations implies that for a Monte Carlo simulation of the ACC and ACF dual representations a strategy for a partial resummation needs to be found. Nevertheless, for SU(2) we find that, when considering a coupled strong coupling hopping expansion, negative terms only appear at $\mathcal{O}\big(\beta^{4}\big)$ or $\mathcal{O}\big((\frac{1}{2m})^{4} \beta^{3} \big)$ and the leading orders are free of the sign problem.

For SU(3) we discussed the strong coupling limit, and found that in our worldline representation strong coupling QCD is a gas of free staggered fermion loops, which describe baryons, that are embedded in a background of monomers, dimers and local link loops. We think that the interpretation of the strong coupling baryon loops as free staggered fermions could open the possibility of updating our form of strong coupling QCD with the fermion bags approach. 

Summarizing, we developed two methods that can be employed for the dualization of non-abelian lattice field theories. Unfortunately, the application of those methods to QCD leads to a worldlines and worldsheets representation that has a sign problem. However, both the ACC and the ACF dualization methods are very general and could be successful in overcoming the sign problem of other non-abelian field theories, as in the case of the SU(2) principal chiral model.

\appendix
\cleardoublepage
\chapter{Composite boson bags in strong coupling QED \label{cha:qed}}

At the beginning of this year yet another reformulation of strong coupling QCD with one flavor of staggered fermions was derived in \cite{Gattringer:2018mrg}. In this publication the author shows that the baryonic degrees of freedom in the QCD path integral are independent of the gauge fields. Moreover, he is able to separate the baryonic contributions from the remaining quark and diquark terms in the partition function. Finally he demonstrates that, in the strong coupling limit, the partition function completely factorizes into so-called \textit{baryons bags} (space-time regions in which the dynamics is given by free baryons) and a complementary domain (where the dynamical degree of freedoms are monomers and dimers of quarks and diquarks). As a natural continuation of this work, we aim to apply the baryon bag idea to strong coupling QCD with two flavors. 

In this appendix we present an intermediate project we studied as a toy model: strong coupling QED with two flavors of staggered fermions of opposite charge. In what follows we will show that in the case of QED the path integral factorizes into composite boson bags (CBB), where composite boson modes of two fermions propagate freely, and a complementary domain with monomers and dimers for the two flavors of fermions. For illustration one may think of the two flavors as "electrons" and "protons" and the composite bosons as "hydrogens". We will partly use this nomenclature in this appendix.

Before coming to the concrete discussion of the model we would like to remark that the idea that fermionic degrees of freedom can be described as being contained inside dynamically determined space-time regions was firstly introduced in \cite{PhysRevD.82.025007} and developed into the so-called \textit{fermion bags} approach. This method constitutes a very powerful tool for treating the fermion sign problem for some fermionic lattice field theories \cite{PhysRevLett.108.140404,PhysRevD.85.091502, PhysRevD.86.021701, Chandrasekharan2013, PhysRevD.88.021701, PhysRevD.91.065035, PhysRevD.93.081701, Ayyar2016, PhysRevD.96.114502, PhysRevD.97.054501}. Among those works we outline \cite{Chandrasekharan2011}, in which the fermion bag approach was applied to strong coupling QED with one flavor of Wilson fermions. 

\section{Gauge integral at strong coupling}
We consider strong coupling lattice QED with two flavors of staggered fermions with opposite charge. The corresponding action is given by
\begin{align}
	S[\overline{e},e,\overline{p},p,U]\ &= \ \sum_{x} \bigg[ 2m_{e} \, \overline{e}_{x} e_{x} \, + \, \sum_{\nu} \gamma_{x,\nu} \Big( \overline{e}_{x} \,  U_{x,\nu} \, e_{x + \hat{\nu}} \, - \,  \overline{e}_{x + \hat{\nu}} \, U_{x,\nu}^{\, \star} \, e_{x} \Big) \bigg] \nonumber \\
	&\, + \ \sum_{x} \bigg[ 2m_{p} \, \overline{p}_{x} p_{x} + \sum_{\nu} \gamma_{x,\nu} \Big( \overline{p}_{x} \, U_{x,\nu}^{\, \star} \, p_{x + \hat{\nu}} \, - \, \overline{p}_{x + \hat{\nu}} \, U_{x,\nu} \, p_{x} \Big) \bigg] \, .
	\label{eq:qed_action}
\end{align}
The fermionic degrees of freedom are the one-component Grassmann variables $e_{x}$, $\overline{e}_{x}$ for the electrons and $p_{x}$, $\overline{p}_{x}$ for the protons. They obviously have opposite charge, as is reflected in the expression of the action $S[\overline{e},e,\overline{p},p,U]$ where the role of the link variable $U_{x,\nu}$ and its complex conjugate $U_{x,\nu}^{\, \star}$ is interchanged for the forward and backward propagation of the electron and the proton. This also ensures overall electric neutrality of the two-flavor system as required by Gauss’ law. The gauge degrees of freedom are the link variables $U_{x,\nu} = e^{\,i\varphi_{x,\nu}} \in U(1)$. They live on the links $(x,\nu)$ of a four-dimensional lattice $\Lambda$ with volume $N_{s}^{3} \times N_{t}$ and they satisfy periodic boundary conditions in all directions. For the fermions, instead, we impose anti-periodic boundary conditions in the time ($\nu=4$) direction, and periodic boundary conditions for the spatial directions ($\nu=1,2,3$). In Eq.~(\ref{eq:qed_action}) $m_{e}$ and $m_{p}$ are the electron's and proton's bare masses respectively, while $\gamma_{x,\nu}$ are the usual staggered sign factors defined as
\begin{equation}
	\label{eq:qed_staggered}
	\gamma_{x,1} = 1\, , \qquad \gamma_{x,2} = (-1)^{x_{1}}\, ,  \qquad \gamma_{x,3} = (-1)^{x_{1}+x_{2}}\, ,
	\qquad \gamma_{x,4} = (-1)^{x_{1}+x_{2}+x_{3}} \, .
\end{equation}

The partition function of the system is obtained as
\begin{align}
	Z \, &= \, \int \! D[\overline{e},e] D[\overline{p},p] \int \! D[U] \, e^{\,S[\overline{e},e,\overline{p},p,U]} \nonumber \\
	&= \, \int \! D[\overline{e},e] D[\overline{p},p] \, \prod_{x} e^{\,2 m_{e}\, \overline{e}_{x} e_{x}} \, e^{\,2 m_{p}\, \overline{p}_{x} p_{x}} \int \! D[U] \, \prod_{x,\nu} L_{x,\nu}[\overline{e},e,\overline{p},p,U]\, ,
	\label{eq:qed_partition}
\end{align}
where $\int \! D[\overline{e},e] D[\overline{p},p]$ are the product of Grassmann measures at all sites
\begin{equation}
	\int \! D[\overline{e},e] D[\overline{p},p] \, = \, \prod_{x} \int dp_{x} d\overline{p}_{x} de_{x} d\overline{e}_{x} \, ,
	\label{eq:qed_grassmannmeasure}
\end{equation}
while  $\int \! D[U]$ is the product of U(1) Haar measures at all links
\begin{equation}
	\int \! D[U] \, = \, \prod_{x,\nu} \, \int_{U(1)} dU_{x,\nu} \, = \, \prod_{x,\nu} \int_{0}^{2\pi} \dfrac{d\varphi_{x,\nu}}{2\pi}  \, .
\label{eq:qed_haarmeasure}
\end{equation}
In the second step of Eq.~(\ref{eq:qed_partition}) we used the fact that every term in the action (\ref{eq:qed_action}) is a commuting Grassmann bilinear to factorize the Boltzmann weight $e^{\,S[\overline{e},e,\overline{p},p,U]}$ into local factors. Moreover, we defined the link terms
\begin{equation*}
	L_{x,\nu}[\overline{e},e,\overline{p},p,U] \, = \, e^{\,\gamma_{x,\nu}\, \overline{e}_{x}\, U_{x,\nu}\, e_{x + \hat{\nu}}} \, e^{-\gamma_{x,\nu}\, \overline{e}_{x + \hat{\nu}}\, U_{x,\nu}^{\star}\, e_{x}} \, e^{\,\gamma_{x,\nu} \, \overline{p}_{x} \, U_{x,\nu}^{\star} \, p_{x + \hat{\nu}}} \, e^{-\gamma_{x,\nu} \, \overline{p}_{x + \hat{\nu}} \, U_{x,\nu} \, p_{x}} \, .
\end{equation*}

In order to separate the hydrogen terms (composite boson terms) in the QED path integral we start by expanding the Boltzmann factors for the mass terms $e^{\,2 m_{e}\, \overline{e}_{x} e_{x}}$ and $e^{\,2 m_{p}\, \overline{p}_{x} p_{x}}$. In the expansions we exploit the nilpotency of the Grassmann variables thus terminating the power series after the linear terms. We also drop all space-time indices for brevity.
\begin{align}
	e^{\,2 m_{e}\, \overline{e} e} \, e^{\,2 m_{p}\,\overline{p} p}& = \big[ 1 + 2 m_{e}\, \overline{e} e\big] \big[ 1 + 2 m_{p}  \, \overline{p} p\big] = 1 + 2 m_{e} \, \overline{e} e + 2 m_{p} \, \overline{p} p + 4 m_{e} m_{p} \, \overline{e} e \, \overline{p} p \nonumber \\
	& = \big[1 + 4 m_{e} m_{p} \, \overline{e} e \, \overline{p} p \big] \big[1 + 2 m_{e} \, \overline{e} e + 2 m_{p} \, \overline{p} p \big] \nonumber \\
	&= \exp\big(2 M \, \overline{H} H\big) \sum_{s^{(e)} = 0}^{1} \big[2m_{e} \, \overline{e} e\big]^{s^{(e)}} \sum_{s^{(p)} = 0}^{1-s^{(e)}} \big[2m_{p} \, \overline{p} p\big]^{s^{(p)}} \, .
	\label{eq:qed_mass}
\end{align}
In the second line we used once more the nilpotency of the Grassmann variables to separate the hydrogen contribution to the mass terms, which we collected in the first factor in brackets in the second line. We then re-exponentiated this factor and used the identities $2M = 4 m_{e} m_{p}$ and $\overline{e} e \overline{p} p = \overline{e} \overline{p} p e = \overline{H} H$, where $H_{x}$ and $\overline{H}_{x}$ are the hydrogen fields defined as
\begin{equation}
	\label{eq:qed_hydrogen}
	H_{x} \, \equiv \, p_{x} e_{x} \, , \qquad \qquad \overline{H}_{x} \, \equiv \, \overline{e}_{x} \overline{p}_{x}\, .
\end{equation}
In the last step of Eq.~(\ref{eq:qed_mass}) we also introduced the monomer variables $s_{x}^{(e)} \in \{0,1\}$ for the mass term of the electron fields and $s_{x}^{(p)} \in \{0,1-s_{x}^{(e)}\}$ for the proton's mass term. As we will see later, this rewriting will turn out to be useful for solving the Grassmann integrals in (\ref{eq:qed_partition}). Notice also that the monomer variables for the proton fields are constrained by the value of the monomer variables for the electron fields at the same site. This is because the mixed term, which would correspond to setting $s_{x}^{(e)} = s_{x}^{(p)} = 1$, was already factorized into the hydrogen contribution.

A second important step to completely separate the hydrogens in the path integral of QED consists in solving the link integral in Eq.~(\ref{eq:qed_partition}), which has the following form (we suppress the link indices to shorten the notation, and we omit even powers of the staggered sign factor, since $\gamma^{2n} = +1$):
\begin{align}
	\int \! dU \, &e^{\,\gamma \,\overline{e}_{x} U e_{y}} \, e^{-\gamma \, \overline{e}_{y} U^{\star} e_{x}} \, e^{\,\gamma \, \overline{p}_{x} U^{\star} p_{y}} \, e^{-\gamma \, \overline{p}_{y} U p_{x}} \nonumber \\
	& = \int \! dU \, (1 + \gamma \, \overline{e}_{x} U e_{y}) \, (1 - \gamma \, \overline{e}_{y} U^{\star} e_{x}) \, (1 + \gamma \, \overline{p}_{x} U^{\star} p_{y}) \, (1 - \gamma \, \overline{p}_{y} U p_{x}) \nonumber \\[0.2em]
	& = 1 - \overline{e}_{x} e_{y} \overline{e}_{y} e_{x} -  \overline{p}_{x} p_{y} \overline{p}_{y} p_{x} +  \overline{e}_{x} e_{y} \overline{p}_{x} p_{y} + \overline{e}_{y} e_{x} \overline{p}_{y} p_{x} + \overline{e}_{x} e_{y} \overline{e}_{y} e_{x} \overline{p}_{x} p_{y} \overline{p}_{y} p_{x}\nonumber \\[0.5em]
	& = \big[1 +  \overline{e}_{x} \overline{p}_{x} p_{y} e_{y} + \overline{e}_{y} \overline{p}_{y} p_{x} e_{x} + \overline{e}_{x} \overline{p}_{x} p_{y} e_{y} \overline{e}_{y}  \overline{p}_{y} p_{x} e_{x} \big] 
	\big[1 + \overline{e}_{x} e_{x} \overline{e}_{y} e_{y} + \overline{p}_{x} p_{x} \overline{p}_{y} p_{y} \big] \nonumber \\
	& = e^{\overline{H}_{x} H_{y} + \overline{H}_{y} H_{x}} 
	\sum_{d^{(e)} = 0}^{1} \big[\overline{e}_{x} e_{x} \overline{e}_{y} e_{y}\big]^{d^{(e)}} \sum_{d^{(p)} = 0}^{1-d^{(e)}} \big[\overline{p}_{x} p_{x} \overline{p}_{y} p_{y}\big]^{d^{(p)}} \, .
	\label{eq:qed_link}
\end{align}
In the first step we Taylor expanded each exponential. The only terms of the resulting product of binomials that survive the Haar integration are the ones  in which the link variable $U$ and its complex conjugate $U^{\star}$ have the same exponent. In fact, the link integrals are now of the type
\begin{equation}
	\int_{U(1)} \! dU \, (U)^{n} (U^{\star})^{m} \, = \,
	\int_{0}^{2\pi}\! \dfrac{d\varphi}{2 \pi} \, e^{i \varphi (n-m)} \, = \, \delta_{n,m} \, .
\end{equation} 
In the third step of (\ref{eq:qed_link}) we have separated the propagation terms for the hydrogen from the electron and proton dimer contributions, using again the nilpotency of the Grassmann variables. Subsequently, we re-exponentiated the hydrogen contribution, that now has the form of a Boltzmann weight for the forward and backward hydrogen hops. Finally, we introduced the dimer variables $d_{x,\nu}^{\,(e)} \in \{0,1\}$ for the electron fields and $d_{x,\nu}^{\,(p)} \in \{0,1-d_{x,\nu}^{\,(e)}\}$ for the proton fields. We again constrain the proton dimer variable with the value of the electron dimer variable at the same link since the configuration $d_{x,\nu}^{\,(e)} = d_{x,\nu}^{\,(p)} = 1$ is already taken into account in the hydrogen contribution. 

Putting things together we obtain the following expression for the partition sum:
\begin{equation}
Z \, = \, \int \! D[\overline{e},e] D[\overline{p},p] \, e^{\,S_{H}[\overline{H},H]} \, W_{MD}[\overline{e},e,\overline{p},p] \, .
\label{eq:qed_partition2}
\end{equation}
In this form the Boltzmann weight $e^{\,S_{H}[\overline{H},H]}$ factorizes the contributions of the hydrogens, which are now completely separated from the electron's and proton's monomer and dimer contributions. The hydrogen action
\begin{equation}
	S_{H}[\overline{H},H] = \sum_{x} \Big[2 M \,\overline{H}_{x} H_{x} + \sum_{\nu} \big( \overline{H}_{x} H_{x+\hat{\nu}} + \overline{H}_{x+\hat{\nu}} H_{x}\big)\Big] \, 
	\label{eq:qed_hydrogenaction}
\end{equation}
has a structure similar to a bosonic action, where backward and forward hopping terms have the same sign. Nevertheless, even though the fields $H_{x}$ and $\overline{H}_{x}$ are indeed bosonic, they still have to satisfy Pauli's exclusion principle, since their fundamental constituents are fermions (see (\ref{eq:qed_commutation}) -- (\ref{eq:qed_Hbarnilpotent}) below for their algebraic relations). $W_{MD}[\overline{e},e,\overline{p},p]$ collects the monomer and dimer contributions of the electrons and protons to the partition sum:
\begin{align}
	W_{MD}[\overline{e},e,\overline{p},p] \, = \, &\sum_{\{s,d\}} \prod_{x} \, \big[2m_{e} \, \overline{e}_{x} e_{x}\big]^{s^{(e)}_{x}}\, \big[2m_{p} \, \overline{p}_{x} p_{x}\big]^{s^{(p)}_{x}} \nonumber \\
	&\, \, \ \times \prod_{x,\nu}\, \big[\overline{e}_{x} e_{x} \overline{e}_{x+\hat{\nu}} e_{x+\hat{\nu}}\big]^{d_{x,\nu}^{\,(e)}} \, \big[\overline{p}_{x} p_{x} \overline{p}_{x+\hat{\nu}} p_{x+\hat{\nu}}\big]^{d_{x,\nu}^{\,(p)}} \, ,
	\label{eq:qed_wmd}
\end{align}
where we used the notation
\begin{equation}
	\sum_{\{s,d\}} = \left[\prod_{x} \sum_{s^{(e)}_{x} = 0}^{1} \sum_{s^{(p)}_{x} = 0}^{1-s^{(e)}_{x}}\right] \left[\prod_{x,\nu} \sum_{d^{\,(e)}_{x,\nu} = 0}^{1} \sum_{d^{\,(p)}_{x,\nu} = 0}^{1-d^{\,(e)}_{x,\nu}}\right]
	\label{eq:qed_sumsd}
\end{equation}
to denote the sum over the configurations of electron's and proton's monomers and dimers.

\section{Factorization of the Grassmann integral}

Having completely integrated out the gauge fields we now discuss the Grassmann integration, starting with the hydrogen contributions. Firstly, notice that the Grassmann integrals in Eq.~(\ref{eq:qed_partition2}) are non-vanishing only if each Grassmann variable $\overline{e}_{x}$ ,$e_{x}$ and $\overline{p}_{x}$, $p_{x}$ appears exactly once. When this condition is fulfilled we say that the Grassmann integral is \textit{saturated}. It is then clear that hydrogen variables saturate the Grassmann integrals on the sites they occupy. In fact, the expansion of the hydrogen Boltzmann weight $e^{\,S_{H}[\overline{H},H]}$ results in terms (hydrogen monomers, dimers and non-intersecting loops) that already contain both the flavors of staggered fermions we are considering (compare with the definition (\ref{eq:qed_hydrogen}) for the hydrogen fields $\overline{H}_{x}$ and $H_{x}$). Therefore, if we denote with $\mathcal{H}_{i}$ the collection of sites that are either occupied by an hydrogen monomer, are the endpoint of an hydrogen dimer, or are run through by an hydrogen loop, we then find that in the region defined by $\mathcal{H}_{i}$ all the Grassmann integrals are saturated by the hydrogen terms. We refer to these regions as \textit{hydrogen bags}. Additionally, we can define the union $\mathcal{H}$ of all bags and the complementary domain $\overline{\mathcal{H}}$
\begin{equation}
	\mathcal{H} = \cup_{i} \mathcal{H}_{i} \, , \qquad \overline{\mathcal{H}} = \Lambda/\mathcal{H}\, .
\label{eq:qed_complementary}
\end{equation}
The complementary region $\overline{\mathcal{H}}$ contains the electron's and proton's monomer and dimer contributions described by the weight factor $W_{MD}[\overline{e},e,\overline{p},p]$. Now, recall that the hydrogens $H_{x}$ and $\overline{H}_{x}$ are nilpotent bosons. In other words, they satisfy the following properties:
\begin{gather}
\label{eq:qed_commutation}
	[H_{x},H_{y}] = 0\, , \qquad [H_{x},\overline{H}_{y}] = 0\, , \qquad [\overline{H}_{x},\overline{H}_{y}] = 0 \, ;\\
\label{eq:qed_Hnilpotent}
	H_{x} H_{x} = 0\, , \qquad H_{x} e_{x} = 0\, , \qquad H_{x} p_{x} = 0\, ; \\
\label{eq:qed_Hbarnilpotent}
	 \overline{H}_{x} \overline{H}_{x} = 0\, , \qquad \overline{H}_{x} \overline{e}_{x} = 0\, , \qquad \overline{H}_{x} \overline{p}_{x} = 0\, .
\end{gather}
Eqs.~(\ref{eq:qed_Hnilpotent}) and (\ref{eq:qed_Hbarnilpotent}) imply that the two regions $\mathcal{H}$ and $\overline{\mathcal{H}}$ do not mix in the strong coupling limit. As a consequence, we can factorize the partition function in the following way
\begin{equation}
	Z \, = \, \sum_{\{\mathcal{H}\}} \left[\prod_{i} Z_{\mathcal{H}_{i}}\right] \times Z_{\overline{\mathcal{H}}} \, .
	\label{eq:qed_partitionfactorized}
\end{equation}
The sum runs over all the possible ways of organizing the lattice into sets of hydrogen bags $	\mathcal{H} = \cup_{i} \mathcal{H}_{i}$. A single hydrogen bag causes a contribution of
\begin{equation}
	Z_{\mathcal{H}_{i}} \, = \prod_{x \in \mathcal{H}_{i}} \int \! dH_{x} d\overline{H}_{x} \, \exp \Big(\sum_{x,y} \overline{H}_{x} D_{x,y}^{(i)} H_{y} \Big) \, ,
	\label{eq:qed_partitionh}
\end{equation}
where we reorganized the canonical Grassmann integrations inside the region $\mathcal{H}_{i}$ as follows:
\begin{equation}
\prod_{x \in \mathcal{H}_{i}} \int dp_{x} d\overline{p}_{x} de_{x} d\overline{e}_{x} \, = \, \prod_{x \in \mathcal{H}_{i}} \int dp_{x} de_{x} d\overline{e}_{x} d\overline{p}_{x} \, = \, \prod_{x \in \mathcal{H}_{i}} \int dH_{x} d\overline{H}_{x} \, ,
\end{equation}
with the hydrogen measures defined as
\begin{equation}
	dH_{x} \, \equiv \, de_{x} dp_{x} \, , \qquad d\overline{H}_{x} \, \equiv \, d\overline{p}_{x} d\overline{e}_{x} \, .
	\label{eq:qed_hydrogenmeasure}
\end{equation}
In the exponent of the exponential in Eq.~(\ref{eq:qed_partitionh}) we introduced the operator
\begin{equation}
	D_{x,y}^{(i)} = \theta_{x}^{(i)}2M \delta_{x,y} + \sum_{\nu} \theta_{x,\nu}^{(i)}\big[ \delta_{x+\hat{\nu},y} + \delta_{x,y+\hat{\nu}} \big] \, ,
	\label{eq:qed_hydrogenoperator}
\end{equation}
with the site and link support functions on the hydrogen bag $\mathcal{H}_{i}$ given by
\begin{equation}
	\theta_{x}^{(i)} = 
	\left\{ \begin{array}{l}
	1 \quad \text{if } x \in \mathcal{H}_{i} \\
	0 \quad \text{if } x \notin \mathcal{H}_{i} \\
	\end{array}\right. \, , \qquad 
	\theta_{x,\nu}^{(i)} = 
	\left\{ \begin{array}{l}
	1 \quad \text{if } (x,\nu) \in \mathcal{H}_{i} \\
	0 \quad \text{if } (x,\nu) \notin \mathcal{H}_{i} \\
	\end{array}\right. \, .
	\label{eq:qed_supportfunctions}
\end{equation}
The partition function of the complementary region $\overline{\mathcal{H}}$ is given by
\begin{align}
	Z_{\overline{\mathcal{H}}} \, = \, \sum_{\{s,d\, ||\, \overline{\mathcal{H}}\}} \, \prod_{x\in \overline{\mathcal{H}}} \, \int \! dp_{x} d\overline{p}_{x} de_{x} &d\overline{e}_{x} \prod_{x} \, \big[2m_{e} \, \overline{e}_{x} e_{x}\big]^{s^{(e)}_{x}}\, \big[2m_{p} \, \overline{p}_{x} p_{x}\big]^{s^{(p)}_{x}} \nonumber \\
	&\, \ \times \prod_{x,\nu}\, \big[\overline{e}_{x} e_{x} \overline{e}_{x+\hat{\nu}} e_{x+\hat{\nu}}\big]^{d_{x,\nu}^{\,(e)}} \, \big[\overline{p}_{x} p_{x} \overline{p}_{x+\hat{\nu}} p_{x+\hat{\nu}}\big]^{d_{x,\nu}^{\,(p)}}.
	\label{eq:qed_partitionc}
\end{align}
$Z_{\overline{\mathcal{H}}}$ is a sum over all the configurations of electron's and proton's monomers and dimers that saturate the Grassmann integrals in the complementary region $\overline{\mathcal{H}}$:
\begin{equation}
	\sum_{\{s,d\, ||\, \overline{\mathcal{H}}\}} \, = \, \left[\, \prod_{x \in \overline{\mathcal{H}}} \, \sum_{s^{(e)}_{x} = 0}^{1} \, \sum_{s^{(p)}_{x} = 0}^{1-s^{(e)}_{x}}\, \right] \left[\,\prod_{(x,\nu) \in \overline{\mathcal{H}}}\, \sum_{d^{\,(e)}_{x,\nu} = 0}^{1} \,\sum_{d^{\,(p)}_{x,\nu} = 0}^{1-d^{\,(e)}_{x,\nu}}\,\right] \, .
	\label{eq:qed_sumsdcomplementary}
\end{equation}
In Eq.~(\ref{eq:qed_partitionc}) all the fermion variables are already in the canonical order for the Grassmann integration. Monomers come with weight factors $2m_{e}$ for the electrons and $2m_{p}$ for the protons. Dimers account for a factor $+1$. The condition of saturation of the Grassmann integral can be implemented by means of the monomer-dimer constraint
\begin{equation}
	C_{MD}[s,d] = \prod_{x \in \overline{\mathcal{H}}}\, \delta \Big( 1 - \big[s_{x}^{(e)} + \sum_{\nu}( d_{x,\nu}^{\, (e)} + d_{x - \hat{\nu},\nu}^{\, (e)} )\big]\Big)	 \,\delta \Big( 1 - \big[s_{x}^{(p)} + \sum_{\nu}( d_{x,\nu}^{\, (p)} + d_{x - \hat{\nu},\nu}^{\, (p)})\big]\Big) \, ,
	\label{eq:qed_cmd}
\end{equation}
where we use the notation $\delta(n) \equiv \delta_{n,0}$ for the Kronecker deltas. Then, we can rewrite (\ref{eq:qed_partitionc}) in the following simple form
\begin{equation}
	Z_{\overline{\mathcal{H}}} \, = \, \sum_{\{s,d\, ||\, \overline{\mathcal{H}}\}} \, \big(2m_{e} \big)^{\mathcal{N}(m_{e})} \big(2m_{p} \big)^{\mathcal{N}(m_{p})} \, C_{MD}[s,d]\, ,
	\label{eq:qed_partitioncdual}
\end{equation}
where $\mathcal{N}(m_{e})$ and $\mathcal{N}(m_{p})$ denote the total number of electron and proton monomers respectively. The partition function $Z_{\overline{\mathcal{H}}}$ sums over all the possible configurations of the dual variables $\{s,d\}$. The allowed configurations must saturate the sites of the complementary domain $\overline{\mathcal{H}}$ by means of electron and proton monomers and dimers. This condition is enforced by the constraint $C_{MD}[s,d\,]$. Additionally, the configurations with $s_{x}^{(e)} = s_{x}^{(p)} = 1$ (hydrogen monomer), as well as with $d_{x,\nu}^{\,(e)} = d_{x,\nu}^{\,(p)} = 1$ (hydrogen dimer) are not allowed in the complementary region. This condition is already implemented in the definition of the sum (\ref{eq:qed_sumsdcomplementary}), where we restrict the value of the proton monomers and dimers to be $s_{x}^{(p)} \in \{0,1-s_{x}^{(e)}\}$ and $d_{x,\nu}^{\,(p)} \in \{0,1-d_{x,\nu}^{\,(e)}\}$.

\section{CBB contributions as permanents}

To obtain the composite boson bag contributions (\ref{eq:qed_partitionh}) to the partition function $Z$ we have to solve integrals of the form
\begin{equation}
I[D] \, = \int \! \prod_{l = 1}^{N} dH_{l} d\overline{H}_{l} \, \exp \Big(\sum_{i,j = 1}^{N} \overline{H}_{i} D_{i,j} H_{j} \Big) \, .
\label{eq:qed_hydrogenintegral}
\end{equation}
If $H_{x}$ and $\overline{H}_{x}$ were Grassmann variables, the result of this integral could be trivially determined as $\det D$ (compare Sec.~\ref{sec:teo_grassmann}). However, we know that $H_{x}$ and $\overline{H}_{x}$ are commuting nilpotent variables, as they obey the properties (\ref{eq:qed_commutation}) -- (\ref{eq:qed_Hbarnilpotent}). Therefore we must explicitly compute the integral based on the algebra (\ref{eq:qed_commutation}) -- (\ref{eq:qed_Hbarnilpotent}).

Before doing so, let us briefly discuss the rules for the integration over the hydrogen measures $dH_{x}$ and $d\overline{H}_{x}$ we defined in Eq.~(\ref{eq:qed_hydrogenmeasure}). They obey the same commutation relations as the hydrogen fields $H_{x}$ and $\overline{H}_{x}$ 
\begin{gather}
\label{eq:qed_commutation2}
[dH_{x},dH_{y}] = 0\, , \qquad [dH_{x},d\overline{H}_{y}] = 0\, , \qquad [d\overline{H}_{x},d\overline{H}_{y}] = 0 \, ,
\end{gather}
and, together with the hydrogen fields, they obey the Grassmann integration rules
\begin{gather}
\label{eq:qed_integrationrules}
	\int dH_{x}  \, = \, 0 \, , \qquad 	\int d\overline{H}_{x}  \, = \, 0 \, , \qquad \int dH_{x} H_{x} \, = \, 1 \, , \qquad \int d\overline{H}_{x} \overline{H}_{x}  \, = \, 1 \,.
\end{gather}

We can now proceed with the computation of the integral (\ref{eq:qed_hydrogenintegral}):
\begin{equation}
	I[D] \, 
	= \int \! \prod_{l = 1}^{N} dH_{l} d\overline{H}_{l} \, \prod_{i=1}^{N}\exp \Big(\overline{H}_{i} \sum_{j} D_{i,j} H_{j} \Big) \, 
	= \, \int \! \prod_{l = 1}^{N} dH_{l} d\overline{H}_{l} \, \prod_{i=1}^{N} \Big(1 + \overline{H}_{i} \sum_{j} D_{i,j} H_{j} \Big) \,.
\label{eq:qed_hydrogenintegral2}
\end{equation}
In the first step we used the fact that all the terms in the exponent of Eq.~(\ref{eq:qed_hydrogenintegral}) commute and brought the sum over the $i$ index down from the exponential as a product. Then we exploited the nilpotency of the hydrogen variables $\overline{H}_{i}$ and wrote the product of exponentials as a product of binomials. Now, from the integration rules (\ref{eq:qed_integrationrules}) follows that the only terms of the product of binomials that survive after the integration are the ones in which all the hydrogen variables appear exactly once:
\begin{align}
	I[D] \, &
	= \, \int \! \prod_{l = 1}^{N} dH_{l} d\overline{H}_{l} \, \prod_{i=1}^{N} \overline{H}_{i} \sum_{j} D_{i,j} H_{j} \nonumber\\
	& = \, \int \! \prod_{l = 1}^{N} dH_{l} d\overline{H}_{l} \ \overline{H}_{1} D_{1,j_{1}} H_{j_1} \,\overline{H}_{2} D_{2,j_{2}} H_{j_2} \dots \overline{H}_{N} D_{N,j_{N}} H_{j_N}\, ,
\label{eq:qed_hydrogenintegral3}
\end{align}
where a sum over repeated indices $j_{k}$, $k=1,\dots,N$ is understood. Then, using the commutation relations (\ref{eq:qed_commutation}) and (\ref{eq:qed_commutation2}) we obtain
\begin{align}
I[D] \, = \, \int \! \prod_{l = 1}^{N} dH_{l} d\overline{H}_{l} \ \overline{H}_{l} H_{l}  \sum_{\sigma \in S_{N}} D_{1,\sigma(1)}  D_{2,\sigma(2)}  \dots D_{N,\sigma(N)} \, = \, \perm D \, ,
\label{eq:qed_hydrogenintegral4}
\end{align}
where the sum extends over all elements $\sigma$ of the symmetric group $S_{N}$, i.e., over all permutations of the numbers $1, 2, \dots, N$. In Eq.~(\ref{eq:qed_hydrogenintegral4}) we used the definition of the \textit{permanent} of a matrix $A$ \cite{minc_marcus_1984}:
\begin{equation}
	\perm A \equiv \sum_{\sigma \in S_{N}} A_{1,\sigma(1)}  A_{2,\sigma(2)}  \dots A_{N,\sigma(N)} \, .
\end{equation}
Summarizing, we obtained that each hydrogen bag gives a contribution
\begin{equation}
Z_{\mathcal{H}_{i}} \, = \perm D^{(i)} \, .
\label{eq:qed_partitionhdual}
\end{equation}

Putting together the results we obtained for the hydrogen bags and the complementary domain, the full partition function is given by
\begin{equation}
Z \, = \, \sum_{\{\mathcal{H}\}} \prod_{i} \perm D^{(i)} \times Z_{\overline{\mathcal{H}}} \, .
\label{eq:qed_partitiondual}
\end{equation}
In the CBB formulation (hydrogen bags formulation) the partition function is a sum over all the possible configurations of the hydrogen bags. Each hydrogen bag $\mathcal{H}_{i}$ contributes with a weight factor $\perm D^{(i)}$, given by the permanent of the free boson operator (\ref{eq:qed_hydrogenoperator}). The remaining part of the lattice, i.e., the complementary domain $\overline{\mathcal{H}}$, has to be filled with monomers and dimers of electrons and protons. $Z_{\overline{\mathcal{H}}}$ sums over the allowed configurations of the monomers ($s_{x}^{(e)} \in \{0,1\}$ and $s_{x}^{(p)} \in \{0,1-s_{x}^{(e)}\}$) and dimers ($d_{x,\nu}^{\,(e)} \in \{0,1\}$ and $d_{x,\nu}^{\,(p)} \in \{0,1-d_{x,\nu}^{\,(e)}\}$) variables, and accounts a weight factor $2m_{e}$ ($2m_{p}$) for each electron (proton) monomer. 

We conclude this appendix with remarking that the permanent of a matrix with non-negative entries can be computed with a polynomial-time approximation algorithm \cite{Jerrum:2004:PAA:1008731.1008738} and thus makes the CBB formulation an interesting candidate for an efficient Monte Carlo simulation.

\bib

\end{document}